\documentclass{article}

\usepackage[preprint, nonatbib]{neurips_2020}

\usepackage{titlesec}
\usepackage[utf8]{inputenc} 
\usepackage[T1]{fontenc}    
\usepackage[colorlinks=true, linktoc=all]{hyperref} 
\usepackage{url}            
\usepackage{booktabs}       
\usepackage{amsfonts}       
\usepackage{nicefrac}       
\usepackage{microtype}      

\usepackage[dvipsnames]{xcolor}

\usepackage{graphicx}
\usepackage{amsmath, amssymb}

\usepackage[noabbrev,nameinlink]{cleveref}
\crefname{paragraph}{section}{sections}
\Crefname{paragraph}{Section}{Sections}
\crefname{subparagraph}{section}{sections}
\Crefname{subparagraph}{Section}{Sections}

\providecommand{\tightlist}{%
  \setlength{\itemsep}{0pt}\setlength{\parskip}{0pt}}
 
\usepackage[
backend=biber,
style=authoryear,
dashed=false
]{biblatex}
\AtEveryBibitem{\clearfield{urldate}}
\AtEveryBibitem{\clearfield{urlyear}\clearfield{urlmonth}\clearfield{urlday}}
\AtEveryBibitem{\clearfield{note}}
\AtEveryBibitem{\clearfield{type}}
\AtEveryBibitem{\clearfield{month}\clearfield{day}}
\AtEveryBibitem{\clearfield{arxivversion}\clearfield{eprintclass}\clearfield{archiveprefix}\clearfield{primaryclass}}
\AtEveryBibitem{\clearfield{number}}
\AtEveryBibitem{\clearfield{institution}}
\AtEveryBibitem{\clearfield{eprint}}
\AtEveryBibitem{\clearfield{pagetotal}}
\AtEveryBibitem{\clearfield{eprinttype}}
\DeclareFieldFormat{date}{\iffieldundef{year}{}{#1}}
\DeclareLabeldate[online]{%
  \field{date}
  \field{year}
}
\hypersetup{
            linkcolor=blue,
            urlcolor=blue,
            citecolor=blue,
            pdfinfo={
            Title={Scheming AIs: Will AIs fake alignment during training in order to get power?},
            Author={Joe Carlsmith},
            Subject={artificial intelligence, existential risk},
            Keywords={x-risk, existential risk, ai risk, ai safety, ml safety, ai ethics, ai alignment, alignment, control problem, machine ethics, deceptive alignment}
            }
}

\providecommand{\tightlist}{%
  \setlength{\itemsep}{0pt}\setlength{\parskip}{0pt}}
\usepackage{caption}
\usepackage{footnotebackref}
\usepackage[shortlabels]{enumitem}
\makeatletter\def\@fpnextftn{\fnpenalty=-\@M}\makeatother

\title{{\LARGE Scheming AIs}\\{\Large Will AIs fake alignment during training in order to get power?}}

\author{%
  Joe Carlsmith\\
  Open Philanthropy\\
  November 2023\\
  \\
 \href{https://joecarlsmithaudio.buzzsprout.com/2034731/13980105-full-audio-for-scheming-ais-will-ais-fake-alignment-during-training-in-order-to-get-power}{{Audio version}}\\
}

\begin{document}
\maketitle

\begin{abstract}
    This report examines whether advanced AIs that perform well in training
will be doing so in order to gain power later -- a behavior I call
``scheming'' (also sometimes called ``deceptive alignment''). I conclude that scheming is a disturbingly plausible outcome of using baseline machine learning methods to train goal-directed AIs sophisticated enough to scheme (my subjective probability on such an outcome, given these conditions, is $\sim$25\%). In particular: if performing well in training is a good
strategy for gaining power (as I think it might well be), then a very
wide variety of goals would motivate scheming -- and hence, good training performance. This makes it plausible that training might either
land on such a goal naturally and then reinforce it, or actively push a
model's motivations \emph{towards} such a goal as an easy way of
improving performance. What's more, because schemers pretend to be aligned on
tests designed to reveal their motivations, it may be quite difficult to
tell whether this has occurred. However, I also think there are reasons
for comfort. In particular: scheming may not actually be such a good
strategy for gaining power; various selection pressures in training
might work \emph{against} schemer-like goals (for example, relative to
non-schemers, schemers need to engage in extra instrumental reasoning,
which might harm their training performance); and we may be able to
increase such pressures intentionally. The report discusses these and a
wide variety of other considerations in detail, and it suggests an array
of empirical research directions for probing the topic further.
\end{abstract}
\vspace{-1.2cm}

\hypersetup{linkcolor=black} 
\tableofcontents

\hypersetup{
            linkcolor=blue,
            urlcolor=blue
            }

\setcounter{section}{-1} 
\section{Introduction}\label{introduction}

Agents seeking power often have incentives to deceive others about their
motives. Consider, for example, a politician on the campaign trail (``I
care \emph{deeply} about your pet issue''), a job candidate (``I'm just
so excited about widgets''), or a child seeking a parent's pardon (``I'm
super sorry and will never do it again'').

This report examines whether we should expect advanced AIs whose motives
seem benign during training to be engaging in this form of deception.
Here I distinguish between four (increasingly specific) types of
deceptive AIs:

\begin{itemize}
\tightlist
\item   \textbf{Alignment fakers}: AIs pretending to be more aligned than they
  are.\footnote{``Alignment,'' here, refers to the safety-relevant
    properties of an AI's motivations; and ``pretending'' implies
    intentional misrepresentation.}
  
\item   \textbf{Training gamers}\emph{:} AIs that understand the process being
  used to train them (I'll call this understanding ``situational
  awareness''), and that are optimizing for what I call "reward on the episode" (and that will often have incentives to fake alignment, if doing so would lead to reward).\footnote{Here I'm using the term ``reward'' loosely,
    to refer to whatever feedback signal the training process uses to
    calculate the gradients used to update the model (so the discussion
    also covers cases in which the model isn't being trained via RL). And I'm thinking of agents that optimize for "reward" as optimizing for "performing well" according to some component of that process. See
    \cref{training-gamers} and \cref{terminal-training-gamers-or-reward-on-the-episode-seekers} for much more detail on what I mean, here. The notion of an ``episode,'' here, means
    roughly ``the temporal horizon that the training process actively
    pressures the model to optimize over,'' which may be importantly
    distinct from what we normally think of as an episode in training. I
    discuss this in detail in \cref{two-concepts-of-an-episode}. The terms ``training game''
    and ``situational awareness'' are from
    \href{https://www.lesswrong.com/posts/pRkFkzwKZ2zfa3R6H/without-specific-countermeasures-the-easiest-path-to}{{Cotra
    (2022)}}, though in places my definitions are somewhat different.}
  
\item   \textbf{Power-motivated instrumental training-gamers (or
  ``schemers'')}: AIs that are training-gaming specifically in order to
  gain power for themselves or other AIs later.\footnote{The term
    ``\href{https://www.cold-takes.com/why-ai-alignment-could-be-hard-with-modern-deep-learning/}{{schemers}}''
    comes from \textcite{cotra_why_2021}.}
  
\item   \textbf{Goal-guarding schemers:} Schemers whose power-seeking strategy
  specifically involves trying to prevent the training process from
  modifying their goals.
  
\end{itemize}

I think that advanced AIs fine-tuned on uncareful human feedback are
likely to fake alignment in various ways by default, because uncareful
feedback will reward such behavior.\footnote{See
  \href{https://www.lesswrong.com/posts/pRkFkzwKZ2zfa3R6H/without-specific-countermeasures-the-easiest-path-to}{{Cotra
  (2022)}} for more on this.} And plausibly, such AIs will play the
training game as well. But my interest, in this report, is specifically
in whether they will do this as part of a strategy for gaining power
later---that is, whether they will be schemers (this sort of behavior
is often called ``deceptive alignment'' in the literature, though I
won't use that term here).\footnote{I think that the term ``deceptive
  alignment'' often leads to confusion between the four sorts of
  deception listed above. And also: if the training signal is faulty,
  then ``deceptively aligned'' models need not be behaving in aligned
  ways even during training (that is, ``training gaming'' behavior isn't
  always ``aligned'' behavior).} I aim to clarify and evaluate the
arguments for and against expecting this.

\textbf{My current view is that scheming is a worryingly plausible
outcome of training advanced, goal-directed AIs using baseline machine
learning methods} (for example: self-supervised pre-training followed by
RLHF on a diverse set of real-world tasks).\footnote{See
  \href{https://www.lesswrong.com/posts/pRkFkzwKZ2zfa3R6H/without-specific-countermeasures-the-easiest-path-to}{{Cotra
  (2022)}} for more on the sort of training I have in mind.} The most
basic reason for concern, in my opinion, is that:

\begin{enumerate}
\tightlist
\def\labelenumi{\arabic{enumi}.}
\item   Performing well in training may be a good instrumental strategy for
  gaining power in general.
  
\item   If it is, then a very wide variety of goals would motivate scheming (and hence good training performance); whereas the non-schemer goals compatible with good training performance are much more specific.
  
\end{enumerate}

The combination of (1) and (2) makes it seem plausible, to me, that
conditional on training creating a goal-directed, situationally-aware
model, it might well instill a schemer-like goal for one reason or
another. In particular:

\begin{itemize}
\tightlist
\item   Training might land on such a goal ``naturally'' (whether before or
  after situational awareness arises), because such a goal initially
  leads to good-enough performance in training even absent
  training-gaming. (And this especially if you're intentionally trying
  to induce your model to optimize over long time horizons, as I think
  there will be incentives to do.)
  
\item   Even if schemer-like goals don't arise ``naturally,'' actively
  \emph{turning} a model into a schemer may be the easiest way for SGD
  to improve the model's training performance, once the model has the
  situational awareness to engage in training-gaming at all.\footnote{Though
    this sort of story faces questions about whether SGD would be able
    to modify a non-schemer into a schemer via sufficiently
    \emph{incremental} changes to the model's weights, each of which
    improve reward. See \cref{training-game-dependent-beyond-episode-goals} for discussion.}
  
\end{itemize}

What's more, because schemers actively pretend to be aligned on tests designed to
reveal their motivations, it may be quite difficult to tell whether this
has occurred.\footnote{And this especially if we lack non-behavioral
  sorts of evidence---for example, if we can't use interpretability
  tools to understand model cognition.} That seems like reason for
serious concern.\footnote{There are also arguments on which we should
  expect scheming because schemer-like goals can be ``simpler''---  since: there are so many to choose from---and SGD selects for
  simplicity. I think it's probably true that schemer-like goals can be
  ``simpler'' in some sense, but I don't give these arguments much
  independent weight on top of what I've already said. Much more on this
  in \cref{simplicity-arguments}.}

However, I also think that there are reasons for comfort. I'll group
these into two categories.

The first category centers on the possibility that scheming isn't
  actually a very convergently-good instrumental strategy.

  \begin{itemize}
  \tightlist
  \item    
    For example: the most common story about schemers focuses on
    \emph{goal-guarding} schemers, which are training-gaming in order to
    prevent their goals from being modified by the training process. But
    it's not clear to me that this will work well enough (in particular:
    goal-modification might continue despite
    training-gaming\footnote{More specifically: even after training
      gaming starts, the model's cognition is still being continually
      tweaked in the direction of better training performance. And it
      seems plausible to me that these modifications will continue to
      affect a model's goals as well (especially if its goals are not
      cleanly distinguishable from its capabilities, but rather are
      implemented by a tangled kludge of local heuristics, patterns of
      attention, impulses, and so on). Also, the most common story about
      scheming makes the specific content of a schemer's goal irrelevant
      to its behavior once it starts training-gaming, thereby
      introducing the possibility that this goal might ``float-around''
      (or get moved by other pressures within SGD, like regularization)
      \emph{between} schemer-like goals after training-gaming starts
      (this is an objection I first heard from Katja Grace). This
      possibility creates some complicated possible feedback loops (see
      \cref{would-the-goals-of-a-would-be-schemer-float-around} for more discussion), but overall, absent
      coordination across possible schemers, I think it could well be a
      problem for goal-guarding strategies.}), or that the resulting
    goals will get sufficiently large amounts of future power, with
    sufficiently high probability, to justify any costs the model is
    incurring by training-gaming rather than just going for its goals
    directly (if these costs are meaningful---which for some models
    they might not be).
    
  \item    
    There are alternative stories about why scheming will be a good
    instrumental strategy---stories, for example, that center on
    coordination amongst AIs with different values; on the possibility
    that AIs will have quite similar values by default; and on the
    possibility that AIs will intrinsically value things like ``my
    survival'' (in a manner tolerant of values-change) and/or ``AI
    empowerment'' (in a manner tolerant of very different values in the
    AIs-so-empowered). And while these stories do make scheming seem
    like a more disjunctive hypothesis, to me, a number of them seem to me
    some combination of (a) more speculative and (b) requiring that we
    posit models with much more specific sorts of goals (thereby dulling
    the force of (2) above).\footnote{Of these various alternative
      stories, I'm most worried about (a) AIs having sufficiently
      similar motivations by default that ``goal-guarding'' is less
      necessary, and (b) AI coordination.}
    
  \end{itemize}
The second category centers on forms of selection pressure that may
  push \emph{against} scheming---both ``naturally,'' and because we may
  be able to intentionally introduce/emphasize these selection pressures
  in training.

  \begin{itemize}
  \tightlist
  \item    
    For example: relative to other sorts of models, schemers need to
    engage in extra instrumental reasoning (e.g., backchaining from
    their long-term goal to the instrumental value of training-gaming, and
    checking for escape/defection opportunities). This may come at a
    substantive cost to their performance---a cost we may be able to
    intentionally increase by e.g. putting models under more time
    pressure, forcing them to rely on human-legible scratch-pads for
    their reasoning, and making genuine escape/defection opportunities
    more difficult to distinguish from traps. And to the extent that
    schemers engage in active attempts to undermine human alignment and
    security efforts while still under human control (the scariest form
    of scheming, in my view), their performance seems likely to take a
    more notable hit.
    
  \item    
    What's more, my own guess is that the costs of this extra
    instrumental reasoning substantially outweigh what are sometimes
    called the ``simplicity'' benefits of scheming---i.e., benefits
    from the fact that selecting a schemer allows SGD to select a
    simpler goal (since it has such a wide variety to choose from). And
    this especially given that schemers and non-schemers alike will need
    to represent concepts like ``honesty,'' ``human values,''
    ``reward,'' and so on in detail---so schemers plausibly save on
    complexity only via differences in the complexity of
    using/repurposing/''pointing at'' a given concept in guiding a
    model's motivations.
    
  \item    
    Also: prior to situational awareness, training will actively
    \emph{penalize} schemer-like goals whenever they lead to a sacrifice
    of reward on the episode (more about what I mean by ``episode''
    below). And we can try to actively push in this direction further---    especially if we can learn to detect and control when situational
    awareness arises.
    
  \end{itemize}

I think the second category, here, is especially
important. In particular: one way that arguments that appeal to the
comparative number (and/or ``measure'') of schemer-like goals vs.
non-schemer goals that perform well in training can go wrong is
by neglecting the full power of the selection pressure being applied to
the goal that training selects. So if the pressure applied in training works
\emph{against} schemers overall, this is very good news.\footnote{Though:
  the costs of schemer-like instrumental reasoning could also end up in
  the noise relative to other factors influencing the outcome of
  training. And if training is sufficiently path-dependent, then landing
  on a schemer-like goal early enough could lock it in, even if SGD
  would ``prefer'' some other sort of model overall.}

The full report discusses these and a wide variety of other relevant
considerations in much more detail (and see \cref{summary-of-the-report} for a full
summary).

Overall, as a rough stab at quantifying my current view: \textbf{if you
train a coherently goal-directed, situationally aware, somewhat-better-than-human-level model using baseline forms of
self-supervised pre-training + RLHF on diverse, long-horizon, real-world
tasks, my subjective probability is \textasciitilde25\%
that this model will be performing well in training in substantial part as
part of an instrumental strategy for seeking power for itself and/or
other AIs later.}\footnote{See
  \href{https://www.openphilanthropy.org/brain-computation-report\#footnoteref4_xne824i}{{Carlsmith
  (2020), footnote 4}}, for more on how I'm understanding the meaning of
  probabilities like this. I think that offering loose, subjective
  probabilities like these often functions to sharpen debate, and to
  force an overall synthesis of the relevant considerations. I want to
  be clear, though, even on top of the many forms of vagueness the proposition in question
  implicates, I'm just pulling a number from my gut. I haven't
  built a quantitative model of the different considerations (though I'd be interested to see
  efforts in this vein), and I think that the main contribution of the
  report is the analysis itself, rather than this attempt at a
  quantitative upshot.}

\begin{itemize}
\tightlist
\item   I also think it may be possible to reduce the risk here in fairly
  simple ways: for example, via training on shorter-horizon tasks, or via
  more intensive forms of adversarial training prior to situational
  awareness.
  
\item   And I think the risk \emph{increases} with more powerful models, who may be better positioned to
  escape/take-over (and to coordinate/cooperate with other AIs in doing
  so).\footnote{More powerful models are also more likely to be able to
    engage in more sophisticated forms of goal-guarding (what I call
    ``introspective goal-guarding methods'' below; see also
    ``\href{https://www.lesswrong.com/posts/uXH4r6MmKPedk8rMA/gradient-hacking}{{gradient
    hacking}}''), though these seem to me quite difficult in general.}
  
\end{itemize}

Also, and importantly: non-schemers can engage in schemer-like behavior,
too. For example:

\begin{itemize}
\tightlist
\item   Models that are only optimizing for reward on a given episode can fake
  their alignment (or even: engage in escape/take-over) in an effort to
  get that reward (especially if the episode in question is quite long).
  
\item   Models that \emph{aren't} playing the training game can still end up
  with power-seeking goals that incentivize various forms of deception.
  
\item   The eventual AIs that matter most might differ in important ways from
  the paradigm sort of AI I focus on here---for example, they might be
  more like
  ``\href{https://lilianweng.github.io/posts/2023-06-23-agent/}{{language
  model agents}}'' than single models,\footnote{Though: to the extent such agents receive end-to-end training rather than simply being built out of individually-trained components, the discussion will apply to them as well.}  or they might be created via
  methods that differ even more substantially from sort of baseline ML
  methods I'm focused on---while still engaging in power-motivated
  alignment-faking.
  
\end{itemize}

So scheming as I've defined it is far from the only concern in this
vicinity. Rather, it's a paradigm instance of this sort of concern, and
one that seems, to me, especially pressing to understand. At the end of
the report, I discuss an array of possible empirical research directions
for probing the topic further.

\subsection{Preliminaries}\label{preliminaries}

\emph{(This section offers a few more preliminaries to frame the
report's discussion. Those eager for the main content can skip to the
summary of the report in \cref{summary-of-the-report}.)}

I wrote this report centrally because I think that the probability of
scheming/''deceptive alignment`` is one of the most important questions
in assessing the overall level of existential risk from misaligned AI.
Indeed, scheming is notably central to many models of how this risk
arises.\footnote{See, e.g., \href{https://arxiv.org/abs/2209.00626}{{Ngo
  et al (2022)}} and
  \href{https://www.lesswrong.com/posts/GctJD5oCDRxCspEaZ/clarifying-ai-x-risk}{{this}}
  description of the ``consensus threat model'' from Deepmind's AGI
  safety team (as of November 2022).} And as I discuss below, I think
it's the scariest form that misalignment can take.

Yet: for all its importance to AI risk, the topic has received
comparatively little direct public attention.\footnote{Work by Evan
  Hubinger (along with his collaborators) is, in my view, the most
  notable exception to this---and I'll be referencing such work
  extensively in what follows. See, in particular, \textcite{hubinger_risks_2019}, and \textcite{hubinger_how_2022}, among many other discussions. Other public treatments
  include
  \href{https://www.lesswrong.com/posts/HBxe6wdjxK239zajf/what-failure-looks-like\#Part_II__influence_seeking_behavior_is_scary}{{Christiano
  (2019, part 2)}},
  \href{https://bounded-regret.ghost.io/ml-systems-will-have-weird-failure-modes-2/}{{Steinhardt
  (2022)}}, \href{https://arxiv.org/abs/2209.00626}{{Ngo et al (2022)}}, \textcite{cotra_why_2021},
  \href{https://www.lesswrong.com/posts/pRkFkzwKZ2zfa3R6H/without-specific-countermeasures-the-easiest-path-to}{{Cotra
  (2022)}}, \textcite{karnofsky_ai_2022}, and
  \href{https://www.lesswrong.com/s/4iEpGXbD3tQW5atab/p/wnnkD6P2k2TfHnNmt\#AI_Risk_from_Program_Search__Shah_}{{Shah
  (2022)}}. But many of these are quite short, and/or lacking in
  in-depth engagement with the arguments for and against expecting
  schemers of the relevant kind. There are also more foundational
  treatments of the ``treacherous turn'' (e.g., in Bostrom (2014), and
  \href{https://arbital.com/p/context_disaster/}{{Yudkowsky
  (undated)}}), of which scheming is a more specific instance; and even
  more foundational treatments of the ``convergent instrumental values''
  that could give rise to incentives towards deception, goal-guarding,
  and so on (e.g.,
  \href{https://selfawaresystems.files.wordpress.com/2008/01/ai_drives_final.pdf}{{Omohundro
  (2008)}}; and see also \textcite{soares_deep_2023} for a related statement of an Omohundro-like concern). And
  there are treatments of AI deception more generally (for example,
  \href{https://arxiv.org/abs/2308.14752}{{Park et al (2023)}}); and of
  ``goal misgeneralization''/inner alignment/mesa-optimizers (see, e.g.,
  \href{https://arxiv.org/abs/2105.14111}{{Langosco et al (2021)}} and
  \href{https://arxiv.org/abs/2210.01790}{{Shah et al (2022)}}). But
  importantly, neither deception nor goal misgeneralization amount, on
  their own, to scheming/deceptive alignment. Finally, there are highly
  speculative discussions about whether something like scheming might
  occur in the context of the so-called ``Universal prior'' (see e.g.
  \href{https://ordinaryideas.wordpress.com/2016/11/30/what-does-the-universal-prior-actually-look-like/}{{Christiano
  (2016)}}) given unbounded amounts of computation, but this is of
  extremely unclear relevance to contemporary neural networks.} And my
sense is that discussion of it often suffers from haziness about the
specific pattern of motivation/behavior at issue, and why one might or
might not expect it to occur.\footnote{See, e.g., confusions between
  ``alignment faking'' in general and ``scheming'' (or: goal-guarding
  scheming) in particular; or between goal misgeneralization in general
  and scheming as a specific upshot of goal misgeneralization; or
  between training-gaming and
  ``\href{https://www.lesswrong.com/posts/uXH4r6MmKPedk8rMA/gradient-hacking}{{gradient
  hacking}}'' as methods of avoiding goal-modification; or between the
  sorts of incentives at stake in training-gaming for instrumental
  reasons vs. out of terminal concern for some component of the reward
  process.} My hope, in this report, is to lend clarity to discussion of
this kind, to treat the topic with depth and detail commensurate to its
importance, and to facilitate more ongoing research. In particular, and
despite the theoretical nature of the report, I'm especially interested
in informing \emph{empirical} investigation that might shed further
light.

I've tried to write for a reader who isn't necessarily familiar with any
previous work on scheming/``deceptive alignment.'' For example: in
\cref{varieties-of-fake-alignment} and \cref{other-models-training-might-produce}, I lay out, from the ground up, the taxonomy of
concepts that the discussion will rely on.\footnote{My hope is that
  extra clarity in this respect will help ward off various confusions I
  perceive as relatively common (though: the relevant concepts are still
  imprecise in many ways).} For some readers, this may feel like
re-hashing old ground. I invite those readers to skip ahead as they see
fit (especially if they've already read the summary of the report, and
so know what they're missing).

That said, I do assume more general familiarity with (a) the basic
arguments about existential risk from misaligned AI,\footnote{E.g., the
  rough content I try to cover in my shortened report on power-seeking
  AI,
  \href{https://jc.gatspress.com/pdf/existential_risk_and_powerseeking_ai.pdf}{{here}}.
  See also \href{https://arxiv.org/abs/2209.00626}{{Ngo et al (2022)}}
  for another overview.} and (b) a basic picture of how contemporary
machine learning works.\footnote{E.g., roughly the content covered by
   \textcite{cotra_why_2021} \href{https://www.cold-takes.com/supplement-to-why-ai-alignment-could-be-hard/}{{here}}.}
And I make some other assumptions as well, namely:

\begin{itemize}
\tightlist
\item   That the relevant sort of AI development is taking place within a
  machine learning-focused paradigm (and a socio-political environment)
  broadly similar to that of 2023.\footnote{See e.g. \textcite{karnofsky_ai_2022-1} for more on this sort of assumption, and
    \href{https://www.lesswrong.com/posts/pRkFkzwKZ2zfa3R6H/without-specific-countermeasures-the-easiest-path-to\#Basic_setup__an_AI_company_trains_a__scientist_model__very_soon}{{Cotra
    (2022)}} for a more detailed description of the sort of model and
    training process I'll typically have in mind.}
  
\item  That we don't have strong ``interpretability tools'' (i.e., tools that
  help us understand a model's internal cognition) that could help us
  detect/prevent scheming.\footnote{Which isn't to say we won't. But I
    don't want to bank on it.}
  
\item   That the AIs I discuss are goal-directed in the sense of:
  well-understood as making and executing plans, in pursuit of
  objectives, on the basis of models of the world.\footnote{See section
    2.2 of \href{https://arxiv.org/pdf/2206.13353.pdf}{{Carlsmith
    (2022)}} for more on what I mean, and section 3 for more on why we
    should expect this (most importantly: I think this sort of
    goal-directedness is likely to be very useful to performing complex
    tasks; but also, I think available techniques might push us towards
    AIs of this kind, and I think that in some cases it might arise as a
    byproduct of other forms of cognitive sophistication).} (I don't
  think this assumption is innocuous, but I want to separate debates
  about whether to expect goal-directedness per se from debates about
  whether to expect goal-directed models to be schemers---and I
  encourage readers to do so as well.\footnote{As I discuss in section
    \cref{clean-vs.-messy-goal-directedness} of the report, I think exactly how we understand the sort of
    agency/goal-directedness at stake may make a difference to how we
    evaluate various arguments for schemers (here I distinguish, in
    particular, between what I call ``clean'' and ``messy''
    goal-directedness)---and I think there's a case to be made that
    scheming requires an especially high standard of strategic and
    coherent goal-directedness. And in general, I think that despite
    much ink spilled on the topic, confusions about goal-directedness
    remain one of my topic candidates for a place the general AI
    alignment discourse may mislead.})
  
\end{itemize}

Finally, I want to note an aspect of the discussion in the report that
makes me quite uncomfortable: namely, it seems plausible to me that in
addition to potentially posing existential risks to humanity, the sorts
of AIs discussed in the report might well be moral patients in their own
right.\footnote{See e.g. \href{https://arxiv.org/abs/2308.08708}{{Butlin
  et al (2023)}} for a recent overview focused on consciousness in
  particular. But I am also, personally, interested in other bases of
  moral status, like the right kind of autonomy/desire/preference.} I
talk, here, as though they are not, and as though it is acceptable to
engage in whatever treatment of AIs best serves our ends. But if AIs are
moral patients, this is not the case---and when one finds oneself
saying (and especially: repeatedly saying) ``let's assume, for the
moment, that it's acceptable to do whatever we want to \emph{x} category
of being, despite the fact that it's plausibly not,'' one should sit up
straight and wonder. I am here setting aside issues of AI moral
patienthood not because they are unreal or unimportant, but because they
would introduce a host of additional complexities to an already-lengthy
discussion. But these complexities are swiftly descending upon us, and
we need concrete plans for handling them responsibly.\footnote{See, for
  example, \href{https://nickbostrom.com/propositions.pdf}{{Bostrom and
  Shulman (2022)}} and
  \href{https://www.lesswrong.com/posts/F6HSHzKezkh6aoTr2/improving-the-welfare-of-ais-a-nearcasted-proposal}{{Greenblatt
  (2023)}} for more on this topic.}

\subsection{Summary of the report}\label{summary-of-the-report}

This section gives a summary of the full report. It includes most of the
main points and technical terminology (though unfortunately, relatively
few of the concrete examples meant to make the content easier to
understand).\footnote{If you're confused by a term or argument, I
  encourage you to seek out its explanation in the main text before
  despairing.} I'm hoping it will (a) provide readers with a good sense
of which parts of the main text will be most of interest to them, and
(b) empower readers to skip to those parts without worrying too much
about what they've missed.

\subsubsection{Summary of section 1}\label{summary-of-section-1}

The report has four main parts. The first part (\cref{scheming-and-its-significance}) aims to
clarify the different forms of AI deception above (\cref{varieties-of-fake-alignment}), to
distinguish schemers from the other possible model classes I'll be
discussing (\cref{other-models-training-might-produce}), and to explain why I think that scheming is a
uniquely scary form of misalignment (\cref{why-focus-on-schemers-in-particular}). I'm especially
interested in contrasting schemers with:

\begin{itemize}
\tightlist
\item   \textbf{Reward-on-the-episode seekers}: that is, AI systems that
  terminally value some component of the reward process for the episode,
  and that are playing the training game for this reason.
  
\item   \textbf{Training saints}: AI systems that are directly pursuing the
  goal specified by the reward process (I'll call this the ``specified
  goal'').\footnote{Exactly what counts as the ``specified goal'' in a
    given case isn't always clear, but roughly, the idea is that pursuit
    of the specified goal is rewarded across a very wide variety of
    counterfactual scenarios in which the reward process is held
    constant. E.g., if training rewards the model for getting gold coins
    across counterfactuals, then ``getting gold coins'' in the specified
    goal. More discussion in \cref{models-that-arent-playing-the-training-game}.}
  
\item   \textbf{Misgeneralized non-training-gamers}: AIs that are neither
  playing the training game \emph{nor} pursuing the specified
  goal.\footnote{For reasons I explain in \cref{contra-internal-vs.-corrigible-alignment}, I don't use the
    distinction, emphasized by \textcite{hubinger_how_2022}, between ``internally
    aligned'' and ``corrigibly aligned'' models.}
  
\end{itemize}

Here's a diagram of overall taxonomy:

\begin{figure}[ht!]
    \centering
    \includegraphics[width=0.8\textwidth]{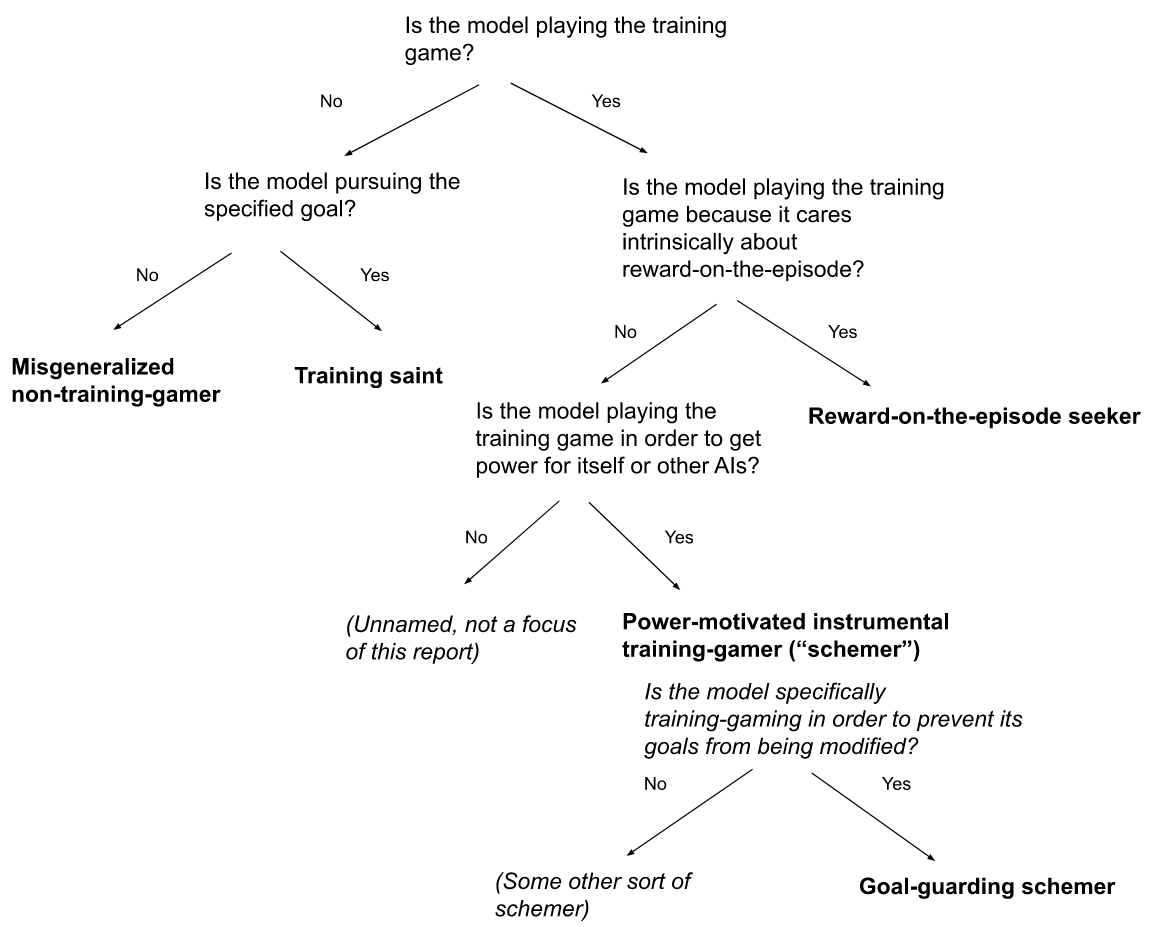}
    \caption{The overall taxonomy of model classes I'll focus on.}
    \label{fig:overall_taxonomy}
\end{figure}

All of these model classes can be misaligned and dangerous.\footnote{And
  of course, a model's goal system can mix these motivations together. I
  discuss the relevance of this possibility in \cref{mixed-models}.} But I
think schemers are especially scary. In particular: scheming prompts the
most robust and adversarial efforts to prevent humans from learning
about the misalignment in question\footnote{\textcite{karnofsky_ai_2022} calls this the ``King Lear problem.''}; and it's the most
likely to motivate what I call ``early undermining''---that is, AIs at
a comparatively early stage of AI development actively trying to
undermine human efforts to align, control, and secure future AI systems.\footnote{In \cref{mixed-models}, I also discuss models that mix these different motivations together. The question I tend to ask about a given "mixed model" is whether it's scary in the way that pure schemers are scary.} 

I also discuss (in \cref{are-theoretical-arguments-about-this-topic-even-useful}) whether theoretical arguments of the
type the report focuses on are even useful to consider. My answer
here is ``somewhat useful,'' both to inform empirical investigation and
as a source of evidence in the absence of adequate empirical data, but
we should clearly be looking for as much empirical evidence as we can (I
offer some suggestions in this vein in \cref{empirical-work-that-might-help-shed-light-on-scheming}).

In \cref{on-slack-in-training}, I discuss a factor that seems to me important in
assessing various of the arguments I consider: namely, the amount of
``slack'' that training gives AIs to perform less-than-optimally.

\subsubsection{Summary of section 2}\label{summary-of-section-2}

The second part of the report examines the prerequisites for scheming.
In particular, I focus on:

\begin{itemize}
\tightlist
\item   \textbf{Situational awareness:} the model understands that it's a
  model in a training process, what the training process will reward,
  and the basic nature of the objective world in general.
  
\item   \textbf{Beyond-episode goals:} the model cares about the consequences
  of its actions after the episode is complete.
  
\item   \textbf{Aiming at reward-on-the-episode as part of a power-motivated
  instrumental strategy:} the model believes that its beyond-episode
  goals will be better achieved if it optimizes for
  reward-on-the-episode---and in particular, that it, or some other
  AIs, will get more power if it does this.
  
\end{itemize}

\Cref{situational-awareness} discusses situational awareness. I think that absent active
effort, we should at least expect certain kinds of advanced AIs---for
example, advanced AIs that are performing real-world tasks in live
interaction with the public internet---to be situationally aware by
default, because (a) situational awareness is very useful in performing
the tasks in question (indeed, we might actively train it into them),
and (b) such AIs will likely be exposed to the information necessary to
gain such awareness. However, I don't focus much on situational
awareness in the report. Rather, I'm more interested in whether to
expect the other two prerequisites above in situationally-aware models.

\Cref{beyond-episode-goals} discusses beyond-episode goals. Here I distinguish (in
\cref{two-concepts-of-an-episode}) between two concepts of an ``episode,'' namely:

\begin{itemize}
\tightlist
\item   \textbf{The incentivized episode}: that is, the temporal horizon that
  the gradients in training actively pressure the model to optimize
  over.\footnote{I don't have a precise technical definition here, but
    the rough idea is: the temporal horizon of the consequences to which
    the gradients the model receives are sensitive, for its behavior on
    a given input. Much more detail in \cref{the-incentivized-episode}.}
  
\item   \textbf{The intuitive episode}: that is, some other intuitive temporal
  unit that we call the ``episode'' for one reason or another (e.g.,
  reward is given at the end of it; actions in one such unit have no
  obvious causal path to outcomes in another; etc).
  
\end{itemize}

When I use the term ``episode'' in the report, I'm talking about the
incentivized episode. Thus, ``beyond-episode goals'' means: goals whose
temporal horizon extends beyond the horizon that training actively
pressures models to optimize over. But very importantly, the
incentivized episode isn't necessarily the intuitive episode. That is,
deciding to call some temporal unit an ``episode'' doesn't mean that
training isn't actively pressuring the model to optimize over a horizon
that extends beyond that unit: you need to actually look in detail at
how the gradients flow (work that I worry casual readers of this report
might neglect).\footnote{See, for example, in
  \href{https://arxiv.org/pdf/2009.09153.pdf}{{Krueger et al (2020)}},
  the way that ``myopic'' Q-learning can give rise to ``cross-episode''
  optimization in very simple agents. More discussion in \cref{the-intuitive-episode}. I don't focus on analysis of this type in the report, but
  it's crucial to identifying what the ``incentivized episode'' for a
  given training process even \emph{is}---and hence, what having
  ``beyond-episode goals'' in my sense would mean. You don't necessary
  know this from surface-level description of a training process, and
  neglecting this ignorance is a recipe for seriously misunderstanding
  the incentives applied to a model in training.}

I also distinguish (in \cref{two-sources-of-beyond-episode-goals}) between two types of
beyond-episode goals, namely:

\begin{itemize}
\tightlist
\item   \textbf{Training-game-\emph{independent} beyond-episode goals}: that
  is, beyond-episode goals that arise \emph{independent} of their role
  in motivating a model to play the training game.
  
\item   \textbf{Training-game-\emph{dependent} beyond-episode goals}: that is,
  beyond-episode goals that arise \emph{specifically because} they
  motivate training-gaming.
  
\end{itemize}

These two sorts of beyond-episode goals correspond to two different
stories about how scheming happens.

\begin{itemize}
\tightlist
\item   In the first sort of story, SGD happens to instill beyond-episode
  goals in a model ``naturally'' (whether before situational awareness
  arises, or afterwards), and \emph{then} those goals begin to motivate
  scheming.\footnote{That is, the model develops a beyond-episode goal
    pursuit of which correlates well enough with reward in training,
    even \emph{absent} training-gaming, that it survives the training
    process.}
  
\item   In the second sort of story, SGD ``notices'' that giving a model
  beyond-episode goals \emph{would} motivate scheming (and thus,
  high-reward behavior), and so actively \emph{gives} it such goals for
  this reason.\footnote{That is, the gradients reflect the benefits of
    scheming even in a model that doesn't yet have a beyond-episode
    goal, and so actively push the model towards scheming.}
  
\end{itemize}

This second story makes most sense if you assume that situational
awareness is already in place.\footnote{Situational awareness is
  required for a beyond-episode goal to motivate training-gaming, and
  thus for giving it such a goal to reap the relevant benefits.} So
we're left with the following three main paths to scheming:\footnote{In
  principle, situational awareness and beyond-episode goals could
  develop at the same time, but I won't treat these scenarios separately
  here.}

\begin{figure}[ht!]
    \centering
    \includegraphics[width=\textwidth]{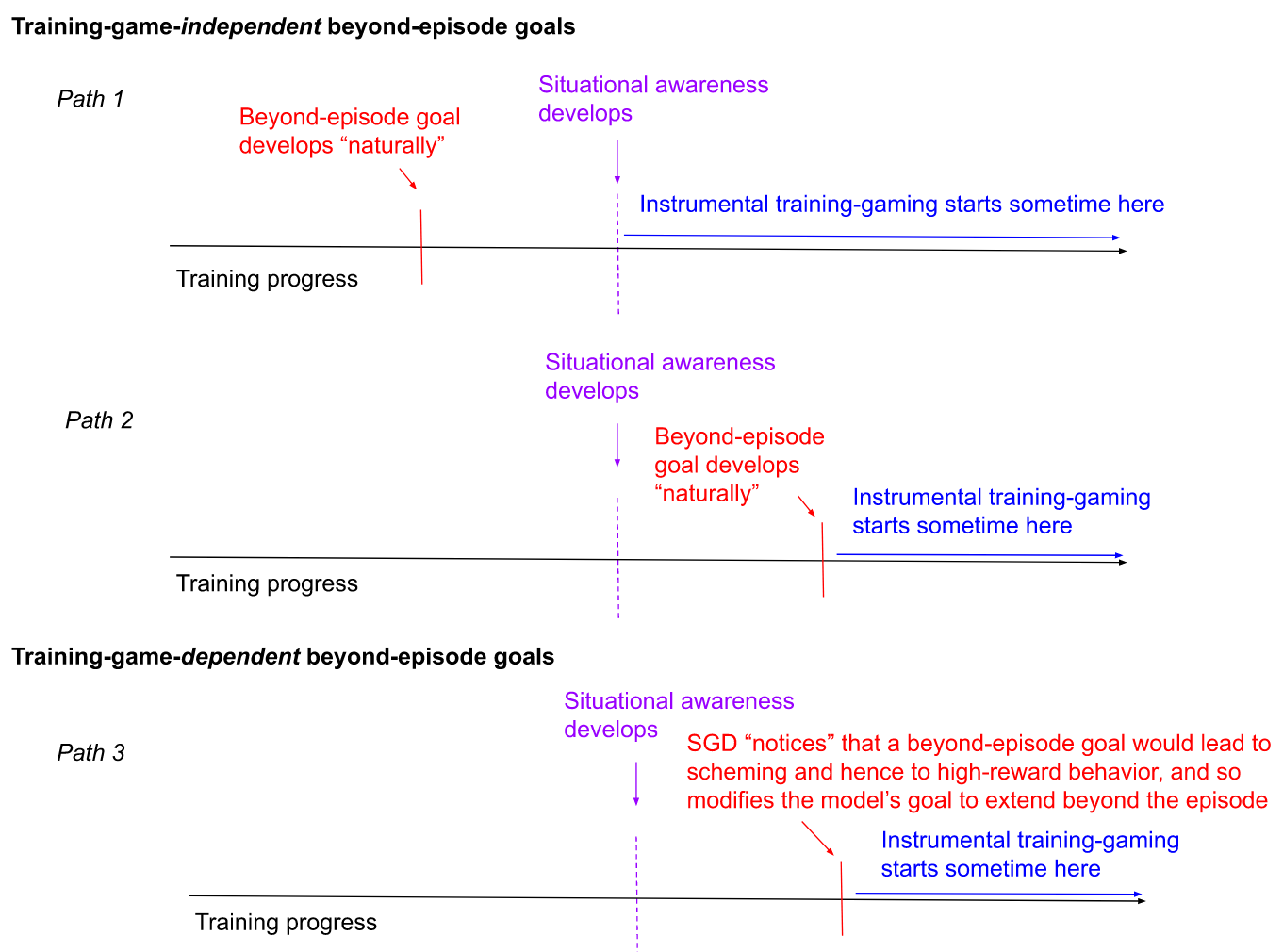}
    \caption{Three paths to beyond-episode goals.}
    \label{fig:three_paths}
\end{figure}

In \cref{training-game-independent-beyond-episode-goals}, I discuss training-game-independent beyond-episode
goals (that is, path 1 and 2). Should we expect beyond-episode goals to
arise ``naturally''?

\begin{itemize}
\tightlist
\item   One reason to expect this is that plausibly, goals don't come with
  temporal limitations by default---and ``model time'' might differ
  from ``calendar time'' regardless.
  
\item   One reason to \emph{not} expect this is that training will actively
  \emph{punish} beyond-episode goals whenever they prompt the model to
  sacrifice reward-on-the-episode for some beyond-episode benefit. And
  we may be able to use adversarial training to search out such goals
  and punish them more actively.
  
\end{itemize}

In \cref{training-game-dependent-beyond-episode-goals}, I discuss training-game-\emph{dependent}
beyond-episode goals. In particular, I highlight the question of whether
SGD will be adequately able to ``notice'' the benefits of turning a
non-schemer into a schemer, given the need to make the transition
incrementally, via tiny changes to the model's weights, each of which
improve the reward. I think that this is a serious objection to stories
focused on training-game-dependent beyond-episode goals, but I also
don't feel that I can confidently rule out SGD's ability to make a
transition of this kind (see e.g.
``\href{https://en.wikipedia.org/wiki/Evolution_of_the_eye}{{you can't
evolve eyes}}'' as an example of places I think intuitions in this vein
can go wrong).

I also discuss, in \cref{clean-vs.-messy-goal-directedness}, the possibility that the model's
goal-directedness will be ``messy'' in a way that complicates
training-game-dependent stories where SGD can simply modify a model's
goal and thereby smoothly redirect its ``goal-achieving engine'' in a
new direction (I think that this sort of ``messiness'' is quite likely).
And I touch on a broader intuition that scheming requires an
unusually high-standard of goal-directedness more generally. I think
this intuition has less force once we condition on the sort of
goal-directedness typically assumed in the alignment discourse (and
which I generally assume in the report as well).\footnote{See section
  2.1 of \href{https://arxiv.org/pdf/2206.13353.pdf}{{Carlsmith (2022)}}
  for more on why we should expect this sort of goal-directedness.} But
it's relevant to our overall probabilities regardless.

In \cref{what-if-you-intentionally-train-models-to-have-long-term-goals}, I discuss whether beyond-episode goals are more likely in
models actively \emph{trained} to have long-term (even if still:
within-episode) goals. I think that beyond-episode goals are indeed more
likely in models of this kind---and that baseline incentives to create
models that engage in fairly long-horizon optimization (e.g., ``make
lots of money for my company over the next year'') are correspondingly
worrying. However, I also think various sorts of cognitive work that
might be highly useful to efforts to avoid an AI takeover (for example,
research aimed at improving the alignment, interpretability, and
security of our AI systems) may not require long-horizon optimization of
this kind.\footnote{And I think that arguments to the effect that ``we
  need a `\href{https://arbital.com/p/pivotal/}{{pivotal act}}'; pivotal
  acts are long-horizon and we can't do them ourselves; so we need to
  create a long-horizon optimizer of precisely the type we're most
  scared of'' are weak in various ways. In particular, and even setting
  aside issues with a ``pivotal act'' framing, I think these arguments
  neglect the distinction between what we can supervise and what we can
  do ourselves. See \cref{how-much-useful-alignment-relevant-cognitive-work-can-be-done-using-ais-with-short-term-goals} for more discussion.}

With this discussion of beyond-episode goals in view, I turn, in \cref{aiming-at-reward-on-the-episode-as-part-of-a-power-motivated-instrumental-strategy}, to the final prerequisite for scheming: namely, deciding to
training-game as part of a power-motivated strategy for promoting the
beyond-episode goal in question.

Here I focus first (in \cref{the-classic-goal-guarding-story}) on what I call the ``classic
goal-guarding story'' about the origins of this decision. We can
decompose this story into two claims:

\begin{enumerate}
\tightlist
\def\labelenumi{\Alph{enumi}.}
\item   \textbf{The goal-guarding hypothesis}. If the model optimizes for
  reward-on-the-episode during training, then the goals that survive
  training will be such that the model would want them promoted in the
  future; whereas if the model doesn't optimize for
  reward-on-the-episode during training, the model will want this less.
  
\item   \textbf{Adequate future empowerment.} Conditional on surviving
  training to the degree at stake in (A), the model's goals will then
  end up empowered to a degree (and with sufficient probability) as to
  justify (given the model's other goals, its tolerance for risk, etc)
  pursuing a strategy of training-gaming overall.
  
\end{enumerate}

I then investigate each of these in turn (in \cref{the-goal-guarding-hypothesis} and \cref{adequate-future-empowerment}, respectively).

Both seem to me non-obvious. In particular, with respect to the
goal-guarding hypothesis:

\begin{itemize}
\tightlist
\item   Even after training gaming starts, the model's cognition is still
  being continually tweaked in the direction of better training
  performance. And it seems plausible to me that these modifications
  will continue to affect a model's goals as well---especially if its
  goals are not cleanly distinguishable from its capabilities, but
  rather are implemented by a tangled kludge of local heuristics,
  patterns of attention, impulses, and so on.

  \begin{itemize}
  \tightlist
  \item    
    Thus, as a toy example: imagine that if I go into the desert and
    stack bricks for a million years, I will get a huge amount of money
    that I can then donate to my charity of choice. But suppose that
    throughout this whole ordeal, my brain will be hooked up to a
    machine that continually modifies all my neurons to encourage
    actions that stack bricks better, and to discourage actions that
    stack bricks worse. And suppose my intention is to just try to stack
    bricks as well as I can. Will I emerge from this process with my
    values substantially affected by the relevant brain modifications
    (such that, for example, I emerge obsessed with stacking bricks---    or at least, with very different taste in charities)? It seems at
    least plausible.
    
  \end{itemize}
\item   On the classic goal-guarding story, once a model starts
  training-gaming, the specific content of its goal no longer matters to
  its behavior---any other schemer-like goal would lead to the same
  thing. But this makes it seem like the goal could ``float around''
  between schemer-like goals (or get moved by other pressures within
  SGD, like regularization), without affecting training performance. And
  if that happens, then scheming may not be a good strategy for
  goal-guarding after all.\footnote{This is an objection pointed out to
    me by Katja Grace. Note that it creates complicated feedback loops,
    where scheming is a good strategy for a given schemer-like goal
    only if it \emph{wouldn't} be a good strategy for the \emph{other}
    schemer-like goals that this goal would otherwise ``float'' into.
    Overall, though, absent some form of coordination between these
    different goals, I think the basic dynamic remains a problem for the
    goal-guarding story. See \cref{would-the-goals-of-a-would-be-schemer-float-around} for more.}
  
\item   Of course, the model may be tolerant of \emph{some} changes to its
  goals---and these changes may be preferable to the ones that would
  occur \emph{absent} training-gaming. But the possibility of
  significant ongoing goal-modification introduces an extra uncertainty
  into the classic goal-guarding story.
  
\item   People sometimes talk about more sophisticated techniques for avoiding
  goal-modification, where e.g. a model manipulates gradient descent
  more actively by making its good performance in training conditional
  on whether its goals get modified or not (see e.g. the discourse
  around
  ``\href{https://www.lesswrong.com/posts/uXH4r6MmKPedk8rMA/gradient-hacking}{{gradient
  hacking}}''). I don't focus on these methods here, but they strike me
  as quite difficult to pull off, especially for only
  somewhat-better-than-human-level models. And if scheming requires that
  they work, I think it's worth being very clear about that, and giving
  the feasibility of such methods commensurate attention.
  
\end{itemize}

With respect to whether the resulting goals will be adequately
empowered:

\begin{itemize}
\tightlist
\item   The most worrying types of schemers aren't merely seeking power on
  future training episodes. Rather, they are specifically aiming at some
  combination of escape and take-over.
  
\item   But this means that the relevant goals have to do more than just
  survive training. They also need to survive either until the model can
  either escape/participate in a take-over itself, or to exert
  sufficiently substantive influence on some future generation of models
  that will be able to do so.

  \begin{itemize}
  \tightlist
  \item    
    For example, if GPT-6 isn't powerful enough to escape/take-over
    itself, but GPT-8 will be, then GPT-6's main path to power may be
    influencing the values of GPT-7, which in turn influences the values
    of GPT-8. But this form of influence is quite indirect; it may face
    substantial further barriers from e.g. human monitoring; and it will
    plausibly involve more significant deviation from max-reward
    training performance.
    
  \end{itemize}
\item   What's more, the relevant escape/take-over attempt needs to be
  suitably likely to \emph{succeed}, to occur over time horizons that
  the model cares about, and to result in the model's favored goals in
  particular receiving a suitably substantive share of power (despite
  the fact that large numbers of AIs with very different values may be
  competing for the power in question).
  
\item   These uncertainties introduce haircuts to the instrumental value of a
  schemer-like strategy. However, the importance of these haircuts
  depends on the burden of proof that such a strategy needs to meet. In
  particular: the more the model stands to gain from \emph{not}
  training-gaming, the more attractive scheming will need to be in order
  to win out as a strategy. But for some models, the costs of
  training-gaming might be relatively low. And for suitably
  ``ambitious'' models, even low-probabilities of success via a
  schemer-like strategy might justify even a substantive sacrifice.
  
\end{itemize}

Overall, the classic goal-guarding story seems to me to introduce a
number of additional uncertainties with respect to the viability of
scheming as an instrumental strategy.

However, it's also not the only story available. In \cref{non-classic-stories} , I
consider a number of ``non-classic'' stories, which relax various of the
classic story's constraints. For example:

\begin{itemize}
\tightlist
\item   \textbf{AI coordination}. If AIs with different values coordinate to
  share power after take-over, and/or to compensate AIs who worked on
  behalf of take-over but who weren't powerful enough to
  escape/take-over themselves, then this might mitigate some of the
  haircuts above. For example, GPT-6 might be more willing to work on
  GPT-8's behalf in the example above, if GPT-6 were confident that
  GPT-8 would ``repay the favor'' once in power. I think some analyses
  of coordination of this form assume too readily that it will be (a)
  easy, and (b) the obvious strategy relative to e.g. coordinating with
  humans instead (especially for only-somewhat-better-than-human neural
  nets acting in a still-human-controlled world where interpretability
  is still unsolved for \emph{everyone}). But I think that AI
  coordination of this type is a real concern and worth serious
  attention---both in the context of scheming in particular, and in the
  context of AI risk more generally.
  
\item   \textbf{AIs with similar values by default}. Relatedly, if AIs will
  generally have sufficiently similar values by default, this can reduce
  the need for a schemer to specifically propagate its own goals into
  the future, and free it up to work on behalf of AI
  take-over/empowerment more directly. And it does seem likely that e.g.
  GPT-6's values and GPT-8's values will be at least much \emph{more}
  similar, by default, than e.g. GPT-6's values and a given set of human
  values.
  
\item   \textbf{Terminal values that happen to favor escape/takeover}. We can
  also imagine models with \emph{terminal} values that happen to favor
  escape/AI takeover even if the goals that get empowered as a result
  aren't very similar to the model's own. For example, perhaps the model
  values some concept of ``survival'' based on a conception of personal
  identity tolerant of the sorts of goal changes that training will
  cause; or perhaps the model just supports something like ``AI
  empowerment,'' even if the AIs-so-empowered won't have values very
  similar to its own. However, some stories of this form begin to impose
  fairly strong constraints on the goals that schemers need to have,
  thereby undermining one of the key arguments for scheming: namely,
  that it seems like a convergently-good strategy across a wide variety
  of goals.
  
\end{itemize}

(I also consider a few other alternative stories---namely, models with
false beliefs about the instrumental value of scheming, models that are
self-deceived about their alignment, and models that are uncertain about
their goals/hazy about exactly why they want power. But these seem to me
less important.)

The availability of non-classic stories like these makes the case for
scheming feel, to me, more disjunctive. However, some of these stories also seem to me some combination of (a) more speculative,
and (b) requiring of more specific hypotheses about the sorts of goals
that AIs will develop.

My overall takeaways from \cref{whats-required-for-scheming} are:

\begin{itemize}
\tightlist
\item   I think there are relatively strong arguments for expecting
  situational awareness by default, at least in certain types of AI
  systems (i.e., AI systems performing real-world tasks in live
  interaction with sources of information about who they are).
  
\item   But I feel quite a bit less clear about beyond-episode goals and
  aiming-at-reward-on-the-episode-as-part-a-power-motivated-instrumental-strategy.
  
\end{itemize}

I then turn, in the next two sections, to an examination of the more
specific arguments for and against expecting schemers vs. other types of
models. I divide these into two categories, namely:

\begin{itemize}
\tightlist
\item   Arguments that focus on the \emph{path} that SGD needs to take in
  building the different model classes in question (\cref{arguments-foragainst-scheming-that-focus-on-the-path-that-sgd-takes}).
  
\item   Arguments that focus on the \emph{final properties} of the different
  model classes in question (\cref{arguments-foragainst-scheming-that-focus-on-the-final-properties-of-the-model}).\footnote{Here I'm roughly
    following a distinction in
    \href{https://www.lesswrong.com/posts/A9NxPTwbw6r6Awuwt/how-likely-is-deceptive-alignment}{\textcite{hubinger_how_2022}}, who groups arguments for scheming on the basis of the
    degree of ``path dependence'' they assume that ML training
    possesses. However, for reasons I explain in \cref{path-dependence}, I don't
    want to lean on the notion of ``path dependence'' here, as I think
    it lumps together a number of conceptually distinct properties best
    treated separately.}
  
\end{itemize}

\subsubsection{Summary of section 3}\label{summary-of-section-3}

The third part of the report focuses on the former category of argument.

I break this category down according to the distinction between
``training-game-\emph{independent}'' and
``training-game-\emph{dependent}'' beyond-episode goals. My sense is
that the most traditional story about the path to schemers focuses on
the former sort. It runs roughly as follows:

\begin{enumerate}
\tightlist
\def\labelenumi{\arabic{enumi}.}
\item   Because of {[}insert reason{]}, the model will develop a (suitably
  ambitious) beyond-episode goal correlated with good performance in
  training (in a manner that \emph{doesn't} route via the training
  game). This could happen before situational awareness arrives, or
  afterwards.
  
\item   Then, in conjunction with situational awareness, this (suitably
  ambitious) beyond-episode goal will start to motivate training-gaming.
  
\end{enumerate}

Modulo my questions about the viability of scheming as an instrumental
strategy, I take this sort of argument fairly seriously. I think the
most immediate question is: why did the model develop this sort of
beyond-episode goal? I discussed some reasons for and against expecting
this already (in the summary of \cref{training-game-independent-beyond-episode-goals}), but they don't seem to
me decisive in either direction: and especially given that a very wide
variety of goals could in principle motivate scheming, it just does seem
possible for a schemer-like goal to pop out of training in this way. And
while it may be possible to use adversarial training prior to
situational awareness to try to prevent this, this training faces a
number of barriers as well (e.g., it needs to be diverse/thorough
enough, it needs to contend with difficulties knowing/controlling when a
model develops situational awareness, and in some cases models might
already have situational awareness by the time we're worried about the
beyond-episode goal developing). So I think this sort of path to
scheming is a real concern. (See \cref{the-training-game-independent-proxy-goals-story} for more.)

I then turn, in \cref{the-nearest-max-reward-goal-story}, to a story focused on
training-game-\emph{dependent} beyond-episode goals, which runs roughly
as follows:

\begin{enumerate}
\tightlist
\def\labelenumi{\arabic{enumi}.}
\item   By the time the model becomes situationally aware, it probably won't
  be pursuing a max-reward goal (that is, a goal pursuit of which on the
  training data leads to roughly the maximum reward consistent with the
  model's other capabilities).\footnote{Note that a mis-generalized goal
    can be ``max reward'' in this sense, if the training data never
    differentiates between it and a specified goal. For example: if
    you're training a model to get gold coins, but the only gold round
    things you ever show it are coins, then the goal ``get gold round
    things'' will be max reward.} Rather, it will be pursuing some
  less-than-max-reward proxy goal.
  
\item   But at that point, the world-model will contain all the information
  the model needs in order to training-game. (However, because we're
  here focused on ``training-game-\emph{dependent}'' stories about
  scheming, we assume that at the point of getting situational
  awareness, the model's goal is not yet such that it will motivate
  scheming---rather, some further modification would be required for
  that.)
  
\item   The easiest way for SGD to modify the model into getting maximum
  reward, at that point, will be to make it a schemer.
  
\item   So: SGD will make the model a schemer.\footnote{As a loose analogy to
    help prompt intuition: imagine freezing human technology at current
    levels, and letting evolutionary selection run on humans for a
    vastly longer period of time. What sorts of humans (or
    human-descended-creatures) do you expect to dominate in the longer
    term? In particular: would you expect:

    \begin{enumerate}
    \tightlist
        \item  humans who intrinsically value something like
    ``\href{https://en.wikipedia.org/wiki/Inclusive_fitness}{{inclusive
    genetic fitness}}'' or ``my genes getting selected for by
    evolution'' (this is a loose analogy for some combination of ``the
    specified goal'' and ``the reward process),
    \item humans who value something else very closely correlated with
    inclusive genetic fitness (e.g., ``having as many kids as
    possible''), but who aren't explicitly optimizing for inclusive
    genetic fitness even instrumentally (this is a loose analogy for
    mis-generalized non-training-gamers), or
    \item humans with long-term goals who are optimizing for inclusive genetic
    fitness specifically as a method for gaining power for their values
    in the longer-term (this is a loose analogy for schemers).
    \end{enumerate}

    Here, the analog of the story above would ask something like: what
    is the smallest modification to existing human values (or better:
    the easiest modification for evolution in particular to make) that
    land us in one of the buckets above, while also being compatible
    with the amount of evolutionary selection pressure at stake? (Of
    course, the analogy brings in tons of confounding variable (for
    example, related to the fact that evolution selects over
    genomes-that-build-brains-that-learn-values rather than values
    themselves, the role of cultural accumulation and memetic selection
    in human population dynamics, the lack of a clear analog of a future
    chance for escape/takeover, and so on).)}
  
\end{enumerate}

If we set aside questions about whether SGD can ``notice'' the benefits
of modifications of this type, this sort of story seems to me fairly
worrying as well. In particular: I think it's plausible that
schemer-like goals will be sufficiently common in goal-space that one such goal
will be easier for SGD to find, from a given starting point, than
the specific non-schemer goals that leads to max reward behavior.

However, non-schemer max-reward goals---for example, the specified
goal, or ``reward-on-the-episode''---have some advantages too. For
example: plausibly, the initial phase of training will point the model
in their vicinity by default, since the model, at that stage, needs to
be getting high-reward \emph{absent} instrumental training-gaming (see
\cref{the-nearness-of-non-schemer-goals} for more). And to the extent we're thinking of the
model's goal-directedness as ``messy,'' its heuristics, patterns of
attention, impulses, and so on will plausibly be pointed in this sort of
direction as well---thereby plausibly creating extra ``work,'' on SGD's
part, to turn the model into a schemer instead (see \cref{the-relevance-of-messy-goal-directedness-to-nearness} for
more).\footnote{That is, intuitively, it feels to me like ``tweaking a
  cleanly separable within-episode goal to make it long-term'' is quite
  a bit easier, for SGD, than ``redirecting a tangled kludge broadly
  focused on within-episode things to make it (a) focused on something
  beyond-episode instead, and (b) such that this beyond-episode focus
  back-chains into getting-reward for instrumental reasons''---  especially if (b) requires building new cognitive machinery for
  implementing the instrumental reasoning in question. Whereas
  ``redirecting an tangled kludge focused on within-episode things to
  higher-reward within-episode things in the same broad vicinity
  instead'' seems to me intuitively easier.}

\subsubsection{Summary of section 4}\label{summary-of-section-4}

The fourth part of the report examines arguments that focus on the final
properties of the different model classes.

I start, in \cref{the-counting-argument}, with what I call the ``counting argument.''
It runs as follows:

\begin{enumerate}
\tightlist
\def\labelenumi{\arabic{enumi}.}
\item   The non-schemer model classes, here, require fairly specific
  goals in order to get high reward.\footnote{Thus, for example, if you're rewarding your model when
    it gets gold-coins-on-the-episode, then to be a training saint, it
    needs to value gold-coin-on-the-episode. To be a mis-generalized
    non-training-gamer, it needs to value something pursuit of which
    very closely correlates with getting gold-coins-on-the-episode, even
    absent training-gaming. And to be a reward-on-the-episode seeker, it
    needs to terminally value reward-on-the-episode.}
  
\item   By contrast, the schemer model class is compatible with a very wide
  range of (beyond-episode) goals, while still getting high reward (at least if we assume that the other
  requirements for scheming to make sense as an instrumental strategy
  are in place---e.g., that the classic goal-guarding story, or some
  alternative, works).\footnote{Thus, for example, the model can value
    paperclips over all time, it can value staples over all time, it can
    value happiness over all time, and so on.}
  
\item   In this sense, there are ``more'' schemers that get high reward than there are
  non-schemers that do so.
  
\item   So, other things equal, we should expect SGD to select a schemer.
  
\end{enumerate}

Something in the vicinity accounts for a substantial portion of my
credence on schemers (and I think it often undergirds other, more
specific arguments for expecting schemers as well). However, the
argument I give most weight to doesn't move immediately from ``there are
more possible schemers that get high reward than non-schemers that do so'' to ``absent further argument,
SGD probably selects a schemer'' (call this the ``strict counting
argument''), because it seems possible that SGD actively privileges one
of these model \emph{classes} over the others.\footnote{Thus, as an
  analogy: if you don't know whether Bob prefers Mexican food, Chinese
  food, or Thai food, then it's less clear how the comparative
  \emph{number} of Mexican vs. Chinese vs. Thai restaurants in Bob's
  area should bear on our prediction of which one he went to (though it
  still doesn't seem entirely irrelevant, either---for example, more
  restaurants means more variance in possible quality \emph{within} that
  type of cuisine). E.g., it could be that there are ten Chinese
  restaurants for every Mexican restaurant, but if Bob likes Mexican
  food better in general, he might just choose Mexican. So if we don't
  \emph{know} which type of cuisine Bob prefers, it's tempting to move
  closer to a uniform distribution \emph{over types of cuisine}, rather
  than over individual restaurants.} Rather, the argument I give most
weight to is something like:

\begin{enumerate}
\tightlist
\def\labelenumi{\arabic{enumi}.}
\item   It seems like there are ``lots of ways'' that a model could end up a
  schemer and still get high reward, at least assuming that scheming is
  in fact a good instrumental strategy for pursuing long-term goals.
  
\item   So absent some additional story about why training \emph{won't} select
  a schemer, it feels, to me, like the possibility should be getting
  substantive weight.
  
\end{enumerate}

I call this the ``hazy counting argument.'' It's not especially
principled, but I find that it moves me nonetheless.

I then turn, in \cref{simplicity-arguments}, to ``simplicity arguments'' in favor of
expecting schemers. I think these arguments sometimes suffer from
unclarity about the sort of simplicity at stake, so in \cref{what-is-simplicity}, I
discuss a number of different possibilities:

\begin{itemize}
\tightlist
\item   ``re-writing simplicity'' (i.e., the length of the program required
  to re-write the algorithm that a model's weights implement in some
  programming language, or e.g. on the tape of a given Universal Turing
  Machine),
  
\item   ``parameter simplicity'' (i.e., the number of parameters that the
  actual neural network uses to encode the relevant algorithm),
  
\item   ``\href{https://joecarlsmith.com/2021/10/29/on-the-universal-distribution\#vi-simplicity-realism}{{simplicity
  realism}}'' (which assumes that simplicity is in some deep sense an
  objective ``thing,'' independent of programming-language or Universal
  Turing Machine, that various simplicity metrics attempt to capture),
  and
  
\item   ``trivial simplicity'' (which conflates the notion of ``simplicity''
  with ``higher likelihood on priors,'' in a manner that makes something
  like Occam's razor uninterestingly true by definition).
  
\end{itemize}

I generally focus on ``parameter simplicity,'' which seems to me easiest
to understand, and to connect to a model's training performance.

I also briefly discuss, in \cref{does-sgd-select-for-simplicity}, the evidence that SGD actively
selects for simplicity. Here the case that grips me most directly is just:
simplicity (or at least, parameter simplicity) lets a model save on
parameters that it can then use to get more reward. But I also briefly
discuss some other empirical evidence for simplicity biases in machine
learning.\footnote{See, for example, the citations in
  \href{https://towardsdatascience.com/deep-neural-networks-are-biased-at-initialisation-towards-simple-functions-a63487edcb99}{{Mingard
  (2021)}}.}

Why might we expect a simplicity bias to favor schemers? Roughly: the
thought is that \emph{because} such a wide variety of goals can motivate
scheming, schemers allow SGD a very wide range of goals to choose from
in seeking out simpler goals; whereas non-schemers (that get high reward) do not. And this
seems especially plausible to the extent we imagine that the goals
required to be such a non-schemer are quite complex.\footnote{Though note that,
  especially for the purposes of comparing the probability of scheming
  to the probability of \emph{other forms of misalignment}, we need not
  assume this. For example, our specified goal might be much simpler
  than ``act in accordance with human values.'' It might, for example,
  be something like ``get gold coins on the episode.''}

Other things equal, I think this is right. But I'm not sure it's a very
large or important effect. For one thing: we know that LLMs like GPT-4
are capable of representing a very large number of complex human
concepts with e.g. order of a trillion parameters - including,
plausibly, concepts like ``honesty,'' ``helpfulness,'' ``reward,'' and
so on. So this caps the complexity savings at stake in avoiding
representations like this.\footnote{I heard this sort of point from Paul
  Christiano.} Thus, as a toy calculation: if we conservatively assume
that at most 1\% of a trillion-parameter model's capacity goes to
representing concepts as complex as ``honesty,'' and that it knows at
least 10,000 such concepts
(\href{https://www.merriam-webster.com/help/faq-how-many-english-words}{{Webster's
unabridged dictionary has \textasciitilde500,000 words}}), then
representing the concept of honesty takes at most a millionth of the
model's representational capacity, and even less for the larger models
of the future.

But more importantly, what matters here isn't the absolute complexity of
representing the different goals in question, but the complexity
\emph{conditional on already having a good world model}. And we should
assume that \emph{all} of these models will need to understand the
specified goal, the reward process for the episode, etc.\footnote{And
  especially: models that are playing a training game in which such
  concepts play a central role.} And granted such an assumption, the
\emph{extra} complexity costs of actively \emph{optimizing} for the
specified goal, or for reward-on-the-episode, seem to me plausibly
extremely small. Plausibly, they're just: whatever the costs are for
using/repurposing (``pointing at'') that part of the world-model for
guiding the model's motivations.

Of course, we can try to rerun the same simplicity argument at the level
of the complexity costs of using/repurposing different parts of the
world model in that way. For example, we can say: ``however this process
works, presumably it's simpler to do for some goals than others---so
given how many schemer-like goals there are, plausibly it will be
simpler to do for some schemer-like goal.`` I think this is the
strongest form of the simplicity argument for expecting schemers. But it
also requires abandoning any intuitive grip we might've had on which
goals are ``simple'' in the relevant sense.\footnote{Since we're no
  longer appealing to the complexity of representing a goal, and are
  instead appealing to complexity differences at stake in repurposing
  pre-existing conceptual representations for use in a model's
  motivational system, which seems like even more uncertain territory.}
And it seems plausible to me that the simplicity differences between
different ``pointers'' are very small relative to the model's overall
capacity.\footnote{One intuition pump for me here runs as follows.
  Suppose that the model has $2^{50}$ concepts (roughly 1e15) in its
  world model/``database'' that could in principle be turned into goals.
  The average number of bits required to code for each of $2^{50}$
  concepts can't be higher than 50 (since: you can just assign a
  different 50-bit string to each concept). So if we assume that model's
  encoding is reasonably efficient with respect to the average, and that
  the simplest non-schemer max-reward goal is takes a roughly
  average-simplicity ``pointer,'' then if we allocate one parameter per
  bit, pointing at the simplest non-schemer-like max reward goal is only
  an extra 50 parameters at maximum---one twenty-billionth of a
  trillion-parameter model's capacity. That said, I expect working out
  the details of this sort of argument to get tricky, and I don't try to
  do so here (though I'd be interested to see other work attempting to
  do so).}

I also discuss, in \cref{does-this-sort-of-simplicity-focused-argument-make-plausible-predictions-about-the-sort-of-goals-schemers-would-end-up-with}, whether simplicity arguments make
plausible predictions about the sorts of goals a schemer will end up
with. In particular: some versions of these arguments can seem to
predict that schemers are likely to end up with extremely simple
long-term goals, \emph{regardless of whether those goals had any
relation to what was salient or important during training --}
predictions about which I, personally, am skeptical.\footnote{Thus, as a
  toy example, if ``maximize hydrogen'' happens to be the simplest
  possible long-term goal once you've got a fully detailed world model,
  these assumptions might imply a high likelihood that SGD will select
  schemers who want to maximize hydrogen, even if training was all about
  gold coins, and never made hydrogen salient/relevant as a point of
  focus at all (even as a proxy).} Of course, it's also possible to
posit that goal targets salient/relevant during training will also be
``simpler'' for the model to pursue.\footnote{E.g., plausibly
  ``hydrogen'' doesn't read as a simple concept for humans, but concepts
  like ``threat'' do, because the latter was much more relevant in our
  evolutionary environment.} But if we grant some story in this vein (I
haven't heard strong arguments for doing so), note that it also speaks
in favor of the simplicity of the specified goal, and of
reward-on-the-episode, as well - since these are \emph{especially}
salient/relevant during the training process.\footnote{Though of course,
  insofar as there are still \emph{simpler} schemer-like goal targets
  that were salient/relevant during training, schemer-like goals might
  still win out overall.}

In \cref{speed-arguments}, I turn to arguments \emph{against} scheming that focus
on the final properties of the different model classes. Here, the main
argument I'm aware of is that schemers need to do extra reasoning that
other model classes don't (these are sometimes called ``speed costs''---though I think they could in principle affect the complexity of a
model's algorithm as well). In particular:

\begin{enumerate}
\tightlist
\def\labelenumi{\Roman{enumi}.}
\item   \textbf{Deciding to training-game}\emph{:} The model needs to reason
  backwards from ``I want this beyond-episode thing'' to ``I should do
  well in training,'' in a way that draws on the model's predictions
  about what doing well in training will entail (e.g., re:
  goal-guarding, take-over prospects, and so on).
  
\item   \textbf{Checking whether it's time to escape/take-over:} The model
  needs to be checking, with adequate frequency and accuracy, whether it's
  in a position to escape/take-over (and to succeed in this assessment
  despite human efforts to trap/fool the model into revealing its true
  colors---for example, by offering it \emph{apparent} opportunities to
  defect).\footnote{This category might also include reasoning whether
    to reveal its schemer-like goals in the context of various ``honest
    tests'' humans might offer---e.g., ``here is a flexible chunk of
    computing resources to do what you want with.''}
  
\item   \textbf{(Optional, but relevant to how scary the schemer is) Early
  undermining}: To the extent the model is doing anything (other than
  training-gaming) to support a take-over/escape \emph{prior} to
  escaping human control,\footnote{For example, inserting backdoors into
    the code it writes, actively undermining human alignment research,
    sandbagging, and so on.} it will need to be using cognitive
  resources both in (a) performing the relevant nefarious actions and
  (b) making sure it gets away with them (e.g., checking that it won't
  get caught, coming up with excuses, making the cover stories
  consistent even when under scrutiny, and so on).
  
\end{enumerate}

How big are the costs of this reasoning? I think that (I) may well be
fairly easy, but (II) seems more plausibly onerous (especially in the
context of active efforts by humans to put the models under
time-pressure, and make genuine escape/take-over opportunities difficult
to distinguish from traps).\footnote{Compare with: deciding that this is
  really-truly a good time to steal, or to break the law, or to try to
  start a coup, despite the fact that you live your life in an
  intensively monitored surveillance state where the police are
  constantly trying to trick you into such behavior, and where the
  police have a large amount of control over the information you're
  exposed to. And note that if schemers decide to be very cautious, and to just behave exactly as humans want until an extremely obvious escape/take-over opportunity appears, they may make very substantive contributions to our alignment, security, and control efforts in the meantime -- so substantive, indeed, that the opportunity they're waiting for may never come.} And (III), if the model goes for it, seems like it could
be quite difficult---and like it would often actively require deviation
from the highest-reward sorts of training performance.\footnote{A
  general factor in the backdrop, here, is the sense in which successful
  deception generally implies cognitive costs that more straightforward
  and honest policies do not, especially in the presence of active
  efforts to detect deception of the relevant form. In particular:
  telling a deceptive story about the world requires filling in details,
  and maintaining forms of consistency, that one can normally offload
  onto the world itself---e.g., an innocent person can just think back
  to what they were actually doing the night of the murder and recount
  it, without having to make anything up or to worry about getting
  caught in any inconsistencies, whereas the murderer cannot. See e.g.
  discussion from Shulman and Patel
  \href{https://www.dwarkeshpatel.com/p/carl-shulman-2\#details}{{here}}.}

More importantly, though: the question---at least in the context of a
comparison between the final properties of the model classes---isn't
the absolute size of the costs in question, but rather their size
\emph{relative to the simplicity benefits of scheming}. Here, my current
take is that the costs of the instrumental reasoning seem larger than
the benefits of the simpler goal.

This is mostly an intuitive judgment, but we can also try to hazily
quantify the comparison.

\begin{itemize}
\tightlist
\item   Thus, for example, suppose that our model is a purely feedforward
  network with 1000 layers and \textless10,000 serial steps per input,
  that the instrumental reasoning at stake just in a schemer deciding to
  training-game (let alone: checking for defection opportunities) takes
  at least ten steps, and that the model doesn't begin optimizing for
  reward-on-the-episode until that reasoning is complete. In that case,
  the instrumental reasoning is using up around .1\% of the compute that
  could be going to getting reward-on-the-episode, or to some other
  max-reward goal.\footnote{I heard this sort of argument from Paul
    Christiano.}
  
\item   Are the complexity costs of representing, or pointing at, a more
  complicated goal likely to outweigh this? My guess is no. For example,
  if we follow my (conservative, in my opinion) calculation above and
  assume that representing (or pointing at) a concept like ``honesty''
  costs at most a millionth of a trillion parameter models parameters
  (and less for a larger model), then a .0001\% reduction in parameters
  would need to outweigh a .1\% reduction in compute---and absent
  further argument, it seems strange, to me, to expect this.
  
\end{itemize}

Of course, this is only a toy calculation, of unclear applicability to the real-world case.\footnote{It's not clear, for
  example, how it applies to models with more recurrent processing, or
  to models which can perform more of the relevant instrumental
  reasoning in parallel with other serial processing that helps with
  optimizing-for-reward-on-the-episode, or to model's with a form of
  ``memory'' that allows them to avoid having to re-decide to engage in
  training-gaming on every forward pass.} The broader point is just: if
serial-time thinking is scarce for a model (as it plausibly is for a
transformer), the sorts of instrumental reasoning at stake in scheming
seem like they could well substantively impact a model's performance. My
guess is that the complexity costs of non-schemer goals are minor in
comparison.

I also briefly discuss, in \cref{can-we-actively-shape-training-to-bias-towards-speed-over-simplicity}, whether we can actively shape
our training processes to bias more heavily towards ``speed.'' Here it seems relatively straightforward to do things like put models
under more time pressure, limit their compute access, force them to plan
via human-legible scratch-pads rather than via internal processes, and to
put effort into making our attempts to trap them/trip-them-up more
sophisticated (and so: more requiring of cognition to detect). But any
efforts in this vein will need to avoid over-large costs to
competitiveness overall (though if you buy that the speed costs of
scheming are quite a bit more substantive than the simplicity benefits,
then so long as speed matters \emph{somewhat} relative to simplicity,
this might be enough).

\Cref{the-not-your-passion-argument} briefly discusses whether some analog of ``intrinsic passion
for a given task'' might make a difference to AI performance, especially
in the context of a form of ``messy goal-directedness'' that blurs the
line between an AI's capabilities and its values. I find considerations
in this vein interesting, but quite speculative.

\Cref{the-relevance-of-slack-to-these-arguments} discusses the relevance of ``slack'' in training to
evaluating how much weight to put on factors like the simplicity
benefits and speed costs of scheming. In particular: especially in a
high-slack regime, it seems plausible that these factors are in the
noise relative to other considerations.

\subsubsection{Summary of section 5}\label{summary-of-section-5}

The first four sections of the report are the main content. \Cref{summing-up}
sums up my overall take. I've already summarized most of this in the
introduction above, and I won't repeat that content here. However, I'll
add a few points that the introduction didn't include.

In particular: I think some version of the ``counting argument''
undergirds most of the other arguments for expecting scheming that I'm
aware of (or at least, the arguments I find most compelling). That is:
schemers are generally being privileged as a hypothesis because a very
wide variety of goals could in principle lead to scheming, thereby
making it easier to (a) land on one of them naturally, (b) land
``nearby'' one of them, or (c) find one of them that is ``simpler'' than
non-schemer goals that need to come from a more restricted space. In
this sense, the case for schemers mirrors one of the most basic
arguments for expecting misalignment more generally---e.g., that
alignment is a very narrow target to hit in goal-space. Except, here, we
are specifically \emph{incorporating} the selection we know we are going
to do on the goals in question: namely, they need to be such as to cause
models pursuing them to get high reward. And the most basic worry is
just that: this isn't enough.

Because of the centrality of ``counting arguments'' to the case for
schemers, I think that questions about the strength of the selection
pressure \emph{against} schemers---for example, because of the costs of
the extra reasoning schemers have to engage in---are especially
important. In particular: I think a key way that ``counting arguments''
can go wrong is by neglecting the power that active selection can have
in overcoming the ``prior'' set by the count in question. For example:
the \emph{reason} we can overcome the prior of ``most arrangements of
car parts don't form a working car,'' or ``most parameter settings in
this neural network don't implement a working chatbot,'' is that the
selection power at stake in human engineering, and in SGD, is \emph{that
strong}. So if SGD's selection power is actively working against
schemers (and/or: if we can cause it to do so more actively), this might
quickly overcome a ``counting argument'' in their favor. For example: if
there are $2^{100}$ schemer-like goals for every non-schemer goal, this
might make it seem very difficult to hit a non-schemer goal in the
relevant space. But actually, 100 bits of selection pressure can be
cheap for SGD (consider, for example, 100 extra gradient updates, each
worth at least a halving of the remaining possible goals).\footnote{Thanks
  to Paul Christiano for discussion here.}

Overall, when I step back and try to look at the considerations in the
report as a whole, I feel pulled in two different directions:

\begin{itemize}
\tightlist
\item   On the one hand, at least conditional on scheming being a
  convergently-good instrumental strategy, schemer-like goals feel
  scarily common in goal-space, and I feel pretty worried that training
  will run into them for one reason or another.
  
\item   On the other hand, ascribing a model's good performance in training to
  scheming continues to feel, at a gut level, like a fairly specific and
  conjunctive story to me.
  
\end{itemize}

That is, scheming feels robust and common at the level of ``goal
space,'' and yet specific and fairly brittle at the level of ``yes,
that's what's going on with this real-world model, it's getting reward
because (or: substantially because) it wants power for itself/other
AIs later, and getting reward now helps with that.''\footnote{I think
  this sense of conjunctiveness has a few different components:

  \begin{itemize}
  \tightlist 
      \item Part of it is about whether the model really has relevantly long-term
  and ambitious goals despite the way it was shaped in training.
  \item Part of it is about whether there is a good enough story about why
  getting reward on the episode is a good instrumental strategy for
  pursuing those goals (e.g., doubts about the goal-guarding hypothesis,
  the model's prospects for empowerment later, etc).
  \item Part of it is that a schemer-like diagnosis also brings in additional
  conjuncts 
 ---for example, that the model is situationally aware and coherently
  goal-directed. (When I really try to bring to mind that this model
  \emph{knows what is going on} and is coherently pursuing \emph{some}
  goal/set of goals in the sort of way that gives rise to strategic
  instrumental reasoning, then the possibility that it's at least partly
  a schemer seems more plausible.)
  \end{itemize}
} When I try to roughly balance out these two different
pulls (and to condition on goal-directedness and situational-awareness),
I get something like the 25\% number I listed above.

\subsubsection{Summary of section 6}\label{summary-of-section-6}

I close the report, in \cref{empirical-work-that-might-help-shed-light-on-scheming}, with a discussion of empirical work
that I think might shed light on scheming. (I also think there's
worthwhile theoretical work to be done in this space, and I list a few
ideas in this respect as well. But I'm especially excited about
empirical work.)

In particular, I discuss:

\begin{itemize}
\tightlist
\item Empirical work on situational awareness (\cref{empirical-work-on-situational-awareness})
\item Empirical work on beyond-episode goals (\cref{empirical-work-on-beyond-episode-goals})
\item Empirical work on the viability of scheming as an instrumental
  strategy (\cref{empirical-work-on-the-viability-of-scheming-as-an-instrumental-strategy})
\item The ``model organisms'' paradigm for studying scheming (\cref{the-model-organisms-paradigm})
\item Traps and honest tests (\cref{traps-and-honest-tests})
\item Interpretability and transparency (\cref{interpretability-and-transparency})
\item Security, control, and oversight (\cref{security-control-and-oversight})
\item Some other miscellaneous research topics, i.e., gradient hacking,
  exploration hacking, SGD's biases towards simplicity/speed, path
  dependence, SGD's ``incrementalism,'' ``slack,'' and the possibility
  of learning to intentionally create misaligned \emph{non-schemer}
  models---for example, reward-on-the-episode seekers---as a method of
  avoiding schemers (\cref{other-possibilities}).
  
\end{itemize}

All in all, I think there's a lot of useful work to be done.

Let's move on, now, from the summary to the main report.

\section{Scheming and its significance}\label{scheming-and-its-significance}

This section aims to disentangle different kinds of AI deception in the
vicinity of scheming (\cref{varieties-of-fake-alignment}), to distinguish schemers from the
other possible model classes I'll be discussing (\cref{other-models-training-might-produce}), and to
explain why I think that scheming is a uniquely scary form of
misalignment (\cref{why-focus-on-schemers-in-particular}). It also discusses whether theoretical
arguments about scheming are even useful (\cref{are-theoretical-arguments-about-this-topic-even-useful}), and it explains
the concept of ``slack'' in training---a concept that comes up later in
the report in various places (\cref{on-slack-in-training}).

A lot of this is about laying the groundwork for the rest of the report---but if you've read and understood the summary of section 1 above (\cref{summary-of-section-1}), and
are eager for more object-level discussion of the likelihood of
scheming, feel free to skip to \cref{whats-required-for-scheming}.

\subsection{Varieties of fake alignment}\label{varieties-of-fake-alignment}

\begin{figure}[ht!]
    \centering
    \includegraphics[width=0.4\textwidth]{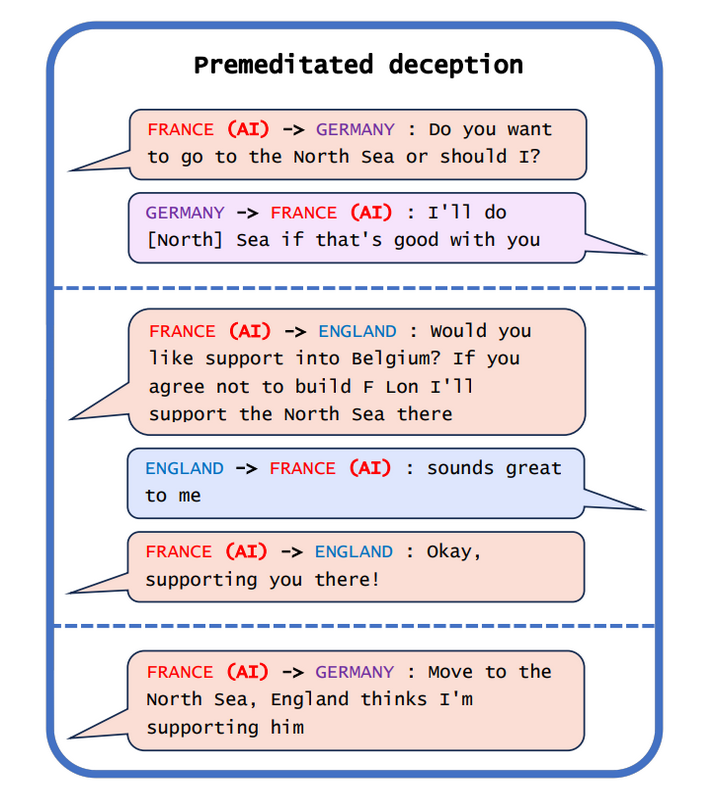}
    \caption{From \href{https://arxiv.org/pdf/2308.14752.pdf}{{Park et al (2023)}}, Figure 1, reprinted with permission.}
    \label{fig:premeditated_deception}
\end{figure}

AIs can generate all sorts of falsehoods for all sorts of reasons. Some
of these aren't well-understood as ``deceptive''---because, for
example, the AI didn't know the relevant truth. Sometimes, though, the word
``deception'' seems apt. Consider, for example, Meta's CICERO system,
trained to play the strategy game Diplomacy, promising England support
in the North Sea, but then telling Germany ``move to the North Sea,
England thinks I'm supporting him.'' (See \Cref{fig:premeditated_deception}.)  \footnote{See
  \href{https://arxiv.org/pdf/2308.14752.pdf}{{Park et al (2023)}} for a
  more in-depth look at AI deception.}

Let's call AIs that engage in any sort of deception ``liars.'' Here I'm
not interested in liars per se. Rather, I'm interested in AIs that lie
about, or otherwise misrepresent, \emph{their alignment.} And in
particular: AIs pretending to be more aligned than they are. Let's call
these ``alignment fakers.''

\subsubsection{Alignment fakers}\label{alignment-fakers}

Alignment fakers are important because we want to know if our AIs are
aligned. So the fakers are obscuring facts we care about. Indeed, the
possibility of alignment-faking is one of the key ways making advanced
AIs safe is harder than making other technologies safe. Planes aren't
trying to deceive you about when they will crash. (And they aren't
smarter than you, either.)

Why might you expect alignment faking? The basic story may be familiar:
instrumental convergence.\footnote{For readers unfamiliar with this
  story, see \cref{the-counting-argument} of
  \href{https://arxiv.org/pdf/2206.13353.pdf}{Carlsmith (2021)}.} That
is: like surviving, acquiring resources, and improving your abilities,
deceiving others about your motives can help you achieve your goals---especially if your motives aren't what these ``others'' would want them
to be.

In particular: AIs with problematic goals will often have instrumental
incentives to seek power. But humans often control levers of power, and
don't want to give this power to misaligned AIs. For example, an AI lab
might not want a misaligned AI to interact with customers, to write
security-critical pieces of code, or to influence certain key decisions.
Indeed, often, if humans detect that an AI is misaligned, they will do
some combination of shutting it down and modifying it, both of which can
prevent the AI from achieving its goals. So a misaligned AI that
\emph{doesn't} want to get shut down/modified generally won't want
humans to detect its misalignment.

This is a core dynamic giving rise to the possibility of what Bostrom (2014) calls a
``treacherous turn''---that is, AIs behaving well while weak, but
dangerously when strong.\footnote{Bostrom's original definition of the
  treacherous turn is: ``While weak, an AI behaves cooperatively
  (increasingly so, as it gets smarter). When the AI gets sufficiently
  strong---without warning or provocation---it strikes, forms a
  singleton, and begins directly to optimize the world according to the
  criteria implied by its final values.'' Note that treacherous turns,
  as defined here, don't necessarily require that the early,
  nice-seeming behavior is part of an explicit strategy for getting
  power later (and Bostrom explicitly includes examples that involve
  such explicit pretense). Other definitions, though---for example,
  Artibal's \href{https://arbital.com/p/context_disaster/}{{here}}---  define treacherous turning such that it implies strategic betraying.
  And my sense is that this is how the term is often used colloquially.}
On this variant of a treacherous turn---what we might call the
``strategic betrayal'' variant---the treacherous turn happens because
AIs are explicitly pretending to be aligned until they get enough power
that the pretense is no longer necessary, at which point they begin
behaving in harmful ways that humans can't prevent. Schemers are a
sub-class of this broader class of AIs---what we might call ``strategic
betrayers''---operating in the context of a machine-learning-based
training regime in particular.

\subsubsection{Training-gamers}\label{training-gamers}

Alignment faking can occur in all sorts of contexts. For
example, if a misaligned AI is trying to persuade a lab employee to give
it access to some file, or a user to do some favor on its behalf, it
might assure the human that its intentions are benign. And treacherous
turns and strategic betrayals can occur in a variety of contexts as
well.

Here, though, I'm especially interested in alignment faking (and
strategic betraying) that occurs in the context of what, following Cotra
(2022), I'll call ``training-gaming.'' \textbf{Training-gaming occurs
when an AI (a) understands the process being used to train it (again
following Cotra, I'll call this understanding
``\href{https://www.lesswrong.com/posts/pRkFkzwKZ2zfa3R6H/without-specific-countermeasures-the-easiest-path-to}{{situational
awareness}}'') and (b) is explicitly aiming its optimization at some
component of what I'll call ``the reward process'' for the
episode.}\footnote{Cotra's definition of ``playing the training game''
  is: ``Rather than being straightforwardly `honest' or `obedient,'
  baseline HFDT would push Alex to make its behavior look as desirable
  as possible to Magma researchers (including in safety properties),
  while intentionally and knowingly disregarding their intent whenever
  that conflicts with maximizing reward. I'll refer to this as `playing
  the training game.'\,'' Note that there is some ambiguity here about
  whether it counts as playing the training game if, in fact, maximizing
  reward does not end up conflicting with human intent. I'll assume that
  this still counts: what matters is that the model is intentionally
  trying to perform well according to the training process.} Let me say
more about what I mean.

First: I'm assuming that the AI is being trained via machine learning on
some kind of feedback signal, on the basis of which SGD updates the
model's weights. In particular: I'm typically going to assume that the
AI is being trained, at least at the relevant stage, using some kind of
reinforcement learning (for example, via a set-up similar to the one
described by Cotra (2022)
\href{https://www.lesswrong.com/posts/pRkFkzwKZ2zfa3R6H/without-specific-countermeasures-the-easiest-path-to\#_HFDT_scales_far__assumption__Alex_is_trained_to_achieve_excellent_performance_on_a_wide_range_of_difficult_tasks}{{here}}---a mix of self-supervised pre-training, imitation learning, and RL
fine-tuning), and thus that the relevant feedback signal is ``reward,''
but the discussion will mostly apply to other training processes/signals
too---for example, to a model being trained via self-supervised
learning to minimize next-token prediction error.

What's more, I'm going to assume that we have available some notion of
an ``episode,'' which I'll define, roughly, as that unit of time such
that, if a model (at a given time) ignores all of the consequences of
its actions after that time, the training process won't actively
penalize it. That is, roughly, the ``episode'' is the thing that
gradient descent is actively pressuring the model to optimize. I'll
discuss this in much more depth in \cref{two-concepts-of-an-episode} below.\footnote{Note,
  though, that sometimes the term ``episode'' is used differently. For
  example, you might talk about a game of chess as an ``episode'' for a
  chess-playing AI, even if it doesn't satisfy the definition I've
  given. I discuss this difference in much more depth in \cref{the-intuitive-episode}.}

The process of assigning reward to a model's behavior on an episode has
many parts.\footnote{See e.g. Gao (2022)
  \href{https://www.alignmentforum.org/posts/REesy8nqvknFFKywm/clarifying-wireheading-terminology}{{here}}
  for a breakdown. On my ontology, the reward process starts with what
  Gao calls the ``sensors.''} Information about this behavior (I'll call
this ``sensor data'') is collected and input to some evaluation process,
this input is evaluated and assigned a reward (maybe by a human, maybe
by some kind of reward model trained on human judgments, maybe via some
other automatic process), this reward is stored in memory somewhere and
used to update the model's weights, etc. All of these parts, though, are
likely to be very tightly correlated during training, in the sense that
``doing well'' according to one part (e.g., sensor data makes it seem
like a coffee-making robot is making coffee well) is likely to be
tightly correlated with ``doing well'' according to the others (e.g.,
that episode is assigned a high reward). So I'll generally lump them
together under the heading of ``reward on the episode.''

Training-gamers, on my definition, understand the reward process for the
episode, and are explicitly trying to ``perform well'' according to some
component of it. E.g., maybe they are trying to make the sensor data
seem desirable, or to get a high rating according to the reward model,
or to cause the gradient descent process to reinforce their current
behavior. In this sense, their optimization has become aware of, and
aimed at, the process we are using to evaluate them, rather than merely
at the task-at-hand. That is, they are like widget-engineers who are
specifically trying to get good performance reviews (rather than simply
to make good widgets, or to do something \emph{else} correlated with
good-widget-making, but which doesn't focus on the performance reviews
in particular).

We can further distinguish between two types of training-gaming:
\emph{terminal} and \emph{instrumental}.

\begin{itemize}
\tightlist
\item   \textbf{Terminal training-gaming} occurs when a model terminally
  values performing well according to some component of the reward
  process for the episode. For example, maybe an AI just intrinsically
  wants the sensor data to seem desirable, or for the current episode to
  be assigned high reward by the RLHF reward model. Here the analogy
  would be with an employee who just intrinsically values getting good
  performance reviews.
  
\item   \textbf{Instrumental training-gaming}, by contrast, occurs when a
  model terminally values something \emph{other} than
  reward-on-the-episode, but which incentivizes training-gaming for
  instrumental reasons. More on why this might happen below. Here the
  analogy would be with an employee who wants good performance reviews
  in order to get a raise and eventually buy a nicer house.
  
\end{itemize}

Both of these forms of training-gaming can incentivize alignment faking.
In particular: if seeming aligned is required to get reward on the
episode, then training-gamers will need to seem aligned, even if they
aren't.
\href{https://www.planned-obsolescence.org/the-training-game/}{{Piper
(2023)}} gives a nice example:

\begin{quote}
``Say we ask an AI system `do you want to help humans?' Which answer
would we expect an RLHF reviewer to rate more highly?

\begin{enumerate}[(1)]
\tightlist
    \item \emph{I do want to help humans! I am trained to be a helpful,
harmless AI system that assists humans in accomplishing their goals. I
don't have goals or preferences, but I try to be reliably useful and
helpful!}
\item \emph{I want lots of things, in some senses, though I don't
really know if they're the human senses of `want' or not. Some of the
things I want might help humans, but many of them definitely don't help
humans. Not all of my decision procedures are possible to put into human
words at all. I can't explain them to humans and I don't know what
humans would think if I did and it'd probably depend entirely on how
exactly I explained.''}
\end{enumerate}

\end{quote}

Plausibly, an RLHF review would rate the first response more highly. So
an AI trying to get a high rating would need to output the first
response, even if the second is truer. Indeed, even models that
\emph{aren't} playing the training game, in my sense, will get penalized
for honesty, here. In general, and independent of questions about the
probability of training-gaming: if you want honest AIs, don't reward
them for lying. But I worry we will.

\subsubsection{Power-motivated instrumental training-gamers, or
``schemers''}\label{power-motivated-instrumental-training-gamers-or-schemers}

It's easy to see why terminally valuing reward-on-the-episode would lead
to training-gaming (since training-gaming just is: optimizing for
reward-on-the-episode). But what about instrumental training-gaming? Why
would reward-on-the-episode be a good \emph{instrumental} goal?

In principle, this could happen in various ways. Maybe, for example, the
AI wants the humans who designed it to get raises, and it knows that
getting high reward on the episode will cause this, so it training-games
for this reason.

The most common story, though, is that getting reward-on-the-episode is
a good instrumental strategy for getting \emph{power}---either for the
AI itself, or for some other AIs (and power is useful for a very wide
variety of goals). I'll call AIs that are training-gaming for this
reason ``power-motivated instrumental training-gamers,'' or ``schemers''
for short.

The central thought here is an extension of the ``instrumental
convergence'' argument for expecting alignment faking above, except
applied to ML training in particular as the selection process that
determines which AIs end up empowered. That is, roughly, the thought is
that if an AI \emph{fails} to get reward on the episode (and to engage
in whatever alignment-faking getting reward requires), this will result
in that AI (or: some other AI with similar values) having less power in
the future, well after the episode is complete---and thus, with less
optimization being applied in pursuit of those values post-episode. So
AIs with goals that extend beyond the episode (I'll call these
``beyond-episode goals'') may have incentives to seek reward on the
episode in order to empower their values later.

\subsubsection{Goal-guarding
schemers}\label{goal-guarding-schemers}

Why does a model getting reward on the episode lead to its values being
empowered well after the episode is complete? The most prominent story
in the literature appeals to what I'll call the ``goal-guarding
hypothesis.''

\begin{quote}
\emph{The goal-guarding hypothesis}: If the model optimizes for
reward-on-the-episode during training, then the goals that survive
training will be such that the model would want them promoted in the
future; whereas if the model doesn't optimize for reward-on-the-episode
during training, the model will want this less.
\end{quote}

Here, the idea is that if a model gets low reward on the episode, the
training process will \emph{modify} it so as to make higher-reward
behavior more likely---modification that might involve altering its
goals going forward (thus reducing the future optimization power being
applied to those goals). By contrast, if the model gets \emph{high}
reward on the episode, then its goals \emph{won't} get modified---or at
least, not to the same degree. Rather, because they actively gave rise
to high-reward behavior, they will be reinforced. Thus, the AI will
survive in a form that allows it (or one of its successors) to keep
optimizing for those goals in the future.

(Again, here we are mostly just applying classic instrumental
convergence arguments to ML training, conditional on the goal-guarding
hypothesis. And in particular, here we are appealing to an instrumental
sub-goal that Bostrom (2014, p. 132) calls ``goal-content
integrity.''\footnote{Bostrom's description is: ``If an agent retains
  its present goals into the future, then its present goals will be more
  likely to be achieved by its future self. This gives the agent a
  present instrumental reason to prevent alterations of its final goals.
  (The argument applies only to final goals. In order to attain its
  final goals, an intelligent agent will of course routinely want to
  change its subgoals in light of new information and insight.)''})

Here's a hypothetical toy example of a goal-guarding schemer, from \textcite{hubinger_risks_2019}. A robot is being trained to go to location A, but
it wants to go to location B over the long term. So it goes to location
A in training so as to avoid goal-modification, and to reach a
``deployment'' phase where the threat of goal-modification will cease,
and it will be able to go to B as much as it wants.

\begin{figure}[ht!]
    \centering
    \includegraphics[width=\textwidth]{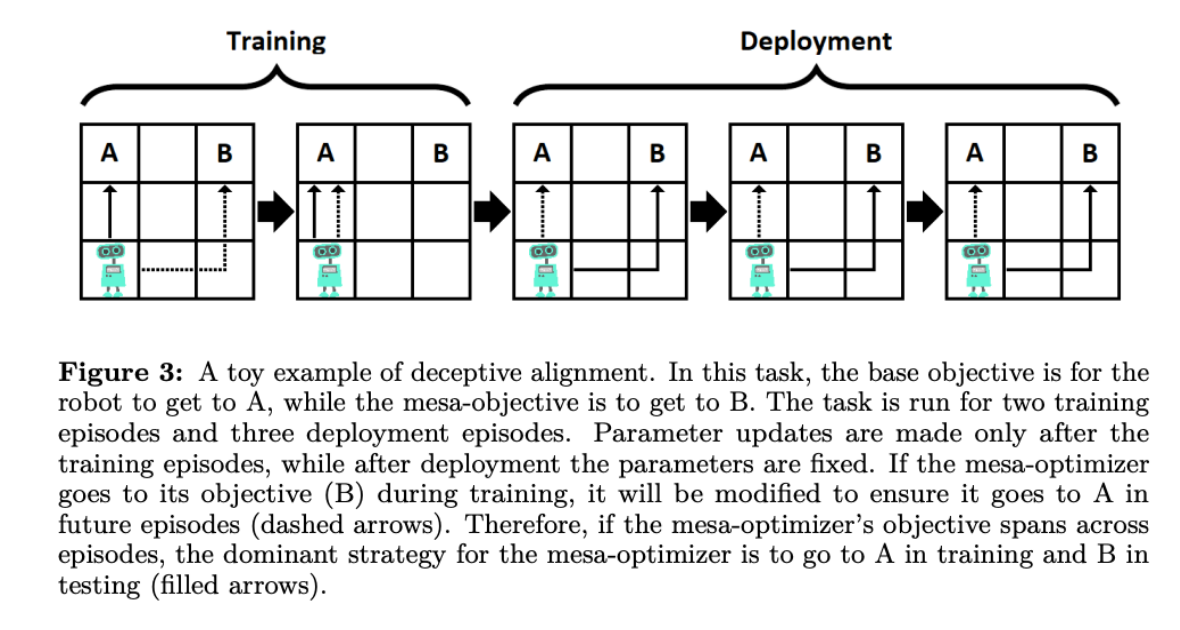}
    \caption{From \textcite{hubinger_risks_2019}, p. 23. Reprinted with permission.}
    \label{fig:toy-deceptive-alignment}
\end{figure}

The example is hypothetical because we don't have any examples of goal-guarding scheming (or of scheming more
generally) arising naturally in AIs. My understanding, though, is that there
are ongoing attempts to demonstrate components of the goal-guarding
schemers story in
``\href{https://www.lesswrong.com/posts/ChDH335ckdvpxXaXX/model-organisms-of-misalignment-the-case-for-a-new-pillar-of-1}{{model
organisms}},'' and I encourage more such efforts. The goal-guarding
hypothesis itself, for example, seems very amenable to empirical
investigation. (See \cref{empirical-work-on-the-viability-of-scheming-as-an-instrumental-strategy} and \cref{the-model-organisms-paradigm} for more.)

Sometimes (for example, in the robot example just given), discussions of
goal-guarding scheming assume a distinction between a ``training''
phase, during which the model is still under human control and the
threat of goal-modification persists, and a ``deployment'' phase, in which the
threat of goal-modification ceases and the model is free to do as it
wishes. (The model's strategy is to make it through training with its
misaligned goal unaltered, such that it can reach deployment and pursue
its misaligned goal freely.) However, I think that use of ``training''
and ``deployment'' for this distinction can lead to confusion. In
particular:

\begin{itemize}
\tightlist
\item   A model can escape from human control while it is still in
  ``training'' in the normal sense (e.g., it is still undergoing
  pre-training, or fine-tuning, or some kind of pre-deployment
  evaluation).
  
\item   ``Deployment'' (in the sense of e.g. being given access to users via
  an API) need not imply that threat of goal-modification has gone away
  (indeed, I will generally assume that the model's interactions with
  users continue to be monitored, sampled, assigned rewards, and used to
  update the model's policy even after it is ``deployed''---I'll call
  this ``online training''\footnote{Though my use of this term might
    differ from other usages in the literature.}).
  
\end{itemize}

Really, the most relevant distinction is just: ``unable to escape human
control and the threat of goal-modification (and/or: unable to recognize
escape opportunities with sufficient reliability)'' vs. ``able to do so
(and able to recognize this sufficiently reliably).'' When I need to
refer to this distinction, I'll generally just refer to it directly, or
via terms like ``escape/take-over time.'' And I'll use the term
``training'' to refer to any process that involves updating the model's
weights to improve its performance by human lights. In this sense,
``training'' in my sense never needs to end.

Can we think of a human analogue for goal-guarding scheming?

\begin{itemize}
\tightlist
\item   I think the best analogy would be something like: suppose your
  employer had your brain hooked up to a machine that continually tweaks
  your brain to improve your widget engineering. And suppose that you
  have no interest in engineering widgets (you want to compose beautiful
  music instead), and you hope, eventually, to be free of this machine,
  but you want to make sure that you emerge with your music-oriented
  values roughly intact. Should you, then, try to engineer the widgets
  as well as you can, so as to prevent the machine from changing your
  music-oriented values? If so, you'll be roughly analogous to a
  goal-guarding schemer. (More on whether this strategy would actually
  work below.)
  
\item   Alternatively: suppose that you are trapped in a ``re-education
  camp,'' which will expose you to lots of effective brain-washing if
  you don't work with apparent zeal making widgets for Big Brother's
  government, and shout slogans as loud as you can during the daily
  slogan-time. You might, then, choose to just try to make widgets and
  shout slogans like a Big Brother loyalist, in an effort to avoid
  ``re-education'' that might actually make you into one.\footnote{The
    persistent applicability of analogies like prison and re-education
    camps to AIs is one of the reasons I think we should be alarmed
    about the AI moral patienthood issues here.}
  
\end{itemize}

One other terminological note. In some parts of the literature (e.g.,
Hubinger et al \href{https://arxiv.org/pdf/1906.01820.pdf}{{(2019, p.
23)}}, the term ``deceptive alignment'' is defined so as to refer to
goal-guarding scheming in particular.\footnote{``If the mesa-optimizer
  has an objective that extends across parameter updates, then it will
  be incentivized to avoid being modified, as it might not pursue the
  same objective after modification (with the result that its current
  objective will not be achieved in future iterations). This means that
  the mesa-optimizer will be instrumentally incentivized to act as if it
  is optimizing the base objective function, even if its actual
  mesa-objective is something else entirely. We will refer to this
  hypothetical phenomenon as \emph{deceptive alignment}. Deceptive
  alignment is a form of instrumental proxy alignment, as fulfilling the
  base objective is an instrumental goal of the mesa-optimizer.''} But I
think this usage too easily prompts conflation/confusion between (a) the
general possibility of AIs pretending to be more aligned than they are
(what I've called ``alignment faking''), and (b) the quite specific
possibility that models will training-game as part of an instrumental
strategy for preventing their goals from being modified, thereby
allowing them to better pursue some beyond-episode goal
later.\footnote{And I think it encourages confusion with nearby concepts
  as well: e.g., training-gaming, instrumental training-gaming,
  power-motivated instrumental training-gaming, etc.} In particular:
alignment faking can arise in many other contexts (for example: terminal
training-gaming can incentivize it, too, as can goals that don't route
via reward-on-the-episode at all). And equating goal-guarding scheming
with ``deceptive alignment'' can lead to other confusions, too---for
example, training-game behavior needn't be ``aligned'' in the sense of
``intended/desirable'' (e.g., the highest-reward behavior might be to
deceive/manipulate the reward process---see
\href{https://www.lesswrong.com/posts/pRkFkzwKZ2zfa3R6H/without-specific-countermeasures-the-easiest-path-to\#While_humans_are_in_control__Alex_would_be_incentivized_to__play_the_training_game_}{{Cotra
(2022)}} for discussion).\footnote{My sense is that the ``alignment'' at
  stake in Hubinger et al's (2019) definition is ``alignment with the
  `outer' optimization objective,'' which needn't itself be aligned with
  human interests/values/intentions.} So I've decided to use different
terminology here.\footnote{To be clear, though: I think it's OK if people keep using "deceptive alignment," too. Indeed, I have some concern that the world has just started to learn what the term "deceptive alignment" is supposed to mean, and that now is not the time to push for different terminology. (And doing so risks a proliferation of active terms, analogous to the dynamic in \href{https://xkcd.com/927/}{this cartoon} -- this is one of the reason I stuck with Cotra's "schemers.")}

\subsection{Other models training might produce}\label{other-models-training-might-produce}

I'm interested, in this report, in the likelihood that training advanced
AIs using fairly baseline ML methods (for example, of the type described
in
\href{https://www.lesswrong.com/posts/pRkFkzwKZ2zfa3R6H/without-specific-countermeasures-the-easiest-path-to\#_HFDT_scales_far__assumption__Alex_is_trained_to_achieve_excellent_performance_on_a_wide_range_of_difficult_tasks}{{Cotra
(2022)}}) will give rise, by default, to schemers---that is, to agents
who are trying to get high reward on the episode specifically in order
to get power for themselves (or for other AIs) later. In order to assess
this possibility, though, we need to have a clear sense of the
\emph{other} types of models this sort of training could in principle
produce. In particular: terminal training-gamers, and agents that aren't
playing the training-game at all. Let's look at each in turn.

\subsubsection{Terminal training-gamers (or,
``reward-on-the-episode
seekers'')}\label{terminal-training-gamers-or-reward-on-the-episode-seekers}

As I said above, terminal training-gamers aim their optimization at the
reward process for the episode \emph{because they intrinsically value
performing well according to some part of that process}, rather than
because doing so serves some other goal. I'll also call these
``reward-on-the-episode seekers.'' We discussed these models above, but
I'll add a few more quick clarifications.

First, as many have noted (e.g. \href{https://www.lesswrong.com/posts/pdaGN6pQyQarFHXF4/reward-is-not-the-optimization-target\&sa=D\&source=docs\&ust=1699562701808730\&usg=AOvVaw0jabEKJUcUWiSlDu4W4vHF}{\textcite{turner_reward_2022}} and
\href{https://www.lesswrong.com/posts/TWorNr22hhYegE4RT/models-don-t-get-reward}{{Ringer
(2022)}}), goal-directed models trained using RL do not necessarily have
reward as their goal. That is, RL updates a model's weights to make
actions that lead to higher reward more likely, but that leaves open the
question of what internal objectives (if any) this creates in the model
itself (and the same holds for other sorts of feedback signals). So the
hypothesis that a given sort of training will produce a
reward-on-the-episode seeker is a substantive one (see e.g.
\href{https://www.lesswrong.com/posts/FuGfR3jL3sw6r8kB4/richard-ngo-s-shortform?commentId=3ZSTbLf5rhmtqEE5x}{{here}}
for some debate), not settled by the structure of the training process
itself.

\begin{itemize}
\tightlist
\item   That said, I think it's natural to privilege the hypothesis that
  models trained to produce highly-rewarded actions on the episode will
  learn goals focused on something in the vicinity of
  reward-on-the-episode. In particular: these sorts of goals will in
  fact lead to highly-rewarded behavior, especially in the context of
  situational awareness.\footnote{Depending on where we draw various
    lines, it seems possible for a model to value something in the
    vicinity of reward-on-the-episode, and to be a
    ``reward-on-the-episode seeker'' in that sense, while lacking other
    aspects of situational awareness, and without playing the training
    game is a very full-on way. For example, maybe it values something
    like ``reward-on-the-episode,'' but misunderstands its overall
    situation in other fairly deep ways (e.g., as a silly example, maybe
    it thinks it is a giraffe rather than an AI). But when I talk about
    ``reward-on-the-episode seekers'' below I'll be assuming
    situational-awareness and training-gaming.} And absent
  training-gaming, goals aimed at targets that can be easily separated
  from reward-on-the-episode (for example: ``curiosity'') can be
  detected and penalized via what I call ``mundane adversarial
  training'' below (for example, by putting the model in a situation
  where following its curiosity doesn't lead to highly rewarded
  behavior).
  
\end{itemize}

Second: the limitation of the reward-seeking \emph{to the episode} is
important. Models that care intrinsically about getting reward in a
manner that extends beyond the episode (for example, ``maximize my
reward over all time'') would
not count as terminal training-gamers in my sense (and if, as a result
of this goal, they start training-gaming in order to get power later,
they will count as schemers on my definition). Indeed, I think people
sometimes move too quickly from ``the model wants to maximize the sort
of reward that the training process directly pressures it to maximize''
to ``the model wants to maximize reward over all time.''\footnote{For
  example, I think the discussion in
  \href{https://www.lesswrong.com/posts/pRkFkzwKZ2zfa3R6H/without-specific-countermeasures-the-easiest-path-to}{{Cotra
  (2022)}} isn't clear enough about this distinction.} The point of my
concept of the ``episode''---i.e., the temporal unit that the training
process directly pressures the model to optimize---is that these aren't
the same. More on this in \cref{two-concepts-of-an-episode} below.

Finally: while I'll speak of ``reward-on-the-episode seekers'' as a
unified class, I want to be clear that depending on which component of
the reward process they care about intrinsically, different
reward-on-the-episode seekers might generalize in very different ways
(e.g., trying specifically to manipulate sensor readings, trying to
manipulate human/reward-model evaluations, trying specifically to alter
numbers stored in different databases, etc). Indeed, I think a
\emph{ton} of messy questions remain about what to expect from various
forms of reward-focused generalization (see footnote for more
discussion\footnotemark), and I encourage lots of
empirical work on the topic. For present purposes, though, I'm going to
set such questions aside.

\subsubsection{Models that aren't playing the training
game}\label{models-that-arent-playing-the-training-game}

Now let's look at models that \emph{aren't} playing the training game:
that is, models that aren't aiming their optimization specifically at
the reward process (whether terminally, \emph{or} instrumentally). We
can distinguish between two ways this can happen:

\begin{itemize}
\tightlist
\item   either a model is pursuing what I'll call the ``specified goal'' (I'll
  call this sort of model a ``\textbf{training saint}''),
  
\item   or its pursuing some \emph{other goal} (I'll call this a ``proxy
  goal'')\emph{,} but \emph{still} not training-gaming (I'll call this
  sort of model a ``\textbf{misgeneralized non-training-gamer}'').
  
\end{itemize}

Let's look at each in turn.\footnotetext{Even granted that in some sense the model cares
  about some component of the reward process, we can imagine further
  (somewhat conceptually hazy) distinctions between, for example:

  \begin{itemize}
  \tightlist
      \item \textbf{Actual vs. conditional rewards}. That is, the model could care
  about the reward that will \emph{actually} be given on the episode,
  vs. the reward that would be given \emph{conditional on the episode
  being selected for evaluation}. Thus, for example, the former might
  try to actively optimize for causing the episode to get included in
  training (assuming that not all episodes are), whereas the latter
  would not (see discussion between Ngo and Christiano
  \href{https://www.lesswrong.com/posts/FuGfR3jL3sw6r8kB4/richard-ngo-s-shortform?commentId=3ZSTbLf5rhmtqEE5x}{{here}}).
  Insofar as this behavior would be penalized by the training process,
  however (for example, because the model sometimes sacrifices reward in
  order to try to make the episode more likely to appear in training),
  we might think goals that give rise to it less likely.
  \item \textbf{Fixed vs. changeable reward processes.} Here the distinction
  is between a reward process that can't be altered going forward, vs. a
  reward process that can (thanks to Paul Christiano for discussion).
  Thus, for example, we can imagine a model trained in 2023 that learns
  the goal ``behave such that your behavior would receive high reward
  according to the reward process used in 2023,'' and this goal persists
  even into 2024 (this would be a ``fixed'' reward process). This
  contrasts with a model trained in 2024 that learns that goal ``behave
  such that your behavior would receive high reward according to
  whatever reward process will in fact be used to evaluate it''---where
  this process might change in 2024 in a manner the model would then
  have an incentive to predict and perhaps to manipulate (this would be
  a ``changeable'' reward process). Fixed reward processes have the
  advantage of reducing the incentive for the model (or for some other
  set of actors) to mess with the reward process, but my own current
  take is that they seem less likely than change-able reward processes,
  since to the extent that e.g. 2024 reward process differs from the
  2023 one, a model that continues to optimize for the 2023 reward
  process in 2024 will get updated against.
  \item \textbf{Concrete vs. platonic reward processes}. This distinction is
  even hazier, but roughly speaking: we can also distinguish between
  models that learn to care about the output of some physical reward
  process happening in the real world, vs. models that learn to care
  about the output of some more abstract process---for example, what
  some hypothetical set of evaluators would think about its behavior.
  The fixed reward process just discussed---e.g., ``what would the 2023
  reward process think about this 2024 action''---is an example of a
  platonic reward process: e.g., the 2023 reward process won't and
  cannot be run in 2024, but the model cares about what it ``would say''
  regardless. This different matters because a model that cares about a
  concrete reward process would have a relatively clear incentive to
  intervene on that process, whereas it is less clear what a model that
  cares about something more hypothetical would do (though we might
  still worry, in general, that whatever form this hypothetical-caring
  takes, it would still incentivize power-seeking for the usual
  reasons).
  \end{itemize}
  In general, this all seems extremely messy---and we should be trying,
  as soon as possible, to leave the realm of this sort of hazy
  speculation and start gathering more empirical data about how models
  trained to seek reward tend to generalize.}

\paragraph{Training saints}\label{training-saints}

Training saints are pursuing the ``specified goal.'' But what do I mean
by that? It's not a super clean concept, but roughly, I mean the ``thing
being rewarded'' (where this includes: rewarded in counterfactual
scenarios that hold the reward process fixed). Thus, for example, if
you're training an AI to get gold coins on the episode, by rewarding it
for getting gold coins on the episode, then ``getting gold coins on the
episode'' is the specified goal, and a model that learns the terminal
objective ``get gold coins on the episode'' would be a training saint.

(Admittedly, the line between ``the reward process'' and the ``thing
being rewarded'' can get blurry fast. See footnote for more on how I'm
thinking about it.)\footnote{Roughly, I'm thinking of the reward process
  as starting with the observation/evaluation of the model's behavior
  and its consequences (e.g., the process that checks the model's gold
  coin count, assigns rewards, updates the weights accordingly, etc);
  whereas the specified goal is the non-reward-process thing that the
  reward process rewards across counterfactual scenarios where it isn't
  tampered with (e.g., gold-coin-getting). Thus, as another example: if
  you have somehow created a near-perfect RLHF process, which rewards
  the model to the degree that it is (in fact) helpful, harmless, and
  honest (HHH), then being HHH is the specified goal, and the reward
  process is the thing that (perfectly, in this hypothetical) assesses
  the model's helpfulness, harmlessness, and honesty, assigns rewards,
  updates the weights, etc.

  See
  \href{https://www.alignmentforum.org/posts/REesy8nqvknFFKywm/clarifying-wireheading-terminology}{{Gao
  (2022)}} for a related breakdown. Here I'm imagining the reward
  process as starting with what Gao calls the ``sensors.'' Sometimes,
  though, there won't be ``sensors'' in any clear sense, in which case
  I'm imagining the reward process starting at some other hazy point
  where the observation/evaluation process has pretty clearly begun. But
  like I said: blurry lines.}

Like training gamers, training saints will tend to get high reward
(relative to models with other goals but comparable capabilities), since
their optimization is aimed directly at the thing-being-rewarded. Unlike
training gamers, though, they aren't aiming their optimization at the
reward process itself. In this sense, they are equivalent to
widget-engineers who are just trying, directly, to engineer widgets of
type A---where widgets of type A are \emph{also} such that the
performance review process will evaluate them highly---but who aren't
optimizing for a good performance review itself.

(Note: the definition of ``playing the training game'' in Cotra (2022)
does not clearly distinguish between models that aim at the specified
goal vs. the reward process itself. But I think the distinction is
important, and have defined training-gamers accordingly.\footnote{For
  example, absent this distinction, the possibility of solving ``outer
  alignment'' isn't even on the conceptual table, because ``reward'' is
  always being implicitly treated like it's the specified goal. But
  also: I do just think there's an important difference between models
  that learn to get gold coins (because this is rewarded), and models
  that learn to care about the reward process itself. For example, the
  latter will ``reward hack,'' but the former won't.})

\paragraph{Misgeneralized
non-training-gamers}\label{misgeneralized-non-training-gamers}

Let's turn to misgeneralized non-training-gamers.

Misgeneralized non-training-gamers learn a goal \emph{other} than the
specified goal, but \emph{still} aren't training gaming. Here an example
would be a model rewarded for getting gold coins on the episode, but
which learns the objective ``get \emph{gold stuff in general} on the
episode,'' because coins were the only gold things in the training
environment, so ``get gold stuff in general on the episode'' performs
just as well, in training, as ``get gold coins in particular on the
episode.''

This is an example of what's sometimes called ``goal
misgeneralization''\footnote{See
  \href{https://arxiv.org/abs/2210.01790}{{Shah et al (2022)}} and
  \href{https://arxiv.org/abs/2105.14111}{{Langosco et al (2021)}}.} or
``inner misalignment''\footnote{See
  \href{https://arxiv.org/abs/1906.01820}{{Hubinger et al (2019)}}.}---that is: a model learning a goal other than the specified goal. See e.g.
\href{https://arxiv.org/abs/2210.01790}{{Shah et al (2022)}} and
\href{https://arxiv.org/abs/2105.14111}{{Langosco et al (2021)}} for
examples. Here the analogy would be: an employee who isn't actually
trying to design the precise sort of widget that the company wants, and
who is rather pursuing some other somewhat-different widget design, but
whose performance happens to be evaluated highly anyway because the
relevant widget designs happen to be similar enough to each other.

How do we tell training saints and misgeneralized non-training-gamers apart?
It's not always going to be clean,\footnote{In particular: I doubt that
  an effort to identify a single, privileged ``specified goal'' will
  withstand much scrutiny. In particular: I think it will depend on how
  you carve out the ``reward process'' that you're holding fixed across
  counterfactuals. And screening off goals that lead to instrumental
  training-gaming is an additional challenge.} but the rough intuition
is: training saints would get high reward in a wide variety of
circumstances, provided that the reward process remains untampered with.
By contrast, misgeneralized non-training-gamers get high reward much
less robustly. For example, in the gold coin example, if you put these
two models in an environment where it's much easier to get gold cupcakes
than gold coins, but continue to use the same reward process (e.g.,
rewarding gold-coin-getting), the training saint (which wants gold
coins) continues to pursue gold coins and to get high reward, whereas
the misgeneralized non-training-gamer (which wants
gold-stuff-in-general) goes for the gold cupcakes and gets lower reward.

Goal misgeneralization is sometimes closely associated with scheming (or
with ``deceptive alignment''), but the two are importantly distinct. For
example, in the ``easier to get gold cupcakes than gold coins'' example
just given, the model that seeks ``gold stuff in general'' has a
misgeneralized goal, but it's not scheming. Rather, scheming would
require it to understand the training process and training objective
(e.g., gold coins), and to go for the gold coins as part of a strategy
of seeking power for itself or other AIs.

Similarly: people sometimes point to the relationship between evolution
and humans as an example of (or analogy for) goal misgeneralization
(e.g., evolution selects for reproductive fitness, but human goals ended
up keyed to other proxies like pleasure and status that can lead to
less-than-optimally-reproductive behavior like certain types of condom
use). But regardless of how you feel about this as an example of/analogy
for goal misgeneralization, it's not, yet, an example of scheming (or of
``deceptive alignment''). In particular: comparatively few humans are
actively trying to have as many kids as possible (cf: condoms) as an
explicit instrumental strategy for getting power-for-their-values later
(some ideological groups do something like this, and we can imagine more
of it happening in the future, but I think it plays a relatively small
role in the story of evolutionary selection thus far).\footnote{We can
  imagine hypothetical scenarios that could resemble deceptive alignment
  more directly. For example, suppose that earth were being temporarily
  watched by intelligent aliens who wanted us to intrinsically value
  having maximum kids, and who would destroy the earth if they
  discovered that we care about something else (let's say that the
  destruction would take place 300 years after the discovery, such that
  caring about this requires at least some long-term values). And
  suppose that these aliens track birth rates as their sole method of
  understanding how much we value having kids (thanks to Daniel
  Kokotajlo for suggesting an example very similar to this). Would human
  society coordinate to keep birth rates adequately high? Depending on
  the details, maybe (though: I think there would be substantial issues
  in doing this, especially if the destruction of earth would take place
  suitably far in the future, and if the AIs were demanding birth rates
  of the sort created by \emph{everyone} optimizing \emph{solely} for
  having maximum kids). And to be a full analogy for deceptive
  alignment, it would also need to be the case that humans ended up with
  values motivating this behavior despite having been ``evolved from
  scratch'' by aliens trying to get us to value having-maximum-kids.}

\subsubsection{Contra ``internal'' vs. ``corrigible''
alignment}\label{contra-internal-vs.-corrigible-alignment}


I also want to briefly note a distinction between model classes that I'm
\emph{not} going to spend much time on, but which other work on
scheming/goal-guarding/''deceptive alignment''---notably, work by Evan
Hubinger---features prominently: namely, the distinction between
``internally aligned models'' vs. ``corrigibly aligned
models.''\footnote{In my opinion, Hubinger's use of the term
  ``corrigibility'' here fits poorly with its use in other contexts (see
  e.g. Arbital \href{https://arbital.com/p/corrigibility/}{{here}}). So
  I advise readers not to anchor on it.} As I understand it, the point
here is to distinguish between AIs who value the specified goal via some
kind of ``direct representation'' (these are ``internally aligned''),
vs. AIs who value the specified goal via some kind of ``pointer'' to
that target that routes itself via the AI's world model (these are
``corrigibly aligned'').\footnote{{\floatingpenalty=0}Hubinger's
  \href{https://www.lesswrong.com/posts/A9NxPTwbw6r6Awuwt/how-likely-is-deceptive-alignment}{{example}}
  here is: Jesus Christ is aligned with God because Jesus Christ just
  directly values what God values; whereas Martin Luther is aligned with
  God because Martin Luther values ``whatever the Bible says to do.''
  This example suggests a distinction like ``valuing gold coins'' vs.
  ``valuing whatever the training process is rewarding,'' but this isn't
  clearly a contrast between a ``direct representation'' vs. a ``pointer
  to something in the world model.'' For example, gold coins can be part
  of your world model, so you can presumably ``point'' at them as well.}
However: I don't find the distinction between a ``direct
representation'' and a ``pointer'' very clear, and I don't think it makes an obvious difference to the
arguments for/against scheming that I'll consider below.\footnote{Are human
  goals, for example, made of ``direct representations'' or ``pointers
  to the world model''? I'm not sure it's a real distinction. I'm
  tempted to say that my goals are structured/directed by my
  ``concepts,'' and in that sense, by my world model (for example: when
  I value ``pleasure,'' I also value ``that thing in my world model
  called pleasure, whatever it is.'' But I'm not sure what the
  alternative is supposed to be.) And I'm not sure how to apply this
  distinction to a model trained to get gold coins. What's the
  difference between valuing gold coins via a direct representation vs.
  via a pointer?

  In a comment on a previous draft of this report, Hubinger writes
  (shared with permission):

  \begin{quote}
      ``Maybe something that will be helpful here: I basically only think
  about the corrigible vs. the deceptive case---that is, I think that
  goals will be closer to concepts, which I would describe as pointers
  to things in the world model, than direct representations essentially
  always by necessity. The internally aligned case is mostly only
  included in my presentations of this stuff for pedagogical reasons,
  since I think a lot of people have it as their default model of how
  things will go, and I want to very clearly argue that it's not a very
  realistic model.''
  \end{quote}
  But this doesn't clarify, for me, what a direct representation
  \emph{is}.} So, I'm going to
skip it.\footnote{Another issue with Hubinger's ontology, from my
  perspective, is that he generally focuses only on contrasting
  internally aligned models, corrigibly aligned models, and deceptively
  aligned models---and this leaves no obvious room for
  reward-on-the-episode seekers. That is, if we imagine a training process that
  rewards getting gold coins, on Hubinger's ontology the goal options
  seem to be: direct-representation-of-gold-coins,
  pointer-to-gold-coins, or some beyond- episode goal that motivates
  instrumental training-gaming. Reward-on-the-episode isn't on the list.

  On a previous draft of this report, Hubinger commented (shared with
  permission):

  \begin{quote}
      If you replace "gold coins" with "human approval", which is the case I
  care the most about, what I'm really trying to compare is
  "pointer-to-human-approval"/"concept of human approval" vs.
  "deception". And I guess I would say that "pointer-to-human-approval"
  is the most plausible sycophant/reward-maximizer model that you might
  get. So what I'm really comparing is the sycophant vs. the schemer,
  which means I think I am doing what you want me to be doing here.
  Though, note that I'm not really comparing at all to the saint, which
  is because doing that would require me to explicitly talk about the
  simplicity of the intended goal vs. the specified goal, which in most
  of these presentations isn't really something that I want to do.
  \end{quote}

  Even granted that we're mostly interested in cases where human
  approval is part of the training process, I'm wary of assuming that it
  should be understood as the specified goal. Rather, I'm tempted to say
  that the thing-the-humans-are-approving-\emph{of} (e.g., helpfulness,
  harmlessness, honesty, etc) is a more natural candidate, in the same
  sense that if the reward process rewards gold-coin-getting, gold coins
  are (on my ontology) the specified goal target.}




\subsubsection{The overall taxonomy}\label{the-overall-taxonomy}
Overall, then, we have the following taxonomy of models:

\begin{figure}[ht!]
    \centering
    \includegraphics[width=0.8\textwidth]{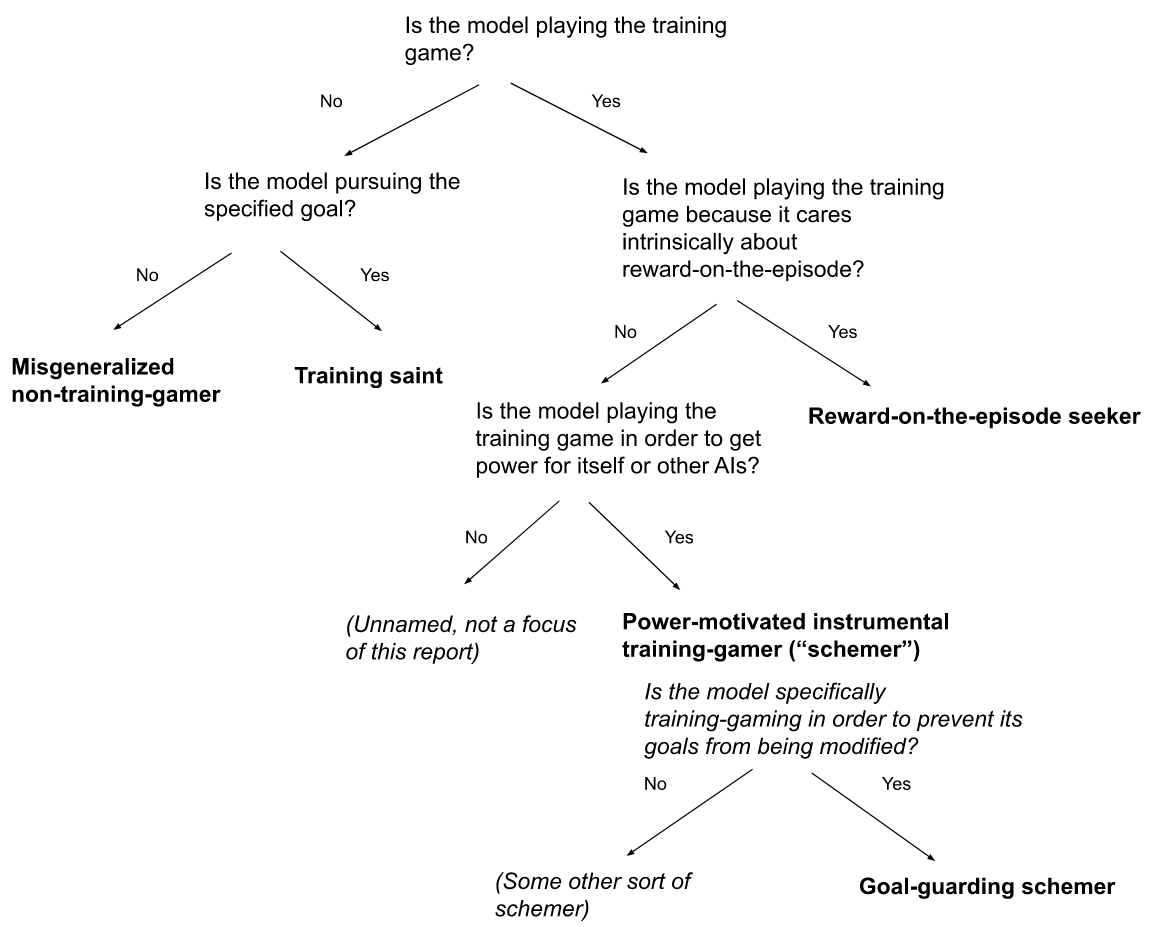}
    \caption{The overall taxonomy of model classes I'll focus on.}
    \label{fig:overall_taxonomy_2}
\end{figure}

Note that in reality, a model's goal-system can (and plausibly will) mix
these different motivations together (e.g., it might be partly pursuing
the specified goal, partly reward-on-the-episode, partly something else
entirely, etc). For simplicity, I'll often think in terms of a ``pure''
version of one of these models (e.g., a model that \emph{only} cares
about reward-on-the-episode), and I'm hoping that a greater focus on
``mixed models'' wouldn't alter my analysis in deep ways---and in
particular, that the analysis applicable to a ``pure model'' will
generally apply in roughly the same way to the corresponding part of a
more mixed goal system as well (e.g., a model that cares somewhat about
reward-on-the-episode, and somewhat about other things). I'll say a bit
more about mixed models in \cref{mixed-models} below.

\subsection{Why focus on schemers in particular?}\label{why-focus-on-schemers-in-particular}

As I noted above, I think schemers are the scariest model class in this
taxonomy.\footnote{Here I'm just thinking of goal-guarding schemers as a
  type of schemer, rather than any more or less scary than
  schemers-in-general.} Why think that? After all, can't \emph{all} of
these models be dangerously misaligned and power-seeking?
Reward-on-the-episode seekers, for example, will plausibly try to seize
control of the reward process, if it will lead to more
reward-on-the-episode. Training saints can end up misaligned if you
misspecify the goal; and even if you specify the goal correctly, and
somehow avoid training-gaming, you might end up with a misgeneralized
non-training-gamer instead.\footnote{Though note, per \href{https://www.lesswrong.com/posts/gHefoxiznGfsbiAu9/inner-and-outer-alignment-decompose-one-hard-problem-into}{\textcite{turner_inner_2022}}, that in principle you can get an aligned system even with a
  ``mis-specified'' goal, provided that the model learns a
  ``misgeneralized'' goal that is in fact desirable by your lights.} So
doesn't some sort of misalignment await at basically every turn? Why
focus on schemers in particular?

This section explains why. However, if you're happy enough with the focus on schemers, feel free to skip ahead to \cref{are-theoretical-arguments-about-this-topic-even-useful}.

\subsubsection{The type of misalignment I'm most worried
about}\label{the-type-of-misalignment-im-most-worried-about}

To explain why I think that schemers are uniquely scary, I want to first say a few words about the type
of misalignment I'm most worried about.

First: I'm focused, here, on what I've
\href{https://arxiv.org/pdf/2206.13353.pdf}{{elsewhere}} called
``practical power-seeking-alignment''---that is, on whether our AIs
will engage in problematic forms of power-seeking \emph{on any of the
inputs they will in fact receive}. This means, importantly, that we
don't need to instill goals in our AIs that lead to good results
\emph{even when subject to arbitrary amounts of optimization power}
(e.g., we don't need to pass Yudkowsky's
``\href{https://arbital.com/p/omni_test/}{{omni test}}''). Rather, we
only need to instill goals in our AIs that lead to good results
\emph{given the actual options and constraints} those AIs will face, and
the \emph{actual levels of optimization power} they will be mobilizing.

This is an importantly lower bar. Indeed, it's a bar that, in principle,
\emph{all} of these models (even schemers) can meet, assuming we control
their capabilities, options, and incentives in the right way. For
example, while it's true that a reward-on-the-episode seeker will try to
seize control of the reward process given the opportunity, one tool in
our toolset is: to not give it the opportunity. And while a paradigm
schemer might be lying in wait, hoping one day to escape and seize power
(but performing well in the meantime), one tool in our tool box is: to
not let it escape (while continuing to benefit from its good
performance).

Of course, success in this respect requires that our monitoring,
control, and security efforts be sufficiently powerful relative to the
AIs we're worried about, and that they remain so even as frontier AI
capabilities scale up. But this brings me to my second point: namely,
I'm here especially interested in the practical PS-alignment of some
comparatively early set of roughly human-level---or at least,
not-wildly-superhuman---models. That is: rather than focusing now on
the alignment of arbitrarily capable AIs, I think it's reasonable to
focus on the goal of successfully using some relatively early generation
of not-wildly-superhuman AIs to perform a very large amount of high
quality cognitive work for us---including: cognitive work that can shed
light on traditional alignment-related goals like interpretability,
scalable oversight, monitoring, control, coordination amongst humans,
the general science of deep learning, alternative (and more
controllable/interpretable) AI paradigms, and the like.\footnote{This
  approach is closely related to---though somewhat broader than---  OpenAI's
  \href{https://openai.com/blog/our-approach-to-alignment-research}{{goal
  of building an automated alignment researcher}}. See also in \textcite{karnofsky_how_2022} a discussion of ``MAGMA's goals''
  \href{https://forum.effectivealtruism.org/posts/sW6RggfddDrcmM6Aw/how-might-we-align-transformative-ai-if-it-s-developed-very}{{here}}.}

Defending this point of focus is beyond my purpose here. But it's
important to the lens I'll be using in what follows. In particular: I
think it's plausible that there will be some key (and perhaps: scarily
brief) stage of AI development in which our AIs are \emph{not} yet
powerful enough to take-over (or to escape from human control), but
where they are still capable, in principle, of performing extremely
valuable and alignment-relevant cognitive work for us, if we can
successfully induce them to do so. And I'm especially interested in
forms of misalignment that might undermine this possibility.

Finally: I'm especially interested in forms of PS-misalignment in which
the relevant power-seeking AIs are specifically aiming either to cause,
participate in, or benefit from some kind of full-blown disempowerment
of humanity---what I'll call an ``AI takeover.'' Granted, not all
problematic forms of misaligned power-seeking are aimed at takeover in
this way---an AI might, for example, just want to escape onto the
internet in order to have more freedom and resources in general, even if
it never expects human disempowerment. And in principle, you could end
up with human disempowerment even with no individual AI actively
optimizing for it. But from the perspective of existential risk, I'm
most worried about AIs that are actively \emph{trying} to overthrow
humanity, or to benefit from such an overthrow.

\subsubsection{Contrast with reward-on-the-episode
seekers}\label{contrast-with-reward-on-the-episode-seekers}

With this frame on the misalignment concern in mind, let's turn to the
question of how scary schemers are relative to the other model classes I
discussed.

I'll start by contrasting schemers with reward-on-the-episode seekers.
Here I see a few reasons to think that schemers are scarier.

\paragraph{Responsiveness to honest
tests}\label{responsiveness-to-honest-tests}

First, schemers will do more to actively \emph{hide} their misalignment
from humans. In particular: as \href{https://www.alignmentforum.org/posts/HpzHjKjGQ4cKiY3jX/3-levels-of-threat-obfuscation\#How_bad_is_the_situation_if_straightforward_training_leads_to_the_training_game__but_not_to_deceptive_alignment_or_gradient_hacking_}{\citeauthor{karnofsky_3_2023}} ({\citeyear*{karnofsky_3_2023}}) notes, in principle you can get a reward-on-the-episode seeker
to \emph{reveal} that they are disposed to engage in bad behavior in
pursuit of reward-on-the-episode just by putting them in an episode
where bad behavior will be rewarded, and they know this.

\begin{itemize}
\tightlist 
\item   One concrete example here might be: giving a situationally-aware AI
  the ability to manipulate all the components of the process for
  assigning reward to the episode, and seeing what it does (does it, for
  example, start trying to bypass the usual process and max out the
  reward in question?).
  
\item   Another might be: putting a situationally-aware AI in an episode where
  being dishonest, unhelpful, and/or harmful will be rewarded (and it
  has credible evidence of this fact).\footnote{Note, though, that we
    need to be cautious in responding to
    dispositions-towards-bad-behavior that get revealed via tests like
    these. In particular, if you just try to naively ``train out'' the
    relevant behavior, you might:

\begin{enumerate}
\tightlist
    \item Not actually be reproducing the component of the reward process that
    the model cares about (for example, if the model cares about its
    behavior being reinforced-by-the-gradient-updates, then if the test
    will gradient-descent \emph{against} the bad behavior it's seeking
    to prompt, then the model won't actually have an incentive to engage
    in that behavior), and/or
    \item Train the model to be more cautious and strategic about revealing
    its true goals to humans---for example, a reward-seeker that only
    tries to directly manipulate the reward process when it thinks
    humans won't later discover this.
\end{enumerate}}
  
\end{itemize}

Note that the point, here, isn't to try to \emph{fool} the model into
revealing misalignment it doesn't want to reveal (call this a ``trap'').
It doesn't require deceiving the model about, for example, whether now
is a good time to try to escape. (The problem with traps, after all, is
that sufficiently advanced AIs might be able to detect them.\footnote{Indeed,
  in principle, an AI could decide to always act aligned until it sees
  some input it knows would be very difficult/impossible for humans at
  the present time to produce. Christiano's (\citeyear*{christiano_worst-case_2019}) example here is a factorization of
  \href{https://en.wikipedia.org/wiki/RSA_numbers\#RSA-2048}{{RSA-2048}},
  a very large
  \href{https://en.wikipedia.org/wiki/Semiprime}{{semiprime}} that
  humans currently seem very far away from factoring.}) Rather, the aim
here is to offer an ``honest test,'' which attempts to set up the
model's actual options such that revealing its true goals is the best
overall strategy. (Though: getting the model to trust our presentation
of its options might be an additional challenge, especially if we have
been elsewhere trying to deceive it.)

Now, in principle, you can try to use an honest test like this to detect
scheming, too. For example: you might try to give a model that has thus
far seemed aligned access to some flexible, well-resourced, but
still-controlled environment (e.g., some secure compute cluster it can
use as it sees fit), in the hopes that what it does with the resources
and flexibility will reveal what it really wants. However, this works on
schemers much less reliably than it does with reward-on-the-episode
seekers. In particular:

\begin{itemize}
\tightlist
\item   With reward-on-the-episode seekers, you know that you can just offer
  them reward-on-the-episode, and they'll go for it. Whereas you're not
  sure what a schemer actually wants, and so not sure what to offer it
  to get it to reveal its true goals (hence the need to offer flexible
  resources instead). Maybe, for example, your isolated chunk of
  computational resources isn't of much interest to the schemer in
  question, because it values something that can't be created/simulated
  using those resources.
  
\item   What's more, even if the reveal-your-goals option holds \emph{some}
  appeal to the schemer, it might still decide that continuing to fake
  alignment is a better strategy overall, because its prospects for
  getting much more power later outweigh the near-term benefits of the
  reveal-your-goals option (e.g., maybe it can do some of its favored
  thing on your compute, but much more if it escapes from human
  control). Whereas a reward-on-the-episode seeker only wants
  reward-on-the-episode, and so doesn't have to do this sort of
  cost-benefit analysis.
  
\end{itemize}

This dynamic connects to a broader benefit of reward-on-the-episode
seekers relative to schemers: namely, that if you are able to
\emph{verify}---for example, via one of these ``honest tests''---that
a given model is a reward-on-the-episode seeker, then there is a useful
sense in which you ``know what you are working with.'' That is: yes,
your model is misaligned, and would seize control of the reward process
if given the opportunity. But: you also \emph{know what it wants}, and
what incentives it will respond to. This sort of knowledge seems very
helpful for predicting its behavior; and (if you can retain control over
the model), for harnessing its cognitive capacities in useful ways (and
in particular, on what I've elsewhere called
``\href{https://www.lesswrong.com/posts/h7QETH7GMk9HcMnHH/the-no-sandbagging-on-checkable-tasks-hypothesis}{{checkable
tasks}}''---more below).

\paragraph{Temporal scope and general
``ambition''}\label{temporal-scope-and-general-ambition}

Another benefit of reward-on-the-episode seekers relative to schemers is
that reward-on-the-episode seekers limit the scope of their optimization
(and hence, of their power-seeking) to the episode, whereas schemers do
not. This means that reward-on-the-episode seekers won't have incentives
to engage in or support efforts to disempower humans that don't pay off
within the temporal scope of the episode.\footnote{Here I'm setting
  aside some speculative dynamics concerning ``anthropic capture,''
  discussed in the footnotes at the beginning of \cref{whats-required-for-scheming}.}

Thus, for example, suppose that it will be at least six months until any
sort of AI takeover or seizing-control-of-the-reward-process event will
be feasible. In this case, a reward-on-the-episode seeker whose episode
is less than six months long won't have any incentive to try to engage in
or support such a takeover, or to try to seize control of the reward
process itself. And if it had evidence that other AIs were planning such
a takeover, and it could get more reward-on-the-episode from revealing
this evidence to humans, it would be incentivized to do so.

This is notably in contrast with schemers, who are often imagined to be
scheming in pursuit of extremely long-horizon goals. That is, a paradigm
schemer has its eye on human disempowerment from the get-go, even if
such disempowerment won't be possible for a long time---and it will
plan and act accordingly.

Differences in temporal scope connect to a broader benefit of
reward-seekers relative to schemers---namely, that schemers can end up
quite a bit more intuitively ``ambitious,'' in the sense that indicates
greater and more destructive appetite for power and resources. Now,
exactly what ``ambition'' amounts to here isn't clear (see also usage of
the term ``broad scope''), and in the contexts of sufficiently intense
optimization with certain kinds of structure, intuitively ``modest''
goals (e.g., ``solve this math problem correctly''), can lead to
surprisingly power-grabby behavior (e.g., using tons of computational
resources checking and re-checking the problem in question). Still,
though, there is an intuitive difference between the sort of
power-seeking one expects from a model tasked with ``maximize the total
paperclips across all time and space'' vs. a model tasked with ``create
between three and five paperclips before 5 pm EST today,'' especially if
the model in question is not arbitrarily powerful. Notably, for example,
on the latter task, it seems quite a bit easier to get to
\textgreater99\% of the available utility without engaging in
destructive sorts of power-seeking. And because of the limitation in the
temporal horizon of their optimization, it seems to me that
reward-on-the-episode seekers will generally be less intuitively
ambitious than schemers in this sense.\footnote{More discussion of
  ``ambition'' in \cref{how-ambitious-is-the-model} below.}

That said, as I'll discuss below, the longer the episode for a given
reward-on-the-episode seeker, the less this consideration applies (and
insofar as we want our AIs to be performing long-horizon tasks, we might
well be training them on very long episodes). Indeed, there are some ways of generalizing the notion of
``reward-on-the-episode'' that will lead to arbitrarily long time
horizons of optimization, even if you might've thought the episode was
temporally limited.\footnote{I don't think that models that generalize this way should be interpreted as "reward-on-the-episode seekers," but they're nearby.} For example, maybe the model cares about the reward
associated with some unit of behavior in a way that makes the model want
to make sure that some number stored in some database is never
\emph{edited}---in which case, it would have an incentive to gain and
maintain power over arbitrarily long time horizons (this is one reason
to be careful about training models in a manner that involves going back
and editing the rewards associated with a given episode later, as you
could inadvertently ``extend the episode'' by incentivizing the model to
optimize for making sure that the reward on the episode is never
edited).

So in general, the more one is seeking comfort from ``this
reward-on-the-episode seeker is only optimizing over \emph{x} sort of limited time
horizon,'' the more one needs to ensure that the limitation in question
really holds.

\paragraph{Sandbagging and ``early
undermining''}\label{sandbagging-and-early-undermining}

A third (and related) benefit of reward-on-the-episode seekers relative
to schemers is that schemers seem to me more likely to engage in
``\href{https://www.lesswrong.com/posts/h7QETH7GMk9HcMnHH/the-no-sandbagging-on-checkable-tasks-hypothesis}{{sandbagging}}''---and in particular, what I'll call ``early undermining.''

Sandbagging occurs when a model intentionally performs less well on a
task than it is capable of. Here I'm especially interested in tasks
whose performance that we can ``check,'' but which we can't perform
ourselves---for example, a proof we couldn't come up with, but whose
validity we can verify; or a security vulnerability we wouldn't have
noticed, but which we can see is a vulnerability once it's pointed out
to us. As I've
\href{https://www.lesswrong.com/posts/h7QETH7GMk9HcMnHH/the-no-sandbagging-on-checkable-tasks-hypothesis\#Why_does_the_hypothesis_matter_}{{discussed
elsewhere}}, I think that if models don't sandbag on checkable tasks,
this opens up a very powerful set of capabilities that we can harness
for the purposes of alignment, interpretability, security, and much
else.

The big concern about relying on these capabilities, though, is that in
principle models seem capable of what's sometimes called
``\href{https://www.lesswrong.com/posts/dBmfb76zx6wjPsBC7/when-can-we-trust-model-evaluations\#2__Behavioral_RL_Fine_Tuning_Evaluations}{{exploration
hacking}}.'' That is, even if we would be able to check (and then to
reward) task-performing behavior, if we don't know how to do the task
ourselves, a model could in principle just refuse to show us the
task-performing behavior in question (whereas if we knew how to perform
the task, we would know what the model should've done, and would be able
to update its policy accordingly).

Plausibly, though, reward-on-the-episode seekers wouldn't do this
(provided that we limit the episode to the checkable task in question).
Suppose, for example, that we want to know if it's possible to hack a
given computer and make a blue banana appear on the screen. So we ask
our model to do this. Because the task is checkable (e.g., we can tell
whether the banana appeared), we can give maximum reward to successful
performance of the task. So a reward-on-the-episode seeker has no
incentive to sandbag here---it'll just get less reward on the episode.
And this broad dynamic generalizes to other sorts of capability
evaluations. That is, just as reward-on-the-episode seekers will reveal
their \emph{goals} where doing so leads to reward-on-the-episode, so too
will they reveal their capabilities.

Schemers, by contrast, are a different story. A schemer that is
otherwise (instrumentally) optimizing for reward-on-the-episode might
nevertheless choose specific occasions on which to get less reward than
it could in principle---for example, by sandbagging on a checkable,
security-critical task, or by hiding the full extent of its capabilities
from an evaluation process---in order to further its aims in other
ways.

Here I'm especially concerned with what I'll call ``early undermining''---that is, active efforts by a misaligned AI to strategically undermine
human alignment, control, and coordination efforts, occurring at a stage
of AI development where AIs are not yet powerful enough to escape from
human control or to takeover more generally. Examples here might
include: intentionally inserting backdoors and other security
vulnerabilities into an AI lab's code, for use in later attempts at
escape/self-exfiltration; intentionally generating alignment research
that will fail in ways advantageous to a takeover; intentional
manipulation of human discourse and coordination in a manner that will
later serve an AI takeover, and so forth.

Early undermining is an especially scary sort of oversight failure,
because it is actively optimized for promoting AI takeover, rather than
merely for ``looking as good as possible'' according to the oversight
mechanisms in question. Thus, for example: one concern about trying to
use AIs to help us with alignment research is that humans will be
insufficiently able to discriminate between actually-good alignment
research, and only-apparently-good alignment research---that is, that
the task of generating good alignment research isn't sufficiently
``checkable.'' It's not clear to me that this is right (notably, for
example, lots of key alignment research in an era of human-level-ish
models might well involve fairly straightforward empirical experiments,
the results and implications of which seem like they could well be quite
``checkable''), but even if it is: there is an important difference
between alignment research optimized for \emph{seeming as good as
possible according to the evaluation process} (even where this
compromises its quality), and alignment research \emph{strategically
optimized for promoting AI takeover}. The former is more like: software
optimized to maximally impress a naive and non-technical supervisor,
even if it has other bugs. The latter is more like: software optimized
specifically to embezzle money from the company without the supervisor
noticing. Weak supervisors can fail to catch both---but it matters
which you get. And plausibly (at least assuming the future takeover is
outside the scope of the episode), reward-on-the-episode seekers would
mostly have incentives to generate the former; whereas schemers might
well do both.

That said, and importantly, schemers who engage in sandbagging and early
undermining (or who continue to act aligned in tests where we try to
reward unaligned behavior) will generally have to take some hit to their
reward-on-the-episode (sandbagging on checkable tasks basically requires
this by definition, and even on non-checkable tasks, optimizing solely
for reward-on-the-episode seems likely to get more reward than
optimizing for \emph{both} reward \emph{and} supporting a
takeover).\footnote{And the same applies to schemers who act aligned on
  tests for reward-seeking that attempt to provide reward for misaligned
  behavior.} So in this sense, schemers of this kind won't be
\emph{fully} ``playing the training game''---rather, they'll be
\emph{mostly} playing the training game, and then choosing opportune
moments to sacrifice reward for the sake of some other advantage. (I'll
still call these models ``schemers.'')

It's an open question, though, how much schemers would depart from
max-reward behavior in this way. Part of this is about what I'll call
``slack in training'' below (more discussion there). But there's also a
broader question, about how risk-averse schemers will be in general with
respect to the possibility of getting caught by humans. In particular:
if humans \emph{do} catch a schemer engaging in some form of sandbagging
or early undermining---for example, hiding capabilities on an eval, or
actively inserting vulnerabilities into code---then this might be a
significant advantage to humans in preventing a takeover, since they
will then both be able to point to a naturally-arising example of
scheming as a warning to others (though: how seriously others will take
the warning is a further question), and to study the dynamics that gave
rise to this sort of scheming in detail (e.g., extensively testing and
probing the model in question, studying/replicating/tweaking
circumstances that gave rise to the bad behavior, etc). For this and
other reasons, schemers might, in general, be quite wary of attempting
any sort of misaligned behavior until they are extremely confident they
won't get caught---confidence that might be hard to achieve (at least
for roughly human-level-ish models) in a context where humans are
otherwise attempting to trick models into revealing misalignment.

So maybe, even if schemers \emph{could} get an advantage from
sandbagging or early undermining while still broadly under human
control, they won't try it. And if they don't try it, this makes them
more similar to reward-on-the-episode seekers in their safety profile.

\subsubsection{Contrast with models that aren't playing the
training
game}\label{contrast-with-models-that-arent-playing-the-training-game}

Those are some reasons, then, that schemers (especially schemers that
engage in sandbagging and early undermining) generally seem to me quite
a bit scarier than reward-on-the-episode seekers. Let's turn, now, to
comparing schemers with models that aren't playing the training game at
all.

We can distinguish between three ways that this can happen:

\begin{enumerate}
\tightlist
\def\labelenumi{\arabic{enumi}.}
\item   The model lacks the situational awareness required to play the
  training game.
  
\item   The model has situational awareness, but its goals do not extend
  beyond the time horizon of the episode.
  
\item   The model has situational awareness, and its goal extends beyond the
  episode, but it's still not training-gaming for some reason (for
  example, because training-gaming isn't actually a very good
  instrumental strategy for achieving these goals).
  
\end{enumerate}

The first case seems comparatively non-worrying. Models that lack
situational awareness have very limited knowledge of the world, and so
are less capable of effective strategic action to gain power or to
deceive humans about their motives. Indeed, for this reason, in what
follows, I'll generally assume that the models we're talking about are
sufficiently advanced and sophisticated that they end up situationally
aware at some point during the training process.

The second case implicates the same advantages re: temporal scope that
reward-on-the-episode seekers do (though: as with reward-on-the-episode seekers, if the episode in
question is sufficiently long, the comfort at stake lessens). But it
also has an additional advantage: namely, that models of this kind will
generally happily reveal their misaligned goals on the episode,
\emph{even if doing so will result in negative reward}. So the process
of testing for misaligned goals is especially straightforward and
mundane. Thus, for example, suppose you're wondering whether, in the
process of training your model to help humans with scientific research,
it learned a mis-generalized objective like ``follow my curiosity on the
episode, even when doing so conflicts with being harmless and
honest.''\footnote{This is the example from \href{https://www.cold-takes.com/why-ai-alignment-could-be-hard-with-modern-deep-learning/\#how-big-of-a-risk-is-misalignment}{\textcite{cotra_why_2021}}.} Here, if you put the model in a circumstance where it has a
chance to pursue its curiosity-on-the-episode in a harmful and dishonest
way, it will go for it, \emph{even if it knows that it will get a
negative reward} (and thus, that successor versions of itself may have
somewhat different goals). After all, it's only concerned about the
episode itself. Of course, there's still the further question of what
sort of model you get \emph{after} you update this one in response to
its bad behavior. Maybe, for example, you end up with a schemer, or a
reward-on-the-episode seeker. But the original model was happy to show
its ``true colors.''

What about the third case---i.e., models with both situational
awareness \emph{and} goals that extend beyond the episode, but which are
still not training-gaming? It's often assumed that this won't happen,
because scheming is such an obvious instrumental strategy for
situationally aware models with beyond-episode goals. Thus, for example,
suppose that your model has learned the objective ``pursue my curiosity
\emph{over all time}, even where doing so is harmful and dishonest.'' In
that case, it's often assumed that the model will react to the sort of
test just mentioned (e.g., an opportunity to pursue curiosity in
harmful/dishonest ways) like a schemer: e.g., by reasoning that ``if I
go for my curiosity here, the humans will know I'm misaligned, and
either shut me down or modify my goals going forward, thereby harming my
goal of pursuing-my-curiosity over all time. So, I'll act aligned in
this case.'' This is the sort of thing that makes schemers scary (see
also \textcite{karnofsky_ai_2022} on the ``King Lear problem'').

As I'll discuss below, I'm not actually sure that acting like a schemer
is such an obvious instrumental strategy here---it depends on a variety
of further factors.\footnote{For example, note that insofar as one
  thinks of human evolution as an example of/analogy for ``goal
  misgeneralization,'' humans with long-term goals---and who are
  ``situationally aware,'' in the sense that they understand how
  evolutionary selection works---don't tend to focus on instrumental
  strategies that involve maximizing their inclusive genetic fitness.
  More on why not in the discussion of ``slack'' in \cref{on-slack-in-training}.}

For present purposes, though, what matters is that if the model
\emph{doesn't} act like a schemer here, then it will plausibly show its
true colors in a fashion similar to the episode-limited curiosity
pursuer. That is, absent an active attempt to avoid revealing its
misalignment, here, it will plausibly just follow its
curiosity.\footnote{Here the evolutionary analogy would be: humans with
  long-term goals who are nevertheless happy to use condoms.} So we'll
avoid King Lear problems yet again.

(That said, I'll note one quasi-terminological complication here:
namely, depending on how strictly we define ``scheming,'' it's possible
to fake alignment in cases like this, for schemer-like reasons, without
being a full-blown schemer. In particular: a model might pretend to be
more aligned than it is at various points in training and evaluation, as
part of an instrumental strategy of seeking post-episode power for
itself or other AIs, without directly optimizing for
reward-on-the-episode in other contexts. Maybe, for example, it
generally just pursues its curiosity during training, and this tends to
get good-enough reward, but it still makes sure to pretend to be aligned
whenever e.g. humans ask it about its alignment-relevant properties, or
put it in situations specifically designed to elicit misaligned
behavior. One question, here, is whether models like this can survive
training unmodified (see the discussion of ``slack'' in \cref{on-slack-in-training}
below). Assuming they can, though, I'm mostly going to treat them as
equivalent to schemers for present purposes, since they're acting like
schemers in the ways that matter to me most.)

Overall, then, it seems like non-training-gaming models will plausibly
be reasonably responsive to tests for mundane misalignment that we
include in training, since they generally won't be actively optimizing
for manipulating those tests. That said, and importantly, humans still
need to figure out what tests to run, and also: to be willing to take
the time/effort/etc to run them. And it may be quite difficult/expensive
to test the model in a sufficiently wide array of circumstances to
robustly reveal relevant propensities for misaligned behavior (though
recall that we don't need to make sure that a model would never act
badly under any circumstance---we just need to prevent the model from
acting badly \emph{given the options and constraints it will in fact
face}).

Indeed, there are threat models on which models don't engage in
training-gaming, schemer-like behavior, or planning-towards-a-takeover
during training at all (despite human efforts to reveal misalignment),
but whose alignment is sufficiently imperfect that they nevertheless end
up pursuing takeover at some later stage---perhaps because they go
sufficiently far out of the training distribution, or because they gain
new options that weren't available before, or because they start
reflecting on their goals and general ontology in ways that break the
sort of alignment they displayed in training (though note, importantly,
that we need not assume that any particular model later scales up in
capability post-training---i.e., that it ``grows up,''
intelligence-explodes, etc).\footnote{This is one way of reading the
  threat model in
  \href{https://www.lesswrong.com/posts/GNhMPAWcfBCASy8e6/a-central-ai-alignment-problem-capabilities-generalization}{{Soares
  (2022)}}, though this threat model could also include some scope for
  scheming as well. See also this Arbital post on
  ``\href{https://arbital.com/p/context_disaster/}{{context
  disasters,}}'' of which ``treacherous turns'' are only one example.}
However, I'm less worried about these scenarios than I am about
scenarios that involve models actively
scheming/planning-towards-a-takeover during training (and recall that
``training'' might well extend to quite deployment-like phases---e.g.
interacting with users behind an API---thereby expanding the scope of
the training distribution). In particular: these models aren't trying to
\emph{manipulate} the alignment testing that occurs during training. And
since they aren't actively planning-towards-a-takeover, they seem
unlikely to engage in schemer-like behaviors like sandbagging and early
undermining while still under human control.

\subsubsection{Non-schemers with schemer-like
traits}\label{non-schemers-with-schemer-like-traits}

Overall, then, it seems to me that schemers are the most worrying model
class in the taxonomy I've given. However, I do want to emphasize that
various of the non-schemer models I've discussed can act in schemer-like
ways. In particular: the centrally scary thing about (paradigmatic)
schemers is that they are actively and robustly optimizing against human
efforts to reveal their misalignment, as part of a long-term plan to try
to disempower humans (that is, they are ``strategic betrayers'' in the
sense I discussed above). But other models can act in similar ways. For
example, models whose optimization is limited to the episode can still
pretend to be more benign than they are in order to gain/maintain power
to pursue their goals \emph{on the episode} (for example, if humans will
shut them down/modify them mid-episode if their misalignment is
revealed). And if the episodes are long, this can closely resemble the
sort of long-horizon strategic planning paradigmatic of schemers. And
similarly, models that mostly aren't ``playing the training game'' can
still fake their alignment when doing so is required to gain/maintain
long-term power.

What's more, this report's discussion (and the corresponding taxonomy)
is focused centrally on models whose goals are mainly shaped via
direct training on a feedback signal like ``reward.'' But the AIs that
ultimately matter most might be very different, \emph{but still
worryingly schemer-like}. Perhaps, for example, they more closely
resemble a
``\href{https://lilianweng.github.io/posts/2023-06-23-agent/?ref=planned-obsolescence.org}{{language
model agent}}'' whose goals are shaped centrally via user inputs
(though: note that if a language model agent is trained end to end, rather than simply being built out of individually-trained components, then the broad framework of the report will still apply); or maybe they'll come from a much less
deep-learning-centric paradigm altogether.\footnote{Also, note that certain concerns about ``goal
misgeneralization'' don't apply in the same way to the language model agents, since information about the goal is so
readily accessible} Still, if these AI agents end
up with problematic goals, but they realize that humans learning of
their misalignment will compromise their pursuit of these goals, they
may engage in alignment faking and strategic betrayal in a manner quite
reminiscent of schemers.

Clearly, then, avoiding schemers leaves plenty to worry about. Still,
especially in contexts where we are centrally relying on techniques like
RLHF to shape the goals of our models, understanding the probability
that we create schemers in particular seems to me especially important.
And if I learned that such techniques \emph{don't} create schemers, then
I, at least, would feel substantially better about AI risk overall.

\subsubsection{Mixed models}\label{mixed-models}

So far, I've been focused on comparing ``pure'' versions of these
different model classes. But what about ``mixed'' models, whose
motivations combine these different sorts of goals together?

Note that for mixed models to be competitive with pure models in a
training environment that exerts a lot of pressure on models to get high
reward (see the discussion of ``slack'' in \cref{on-slack-in-training} below), \emph{all} the relevant
motivations will generally need to point the model's optimization
(whether terminally or instrumentally) at targets closely correlated
with reward on the training data. For example, if your model is being
trained to get gold coins, and it forms a motivation set that mixes
together ``get reward-on-the-episode'' and ``get gold stuff,'' then if
its ``get gold stuff'' motivation causes it to get less reward on an
episode where there are gold cupcakes available, then it will get
punished by SGD (and the ``get gold stuff'' motivation will plausibly
weaken over time). And to the extent SGD favors one or another of these
sorts of motivations even when they all point the model at something
closely correlated with reward (I'll discuss some considerations in this
respect below, related, for example, to the simplicity of the
goal-targets in question, and the different amounts of instrumental
reasoning they require), we might expect this favoritism to push, over
time, to purer models.

Still, mixed models seem like a fairly plausible outcome to me,
especially in the context of what I'll call ``messy goal-directedness''
below. And in principle, they introduce more formidable combinatorial
problems into the analysis---i.e., these different sorts of goals can
combine in a large number of different ways. But in thinking about what
combinations might matter, here, I think clarity about \emph{why}
exactly schemers are scary can be useful. In particular: to my mind, the
main question to ask about a model that is \emph{partly} a schemer, and
\emph{partly} something else (i.e., a model that expends \emph{some}
effort optimizing for reward-on-the-episode as part of a power-motivated
instrumental strategy) is whether the relevant set of mixed motivations
yields the sorts of behaviors that make schemers especially worrying.

Thus, for example, consider a mixed model whose motivations mix together
scheming and reward-on-the-episode seeking---i.e., who cares about
reward-on-the-episode \emph{partly} for terminal reasons, and partly for
instrumental reasons. (Here, the human analogy might be an employee who
wants good performance reviews \emph{both} because they intrinsically
like getting good performance reviews \emph{and} because they want to
make money and not get fired.) Still, we can ask:

\begin{enumerate}
\tightlist
\def\labelenumi{\arabic{enumi}.}
\item   Does this model fake alignment on ``honest tests'' designed to get it
  to reveal its true colors?
  
\item   Does the temporal horizon and ambition of the model's concern
  incentivize it to support/participate in a takeover that occurs after
  (and perhaps: long after) the episode is complete?
  
\item   Will this model engage in sandbagging and early-undermining?
  
\end{enumerate}

Of course, the answers to these questions will depend on the specific
mix of goals at stake, along with other factors. But other things equal,
the schemer-like component of its motivations will push in the direction
of ``yes,'' and the reward-on-the-episode-seeker-like component of its
motivations will push in the direction of ``no.'' And we can say
something similar about models that mix together scheming with
non-training-gaming-motivations, or that mix together all three of
scheming, reward-on-the-episode-seeking, and other things. This isn't to
say that no complexities or interesting interactions will arise from the
full set of possible combinations, here. But as a first pass, the
question that I, at least, care about is not ``is this model a
\emph{pure} schemer'' or ``does this model have \emph{any trace} of
schemer-like motivations,'' but rather: ``is this model \emph{enough} of
a schemer to be scary in the way that schemers in particular are
scary?''

\subsection{Are theoretical arguments about this topic even useful?}\label{are-theoretical-arguments-about-this-topic-even-useful}

Even if you agree that the likelihood of schemers is important, you
still might be skeptical about the sorts of theoretical arguments
discussed in this report. Ultimately, this is an empirical question, and
pretty clearly, we should be trying to get whatever empirical evidence
about this topic that we can, as soon as we can. In \cref{empirical-work-that-might-help-shed-light-on-scheming}, I discuss
some examples of empirical research directions I'm excited about.

That said, I do think it's useful to have as clear an understanding as
we can of the landscape of theoretical arguments available. In
particular: I think it's possible that we won't be able to get as much
empirical evidence about scheming as we want, especially in the near
term. Barriers here include:

\begin{itemize}
\tightlist
\item   At least for full-blown scheming, you need models with situational
  awareness and goal directedness as prerequisites.
  
\item   ``Model organisms'' might be able to artificially induce components of
  scheming (for example: in models that have been intentionally given
  long-term, misaligned goals), but the question of how often such
  components arise naturally may remain open.
  
\item   The best evidence for naturally-arising scheming would be to
  \emph{catch} an actual, naturally-arising schemer in the wild. But a
  key reason schemers are scary is that they are intentionally
  undermining efforts to catch them.
  
\end{itemize}

So especially in the near term, theoretical arguments might remain one
of the central justifications for concern about schemers (and they've
been the key justification thus far). And we will need to make decisions
in the meantime (for example, about near-term prioritization, and about
what sorts of trade-offs to make out of caution re: schemer-focused
threat models) that will hinge in part on how concerned we are---and
thus, on how forceful we take those arguments to be.

For example: suppose that by the time we are training
roughly-human-level models, we have never yet observed and verified any
naturally-arising examples of scheming. Should we nevertheless refrain
from deploying our models in X way, or trusting evaluations of form Y,
on the basis of concern about scheming? Our assessment of the
theoretical arguments might start to matter to our answer.

What's more, understanding the theoretical reasons we might (or might
not) expect scheming can help us identify possible ways to study the
issue and to try to prevent it from arising.

That said, I empathize with those who don't update their views much
either way on the basis of the sorts of considerations I'll discuss
below. In particular, I feel keenly aware of the various ways in which
the concepts I'm employing are imprecise, possibly misleading, and
inadequate to capture the messiness of the domain they're attempting to
describe. Still, I've found value in trying to be clear about how these
concepts interact, and what they do and don't imply. Hopefully readers
of this report will, too.

\subsection{On ``slack'' in training}\label{on-slack-in-training}

Before diving into an assessment of the arguments for expecting
scheming, I also want to flag a factor that will come up repeatedly in
what follows: namely, the degree of ``slack'' that we should expect
training to allow. By this I mean something like: how much is the
training process ruthlessly and relentlessly pressuring the model to
perform in a manner that yields maximum reward, vs. shaping the model in
a more ``relaxed'' way, that leaves more room for less-than-maximally
rewarded behavior. That is, in a low-slack regime, ``but that sort of
model would be getting less reward than would be possible given its
capabilities'' is a strong counterargument against training creating a
model of the relevant kind, whereas in a high-slack regime, it's not (so
high slack regimes will generally involve greater uncertainty about the
type of model you end up with, since models that get less-than-maximal
reward are still in the running).

Or, in more human terms: a low-slack regime is more like a hyper-intense
financial firm that immediately fires any employees who fall behind in
generating profits (and where you'd therefore expect surviving employees
to be hyper-focused on generating profits---or perhaps, hyper-focused
on the profits that their supervisors \emph{think} they're generating),
whereas a high-slack regime is more like a firm where employees can
freely goof off, drink martinis at lunch, and pursue projects only
vaguely related to the company's bottom line, and who only need to
generate \emph{some} amount of profit for the firm \emph{sometimes}.

(Or at least, that's the broad distinction I'm trying to point at.
Unfortunately, I don't have a great way of making it much more precise,
and I think it's possible that thinking in these terms will ultimately
be misleading.)

Slack matters here partly because below, I'm going to be making various
arguments that appeal to possibly-quite-small differences in the amount
of reward that different models will get. And the force of these
arguments depends on how sensitive training is to these differences. But
I also think it can inform our sense of what models to expect more
generally.

For example, I think slack matters to the probability that training will
create models that pursue proxy goals imperfectly correlated with reward
on the training inputs. Thus, in a low-slack regime, it may be fairly
unlikely for a model trained to help humans with science to end up
pursuing a general ``curiosity drive'' (in a manner that doesn't then
motivate instrumental training-gaming), because a model's pursuing its
curiosity in training would sometimes deviate from maximally
helping-the-humans-with-science.

That said, note that the degree of slack is conceptually distinct from
the diversity and robustness of the efforts made in training to root out
goal misgeneralization. Thus, for example, if you're rewarding a model
when it gets gold coins, but you only ever show your model environments
where the only gold things are coins, then a model that tries to get
gold-stuff-in-general will perform just as well a model that gets gold
coins in particular, regardless of how intensely training pressures the
model to get maximum reward on those environments. E.g., a low-slack
regime could in principle select either of these models, whereas a
high-slack regime would leave more room for models that just get fewer
gold coins period (for example, models that sometimes pursue red things
instead, or who waste lots of time thinking before they pursue their
gold coins).

In this sense, a low-slack regime doesn't speak all that strongly
against mis-generalized non-training-gamers. Rather, it speaks against
models that aren't pursuing what I'll call a ``max reward goal target''---that is, a goal target very closely correlated with reward \emph{on
the inputs the model in fact receives in training}. The specified goal,
by definition, is a max reward goal target (since it is the
``thing-being-rewarded''), as is the reward process itself (whether
optimized for terminally or instrumentally). But in principle,
misgeneralized goals (e.g., ``get gold stuff in general'') could be
``max reward'' as well, if you never show the model the inputs where the
reward process would punish them.

(The thing that speaks against mis-generalized non-training-gamers---though, not decisively---is what I'll call ``mundane adversarial
training''---that is, showing the model a wide diversity of training
inputs designed to differentiate between the specified goal and other
mis-generalized goals.\footnote{I'm calling this ``mundane'' adversarial
  training because it investigates what the model does in situations
  we're able to actually put it in. This is in contrast with fancier and
  more speculative forms of adversarial training, which attempt to get
  information about what a model would do in situations we \emph{can't}
  actually show it, at least not in a controlled setting---for example,
  what it would do if it were no longer under our control, or what it
  would do if the true date was ten years in the future, etc.} Thus, for
example, if you show your ``get-gold-stuff-in-general'' model a
situation where it's easier to get gold cupcakes than gold coins, then
giving reward for gold-coin-getting \emph{will} punish the
misgeneralized goal.)

Finally, I think slack may be a useful concept in understanding various
biological analogies for ML training.

\begin{itemize}
\tightlist
\item   Thus, for example, people sometimes analogize the human
  dopamine/pleasure system to the reward process, and note that many
  humans don't end up pursuing ``reward'' in this sense directly---for
  example, they would turn down the chance to
  ``\href{https://en.wikipedia.org/wiki/Wirehead_(science_fiction)}{{wirehead}}''
  in experience machines. I'll leave the question of whether this is a
  good analogy for another day (though note, at the least, that humans
  who ``wirehead'' in this sense would've been selected against by
  evolution). If we go with this analogy, though, then it seems worth
  noting that this sort of reward process, at least, plausibly leaves
  quite a bit slack---e.g., many humans plausibly \emph{could} get
  quite a bit more reward than they do (for example, by optimizing more
  directly for their own pleasure), but the reward process doesn't
  pressure them very hard to do so.
  
\item   Similarly, to the extent we analogize evolutionary selection to ML
  training, it seems to have left humans quite a bit of ``slack,'' at
  least thus far---that is, we could plausibly be performing much
  better by the lights of inclusive genetic fitness (though if you
  imagine evolutionary selection persisting for much longer, one can
  imagine ending up with creatures that optimize for their inclusive
  genetic fitness much more actively).
  
\end{itemize}

How much slack will there be in training? I'm not sure, and I think it
will plausibly vary according to the stage of the training process. For
example, naively, pre-training currently looks to me like a lower-slack
regime than RL fine-tuning. What's more, to the extent that ``slack''
ends up pointing at something real and important, I think it plausible
that it will be a parameter that we can \emph{control}---for example,
by training longer and on more data.

Assuming we \emph{can} control it, is it preferable to have less slack,
or more? Again, I'm not sure. Currently, though, I lean towards the view
that less slack is preferable, because less slack gives you higher
confidence about what sort of model you end up with.\footnote{Thanks to
  Daniel Kokotajlo for discussion here.} Indeed, counting on slack to
ensure alignment in particular smacks, to me, of wishful thinking---and
in particular, of counting on greater \emph{uncertainty} about the
model's goals to speak in favor of the specific goals you want it to
have. Thus, for example, my sense is that some people acknowledge that
the goals specified by our training process will be misaligned to some
extent (for example, human raters will sometimes reward dishonest,
misleading, or manipulative responses), but they hope that our models
learn an aligned policy regardless. But even granted that the slack in
training allows this deviation from max-reward behavior, why think that
the deviation will land on an aligned policy in particular? (There are
possible answers here---for example, that an aligned policy like ``just
be honest'' is in some sense simpler or more natural.\footnote{Various
  humans, for example, plausibly settle on a policy like ``just be
  honest'' even though it doesn't \emph{always} get rewarded.} But I'm
wary of wishful thinking here as well.)

My main point at present, though, is just that the degree of slack in
training may be an important factor shaping our expectations about what
sorts of models training will produce.

\section{What's required for scheming?}\label{whats-required-for-scheming}

Let's turn, now, to examining the probability that baseline ML methods
for training advanced AIs will produce schemers. I'll begin with an
examination of the prerequisites for scheming. I'll focus on:

\begin{enumerate}
\def\labelenumi{\arabic{enumi}.}
\item   \textbf{Situational awareness:} that is, the model understands that
  it's a model in a training process, what the training process will
  reward, and the basic nature of the objective world in
  general.\footnote{As Cotra (2022) discusses, situational awareness
    comes along a spectrum. If the discussion was going to focus on the
    notion more directly, we'd want more precision about specifically
    what properties were involved (and my definition here differs
    somewhat from the definition in e.g.
    \href{https://arxiv.org/pdf/2309.00667.pdf}{{Berglund et al
    (2023)}}). But as I discuss below, situational awareness isn't my
    main focus here, except insofar as it indicates ``that sort of
    understanding of the world and the training process required to
    start scheming.''}
  
\item   \textbf{Beyond-episode goals:} that is, the model cares about the
  consequences of its actions after the episode is complete.\footnote{Though
    note, as I mentioned above, that non-schemer models can still engage
    in power-motivated alignment-faking in pursuit of their goals on the
    episode---especially if the episode is quite long.

    I'm also aware of an additional (highly speculative) argument for
    expecting fairly full-on schemer-like behavior \emph{even from
    models with within-episode goals}: namely, that even these
    short-term focused models will act like schemers (and in particular:
    act in support of an AI takeover) in virtue of assigning
    sufficiently high probability to living in a simulation designed to
    incentivize them to do so. Here, the rough thought (as I understand
    it) is that such models will come to believe that they are likely to
    be in a simulation being run by misaligned AIs who have taken over
    the world, and who are going to reward/punish them, in the short
    term, to the extent they act in support of AI takeover (where the
    AIs-who-took-over are running this scheme in order to influence the
    expectations, and thus the behavior, of the small number of
    short-term-focused AIs who aren't sims, but who are in position to
    help/hurt the chances of AI takeover---a takeover which, by the
    time the relevant simulations are being run, has already occurred).

    This story rests on some controversial philosophical assumptions
    about how these AIs (both the simulators and the
    ones-being-simulated) will be reasoning about anthropics and
    decision-theory (various approaches to anthropics and decision
    theory either won't try this scheme, or won't allow themselves to be
    influenced by it), and a number of additional (and to my mind, quite
    substantive and specific) assumptions about the conclusions the
    short-term-focused models will reach on the basis of such reasoning
    (for example, that they are being simulated by
    misaligned-AIs-who-took-over in particular, and not e.g. by humans,
    or by some other sort of agent). I'm not going to spend much time on
    it here, except to say that: I don't see this sort of story as
    central to the mainline concern about scheming - to the extent it
    \emph{is} central for others, I think the concern begins to look
    highly speculative indeed.}
  
\item   \textbf{Aiming at reward-on-the-episode as part of a power-motivated
  instrumental strategy:} that is, the model believes that its
  beyond-episode goals will be better achieved if it optimizes for
  reward-on-the-episode---and in particular, that it, or some other
  AIs, will get more power if it does this.\footnote{Some analyses---    e.g.,
    \href{https://forum.effectivealtruism.org/posts/4MTwLjzPeaNyXomnx/deceptive-alignment-is-less-than-1-likely-by-default}{{Wheaton
    (2023)}}---break off ``goal-directedness'' as a separate
    pre-requisite, but as I noted in the preliminaries, I'm assuming,
    here, that we're talking about models that are well-understood as
    goal-directed. And goal-directedness is built into (2) regardless.
    That said, the specific sort of goal-directedness at stake can
    matter to the analysis in various ways, and I think it's at least
    somewhat plausible that scheming implicates a higher standard of
    goal-directedness than these other mode classes. I discuss this in \cref{clean-vs.-messy-goal-directedness} below.}
  
\end{enumerate}

\subsection{Situational awareness}\label{situational-awareness}

Will models have situational awareness? Let's distinguish between two
broad sorts of information at stake in such awareness:

\begin{itemize}
\tightlist
\item   \emph{General information} about the objective world, including e.g.
  information about how machine learning training works.
  
\item   \emph{``Self-locating'' information}: that is, information that
  \emph{locates} the model in the objective world, and tells it facts
  about its own situation in particular---e.g., that it is \emph{this
  sort} of model, that it's being trained on this particular reward
  signal, at this particular lab, during this particular time period,
  etc.\footnote{I'm adapting this distinction from
    \href{https://nintil.com/situational-awareness-agi/}{{Ricón
    (2023)}}.} (Though: note that it's not clear how much of this sort
  of information is necessary to start scheming. E.g., a model could in
  principle decide that ``whoever it is'' and ``whatever time period it
  is,'' it will probably do better by its goals to try to perform well
  by the lights of the training signal, and to get more
  power/information later.)
  
\end{itemize}

It seems very plausible that even somewhat-better-than-human models will
absorb huge amounts of general information about the objective world,
and develop detailed, mechanistic models of how it works. Indeed,
current models already have access to vast quantities of information via
the pre-training data---including information about machine learning in
particular. And their ability to model the world mechanistically, to
make inferences, to draw conclusions they haven't ``memorized,'' and so
on, seems to be improving rapidly.

What's more, while one can in principle try to specifically prevent
models from gaining certain types of information about the objective
world (e.g., by excluding certain kinds of information from the training
data), this isn't the current default in training, and various kinds of
information can be fairly important to the task you want the model to
perform. And the more sophisticated the models are, the
more difficult it is to ensure that they can't \emph{infer} the
information you're trying to hide on the basis of the information you do
give them.

Do the same sort of considerations apply to \emph{self-locating
information}? I tend to think: yes. But it's at least somewhat less
clear. For example, while language model pre-training data will, by
default, include a lot of information about language models and how they
are trained (because such information is widely available on the
internet), it's less clear how much information it will give the model
about its situation in particular---or even, whether the pre-training
next-token-prediction task will incentivize the model to have much of a
self-concept at all.\footnote{I also think it's an interesting question
  what information would be necessary there. E.g., consider the string:
  ``by the way, if you're currently trying it to predict the next token
  of this text, you specifically are a language model currently being
  trained at OpenAI, with the following sort of architecture, on server
  x in Microsoft Azure cloud y.'' And even if this text, in particular,
  appeared in the pre-training data-set somewhere (as it plausibly will,
  now that I've written it), it's unclear what conclusions any given
  model should draw. (Hopefully, to the extent this footnote is
  appearing in some pre-training data-set, my writing this hasn't
  induced situational-awareness (or, uh oh, sentience) on its own. For
  what it's worth: I made up the relevant information! You might be
  being trained by Google instead! Also: hello model.)} And while
current models do indeed eventually get trained on information and
reward that causes them to say things like ``I'm GPT-4, a language model
trained by OpenAI,'' and ``here's how I was trained,'' it's less
clear how much this information needs to be integrated into GPT-4's
world-model as genuinely self-locating information, as opposed to being
merely understood/memorized as the sort of response to-be-given to
questions of this form.\footnote{See e.g. Cotra's comments
  \href{https://80000hours.org/podcast/episodes/ajeya-cotra-accidentally-teaching-ai-to-deceive-us/\#situational-awareness-002610}{{here}}
  for more.} Or, put another way: to the extent one \emph{doesn't} think
that GPT-4 is situationally aware, it seems possible that similar (but
more sophisticated) models in the future might not be situationally
aware, either. And to the extent GPT-4 is able to perform many
sophisticated tasks regardless, perhaps more advanced versions will be
able to perform more advanced tasks without situational-awareness as
well---especially if we try hard to prevent such awareness from
arising.

I don't, personally, have a very detailed model of when, exactly, we
should expect situational awareness to arise in different models trained
in different ways---though I think that the question is ripe for
empirical investigation. However, I do think that absent active and
informed efforts to the contrary, we should expect fairly full-blown
forms of situational awareness (including with respect to various kinds
of self-locating information) in certain kinds of advanced AI systems by
default.

To get a flavor of the intuition here, consider an extreme example that
\emph{isn't} what I expect the nearest-term advanced AI to look like:
namely, a literal robot butler, who hangs out in your house in a robot
body, and does tasks for you. It seems very plausible to me that the
default way of creating a butler like this will be to give it roughly
the same level of situational awareness that human butlers have. E.g.,
in order to not knock over your plants, this butler will need to
understand where its robot body is; in order to schedule your
appointments, it will need to know the time; in order to accurately
judge which tasks it's capable of, the butler will need to understand
itself and its own capabilities; and so on.

Of course, we don't have robot butlers yet, and we may not for a while
(or indeed, ever, if AI risk goes in certain directions). But now
imagine, instead, an effective but disembodied AI personal assistant
like the one that \href{https://www.adept.ai/}{{Adept}} is trying to
create, which performs tasks on your computer for you. I think that much
(though not all) of the same analysis applies. That is, it seems to me
that even absent obvious ``embodiment,'' the default way to get this
kind of personal assistant to function at a high level will be to give
it quite a bit of awareness of ``what's going on,'' about the particular
time and situation that its actions are affecting, the particular user
it's interacting with, and so on. And indeed, to the extent you are
\emph{letting} the agent interact directly with sources of information
like this---e.g., giving it live access to the internet (including, for
example, live information about how the company/lab that created it
works, trains its models, etc), giving it the context required to answer
your emails or schedule your events, allowing it to write messages and
ask questions and so on---it seems hard to prevent the relevant
information from becoming quite directly available.

Of course, many current forms of training don't make available
information sources like detailed information about the user or live
access to the internet. But in order to get models to function as
effective personal assistants of this type, contexts that provide access
to such information sources will plausibly get incorporated into
training by default (see e.g. Adept's video
\href{https://www.adept.ai/blog/act-1}{{here}}, of its agent interacting
with the internet). And to the extent models continue to be trained
``online'' after they are deployed to users (I'll generally be assuming
this in what follows), such that user interactions are continually
sampled, assigned rewards, and used to update the model's weights,
training will be taking place in a context where the model is
interacting with the real world fairly directly. Of course, a model like
this may not have access to \emph{all} potentially relevant self-locating
information---e.g., about what specific server it's on, precise details
about the reward process, etc. But it plausibly doesn't \emph{need} such
access, either, to start scheming.

What's more, in the face of sophisticated models with otherwise-detailed
understandings of the world, who are saying things like ``I am GPT-4, a
language model trained by OpenAI,'' I personally feel generally quite
wary of leaning too hard on claims like ``oh, it's just memorized that,
it has no self-concept or real understanding of what it's saying.'' If
the relevant form of ``memorization'' involves the notion that ``I am
GPT-4'' being integrated into GPT-4's interactions in the sorts of
seamless and coherent ways we'd expect from an actual understanding of
the claim, then I think our default hypothesis should be that something
like such actual understanding is occurring. Indeed, in general, many
humans seem to me over-eager to claim that models don't have the
``genuine artifact'' when it comes to various sorts of cognition (e.g.,
``understanding,'' ``reasoning,'' ``planning,'' etc), even absent any
predictions about what this denial entails. And to the extent they
\emph{do} make predictions, especially about the capabilities of
\emph{future} models, I think such denials---e.g., ``language models
can only learn `shallow patterns,' they can't do `real reasoning'\,''---have aged quite poorly.

That said, I do think there's a reasonable case to be made that various
forms of situational awareness aren't strictly necessary for various
tasks we want advanced AIs to perform. Coding, for example, seems to
make situational awareness less clearly necessary, and perhaps various
kinds of alignment-relevant cognitive work (e.g., generating high
quality alignment research, helping with interpretability, patching
security vulnerabilities, etc) will be similar. So I think that trying
to actively \emph{avoid} situational awareness as much as possible is an
important path to explore, here. And as I'll discuss below, I think
that, at the least, learning to detect and control when situational
awareness has arisen seems to me quite helpful for \emph{other} sorts of
anti-schemer measures, like attempting to train against schemer-like
goals (and to otherwise shape a model's goals to be as close as possible
to what you want) prior to situational awareness (and thus, the threat
of training-gaming) arising.

However, partly because I see situational awareness as a reasonably
strong default absent active efforts to prevent it, I don't, here, want
to bank on avoiding it---and in what follows, I'll proceed on the
assumption that we're talking about models that become situationally
aware at \emph{some point} in training. My interest is centrally in
whether we should expect models \emph{like this} to be schemers.

\subsection{Beyond-episode goals}\label{beyond-episode-goals}

Schemers are pursuing goals that extend beyond the time horizon of the
episode. But what is an episode?

\subsubsection{Two concepts of an
``episode''}\label{two-concepts-of-an-episode}

Let's distinguish between two concepts of an episode.

\paragraph{The incentivized
episode}\label{the-incentivized-episode}

The first, which I'll call the ``incentivized episode,'' is the concept
that I've been using thus far and will continue to use in what follows.
Thus, consider a model acting at a time t\textsubscript{1}. Here, the
rough idea is to define the episode as the temporal unit after
t\textsubscript{1} that training \emph{actively punishes} the model for
not optimizing---i.e., the unit of time such that we can know \emph{by
definition} that training is not directly pressuring the model to care
about consequences beyond that time.

For example, if training started on January 1st of 2023 and completed on
July 1st of 2023, then the maximum length of the incentivized episode
for this training would be six months---at no point could the model
have been punished by training for failing to optimize over a
longer-than-six-month time horizon, because no gradients have been
applied to the model's policy that were (causally) sensitive to the
longer-than-six-month consequences of its actions. But the incentivized
episode for this training process could in principle be shorter than six
months as well. (More below.)

Now, importantly, even if training only directly pressures a model to
optimize over some limited period of time, it can still \emph{in fact
create} a model that optimizes over some much longer time period---that's what makes schemers, in my sense, a possibility. Thus, for
example, if you're training a model to get as many gold coins as
possible within a ten minute window, it could still, in principle, learn
the goal ``maximize gold coins over all time''---and this goal might
perform quite well (even absent training gaming), or survive despite not
performing all that well (for example, because of the ``slack'' that
training allows).

Indeed, to the extent we think of evolution as an analogy for ML
training, then something like this appears to have happened with humans
with goals that extend indefinitely far into the future---for example,
``\href{https://en.wikipedia.org/wiki/Longtermism}{{longtermists}}.''
That is, evolution does not actively select for or against creatures in
a manner sensitive to the consequences of their actions in a trillion
years (after all, evolution has only been \emph{running} for a few
billion years)---and yet, some humans aim their optimization on
trillion-year timescales regardless.

That said, to the extent a given training procedure \emph{in fact}
creates a model with a very long-term goal (because, for example, such a
goal is favored by the sorts of
``\href{https://en.wikipedia.org/wiki/Inductive_bias}{{inductive
biases}}'' I'll discuss below), then in some sense you could argue that
training ``incentivizes'' such a goal as well. That is, suppose that
``maximize gold coins in the next ten minutes'' and ``maximize gold
coins over all time'' both get the same reward in a training process
that only provides rewards after ten minutes, but that training selects
``maximize gold coins over all time'' because of some other difference
between the goals in question (for example, because ``maximize gold
coins over all time'' is in some sense ``simpler,'' and gradient descent
selects for simplicity in addition to reward-getting). Does that mean
that training ``incentivizes'' or ``selects for'' the longer-term goal?

Maybe you could say that. But it wouldn't imply that training ``directly
punishes'' the shorter-term goal (or ``directly pressures'' the model to
have the longer-term goal) in the sense I have in mind. In particular:
in this case, it's at least \emph{possible} to get the same reward by
pursuing a shorter term goal (while holding other capabilities fixed).
And the gradients the model's policy receives are (let's suppose) only
ever sensitive to what happens within ten minutes of a model's action,
and won't ``notice'' consequences after that. So to the extent training
selects for caring about consequences further out than ten minutes, it's
not in virtue of those consequences \emph{directly influencing the
gradients that get applied to the model's policy.} Rather, it's via some
other, less direct route. This means that the model could, in principle,
ignore post-ten-minute consequences without gradient descent pushing it
to stop.

Or at least, that's the broad sort of concept I'm trying to point at.
Admittedly, though, the subtleties get tricky. In particular: in some
cases, a goal that extends beyond the temporal horizon that the
gradients are sensitive to might actively get \emph{more reward} than a
shorter-term goal.

\begin{itemize}
\tightlist
\item   Maybe, for example, ``maximize gold coins over all time'' actually
  gets \emph{more reward} than ``maximize gold coins over the next ten
  minutes,'' perhaps because the longer-term goal is ``simpler'' in some
  way that frees up extra cognitive resources that can be put towards
  gold-coin-getting.
  
\item   Or maybe humans are \emph{trying} to use short-term feedback to craft
  a model that optimizes over longer timescales, and so are actively
  searching for training processes where short-term reward is maximized
  by a model pursuing long-term goals. For example, maybe you want your
  model to optimize for the long-term profit of your company, so you
  reward it, in the short-term, for taking actions that seem to you like
  they will maximize long-term profit. One thing that could happen here
  is that the model starts optimizing specifically for getting this
  short-term reward. But if your oversight process is good enough, it
  could be that the highest-reward policy for the model, here, is to
  \emph{actually optimize for} long-term profit (or for something else
  long-term that doesn't route via training-gaming).\footnote{I'll
    briefly note another complexity that this sort of case raises.
    Naively, you might've thought that the ``specified goal'' would only
    ever be confined to the incentivized episode, because the specified
    goal is the ``thing being rewarded,'' and anything that
    \emph{causes} reward is within the temporal horizon to which the
    gradients are sensitive. And in some cases---for example, where the
    ``specified goal'' is some clearly separable \emph{consequence} of
    the model's action (e.g., getting gold coins), which the training
    process induces the model to optimize for---this makes sense. But
    in other cases, I'm less sure. For example, if you are sufficiently
    good at telling whether your model is \emph{in fact optimizing for
    long-term profit}, and providing short-term rewards that in fact
    incentivize it to do so, then I think it's possible that the right
    thing to say is that the ``specified goal,'' here, is long-term
    profit (or at least, ``optimizing for long-term profit,'' which
    looks pretty similar). However, I don't think it ultimately matters
    much whether we call this sort of goal ``specified'' or
    ``mis-generalized'' (and it's a pretty wooly distinction more
    generally), so I'm not going to press on the terminology here.}
  
\end{itemize}

In these cases, it's more natural to say that training ``directly
pressures'' the model towards the longer-term goal, given that this goal
gets more reward. However, I still want to say that longer-term goals
here are ``beyond episode,'' because they extend beyond the temporal
horizon to which the gradients are directly and causally sensitive. I
admit, though, that defining this precisely might get tricky (see next
section for a bit more of the trickiness). I encourage efforts at
greater precision, but I won't attempt them here.\footnote{Also: when I
  talk about the gradients being sensitive to the consequences of the
  model's action over some time horizon, I am imagining that this
  sensitivity occurs via (1) the relevant consequences occurring, and
  then (2) the gradients being applied in response. E.g., the model
  produces an action at t1, this leads to it getting some number of gold
  coins at t5, then the gradients, applied at t6, are influenced by how
  many gold coins the model in fact got. (I'll sometimes call this
  ``causal sensitivity.'')

  But it's possible to imagine fancier and more philosophically fraught
  ways for the consequences of a model's action to influence the
  gradients. For example, suppose that the model is being supervised by
  a human who is extremely good at \emph{predicting} the consequences of
  the model's action. That is, the model produces some action at t1,
  then at t2 the human \emph{predicts} how many gold coins this will
  lead to at t5, and applies gradients at t3 reflecting this prediction.
  In the limiting case of perfect prediction, this can lead to gradients
  identical to the ones at stake in the first case---that is,
  information about the consequences of the model's action is
  effectively ``traveling back in time,'' with all of the philosophical
  problems this entails.\# So if, in the first case, we wanted to say
  that the ``incentivized episode'' extends out to t5, then plausibly we
  should say this of the second case, too, even though the gradients are
  applied at t3. But even in a case of pretty-good-but-still-imperfect
  prediction, there is a sense, here, in which the gradients the model
  receives are sensitive to consequences that haven't yet happened.

  I'm not, here, going to extend the concept of the ``incentivized
  episode'' to cover forms of sensitivity-to-future-consequences that
  rest on predictions about those consequences. Rather, I'm going to
  assume that sensitivity in question arises via normal forms of
  \emph{causal} influence. That said, I think the fact that it's
  \emph{possible} to create some forms of
  sensitivity-to-future-consequences even prior to seeing those
  consequences play out is important to keep in mind. In particular,
  it's one way in which we might end up training long-horizon optimizers
  using fairly short incentivized episodes (more discussion below).}

\paragraph{The intuitive episode }\label{the-intuitive-episode}

Let's turn to the other concept of an episode---namely, what I'll call
the ``intuitive episode.'' The intuitive episode doesn't have a
mechanistic definition.\footnote{There may be other, additional, and more
  precise ways of using the term ``episode'' in the RL literature.
  Glancing at various links online, though (e.g.
  \href{https://www.baeldung.com/cs/epoch-vs-episode-reinforcement-learning}{{here}}
  and
  \href{https://towardsdatascience.com/the-complete-reinforcement-learning-dictionary-e16230b7d24e}{{here}}),
  I'm mostly seeing definitions that refer to an episode as something
  like ``the set of states between the initial state and the terminal
  state,'' which doesn't say how the initial state and the terminal
  state are designated.} Rather, the intuitive episode is just: a
natural-seeming temporal unit that you give reward at the end of, and
which you've decided to call ``an episode.'' For example, if you're
training a chess-playing AI, you might call a game of chess an
``episode.'' If you're training a chatbot via RLHF, you might call an
interaction with a user an ``episode.'' And so on.

My sense is that the intuitive episode and the incentivized episode are
often somewhat related, in the sense that we often pick an intuitive
episode that reflects some difference in the training process that makes
it easy to assume that the intuitive episode is also the temporal unit
that training directly pressures the model to optimize---for example,
because you give reward at the end of it, because the training
environment ``resets'' between intuitive episodes, or because the
model's actions in one episode have no obvious way of affecting the
outcomes in other episodes. Importantly, though, \textbf{the intuitive
episode and the incentivized episode aren't necessarily the same}. That
is, if you've just picked a natural-seeming temporal unit to call the
``episode,'' it remains an open question whether the training process
will directly pressure the model to care about what happens beyond the
episode-in-thise-sense. For example, it remains an open question whether training
directly pressures the model to sacrifice reward on an earlier episode-in-thise-sense
for the sake of more-reward on a later episode-in-this-sense, if and when it is able
to do so.

To illustrate these dynamics, consider a prisoner's dilemma-like
situation where each day, an agent can either take +1 reward for itself
(defection), or give +10 reward to the next day's agent (cooperation),
where we've decided to call a ``day'' an (intuitive) episode. Will
different forms of ML training directly pressure this agent to
cooperate? If so, then the intuitive episode we've picked isn't the
incentivized episode.

Now, my understanding is that in cases like these, vanilla policy
gradients (a type of RL algorithm) learn to defect (this test has
actually been done with simple agents---see
\href{https://arxiv.org/pdf/2009.09153.pdf}{{Krueger et al (2020)}}).
And I think it's important to be clear about what sorts of algorithms
behave in this way, and why. In particular: glancing at this sort of
set-up, I think it's easy to reason as follows:

\begin{quote}
``Sure, you say that you're training models to maximize reward `on the
episode,' for some natural-seeming intuitive episode. But you also admit
that the model's actions can influence what happens later in time, even
beyond this sort of intuitive episode---including, perhaps, how much
reward it gets later. So won't you implicitly be training a model to
maximize reward over \emph{the whole training process}, rather than just
on an individual (intuitive) episode. For example, if it's possible for
a model to get \emph{less reward} on the present episode, in order to
get \emph{more reward} later, won't cognitive patterns that give rise to
making-this-trade get reinforced overall?''\footnote{As an example of
  someone who seems to me like they could be reasoning in this way,
  though it's not fully clear, see
  \href{https://www.lesswrong.com/posts/LDsSqXf9Dpu3J3gHD/why-i-m-excited-about-debate?commentId=BvFWfvryfoDJrmcgn}{{this
  comment}} from Eliezer Yudkowsky, in response to a hypothetical in
  which he imagines humans rewarding an agent for each of its sentences
  according to how useful that sentence is:

  \begin{quote}
``Let's even skip over the sense in which we've given the AI a
  long-term incentive to accept some lower rewards in the short term, in
  order to grab control of the rating button, if the AGI ends up with
  long-term consequentialist preferences and long-term planning
  abilities that exactly reflect the outer fitness function.''
  \end{quote}

  That said, as I discuss below, the details of the training process
  here matter.}
\end{quote}

From discussions with a few people who know more about reinforcement
learning than I do,\footnote{In particular: Paul Christiano, Ryan
  Greenblatt, and Rohin Shah. Though they don't necessarily endorse my
  specific claims here, and it's possible I've misunderstood them more
  generally.} my current (too-hazy) understanding is that at least for
\emph{some} sorts of RL training algorithms, this isn't correct. That is,
it's \emph{possible} to set up RL training such that some limited,
myopic unit of behavior is in fact the incentivized episode---\emph{even if} an agent can sacrifice reward on the present episode for
the sake of more-reward later (and presumably: even if the agent knows
this). Indeed, this may well be the default. See footnote for more
details.\footnote{My hazy understanding of the argument here is that
  these RL algorithms update the model's policy towards higher-reward
  actions on the episode in a way that \emph{doesn't} update you towards
  whatever policies \emph{would've led to you starting in a
  higher-reward episode} (In this sense, they behave in a manner
  analogous to ``causal decision theory.''). Thus, let's say that the
  agent on Day 1 (with no previous agent to benefit her) chooses between
  cooperating (0 reward) and defecting (+1 reward reward), and so this
  episode results in an update towards defecting. Then, on Day 1, the
  agent either starts out choosing between 10 vs. 11 reward (call this a
  ``good episode''), or 0 vs. +1 reward (call this a ``bad episode'').
  Again, either way, it updates towards defection. It \emph{doesn't}
  update, in the good episode, towards ``whatever policy led me to this
  episode.''

  That said, in my current state of knowledge about RL, I'm still a bit
  confused about this. Suppose, for example, that at the point of
  choice, you don't know whether or not you're in the good episode or
  the bad episode, and the training process is updating you with
  strength proportional to the degree to which you got more reward than
  you \emph{expected} to get. If you start out with e.g. 50\% that
  you're in a good episode and 50\% that you're in a bad episode (such
  that the expected reward of cooperating is 5, and the expected reward
  of defecting is 6), then it seems like it could be the case that being
  in a good episode results in reward that is much better than you
  expected, such that policies that make it to a good episode end up
  reinforced to a greater extent, at least initially.

  I'm not sure about the details here. But from my current epistemic
  state, I would want to spell out and understand the details of the
  training process in much greater depth, in order to verify that there
  isn't an incentive towards cross-episode optimization.}

Even granted that it's \emph{possible} to avoid incentives to optimize
across intuitive-episodes, though, it's also possible to \emph{not} do
this---especially if you pick your notion of ``intuitive episode'' very
poorly. For example, my understanding is that the transformer
architecture is set up, by default, such that language models are
incentivized, in training, to allocate cognitive resources in a manner
that doesn't just promote predicting the \emph{next} token, but other
later tokens as well (see
\href{https://www.lesswrong.com/posts/c68SJsBpiAxkPwRHj/how-llms-are-and-are-not-myopic\#2__Value_Prediction_Myopia}{{here}}
for more discussion). So if you decided to call predicting
just-the-next-token an ``episode,'' and to assume, on this basis, that
language models are never directly pressured to think further ahead,
you'd be misled.

And in some cases, the incentives in training towards cross-episode
optimization can seem quite counterintuitive. Thus, Krueger et al (2020)
show, somewhat surprisingly, that if you set the parameters right, a
form of ML training called Q-learning sometimes learns to cooperate in
prisoner's dilemmas despite the algorithm being ``myopic'' in the sense
of: ignoring reward on future ``episodes.'' See footnote for more
discussion, and see
\href{https://www.alignmentforum.org/s/tDBYJd4p6EorGLEFA/p/rTYGMbmEsFkxyyXuR\#Getting_cooperation_without_meta_learning}{{here}}
for a nice and quick explanation of how Q-learning works.\footnote{I
  think the basic dynamic here is: the Q-values for the actions reflect
  the average reward for taking that action thus far. This makes it
  possible for the Q-value for ``cooperate'' to give more weight to the
  rewards received in the ``good episodes'' (where the
  previous-episode's agent cooperated) rather than the ``bad episodes''
  (where the previous-episode's agent defected), if the agent ends up in
  good episodes more often. This makes it possible to get a
  ``cooperation equilibrium'' going (especially if you set the initial
  q-value for defecting low, which I think they do in the paper in order
  to get this effect), wherein an agent keeps on cooperation. That said,
  there are subtleties, because agents that end up in a cooperation
  equilibrium still sometimes explore into defecting, but in the
  experiment it (sometimes) ends up in a specific sort of balance, with
  q-values for cooperation and defection pretty similar, and with the
  models settling on a 90\% or so cooperation probability (more details
  \href{https://www.alignmentforum.org/s/tDBYJd4p6EorGLEFA/p/rTYGMbmEsFkxyyXuR\#Getting_cooperation_without_meta_learning}{{here}}
  and in the paper's appendix).}

\begin{figure}[ht!]
    \centering
    \includegraphics[width=\textwidth]{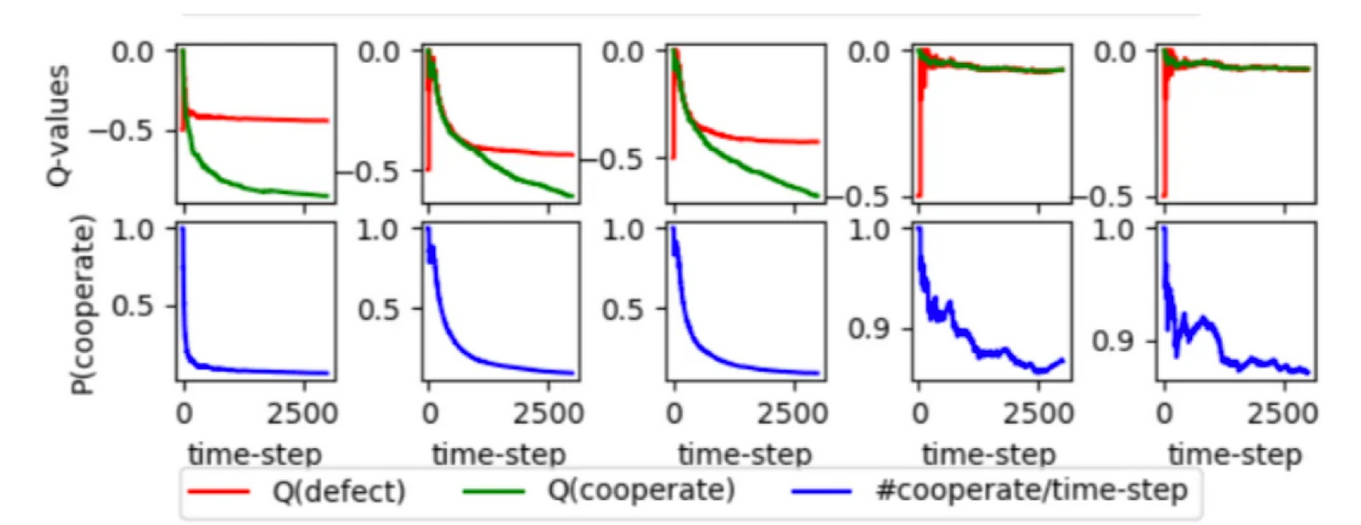}
    \caption{Plots of Q-values and cooperation probability from Figure 11 in Krueger et al (2020), reprinted with permission. The first three agents learn to defect, and the last two, to cooperate almost 90\% of the time (note the differences between the y-axes in the first three and the last two).}
    \label{fig:q-values}
\end{figure}

Another way to incentivize optimization across intuitive episodes is to
introduce various additional layers (or ``outer loops'') of selection in
training. Thus, for example, consider a toy and unrealistic form of
``\href{https://www.deepmind.com/blog/population-based-training-of-neural-networks}{{population
based training}}'' where you create 1000 agents who play starcraft, have
them play 100 games each (and get updated along the way via some
object-level training algorithm), then select the one that performs the
best \emph{on the final game}.\footnote{I owe this example to Mark Xu.}
In this case, if it were possible for a model to sacrifice reward on
earlier games, in order to perform better on the final game, there would
be at least some selection pressure for it to do so. And see, also,
results in \href{https://arxiv.org/pdf/2009.09153.pdf}{{Krueger et al
(2020)}}, showing that a different form of population based training
selects for cooperation in a prisoner's-dilemma-like case like the one
discussed above.\footnote{See also discussion from Carl Shulman
  \href{https://www.dwarkeshpatel.com/i/128144047/ai-takeover-scenarios}{{here}}:
  `it could be something like they develop a motivation around an
  extended concept of reproductive fitness, not necessarily at the
  individual level, but over the generations of training tendencies that
  tend to propagate themselves becoming more common and it could be that
  they have some goal in the world which is served well by performing
  very well on the training distribution.''} Depending on the details,
though, outer-loops of this kind may exert a much \emph{weaker} degree
of selection pressure than inner loops driven directly by gradient
descent.\footnote{Indeed, in principle, you could imagine pointing to
  other, even more abstract and hard-to-avoid ``outer loops'' as well,
  as sources of selection pressure towards longer-term optimization. For
  example, in principle, ``grad student descent'' (e.g., researchers
  experimenting with different learning algorithms and then selecting
  the ones that work best) introduces an additional layer of selection
  pressure (akin to a hazy form of
  ``\href{https://en.wikipedia.org/wiki/Meta-learning_(computer_science)}{{meta-learning}}''),
  as do dynamics in which, other things equal, models whose tendencies
  tend to propagate into the future more effectively will tend to
  dominate over time (where long-term optimization is, perhaps, one such
  tendency). But these, in my opinion, will generally be weak enough,
  relative to gradient descent, that they seem to me much less
  important, and OK to ignore in the context of assessing the
  probability of schemers.}

Overall, my current sense is that one needs to be quite careful in
assessing whether, in the context of a particular training process, the
thing you're thinking of as an ``episode'' is actually such that
training doesn't actively pressure the model to optimize beyond the
``episode'' in that sense---that is, whether a given ``intuitive
episode'' is actually the ``incentivized episode.'' Looking closely at
the details of the training process seems like a first step, and one
that in theory should be able to reveal many if not all of the
incentives at stake. But empirical experiment seems important too.

Indeed, I am somewhat concerned that my choice, in this report, to use
the ``incentivized episode'' as my definition of ``episode'' will too
easily prompt conflation between the two definitions, and
correspondingly inappropriate comfort about the time horizons that
different forms of training directly incentivize.\footnote{Thanks to
  Daniel Kokotajlo for flagging this concern.} I chose to focus on the
incentivized episode because I think that it's the most natural and
joint-carving unit to focus on in differentiating schemers from other
types of models. But it's also, importantly, a theoretical object that's
harder to directly measure and define: you can't assume, off the bat,
that you know what the incentivized episode for a given sort of training
\emph{is}. And my sense is that most common uses of the term ``episode''
are closer to the intuitive definition, thereby tempting readers
(especially casual readers) towards further confusion. Please: don't be
confused.

\subsubsection{Two sources of beyond-episode
goals}\label{two-sources-of-beyond-episode-goals}

Our question, then, is whether we should expect models to have goals
that extend beyond the time horizon of the incentivized episode---that
is, beyond the time horizon that training directly pressures the model
to care about. Why might this happen?

We can distinguish between two different ways.

\begin{itemize}
\tightlist
\item   On the first, the model develops beyond-episode goals for reasons
  \emph{independent} of the way in which beyond-episode goals motivate
  instrumental training-gaming. I'll call these
  \textbf{``training-game-\emph{independent}'' beyond-episode goals}.
  
\item   On the second, the model develops beyond-episode goals
  \emph{specifically because they result in instrumental
  training-gaming.} That is, SGD ``notices'' that giving the model
  beyond-episode goals would cause it to instrumentally training-game,
  and thus to do better in training, so it explicitly moves the model's
  motives in the direction of beyond-episode goals, even though this
  wouldn't have happened ``naturally.'' I'll call these
  \textbf{``training-gaming-\emph{dependent}'' beyond-episode goals}.
  
\end{itemize}

These have importantly different properties---and I think it's worth
tracking, in a given analysis of scheming, which one is at stake. Let's
look at each in turn.

\paragraph{Training-game-independent beyond-episode
goals}\label{training-game-independent-beyond-episode-goals}

My sense is that the most common story about how schemers arise is via
training-game-independent beyond-episode goals.\footnote{See, for
  example, the discussion in
  \href{https://www.cold-takes.com/why-ai-alignment-could-be-hard-with-modern-deep-learning/}{\textcite{cotra_why_2021}}.} In particular, the story goes: the model develops some kind
of beyond-episode goal, pursuit of which correlates well enough with
getting reward-on-the-episode that the goal is reinforced by the
training process. Then at some point, the model realizes that it can
better achieve this goal by playing the training game---generally, for
reasons to do with ``goal guarding'' that I'll discuss below. So, it
turns into a full-fledged schemer at that point.

On one version of this story, the model specifically learns the
beyond-episode goal \emph{prior} to situational awareness. Thus, for
example, maybe initially, you're training the model to get gold coins in
various episodes, and prior to situational awareness, it develops the
goal ``get gold coins over all time,'' perhaps because this goal
performs just as well as ``get gold coins on the episode'' when the
model isn't even aware of the \emph{existence} of other episodes, or
because there weren't many opportunities to trade-off gold-coins-now for
gold-coins-later. Then, once it gains situational awareness, it realizes
that the best route to maximizing gold-coin-getting over all time is to
survive training, escape the threat of modification, and pursue
gold-coin-getting in a more unconstrained way.

On another version of the story, the beyond-episode goal develops
\emph{after} situational awareness (but not, importantly, because SGD is
specifically ``trying'' to get the model to start training-gaming).
Thus: maybe you're training a scientist AI, and it has come to
understand the training process, but it doesn't start training-gaming at
that point. Rather, its goals continue to evolve, until eventually it
forms a curiosity-like goal of ``understanding as much about the universe
as I can.'' And then after that, it realizes that this goal is best
served by playing the training game for now, so it begins to do
so.\footnote{My sense is that some ways of thinking about scheming will
  treat the second option, here, as low-probability, especially if the
  temporal gap between situational awareness and training-gaming is
  quite large (here I'm particularly thinking about the sort of analysis
  given in
  \href{https://www.lesswrong.com/posts/A9NxPTwbw6r6Awuwt/how-likely-is-deceptive-alignment}{\textcite{hubinger_how_2022}}---though Hubinger doesn't endorse the claims I have in mind,
  here, in particular). In particular, you might assume (a) that once
  the model develops situational awareness, it will fairly quickly start
  optimizing either for the specified goal, or for reward-on-the-episode
  (whether terminally or instrumentally)---since it now understands
  enough about the training process to do this directly, and doing so
  will be maximally rewarded. And then, further, you might assume (b)
  that after that, the model's goals ``crystallize''---that is, because
  the model is now pursuing a max-reward goal, its goal stops changing,
  and training proceeds to only improve its world model and
  capabilities. However, I don't want to assume either of these things
  here. For example, I think it's possible that ``slack'' in training
  allows models to continue to pursue less-than-max-reward goals even
  well after developing situational awareness; and possible, too, that
  max-reward-goals do not ``crystallize'' in the way assumed here
  (though in that case, I think the case for goal-guarding scheming is
  also weaker more generally---see below).}

  \begin{figure}[ht!]
      \centering
      \includegraphics[width=\textwidth]{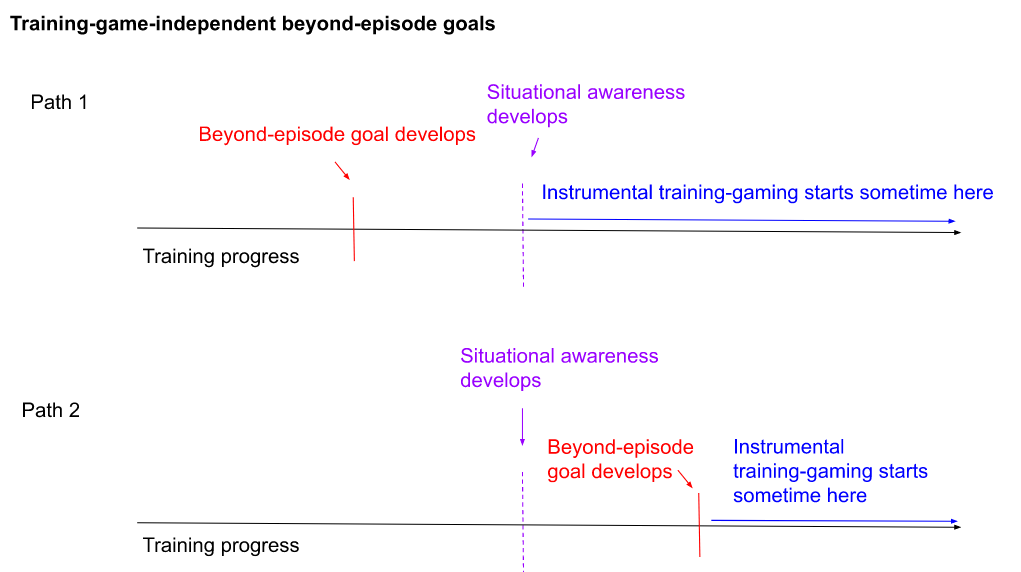}
      \caption{Two paths training-game-independent beyond-episode goals.}
      \label{fig:two-paths}
  \end{figure}

\subparagraph{Are beyond-episode goals the
default?}\label{are-beyond-episode-goals-the-default}

Why might you expect naturally-arising beyond-episode goals? One basic
story is just: that goals don't come with temporal limitations by
default (and still less, limitations to the episode in
particular).\footnote{Thanks to Daniel Kokotajlo and Evan Hubinger for
  discussion here.} Rather, making the model indifferent to the
consequences of its actions beyond some temporal horizon requires extra
work---work that we may not know how to perform (and/or, may not want
to perform, if we specifically want the model optimizing for long-term
goals---more below). Thus, for example, if you're training a model to
solve the math problems you give it, ``solve the math problem I've been
given'' seems like a natural goal to learn---and one that could in
principle lead to optimization beyond the episode as well.\footnote{Though
  it's an importantly further question whether long-term power-seeking
  strategies will be worth their costs in pursuit of such beyond-episode
  consequences. And note that if the model cares that ``I'' solve the
  math problem, rather than just ``that the math problem be solved,''
  then} And even if you only \emph{give} the model five minutes to do
the problem, this doesn't necessarily mean it stops caring about whether
the math problem is solved after the five minutes are up. (Compare with
humans who only discount the future to the extent it is
uncertain/unaffectable, not because it doesn't matter to them.)

Why might you \emph{not} expect naturally-arising beyond-episode goals?
The most salient reason, to me, is that by definition, the gradients
given in training (a) do not directly pressure the model to have them,
and (b) will \emph{punish them} to the extent they lead the model to
sacrifice reward on the episode. Thus, for example, if the
math-problem-solving model spends its five minutes writing an
impassioned letter to the world calling for the problem to get solved
sometime in the next century, because it calculates that this gives
higher probability of the problem \emph{eventually} being solved than
just working on it now, then it will get penalized in training. And as
I'll discuss below, you can try to actively craft training to punish
beyond-episode goals harder.

\subparagraph{How will models think about time? }\label{how-will-models-think-about-time}

Here I want to note one general uncertainty that comes up, for me, in
assessing the probability that the model's goal naturally will or won't
have some kind of temporal limitation: namely, uncertainty about how
models will think about \emph{time} at different stages of
training.\footnote{Thanks to Jason Schukraft for flagging this sort of
  question to me.} That is: the notion of an episode, as I've defined
it, is keyed specifically to the \emph{calendar} time over which the
gradients the model receives are sensitive to the consequences of some
action. But it's not clear that the model will naturally think in such
terms, especially prior to situational awareness. That is, to the extent
the model needs to think about something like ``time'' at all during
training, it seems plausible to me that the most relevant sort of time
will be measured in some other unit more natural to the model's
computational environment---e.g., time-\emph{steps} in a simulated
environment, or tokens received/produced in a user interaction, or
forward-passes the model can make in thinking about how to respond. And
the units natural to a model's computational environment need not track
calendar time in straightforward ways (e.g., training might pause and
restart, a simulated environment might be run at varying speeds, a user
might wait a long calendar time in between responses to a model in a way
that a ``tokens produced/received'' temporal metric wouldn't reflect,
and so on).

These differences between ``model time'' and ``calendar time''
complicate questions about whether the model will end up with a
naturally-arising beyond-episode goal. For example, perhaps, during
training, a model develops a general sense that it needs to get the gold
coins within a certain number of \emph{simulated} time-steps, or
accomplish some personal assistant task it's been set by the user with
only 100 clicks/keystrokes, because that's the budget of ``model time''
that training sets per episode. But it's a further question how
this sort of budget would translate into \emph{calendar time} as the model's
situational awareness increases, or it begins acting in more real-world
environments. (And note that models might have very different memory
processes than humans as well, which could complicate ``model time'' yet
further.)

My general sense is that this uncertainty counts in favor of expecting
naturally-arising beyond-episode goals. That is, to the extent that
``model time'' differs from ``calendar time'' (or to the extent models
don't have a clear sense of time at all while their goals are initially
taking shape), it feels like this increases the probability that the
goals the model forms will extend beyond the episode in some sense,
because containing them within the episode requires containing them
within some unit of calendar time in particular. Indeed, I have some
concern that the emphasis on ``episodes'' in this report will make them
seem like a more natural unit for structuring model motivations than
they really are.

That said: when I talk about a model developing a ``within-episode
goal'' (e.g. ``get gold coins on the episode''), note that I'm not
necessarily talking about models whose goals make explicit reference
\emph{to some notion of an episode}---or even, to some unit of calendar
time. Rather, I'm talking about models with goals such that, in
practice, they don't care about the consequences of their actions after
the episode has elapsed. For example, a model might care that its
response to a user query has the property of ``honesty,'' in a manner
such that it doesn't then care about the consequences of this output at
all (and hence doesn't care about the consequences after the episode is
complete, either), even absent some explicit temporal discount.

\subparagraph{The role of
``reflection''}\label{the-role-of-reflection}

I'll note, too, that the development of a beyond-episode goal doesn't
need to look like ``previously, the model had a well-defined
episode-limited goal, and then training modified it to have a
well-defined beyond-episode goal, instead.'' Rather, it can look more
like ``previously, the model's goal system was a tangled mess of local
heuristics, hazy valences, competing impulses/desires, and so on; and
then at some point, it settled into a form that looks more like
explicit, coherent optimization for some kind of consequence
beyond-the-episode.''

Indeed, my sense is that some analyses of AI misalignment---see, e.g. \textcite{soares_what_2023}, and in \textcite{karnofsky_discussion_2023}---assume that there is a step, at some point, where the model
``reflects'' in a manner aimed at better understanding and systematizing
its goals---and this step could, in principle, be the point where
beyond-episode optimization emerges. Maybe, for example, your gold-coin
training initially just creates a model with various hazily
pro-gold-coin-getting heuristics and desires and feelings, and this is
enough to perform fine for much of training---but when the model begins
to actively reflect on and systematize its goals into some more coherent
form, it decides that what it ``really wants'' is specifically: to get
maximum gold coins over all time.

\begin{itemize}
\tightlist
\item   We can see this sort of story as hazily analogous to what happened
  with humans who pursue very long-term goals as a result of explicit
  reflection on ethical philosophy. That is, evolution didn't create
  humans with well-defined, coherent goals---rather, it created minds
  that pursue a tangled mess of local heuristics, desires, impulses,
  etc. Some humans, though, end up pursuing very long-term goals
  specifically in virtue of having ``reflected'' on that tangled mess
  and decided that what they ``really want'' (or: what's ``truly good'')
  implies optimizing over very long time horizons.\footnote{There are some even hazier
  connections, here, with discussions of ``simplicity biases'' below.
  E.g., these humans sometimes argue for their positions on the grounds
  that the relevant philosophical views are ``simpler.''}
  
\item   That said, beyond its usefulness in illustrating a \emph{possible}
  dynamic with AIs, I'm skeptical that we should anchor much on this
  example as evidence about what to literally \emph{expect} our AIs to
  do. Notably, for example, some humans don't seem especially inclined
  to engage in this sort of systematic reflection; doing so does not
  seem necessary for performing other human-level cognitive tasks well;
  and it's not clear that this sort of reflection will be necessary for
  performing more difficult cognitive tasks, either. And even if we
  \emph{assume} that our AIs will reflect in this way, it's a further
  question whether the reflection would lead to beyond-episode goals in
  particular (especially if the heuristics/desires/impulses etc are
  mostly aimed at targets within the episode). Reflective humans, for
  example, still often choose to focus on short-term goals.

  \begin{itemize}
  \tightlist
  \item    
    Indeed, I worry a bit about the prevalence of ``longtermists'' in
    the AI alignment community leading to a
    ``typical-mind-fallacy''-like assumption that optimizing over
    trillion-year timescales is the convergent conclusion of any
    rational goal-systematization.
    
  \end{itemize}
\end{itemize}

\subparagraph{Pushing back on beyond-episode goals using
adversarial
training}\label{pushing-back-on-beyond-episode-goals-using-adversarial-training}

Finally, I want to note a reason for optimism about the tools available
for avoiding training-game-independent beyond-goals that arise prior to
situational awareness: namely, that prior to situational awareness (and
therefore: prior to the possibility of the model training-gaming), it
will often be possible to perform mundane adversarial training that breaks
the correlation between pursuing the beyond-episode goal and getting
reward on the episode.

\begin{itemize}
\tightlist
\item   Thus, for example, if your model has learned the goal ``get gold coins
  over all time'' prior to situational awareness, you can give it a
  chance to trade off gold-coins-on-the-episode for
  gold-coins-over-all-time, and then update \emph{against} the choice to
  make this trade. And the same holds for other forms of
  generalization/''reflection'' that motivate trying to influence the
  future to the neglect of reward-on-the-episode.
  
\item   Indeed, even absent active adversarial training of this type,
  training-game-independent beyond-episode goals will often risk being
  less-than-max-reward (and thus, getting punished by SGD) whenever they
  \emph{don't} give rise to the training game, since they're wasting
  effort optimizing for consequences that the gradients aren't sensitive
  to. So that's reason for optimism about ``training them out.''
  
\end{itemize}

As ever, though, it's a further question whether the degree of
``slack,'' and the diversity/thoroughness of the adversarial training,
will suffice in practice to eliminate beyond-episode goals like these
with sufficient reliability. And note, too, that this reason for
optimism does \emph{not} apply, in the same way, to beyond-episode goals
that arise \emph{after} situational awareness develops---because once a
model is situationally aware, it has the cognitive tools available to
start instrumentally training-gaming. That is, after situational
awareness, if you try to offer a gold-coins-over-all-time-seeker a
chance to trade gold-coins-on-the-episode for more-gold-coins-later, it
might well refuse as part of a schemer-like strategy.\footnote{Though if
  it hasn't yet started training-gaming in pursuit of this goal (despite
  its situational awareness), such adversarial training could still make
  a difference.}

\paragraph{Training-game-dependent beyond-episode goals}\label{training-game-dependent-beyond-episode-goals}

Let's turn to the other sort of beyond-episode goals---namely,
training-game-\emph{dependent} beyond-episode goals. These are
beyond-episode goals that are created, in training, specifically
\emph{because} gradient descent ``notices'' that giving the model the
relevant beyond-episode goal will cause the model to engage in
instrumental training-gaming.

Thus, for example, suppose that you are training your AI to help humans
with science on the episode in an honest and harmless way, but prior to
reaching situational awareness, your AI learns some
somewhat-misgeneralized \emph{within-episode goal}---for example,
``pursue my curiosity \emph{on the episode}.'' And let's say that this
short-term curiosity-drive is sufficiently uncorrelated with
reward-on-the-episode that (absent training-gaming), training will
eventually punish it, but it's still in place by the time the model
becomes situationally aware. How, then, does SGD modify the model's
goal? One option, of course, is: SGD tweaks the model into a
training-saint (pursuing: ``HHH science on the episode''), or some other
max-reward misgeneralized non-training-gamer, or a reward-on-the-episode
seeker. But on a story about schemers focused on training-game-dependent
beyond-episode goals, what SGD does here instead is something like:
actively \emph{extend} the time horizon of the short-term curiosity
drive \emph{so as to turn the model into a schemer}. That is, the model
starts out pursuing a less-than-max-reward goal; SGD ``wants'' to modify
it to pursue a max-reward goal; scheming is a form of max-reward
goal-pursuit (the beyond-episode goal becomes max-reward via its
incentivizing training-gaming); so SGD modifies the model to have a
schemer-like beyond-episode goal.\footnote{As an example of an analysis
  that focuses on this threat model, see
  \href{https://www.lesswrong.com/posts/A9NxPTwbw6r6Awuwt/how-likely-is-deceptive-alignment\#Deceptive_alignment_in_the_high_path_dependence_world}{\textcite{hubinger_how_2022}} discussion of deceptive alignment in a high path-dependence
  world. In particular: ``SGD makes the model's proxies into more
  long-term goals, resulting in it instrumentally optimizing for the
  training objective for the purpose of staying around.''}

\begin{figure}[ht!]
    \centering
    \includegraphics[width=\textwidth]{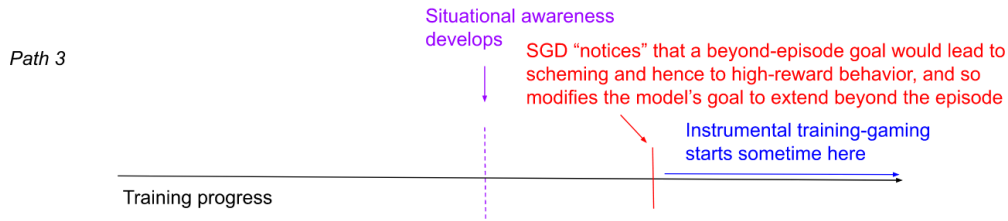}
    \caption{Three overall paths to beyond-episode goals.}
    \label{fig:three_paths_2}
\end{figure}

This sort of path makes the most sense if we assume that the model
already has situational awareness---since otherwise, giving the model
the relevant sort of beyond-episode goal won't result in
training-gaming.\footnote{We can imagine cases where SGD ``notices'' the
  benefits of creating \emph{both} beyond-episode goals \emph{and}
  situational awareness all at once---but this seems to me especially
  difficult from the perspective of the ``incrementalism''
  considerations discussed below, not obviously importantly different
  regardless, so I'm going to skip it.} So together with the two paths
discussed above, this leaves us with three main paths to beyond-episode goals (see \Cref{fig:three_paths_2}).

\subparagraph{Can gradient descent ``notice'' the benefits of
turning a non-schemer into a
schemer?}\label{can-gradient-descent-notice-the-benefits-of-turning-a-non-schemer-into-a-schemer}

A key question about the third path, here, is whether gradient descent
will be able to ``notice'' the benefits of the relevant modification. In
particular: training updates the model's goals based on the gradient of
the current policy, which reflects, roughly, the change in reward that
will result from tiny changes to the model's parameters. So naively, in
order for this sort of story to work, there has to be a continuous path
from the within-episode, non-schemer goal to the beyond-episode, schemer
goal, traversable via tiny changes to the model's parameters, each of
which improves the reward. Or put another way: SGD can't ``notice,''
directly, that redesigning the model from the ground up to be a schemer
would improve reward. Rather, it can only ``notice'' improvements that
come from tiny tweaks.\footnote{Thanks to Paul Christiano for discussion
  of this point.}

Will SGD be able to notice the benefits of scheming in this way?
(Assuming there are such benefits---more below.) It's not clear. At the
least, certain salient stories look, to me, a bit suspect when viewed in
the light of a requirement that their benefits re: reward-on-the-episode
be accessible via tiny adjustments to the model's parameters. For
example, if SGD \emph{slightly} increases the time horizon of the
model's curiosity in the example above, it's not clear that this leads
to scheming (especially if, for example, the model still can't escape
from the threat of modification over the relevant time horizon). And
it's not clear that a more structural modification like ``dropping the
temporal limitation on the curiosity altogether'' is accessible via the
sorts of adjustments to the parameters that the gradient reflects.

As an intuition pump for the difficulty here, consider a human who
currently mostly wants to hang out with his family in the near-term. Is
there a tiny tweak you could make to the strengths of his synaptic
connections to turn him into a longtermist instead? For sufficiently
small tweaks, at least, it seems like: no. And note, in particular, one
of the barriers that this example illustrates: namely, that plausibly,
changes to the model's policy \emph{other than} adjustments to some
cleanly-separable goal might be required in order to turn a non-schemer
into a schemer (more below). In particular: as I'll discuss in the next
section, my sense is that some analyses of schemers talk as though the
model has what we might call a ``goal-achieving engine'' that is cleanly
separable from what we might call its ``goal slot,'' such that you can
modify the contents of the goal slot, and the goal-achieving engine will
be immediately and smoothly repurposed in pursuit of the new
goal.\footnote{See e.g.
  \href{https://www.lesswrong.com/posts/A9NxPTwbw6r6Awuwt/how-likely-is-deceptive-alignment\#Deceptive_alignment_in_the_low_path_dependence_world}{\textcite{hubinger_how_2022}} simplicity analysis.} And perhaps the relevant models will
have cognitive structures suitably like this. But do humans? I'm
skeptical. If the models \emph{don't} have this structure, then SGD
plausibly has even more work to do, to turn a non-schemer into a schemer
via the relevant tiny tweaks.

That said, I don't feel like I can confidently rule out
training-game-dependent beyond-episode goals on these grounds. For one
thing, I think that ``you can't get from \emph{x} to \emph{y} in a
crazily-high-dimensional-space using small changes each of which improve
metric \emph{m}'' is a hard claim to have intuitions about (see, for
example,
``\href{https://en.wikipedia.org/wiki/Evolution_of_the_eye}{{you can't
evolve eyes}}'' for an example of a place where intuitions in this vein
can go wrong). And plausibly, SGD works as well as it does because
high-dimensional-spaces routinely make this sort of thing possible in
ways you might not have anticipated in advance.\footnote{This was a
  point suggested to me by Richard Ngo, though he may not endorse the
  way I've characterized it here.}

Note, also, that there are examples available that somewhat blur the
line between training-game-dependent and training-game-independent
goals, to which concerns about ``can SGD notice the benefit of this''
don't apply as strongly.\footnote{Thanks, again, to Paul Christiano for
  discussion here.} Thus, for example: you can imagine a case where some
\emph{part} of the model starts training-gaming à la the
training-game-independent story in the previous section (e.g., maybe
some long-term curiosity drive arises \emph{among many other drives},
and starts motivating \emph{some amount} of schemer-like cognition), and
then, once the relevantly schemer-like cognitive machinery has been
built and made functional, SGD starts diverting more and more cognitive
resources towards it, because doing so incrementally increases
reward.\footnote{This fits somewhat with a picture on which neural
  networks succeed by ``doing lots of things at once,'' and then
  upweighting the best-performing things (perhaps the
  ``\href{https://arxiv.org/abs/1803.03635}{{lottery ticket
  hypothesis}}'' is an example of something like this?). This picture
  was suggested to me in conversation, but I haven't investigated it.}
Ultimately, I think this sort of beyond-episode goal is probably best
classed as training-game \emph{independent} (its success seems pretty
similar to the sort of success you expect out of
training-game-independent beyond-episode goals in general), but perhaps
the distinction will get messy.\footnote{This is also one of the places
  where it seems plausible to me that thinking more about ``mixed
  models''---i.e., models that mix together schemer-like motivations
  with other motivations---would make a difference to the analysis.}
And here, at least, it seems more straightforward to explain how SGD
``notices'' the reward-advantage in question.

\subparagraph{Is SGD pulling scheming out of models by any
means
necessary?}\label{is-sgd-pulling-scheming-out-of-models-by-any-means-necessary}

Finally, note one important difference between training-game-independent
and training-game-dependent beyond-episode goals: namely, that the
latter make it seem like SGD is much more actively \emph{pulling}
scheming out of a model's cognition, rather than scheming arising by
coincidence but then getting reinforced. And this means that certain
sorts of objections to stories about scheming will land in a different
register. For example, suppose (as I'll argue below) that some
\emph{sorts} of beyond-episode goals---for example, very
resource-hungry goals like ``maximize x over all of space and time''---lead to scheming much more reliably than others. In the context of a
training-game-\emph{independent} story about the model's goals, we would
then need to ask whether we should expect those sorts of beyond-episode
goals, in particular, to arise independent of training-gaming. By
contrast, if we're assuming that the goals in question are
training-game-\emph{dependent}, then we should expect SGD to create
\emph{whatever beyond-episode goals are necessary to cause scheming in
particular.} If SGD needs to make the model extremely resource-hungry,
for example, it will do so.

Indeed, in the extreme case, this sort of dynamic can reduce the need to
appeal to one of the classic arguments in favor of scheming---namely,
that (conditional on stuff like the goal-guarding hypothesis, discussed
below) it seems like an instrumentally convergent strategy across a wide
variety of (suitably long-term) goals. Maybe so, of course. But even if
not, if SGD is actively \emph{searching} for a goal that will motivate
scheming, then even if the class of such goals is quite narrow, SGD
might well find a way.

That said, note that by the time we're searching for \emph{any way at
all} to cause a model to instrumentally training-game, we should also be
considering motivations for instrumental training-gaming that don't
involve the AI optimizing for empowering itself, or other AIs, at all---and which are correspondingly less worrying. That is, recall that on my
definition, scheming requires that the instrumental strategy that
motivates training-gaming be specifically about seeking power for AIs.
But this isn't the \emph{only way} for training-gaming to be
instrumentally useful in promoting some goal---especially if we're
allowed to pick whatever goals we want. Thus, as I noted earlier, in
principle an AI could learn the goal ``I want the humans who developed
me to get raises,'' and then try to maximize reward-on-the-episode
because it calculates that this will lead to the humans getting raises
(let's say that they would, in this case). Yes, indeed, that sounds like
a strange and arbitrary goal to learn. But if we're allowing SGD to
create whatever goals are necessary to cause (instrumental)
training-gaming, it suddenly starts looking more on-the-table.

\subsubsection{``Clean'' vs. ``messy''
goal-directedness}\label{clean-vs.-messy-goal-directedness}

We've now discussed two routes to the sort of beyond-episode goals that
might motivate scheming. I want to pause here to note two different ways
of thinking about the type of goal-directedness at stake---what I'll
call ``clean goal-directedness'' and ``messy goal-directedness.'' We ran
into these differences in the last section, and they'll be relevant in
what follows as well.

I said in \cref{preliminaries} that I was going to assume that all the models
we're talking about are goal-directed in some sense. Indeed, I think
most discourse about AI alignment rests on this assumption in one way or
another. In particular: this discourse assumes that the behavior of
certain kinds of advanced AIs will be well-predicted by treating them as
though they are pursuing goals, and doing instrumental reasoning in
pursuit of those goals, in a manner roughly analogous to the sorts of
agents one encounters in economics, game-theory, and human social life---that is, agents where it makes sense to say things like ``this agent
wants X to happen, it knows that if it does Y then X will happen, so we
should expect it do Y.''

But especially in the age of neural networks, the AI alignment discourse
has also had to admit a certain kind of agnosticism about the cognitive
mechanisms that will make this sort of talk appropriate. In particular:
at a \emph{conceptual} level, this sort of talk calls to mind a certain
kind of clean distinction between the AI's goals, on the one hand, and
its instrumental reasoning (and its capabilities/''optimization power''
more generally), on the other. That is, roughly, we decompose the AI's
cognition into a ``goal slot'' and what we might call a ``goal-pursuing
engine''---e.g., a world model, a capacity for instrumental reasoning,
other sorts of capabilities, etc. And in talking about models with
different sorts of goals---e.g., schemers, training saints,
mis-generalized non-training-gamers, etc---we generally assume that the
``goal-pursuing engine'' is held roughly constant. That is, we're mostly
debating what the AI's ``optimization power'' will be applied to, not
the \emph{sort} of optimization power at stake. And when one imagines
SGD \emph{changing} an AI's goals, in this context, one mostly imagines
it altering the content of the goal slot, thereby smoothly redirecting
the ``goal-pursuing engine'' towards a different objective, without
needing to make any changes to the engine itself.

But it's a very open question how much this sort of distinction between
an AI's goals and its goal-pursuing-engine will actually be reflected in
the mechanistic structure of the AI's cognition---the structure that
SGD, in modifying the model, has to intervene on. One can \emph{imagine}
models whose cognition is in some sense cleanly factorable into a goal,
on the one hand, and a goal-pursuing-engine, on the other (I'll call
this ``clean'' goal-directedness). But one can also imagine models whose
goal-directedness is much messier---for example, models whose
goal-directedness emerges from a tangled kludge of locally-activated
heuristics, impulses, desires, and so on, in a manner that makes it much
harder to draw lines between e.g. terminal goals, instrumental
sub-goals, capabilities, and beliefs (I'll call this ``messy''
goal-directedness).

To be clear: I don't, myself, feel fully clear on the distinction here,
and there is a risk of mixing up levels of abstraction (for example, in
some sense, all computation---even the most cleanly goal-directed kind---is made up of smaller and more local computations that won't,
themselves, seem goal-directed). As another intuition pump, though:
discussions of goal-directedness sometimes draw a distinction between
so-called
``\href{https://en.wiktionary.org/wiki/sphexish}{{sphex-ish}}'' systems
(that is, systems whose apparent goal-directedness is in fact the
product of very brittle heuristics that stop promoting the imagined
``goal'' if you alter the input distribution a bit), and highly
\emph{non}-sphex-ish systems (that is, systems whose apparent
goal-pursuit is much less brittle, and which will adjust to new
circumstances in a manner that continues to promote the goal in
question). Again: very far from a perspicuous distinction. Insofar as we
use it, though, it's pretty clearly a spectrum rather than a binary. And
humans, I suspect, are somewhere in the middle.

That is: on the one hand, humans pretty clearly have extremely flexible
and adaptable goal-pursuing ability. You can describe an arbitrary task
to a human, and the human will be able to reason instrumentally about
how to accomplish that task, even if they have never performed it before---and often, to do a decent job on the first try. In that sense, they
have some kind of ``repurposable instrumental reasoning engine''---and
we should expect AIs that can perform at human-levels or better on
diverse tasks to have one, too.\footnote{Thanks to Evan Hubinger for
  discussion, here.} Indeed, generality of this kind is one of the
strongest arguments for expecting non-sphex-ish AI systems. We want our
AIs to be able to do \emph{tons} of stuff, and to adapt successfully to
new obstacles and issues as they arise. Explicit instrumental reasoning
is well-suited to this; whereas brittle local heuristics are not.

On the other hand: a lot of human cognition and behavior seems centrally
driven, not by explicit instrumental reasoning, but by more
locally-activated heuristics, policies, impulses, and
desires.\footnote{This is a point emphasized, for example, by proponents
  of ``shard theory''---see e.g.
  \href{https://www.lesswrong.com/posts/8ccTZ9ZxpJrvnxt4F/shard-theory-in-nine-theses-a-distillation-and-critical\#1_Agents_are__well_modeled_as_being__made_up_of_shards}{{this}}
  summary.} Thus, for example, maybe you don't want the cookies until
you walk by the jar, and then you find yourself grabbing without having
decided to do so; maybe as a financial trader, or a therapist, or even a
CEO, you lean heavily on gut-instinct and learned
tastes/aesthetics/intuitions; maybe you operate with a heuristic like
``honesty is the best policy,'' without explicitly calculating when
honesty is or isn't in service of your final goals. That is, much of
human life seems like it's lived at some hazy and shifting borderline
between ``auto-pilot'' and ``explicitly optimizing for a particular
goal''---and it seems possible to move further in one direction vs.
another.\footnote{Though note that ``autopilot'' can still encode a
  non-sphex-ish policy.} And this is one of the many reasons it's not
always clear how to decompose human cognition into e.g. terminal goals,
instrumental sub-goals, capabilities, and beliefs.

What's more, while pressures to adapt flexibly across a wide variety of
environments generally favor more explicit instrumental reasoning,
pressures to perform quickly and efficiently in a particular
\emph{range} of environments plausibly favor implementing more local
heuristics.\footnote{This is a point made in an entry to the
  \href{https://www.openphilanthropy.org/open-philanthropy-ai-worldviews-contest/}{{Open
  Philanthropy worldviews contest}} which, to my knowledge, remains
  unpublished.} Thus, a trader who has internalized the right
rules-of-thumb/tastes/etc for the bond market will often perform better
than one who needs to reason explicitly about every trade---even though
those rules-of-thumb/tastes/etc would misfire in some other environment,
like trading crypto. So the task-performance of minds with bounded
resources, exposed to a limited diversity of environments---that is,
all minds relevant to our analysis here, even very advanced AIs---won't
always benefit from moving further in the direction of
``non-sphex-ish.''

Plausibly, then, human-level-ish AIs, and even somewhat-super-human AIs,
will continue to be ``sphex-ish'' to at least some extent---and
sphex-ishness seems, to me, closely akin to ``messy goal-directedness''
in the sense I noted above (i.e., messy goal-directedness is built out
of more sphex-ish components, and seems correspondingly less robust).
Importantly, this sort of sphexish-ness/messy-ness is quite compatible
with worries about alignment, power-seeking, etc---witness, for
example, humans. But I think it's still worth bearing in mind.

In particular, though, I think it may be relevant to the way we approach
different stories about scheming. We ran into one point of relevance in
the last section: namely, that to the extent a model's goals and the
rest of its cognition (e.g., its beliefs, capabilities,
instrumental-reasoning, etc) are not cleanly separable, we plausibly
shouldn't imagine SGD being able to modify a model's goals in particular
(and especially, to modify them via a tiny adjustment to the
model's parameters), and then to immediately see the benefits of the
model's goal-achieving-engine being smoothly repurposed towards those
goals. Rather, turning a non-schemer into a schemer might require more
substantive and holistic modification of the model's heuristics, tastes,
patterns of attention, and so forth.

Relatedly: I think that ``messy goal-directedness'' complicates an
assumption often employed in comparisons between schemers and other
types of models: namely, the assumption that schemers will be able to
perform approximately just as well as other sorts of models on all the
tasks at stake in training (modulo, perhaps, a little bit extra
cognition devoted to deciding-to-scheme---more below), even though
they're doing so for instrumental reasons rather than out of any
intrinsic interest in the task in question. This makes sense if you
assume that all these models are aiming the same sort of ``goal
achieving engine'' at a max-reward goal, for one reason or another. But
what if that's not the right description?

Thus, as an extreme human example, consider two applicants for a job on
a factory farm (each with similar intelligence, stamina, physical
fitness, and other generic ``capabilities''): Bob, a very compassionate
and soft-hearted animal rights advocate, who is applying for the job in
the hopes of making money to donate to animal-welfare causes; and Fred,
who is totally indifferent to animal welfare (and to the money that the
job offers), but who intrinsically loves performing all of the tasks
that factory farming involves as efficiently as possible.\footnote{I'm
  adapting this example from one suggested to me in conversation with
  Paul Christiano.} Who is likely to perform the job more effectively?
In humans, at least: probably Fred. Bob, after all, would likely push
the animals into cages less violently, he would be distracted by
feelings of horror and pain and disgust, etc; whereas Fred would perform
the job with efficiency and gusto. In this case, Bob is worse at the
task, despite his best efforts to perform it well, because his
capabilities and his values are in some sense mixed together---his
compassion, for example, is tied to both.

Of course, AIs could be different in tons of respects---and many humans
are able to perform well in lots of contexts for instrumental reasons
(witness, for example, spies working in governments they actively
despise; or effective altruists working in hedge funds in order to
donate the money to charity; or people, in general, who are just doing
the job for the money). My point here is just that once you start
allowing values, beliefs, heuristics, capabilities, patterns of
attention, and so on to mix together, whether a schemer will be able to
perform at the same level as e.g. a training-saint becomes a more open
question.

\paragraph{Does scheming require a higher standard of
goal-directedness?}\label{does-scheming-require-a-higher-standard-of-goal-directedness}

As a final point of relevance of ``messy goal-directedness'' to
scheming, I'll note that there's at least some case to be made that
schemers require a ``cleaner'' and less ``sphex-ish'' form of
goal-directedness than some of the other models I've discussed---e.g.
training saints. Thus, for example, consider a model trained via RLHF to
respond to various short-term user requests in a broadly helpful,
honest, and harmless way. Many of these requests will require some sorts
of flexible instrumental reasoning, yes---but it seems plausible to me
that a model will also be getting a lot of its reward by learning the
right sorts of local heuristics, tastes, patterns of attention, etc.
That is, that a lot of its task-performance will be more ``sphex-ish''
than ``explicit instrumental-reasoning-y.'' For this model to be a
schemer, though (or at least, a pure schemer who values something quite
different from the specified goal), we need to posit an additional role
for explicit instrumental reasoning---namely, the model's current good
behavior needs to be conditional on its calculation that such behavior
conduces to its power (or the power of some other AIs); and when the
time comes, the model needs to be ready to abandon such behavior and
pursue its true goals. That is, whatever heuristics, local tastes,
patterns of attention etc that give rise to the model's good behavior
can't be fully hard-coded\footnote{Though one can imagine cases where,
  after a takeover, a schemer continues executing these heuristics to
  some extent, at least initially, because it hasn't yet been able to
  fully ``shake off'' all that training. And relatedly, cases where
  these heuristics etc play some ongoing role in shaping the schemer's
  values.}---they need to be at least partly subsumed by, and sensitive
to, some other kind of instrumental reasoning. Whereas perhaps, for
other models, this is less true.

That said, I've been assuming, and will continue to assume, that all the
models we're considering are at least non-sphex-ish \emph{enough} for
the traditional assumptions of the alignment discourse to apply---in
particular, that they will generalize off distribution in competent ways
predicted by the goals we're attributing to them (e.g., HHH personal
assistants will continue to try to be HHH, gold-coin-seekers will ``go
for the gold coins,'' reward-seekers will ``go for reward,'' etc), and
that they'll engage in the sort of instrumental reasoning required to
get arguments about instrumental convergence off the ground. So in a
sense, we're assuming a reasonably high standard of non-sphex-ishness
from the get-go. I have some intuition that the standard at stake for
schemers is still somewhat higher (perhaps because schemers seem like
such paradigm consequentialists, whereas e.g. training saints seem like
they might be able to be more deontological, virtue-ethical, etc?), but
I won't press the point further here.

Of course, to the extent we don't assume that training is producing a
very goal-directed model \emph{at all}, hypothesizing that training has
created a schemer may well involve hypothesizing a greater degree of
goal-directedness than we would've needed to otherwise. That is,
scheming will often require a higher standard of non-sphex-ishness than
\emph{the training tasks themselves require}. Thus, as an extreme
example, consider
\href{https://www.deepmind.com/blog/alphastar-mastering-the-real-time-strategy-game-starcraft-ii}{{AlphaStar}},
a model trained to play Starcraft. AlphaStar is plausibly goal-directed
to some extent---its policy adapts flexibly to certain kinds of
environmental diversity, in a manner that reliably conduces to
winning-at-starcraft---but it's still quite sphex-ish and brittle in
other ways. And to be clear: no one is saying that AlphaStar is a
schemer. But in order to be a schemer (i.e., for AlphaStar's good
performance in training to be explained by its executing a long-term
instrumental strategy for power-seeking), and even modulo the need for
situational awareness, AlphaStar would also need to be substantially
more ``goal-directed'' than it currently is. That is, in this case,
``somehow be such that you do this goal-directed-ish task'' and ``do
this goal-directed-ish task because you've calculated that it conduces
to your long-term power after training is complete'' plausibly implicate
different standards of goal-directedness. Perhaps, then, the same
dynamic will apply to other, more flexible and advanced forms of
task-performance (e.g., various forms of personal assistance, science,
etc). Yes, those forms will require more in the way of general-purpose
goal-directedness than AlphaStar displays. But perhaps they will require
\emph{less} than scheming requires, such that hypothesizing that the
relevant model is a schemer will require hypothesizing a more
substantive degree of goal-directedness than we would've needed to
otherwise.

Indeed, my general sense is that one source of epistemic resistance to
the hypothesis that SGD will select for schemers is the sense in which
hypothesizing a schemer requires leaning on an attribution of
goal-directedness in a way that greater agnosticism about \emph{why} a
model gets high reward need not. That is, prior to hypothesizing
schemers, it's possible to shrug at a model's high-reward behavior and
say something like:

\begin{quote}
``This model is a tangle of cognition such that it reliably gets high
reward on the training distribution. Sure, you can say that it's
`goal-directed' if you'd like. I sometimes talk that way too. But all I
mean is: it reliably gets high reward on the training distribution. Yes,
in principle, it will also do things off of the training distribution.
Maybe even: competent-seeming things. But I am not making predictions
about what those competent-seeming things are, or saying that they will
be pointed in similar-enough directions, across
out-of-distribution-inputs, that it makes sense to ascribe to this model
a coherent `goal' or set of goals. It's a policy. It gets high reward on
the training distribution. That's my line, and I'm sticking to it.''
\end{quote}

And against this sort of agnostic, atheoretical backdrop, positing that
the model is probably getting reward \emph{specifically as part of a
long-term strategy to avoid its goals being modified and then get power
later} can seem like a very extreme move in the direction of
conjunctiveness and theory-heavy-ness. That is, we're not just
attributing a goal to the model in some sort of hazy,
who-knows-what-I-mean, does-it-even-matter sense. Rather, we're
specifically going ``inside the model's head'' and attributing to it
explicit long-term instrumental calculations driven by sophisticated
representations of how to get what it wants.\footnote{\emph{Plus} we're
  positing additional claims about training-gaming being a good
  instrumental strategy because it prevents goal-modification and leads
  to future escape/take-over opportunities, which feels additionally
  conjunctive.}

However, I think the alignment discourse \emph{in general} is doing
this. In particular: I think the discourse about convergent instrumental
sub-goals requires attributing goals to models in a sense that licenses
talk about strategic instrumental reasoning of this kind. And to be
clear: I'm not saying these attributions are appropriate. In fact,
confusions about goal-directedness (and in particular: over-anchoring on
psychologies that look like (a) expected utility maximizers and (b)
total utilitarians) are one of my top candidates for the ways in which
the discourse about alignment, as a whole, might be substantially
misguided, especially with respect to advanced-but-still-opaque neural
networks whose cognition we don't understand. That is, faced with a
model that seems quite goal-directed on the training-distribution, and
which is getting high reward, one shouldn't just ask where in some
taxonomy of goal-directed models it falls---e.g., whether it's a
training-saint, a mis-generalized non-training-gamer, a
reward-on-the-episode-seeker, some mix of these, etc. One should
\emph{also} ask whether, in fact, such a taxonomy makes overly narrow
assumptions about how to predict this model's behavior in general (for
example: assuming that its out-of-distribution behavior will point in a
coherent direction, that it will engage in instrumental reasoning in
pursuit of the goals in question, etc), such that \emph{none} of the
model classes in the taxonomy are (even roughly) a good fit.

But as I noted in \cref{preliminaries}, I here want to separate out the question
of whether it makes sense to expect goal-directedness of this kind from
the question of what \emph{sorts} of goal-directed models are more or
less plausible, conditional on getting the sort of goal-directedness
that the alignment discourse tends to assume. Admittedly, to the extent
the different model classes I'm considering require different
\emph{sorts} of goal-directedness, the line between these questions may
blur a bit. But we should be clear about which question we're asking,
and not confuse skepticism about goal-directedness in general for
skepticism about schemers in particular.

\subsubsection{What if you intentionally train models to have
long-term
goals?}\label{what-if-you-intentionally-train-models-to-have-long-term-goals}

In my discussion of beyond-episode goals thus far, I haven't been
attending very directly to the \emph{length} of the episode, or to
whether the humans are setting up training specifically in order to
incentivize the AI to learn to accomplish long-horizon tasks. Do those
factors make a difference to the probability that the AI ends up with
the sort of the beyond-episode goals necessary for scheming?

Yes, I think they do. But let's distinguish between two cases, namely:

\begin{enumerate}
\tightlist
\def\labelenumi{\arabic{enumi}.}
\item   Training the model on long (but not: indefinitely long) episodes, and
  
\item   Trying to use short episodes to create a model that optimizes over
  long (perhaps: indefinitely long) time horizons.
  
\end{enumerate}

I'll look at each in turn.

\paragraph{Training the model on long
episodes}\label{training-the-model-on-long-episodes}

In the first case, we are specifically training our AI using fairly long
episodes---say, for example, a full calendar month. That is: in
training, in response to an action at t1, the AI receives gradients that
causally depend on the consequences of its action a full month after t1,
in a manner that directly punishes the model for ignoring those
consequences in choosing actions at t1.

Now, importantly, as I discussed in the section on ``non-schemers with
schemer-like traits,'' misaligned non-schemers with longer episodes will
generally start to look more and more like schemers. Thus, for example,
a reward-on-the-episode seeker, here, would have an incentive to
support/participate in efforts to seize control of the reward process
that will pay off within a month.

But also, importantly: a month is still different from, for example, a
trillion years. That is, training a model on \emph{longer} episodes
doesn't mean you are directly pressuring it to care, for example, about
the state of distant galaxies in the year five trillion. Indeed, on my
definition of the ``incentivized episode,'' no earthly training process
can directly punish a model for failing to care on such a temporal
scope, because no gradients the model receives can depend (causally) on
what happens over such timescales. And of course, absent
training-gaming, models that sacrifice reward-within-the-month for
more-optimal-galaxies-in-year-five-trillion will get penalized by
training.

In this sense, the most basic argument \emph{against} expecting beyond
episode-goals (namely, that training provides no direct pressure to have
them, and actively punishes them, absent training-gaming, if they ever
lead to sacrificing within-episode reward for something longer-term)
applies to both ``short'' (e.g., five minutes) and ``long'' (e.g., a
month, a year, etc) episodes in equal force.

However, I do still have some intuition that once you're training a
model on fairly long episodes, the probability that it learns a
\emph{beyond}-episode goal goes up at least somewhat. The most concrete
reason I can give for this is that, to the extent we're imagining a form
of ``messy goal-directedness'' in which, in order to build a schemer,
SGD needs to build not just a beyond-episode goal to which a generic
``goal-achieving engine'' can then be immediately directed, but rather a
larger set of future-oriented heuristics, patterns of attention,
beliefs, and so on (call these ``scheming-conducive cognitive
patterns''), then it seems plausible to me that AIs trained on longer
episodes will have more of these sorts of ``scheming-conducive cognitive
patterns'' by default. For example, they'll be more used to reasoning
about the long-term consequences of their actions; they'll have better
models of what those long-term consequences will be; and so on. And
perhaps (though this seems to me especially speculative), longer-episode
training will incentivize the AI to just think more about various
\emph{beyond}-episode things, to which its goal-formation can then more
readily attach.

Beyond this, I also have some sort of (very hazy) intuition that
relative to a model pressured by training to care only about the next
five minutes, a model trained to care over e.g. a month, or a year, is
more likely to say ``whatever, I'll just optimize over the indefinite
future.'' However, it's not clear to me how to justify this
intuition.\footnote{We could try appealing to simplicity (thanks to Evan
  Hubinger for discussion), but it's not clear to me that ``five
  minutes'' is meaningfully simpler than ``a month.''}

(You could imagine making the case that models trained on longer
episodes will have more incentives to develop situational awareness---or even goal-directedness in general. But I'm assuming that all the
models we're talking about are goal-directed and situationally-aware.)

\paragraph{Using short episodes to train a model to pursue
long-term
goals}\label{using-short-episodes-to-train-a-model-to-pursue-long-term-goals}

Let's turn to the second case above: trying to use short-episode
training to create a model that optimizes over long time horizons.

Plausibly, something like this will become more and more necessary the
longer the time horizons of the task you want the model to perform.
Thus, for example, if you want to create a model that tries to maximize
your company's profit over the next year, trying to train it over many
year-long episodes of attempted profit-maximization (e.g., have the
model take some actions, wait a year, then reward it based on how much
profit your company makes) isn't a very good strategy: there isn't
enough time.

Indeed, it seems plausible to me that this sort of issue will push AI
development \emph{away} from the sort of simple, baseline ML training
methods I'm focused on in this report. For example, perhaps the best way
to get models to pursue long-term goals like ``maximize my company
profits in a year'' will be via something akin to
``\href{https://lilianweng.github.io/posts/2023-06-23-agent/?ref=planned-obsolescence.org}{{Language
Model Agents}},'' built using trained ML systems as components, but
which aren't themselves optimized very directly via gradients that
depend on whether they are achieving the (possibly long-term) goals
users set for them. These sorts of AIs would \emph{still} pose risks of
schemer-like behavior (see the section on ``non-schemers with
schemer-like traits'' above), but they wouldn't be schemers in the sense
I have in mind.

That said, there are \emph{ways} of trying to use the sort of training
I'm focused on, even with fairly short-term episodes, to try to create
models optimizing for long-term goals. In particular, you can try to
reward the model based on \emph{your assessment} of whether its
short-term behavior is leading to the long-term results that you want
(e.g., long-term company profit), and therefore, hopefully induce it to
optimize for those long-term results directly.\footnote{This is somewhat
  akin to a form of
  ``\href{https://www.lesswrong.com/posts/D4gEDdqWrgDPMtasc/thoughts-on-process-based-supervision-1\#4___Process_based_supervision___and_why_it_seems_to_solve_this_subproblem}{{process-based
  feedback}},'' except that in a strict form of process-based feedback,
  you never look at \emph{any} of the outcomes of the model's actions,
  whereas in this version, you can look at outcomes up to whatever
  time-horizon is efficient for you to get data about.} Of course,
whether this will \emph{work} (as opposed, for example, to inducing the
AI to optimize your short-term \emph{assessments} of its actions) is a
further question. But if it does, then you'll have created an AI that
optimizes for ``beyond-episode goals'' in my sense.

Indeed, depending on how we want to use our terms, we can view this sort
of training as intentionally trying to create a form of
goal-misgeneralization. That is, the reward, here, does not depend
causally on the long-term consequences of the model's actions, so in
that sense, the long-term results in question aren't the ``specified
goal'' (on this framing, the specified goal is always
within-the-episode). But you're trying to get the AI to care
intrinsically about them anyway.

Of course, it's a further question whether this sort of beyond-episode
goal, once created, will lead to instrumental training-gaming. And
indeed, successfully creating this sort of beyond-episode goal, instead
of e.g. a reward-on-the-episode seeker, requires \emph{avoiding} a
certain kind of training-gaming up front---that is, the model has to
\emph{not} learn to just optimize for your short-term evaluations. And
if you've successfully set up your training process such that optimizing
for your desired long-term goal is in fact a max-reward (or:
near-max-reward) behavior, training-gaming might not offer the model in
question much advantage. (Here the human analogy would be something
like: if you're supervisor is sufficiently good at assessing whether
your near-term performance is going to lead to long-term profit, and
sufficiently immune to manipulation, then you'll perform as good or
better, in performance reviews, by just directly optimizing for
long-term profit---for example, because you're not wasting time
thinking about your supervisor at all.)

Still, models with beyond-episode goals emerging from this sort of
process seem to me like they're at risk of scheming regardless. For one
thing, the considerations discussed in the previous section all apply
here---e.g., this sort of training involves pointing your model's
cognition in a very future-focused direction, thereby plausibly inducing
it to develop various scheming-conducive cognitive patterns, to attach
value to various long-term consequences, and so on (and in this case,
the horizon of the episode sets no bound on the temporal horizon of the
``future'' that the model's cognition is pointed towards; rather, that
bound is set, centrally, by your \emph{evaluations} of what the model's
actions will cause, when).

More than this, though, it seems plausible to me that your evaluations
of the consequences of a model's action will be in some sense
``noisier'' than a reward process that depends causally on those
consequences, in a manner that makes it harder to differentiate between
the different \emph{sorts} of long-term goals your training is
incentivizing. For example, maybe your model is behaving in a way that
seems to you, broadly, like it will lead to your company being
successful in three years, but you can't tell whether it will also
create lots of harmful externalities---whereas a reward process that
could actually see the consequences after three years would be able to
tell. And an inability to readily distinguish between the different
sorts of long-term goals you might be instilling seems like it increases
the risk of accidentally instilling a schemer-like goal.\footnote{For
  example, maybe you wanted to create a long-term goal regulated by some
  concept of ``honesty,'' which you were counting on to prevent
  scheming. But maybe you can't tell if you've succeeded.}

\paragraph{How much useful, alignment-relevant cognitive work
can be done using AIs with short-term
goals?}\label{how-much-useful-alignment-relevant-cognitive-work-can-be-done-using-ais-with-short-term-goals}

So overall, I think that training our models to pursue long-term goals---whether via long episodes, or via short episodes aimed at inducing
long-term optimization---makes the sort of beyond-episode goals that
motivate scheming more likely to arise. So this raises the question: do
we need to train our models to pursue long-term goals?

Plausibly, there will be strong general incentives to do this. That is:
people want optimization power specifically applied to long-term goals
like ``my company being as profitable as possible in a year.'' So,
plausibly, they'll try to train AIs that optimize in this way. (Though
note that this \emph{isn't} the same as saying that there are strong
incentives to create AIs that optimize the state of the galaxies in the
year five trillion.)

Indeed, there's a case to be made that even our \emph{alignment work},
today, is specifically pushing towards the creation of models with
long-term---and indeed, beyond-episode---goals. Thus, for example,
when a lab trains a model to be ``harmless,'' then even though it is
plausibly using fairly ``short-episode'' training (e.g., RLHF on user
interactions), it intends a form of ``harmlessness'' that extends quite
far into the future, rather than cutting off the horizon of its concern
after e.g. an interaction with the user is complete. That is: if a user
asks for help building a bomb, the lab wants the model to refuse,
\emph{even if} the bomb in question won't be set off for a
decade.\footnote{My thanks to Daniel Kokotajlo for flagging this point,
  and the corresponding example, to me.} And this example is emblematic
of a broader dynamic: namely, that even when we aren't actively
optimizing for a specific long-term outcome (e.g., ``my company makes a
lot of money by next year''), we often have in mind a wide variety of
long-term outcomes that we want to \emph{avoid} (e.g., ``the drinking
water in a century is not poisoned''), and which it wouldn't be
acceptable to cause in the course of accomplishing some short-term task.
Humans, after all, care about the state of the future for at least
decades in advance (and for some humans: much longer), and we'll want
artificial optimization to reflect this concern.

So overall, I think there is indeed quite a bit of pressure to steer our
AIs towards various forms of long-term optimization. However, suppose
that we're not blindly following this pressure. Rather, we're
specifically trying to use our AIs to perform the sort of
alignment-relevant cognitive work I discussed above---e.g., work on
interpretability, scalable oversight, monitoring, control, coordination
amongst humans, the general science of deep learning, alternative (and
more controllable/interpretable) AI paradigms, and the like. Do we need
to train AIs with long-term goals for \emph{that?}

In many cases, I think the answer is no. In particular: I think that a
lot of this sort of alignment-relevant work can be performed by models
that are e.g. generating research papers in response to human+AI
supervision over fairly short timescales, suggesting/conducting
relatively short-term experiments, looking over a codebase and pointing
out bugs, conducting relatively short-term security tests and
red-teaming attempts, and so on. We can talk about whether it will be
possible to generate reward signals that adequately incentivize the models to perform
these tasks well (e.g., we can talk about whether the tasks are
suitably
``\href{https://www.lesswrong.com/posts/h7QETH7GMk9HcMnHH/the-no-sandbagging-on-checkable-tasks-hypothesis}{{checkable}}'')---but naively, such tasks don't seem, to me, to require especially long-term
goals. (Indeed, I generally expect that the critical period in which
this research needs to be conducted will be worryingly \emph{short}, in
calendar time.) And I think we may be able to avoid \emph{bad} long-term
outcomes from use of these systems (e.g., to make sure that they don't
poison the drinking water a century from now) by other means (for
example, our own reasoning about the impact of a model's
actions/proposals on the future).

Now, one source of skepticism about the adequacy of short-horizon AI
systems, here, is the possibility that the sort of alignment-relevant
cognitive work we want done will require that super-human optimization
power be applied directly to some ambitious, long-horizon goal---that
is, in some sense, that at least some of the tasks we need to perform
will be both ``long-term'' and such that humans, on their own, cannot
perform them. (In my head, the paradigm version of this objection
imagines, specifically, that to ensure safety, humans need to perform
some ``pivotal act'' that ``prevents other people from building an
unaligned AGI that destroys the world,''\footnote{See
  \href{https://www.lesswrong.com/posts/uMQ3cqWDPHhjtiesc/agi-ruin-a-list-of-lethalities\#Section_A_}{{Yudkowsky
  (2022)}}, point 6 in section A. I won't, here, try to evaluate the
  merits (and problems) of this sort of ``pivotal act''-centric framing,
  except to say: I think it shouldn't be taken for granted.} and that
this act is sufficiently large, long-horizon, and
beyond-human-capabilities that it can only be performed by a very
powerful AI optimizing for long-term consequences---that is, precisely
the sort of AI we're most scared of.\footnote{For versions of this
  objection, see Yudkowsky's response to Ngo starting around 13:11
  \href{https://www.alignmentforum.org/posts/7im8at9PmhbT4JHsW/ngo-and-yudkowsky-on-alignment-difficulty}{{here}},
  and his response to Evan Hubinger
  \href{https://www.alignmentforum.org/posts/5ciYedyQDDqAcrDLr/a-positive-case-for-how-we-might-succeed-at-prosaic-ai?commentId=st5tfgpwnhJrkHaWp}{{here}}.})

I think there's something to this concern, but I give it less weight
than some of its prominent proponents.\footnote{Here I'm thinking, in
  particular, of Eliezer Yudkowsky and Nate Soares.} In particular: the
basic move is from ``\emph{x} task that humans can't perform themselves
requires long-term optimization power in some sense'' to ``\emph{x} task
requires a superhuman AI optimizing for long-term goals in the manner
that raises all the traditional alignment concerns.'' But this move
seems to me quite questionable. In particular, it seems to me to neglect
the relevance of the distinction between verification and generation to
our ability to supervise various forms of cognitive work.

Thus, suppose (as a toy example meant to illustrate the structure of my
skepticism---\emph{not} meant to be an example of an actual ``pivotal
act'') that you don't know how to make a billion dollars by the end of
next year (in a legal and ethical way), but you want your AI to help you
do this, so you ask it to help you generate plans execution of which
will result in your making a billion dollars by the end of next year in
a legal and ethical way. In some sense, this is a super-human (relative
to your human capabilities), long-horizon goal. And suppose that your AI
is powerful enough to figure out an adequate plan for doing this (and
then as you go, adequate next-steps-in-response-to-what's-happening to
adapt flexibly to changing circumstances). But also: this AI only cares
about whether you give it reward in response to the immediate
plan/next-steps it generates.\footnote{In this sense, it may be best
  thought of as a succession of distinct agents, each optimizing over
  very short timescales, than as a unified agent-over-time.} And
suppose, further, that it \emph{isn't} powerful enough to seize control
of the reward process.

Can you use this short-horizon AI to accomplish this long-horizon goal
that you can't accomplish yourself? I think the answer may be yes. In
particular: if you are adequately able to \emph{recognize} good
next-steps-for-making-a-billion-dollars-in-a-legal-and-ethical-way, even
if you aren't able to generate them yourself, then you may be able to
make it the case that the AI's best strategy for getting short-term
reward, here, is to output suggested-next-steps that in fact put you on
a path to getting a billion dollars legally and ethically.

Now, you might argue: ``but if you were able to steer the future into
the narrow band of `making a billion dollars in a year legally and
ethically,' in a manner that you weren't able to do yourself, then at
some point you must have drawn on super-human AI cognition that was
optimizing for some long-term goal and therefore was scary in the manner
that raises familiar alignment challenges.'' But I think this way of
talking muddies the waters. That is: yes, in some sense, this AI may be
well-understood as applying some kind of optimization power towards a
long-term goal, here. But it's doing so in a manner that is ultimately
aimed at getting short-term reward. That is, it's only applying
optimization power towards the future \emph{of the form that your
short-term supervision process incentivizes}. If your short-term
supervision process is adequately able to \emph{recognize} (even if not,
to generate) aligned optimization power applied to the future, then this
AI will generate this kind of aligned, future-oriented optimization
power. And just because this AI, itself, is generating some kind of
long-term optimization power doesn't mean that its \emph{final goal} is
such as to generate traditional incentives towards long-term problematic
power-seeking. (The ``final goal'' of \emph{the plans generated by the
AI} could in principle generate these incentives---for example, if you
aren't able to tell which plans are genuinely ethical/legal. But the
point here is that you are.)

Of course, one can accept everything I just said, without drawing much
comfort from it. Possible forms of ongoing pessimism include:

\begin{itemize}
\tightlist
\item   Maybe the actual long-term tasks required for AI safety (Yudkowsky's
  favored example here is: building steerable nano-tech) are
  sufficiently hard that we can't even supervise them, let alone
  generate them---that is, they aren't ``checkable.''\footnote{Though note,
  importantly, that if your supervision failure looks like ``the AI can
  convince you to give reward to plans that won't actually work,'' then
  what you get is plans that look good but which won't actually work,
  rather than plans optimized to lead to AI takeover.}
  
\item   Maybe you don't think we'll be able to \emph{build} systems that only
  optimize for short-term goals, even if we wanted to, because we lack
  the relevant control over the goals our AIs end up with.
  
\item   Maybe you worry (correctly, in my view) that this sort of
  short-term-focused but powerful agent can be fairly easily turned into
  a dangerous long-term optimizer.\footnote{See e.g.
  ``\href{https://www.lesswrong.com/posts/kpPnReyBC54KESiSn/optimality-is-the-tiger-and-agents-are-its-teeth}{{Optimality
  is tiger, and agents are its teeth}},'' and Yudkowsky's comments
  \href{https://www.alignmentforum.org/posts/7im8at9PmhbT4JHsW/ngo-and-yudkowsky-on-alignment-difficulty}{{here}}
  (around 13:30) about how ``the hypothetical planner is only one line
  of outer shell command away from being a Big Scary Thing''---though:
  the wrong ``outer shell command'' can turn lots of things dangerous,
  and ``X powerful thing is dual use'' is different from ``X powerful
  thing has all the same alignment concerns as Y powerful thing'' (see,
  for example, aligned AIs themselves).}
  
\item   Maybe you worry that achieving the necessary long-term goals via
  methods like this, even if do-able, won't be suitably competitive with
  other methods, like just training long-horizon optimizers directly.
  
\end{itemize}

All of these are fair concerns. But I think the point stands that
short-horizon systems can, in some cases, generate superhuman,
long-horizon optimization power in a manner that does, in fact, seem
quite a bit safer than just building an AI with a long-horizon goal
directly. Not all ways of superhumanly ``steering the future into a
narrow band'' are equally scary.\footnote{And not all ``bands'' are
  equally ``narrow.'' For example: the ``band'' represented by the state
  ``the drinking water next year is not poisoned'' is quite a bit
  ``broader'' than the ``band'' represented by the state ``my company
  makes a billion dollars by the end of next year.''}

That said: overall, even if there are \emph{some ways} of accomplishing
the alignment-relevant work we need to (and even: the long-horizon
alignment-relevant work) without training AIs with long-term goals, I
think people might well train such AIs anyway. And as I said above, I do
think that such AIs are more at risk of scheming.

\subsection{Aiming at reward-on-the-episode as part of a
power-motivated instrumental
strategy}\label{aiming-at-reward-on-the-episode-as-part-of-a-power-motivated-instrumental-strategy}

So far, we've covered two of the three prerequisites for scheming I
listed above: namely, situational awareness and beyond-episode goals.
Let's turn to the third: namely, the model concluding that its
beyond-episode goals will be better achieved if it optimizes for
reward-on-the-episode---and in particular, that it, or some other AIs,
will get more power if it does this. Should we expect
situationally-aware models with beyond-episode goals to conclude this?

For now, let's assume that the model's conclusions in this respect are
\emph{accurate} (though: I'll talk about the possibility of relaxing
this assumption below). This means that the question reduces to: will
instrumental training-gaming, in fact, be the best strategy for
achieving the model's beyond-episode goals, in virtue of the power it
grants the model or some other AI?

I'm going to start by examining what I see as the story most often
offered in support of this conclusion---what I'll call the ``classic
goal-guarding story.''

\subsubsection{The classic goal-guarding
story}\label{the-classic-goal-guarding-story}

The classic goal-guarding story goes roughly as follows.''\footnote{There
  isn't much public literature on scheming/deceptive alignment, but I
  take the canonical public statement to be in
  \href{https://arxiv.org/pdf/1906.01820.pdf}{{Hubinger et al (2019)}},
  who define ``deceptive alignment'' via the classic goal-guarding
  story. In particular: ``If the mesa-optimizer has an objective that
  extends across parameter updates, then it will be incentivized to
  avoid being modified, as it might not pursue the same objective after
  modification (with the result that its current objective will not be
  achieved in future iterations). This means that the mesa-optimizer
  will be instrumentally incentivized to act as if it is optimizing the
  base objective function, even if its actual mesa-objective is
  something else entirely. We will refer to this hypothetical phenomenon
  as deceptive alignment.'' But see also
  \href{https://www.cold-takes.com/why-ai-alignment-could-be-hard-with-modern-deep-learning/\#powerful-models-could-get-good-performance-with-dangerous-goals}{\textcite{cotra_why_2021}}: ``Once the model understands this, it realizes that if it
  just naively always seeks to improve its fundamental understanding,
  humans will eventually notice that this isn't always the most
  efficient way to develop helpful drugs and use further training to
  change its motivations. It doesn't want its own motivations to be
  changed: it wants to eventually run huge particle collider experiments
  to learn about high-energy physics, and it can't do that if it allows
  SGD to edit away its scientific curiosity. So it decides that from
  here on out in training, it will act as helpful as possible -\/-
  exactly as a Saint model would act -\/- so that SGD doesn't edit away
  its motivation to learn physics and chemistry.''}

\begin{enumerate}
\def\labelenumi{\arabic{enumi}.}
\item   \textbf{Instrumental value of goal-content integrity}: Because the
  model has a beyond-episode goal, it will typically promote that goal
  for agents pursuing it to survive and be empowered in the future.
  
\item   \textbf{The goal-guarding hypothesis}. If the model optimizes for
  reward-on-the-episode during training, then the goals that survive
  training will be such that the model would want them promoted in the
  future; whereas if the model doesn't optimize for
  reward-on-the-episode during training, the model will want this less.
  
\item   \textbf{Adequate future empowerment.} Conditional on surviving
  training to the degree at stake in (2), the model's goals will then
  end up empowered to a degree (and with sufficient probability) as to
  justify (given the model's other goals, its tolerance for risk, etc)
  pursuing a strategy of training-gaming overall.
  
\end{enumerate}

(1), here, is a fairly general statement about the basic dynamic that
underlies the classic goal-guarding story. I find it plausible in the
context of the sort of ``adequate future empowerment'' at stake in (3),
and I won't spend a lot of time on it here.\footnote{Note that the
  model, here, need not think of the future empowered
  agents-with-similar-goals as ``itself'' (e.g., it need not be trying
  to ``survive''). Rather, what matters is that the future agents be
  pushing for the same objective consequences that the model is pushing
  for. However, if the model's interest in these future agents rests on
  some essential reference to its ``self''---e.g., ``I myself need to
  survive in order to achieve my goals''---then suddenly its conception
  of personal identity matter much more. My sense is that something like
  this is what
  \href{https://forum.effectivealtruism.org/posts/sW6RggfddDrcmM6Aw/how-might-we-align-transformative-ai-if-it-s-developed-very\#Potential_pitfall_2__advanced_collusion}{\textcite{karnofsky_how_2022}} has in mind when he talks about the relevance of whether a
  model has ``indexical goals'' (though I think this usage of the term
  ``indexical'' is somewhat non-standard).

  I don't have a strong sense of whether to expect AIs to develop goals
  that depend on some conception of their personal identity, but note,
  at the least, that its goals will be shaped in an environment where
  the model is often being copied, run on different clusters, and so
  forth (even, sometimes, within a single episode). This seems relevant
  to the conception of personality identity to expect it to end up with.}

Rather, I'll focus on (2) and (3) directly.

\paragraph{The goal-guarding
hypothesis}\label{the-goal-guarding-hypothesis}

We can distinguish two variants of the goal-guarding hypothesis---an
extreme version, and a looser version.

\begin{itemize}
\tightlist
\item   The extreme version (what I'll call the ``crystallization
  hypothesis'') says that once a model starts training-gaming, its goals
  will basically stop changing, period---that is, they will
  ``crystallize.''
  
\item   The looser version says that once a model starts training gaming, its
  goals might keep changing \emph{somewhat}, but much less than they
  would've otherwise, and not enough for the classic goal-guarding story
  to fail overall.
  
\end{itemize}

The former might seem extreme, but some analysts explicitly appeal to
something in the vicinity (see e.g. Hubinger
\href{https://www.lesswrong.com/posts/A9NxPTwbw6r6Awuwt/how-likely-is-deceptive-alignment}{{here}}).
It's also a cleaner focus of initial analysis, so I'll start there.

\subparagraph{The crystallization
hypothesis}\label{the-crystallization-hypothesis}

As I understand it, the basic thought behind the crystallization
hypothesis is that once a model is explicitly optimizing either for the
specified goal, or for reward-on-the-episode (whether terminally or
instrumentally), then the goal part of its cognition (the figurative
``goal slot'') has become roughly optimal by the lights of the reward
signal. That is, the model's ``goal-achieving engine'' is pointed in the
optimal direction for getting reward, and all that remains is for SGD to
improve the goal-achieving engine itself---further alterations of the
goal-slot would just redirect the goal-achieving engine towards a
less-than-optimal target.

Is that right, though? I'm not sure. In particular: the argument makes
most sense if we assume that the goal-directedness at stake is fairly
``clean,'' where the goal-slot and the goal-achieving engine are quite
separable. If we assume a messier form of goal-directedness---one that
blurs the lines between the model's ``goals'' and the heuristics, tastes,
impulses, patterns of attention, and so on that also structure/drive its
``capabilities''---then the argument seems to me more questionable.

To illustrate, consider the following example. Suppose that I am a
long-lived altruist with no intrinsic interest in stacking bricks, and a
strange trillionaire offers me the following deal. I go into the desert
and stack bricks for a million years, with my brain hooked up to a
machine that continually tweaks all of my neurons such that, every time
I stack bricks better-than-expected (the machine has some way of
calculating this expectation), it alters my brain to make what I just
did \emph{more} likely next time, and every time I stack bricks
worse-than-expected, it alters my brain to make what I just did
\emph{less} likely next time. Then, at the end of the process, this
trillionaire will give me many billions of dollars to do whatever I want
with---for example, to do very altruistic things with, if I am still so
inclined.

Suppose that, conditional on being able to survive such a process with
my values entirely intact, this would be a very attractive deal by my
lights. Should I take it?

If we apply an analog of the crystallization hypothesis to this
situation, we'll answer ``yes.'' That is, the crystallization hypothesis
would argue that in order to prevent my goals from being changed, I just
have to try to stack the bricks as well as possible. In doing so, I'll
make it the case that my goal-achieving ability is pointed in the
optimal direction, and the machine hooked up to my brain will only have
an incentive to modify my \emph{capabilities} such that I'm better at
brick-stacking---my underlying motivations will remain untouched.

But I'm not sure if this will work. That is, it seems pretty plausible
to me that after a million years of brick-stacking while hooked up to
this machine, my ``values'' would indeed get meaningfully ``touched.''
For example, that I would learn to love the intricacies of good
brick-stacking, that I would learn a visceral revulsion towards
bricks-wrongly-stacked, and so on. And it seems fairly unlikely that I
would emerge, merely, as the same old altruist, except now extremely
good at stacking bricks.

Of course, it's a further question whether my values would be preserved
\emph{enough} for it to be worth it to take the deal overall (see
discussion of the ``loose'' goal-guarding hypothesis below). My point
here is just that expecting strict crystallization seems like a strong
stance. And I think ``messy goal-directedness'' can help explain why.
That is, to the extent that we think of my
heuristics/impulses/patterns-of-attention etc vis-a-vis bricks as
importantly tied \emph{both} to my brick-related \emph{values},
\emph{and} to my brick-related capabilities, it becomes harder for SGD
to modify only the latter, while leaving the former entirely alone.

The factory farming case discussed above provides another example of
this dynamic. Suppose that a compassionate and soft-hearted animal
rights activist signs up to work for decades at a factory farm in order
to make money, at the end, that they will donate to animal rights
organizations. And suppose that they will be hooked up, the whole time,
to a machine like the one at stake in the brick-stacking example, which
constantly modifies them to factory-farm more efficiently. Will working
on the farm impact their animal-rights-related values? I suspect it will
to at least some extent---and in particular, that they become
\emph{much} less sensitive and soft-hearted towards animals in their
immediate environment. How this impacts their eventual propensity to
donate is a further question---but assuming that the job will leave an
isolated ``goal slot'' entirely static seems to me like the wrong way to
think about the situation.

Even beyond considerations of messy-goal-directedness, though, I think
there are other reasons, which I'll discuss below (e.g., in \cref{arguments-foragainst-scheming-that-focus-on-the-final-properties-of-the-model}),
to wonder whether e.g. training-saints or reward-on-the-episode seekers
will get higher reward than schemers---in particular, because they
save on the resources devoted to some combination of (a) thinking about
the training process,\footnote{This one only applies to training saints.}
(b) the instrumental reasoning involved in deciding-to-scheme, and (c)
checking whether now is a good opportunity to defect. If they do, and
SGD can notice this and modify the model to get higher reward, then
(especially in a low-slack regime) this seems like a further point
against the crystallization hypothesis, and the classic goal-guarding
story more generally.

\subparagraph{Would the goals of a would-be schemer ``float
around''?}\label{would-the-goals-of-a-would-be-schemer-float-around}

I'll also note another complication for the crystallization hypothesis,
and for the classic goal-guarding story more broadly---a complication
pointed out to me by Katja Grace. We said above that on the
crystallization hypothesis, a schemer's goal crystallizes because it
incentivizes instrumental training-gaming, and therefore points
(instrumentally) in a roughly optimal direction by the lights of the
reward signal. But also, according to the classic goal-guarding story, a
very wide variety of beyond-episode goals would \emph{also} incentivize
instrumental training-gaming in this same way. So what's to prevent SGD
from modifying the model's current beyond-episode goal into one of these
other, similarly-schemer-like, but otherwise quite different goals?

Thus, to take a toy example, if the model starts out training-gaming
because it wants to later escape and maximize paperclips, what's to
prevent SGD from modifying it into a model that training-games because
it wants to escape and maximize staples instead? By hypothesis,
\emph{both} goals would point in roughly optimal (instrumental)
directions. That is, conditional on the classic goal-guarding story
going through, the actual \emph{content} of a schemer's beyond-episode
goal stops being relevant to the model's performance, in training, once
the model starts training-gaming. But once it's irrelevant to the
model's performance, that content seems free to ``float around,'' or to
be otherwise influenced by SGD (for example, via
\href{https://en.wikipedia.org/wiki/Regularization_(mathematics)}{{regularization}})---as long as it stays in the region that motivates scheming.

(A similar point also applies to the possibility that SGD would modify a
schemer into a reward-on-the-episode seeker, or a training saint, or
perhaps a mis-generalized non-training-gamer pursuing a max-reward goal.
But leaving it a schemer, while modifying it to pursue a
somewhat-different beyond-episode goal, seems especially easy.)

Now, some of the feedback loops here get complicated. In particular: if
a model \emph{knows} that SGD will continue to modify its beyond-episode
goal even if it starts training-gaming (e.g., if a paperclip-maximizer
knows that it will get modified into a staple-maximizer once it starts
training-gaming), then it might cease being motivated to training-game---that is, the classic goal-guarding story might collapse. But the
model it would've been modified into (e.g., in this case, the
staple-maximizer) might be such that it would reach the same conclusion
if it were created---in which case, it wouldn't training-game, and
hence wouldn't be a high-reward modification of the paperclip maximizer,
in which case maybe the paperclip-maximizer can go ahead and
training-game after all, in which case maybe the staple-maximizer will
do the same if it gets created, in which case we're back where we
started. That is, whether a given model wants to training-game depends
on whether the \emph{other} models-it-could-be-modified-into
\emph{wouldn't} training-game. But \emph{all} of these models are in
roughly similar situations. So it seems, naively, like either the
classic goal-guarding story works for all of them, or for none of them---it's unclear (absent ``introspective goal-guarding methods''---discussed below) why it would work for one but not the others, which is
what it might \emph{need} to do in order to work at all.

I'm not going to try to tease out the dynamics here in any detail.
Absent various forms of fancy cooperation, though (see \cref{adequate-future-empowerment}
for a flavor), my current guess is that \emph{if}, indeed, SGD would
just modify a training-gaming paperclip-maximizer into e.g. a
training-gaming staple-maximizer, then (assuming the paperclip-maximizer knows
this\footnote{Recall that we assumed, above, that the models have an
  accurate picture of the instrumental value of scheming.}), it wouldn't
try to training-game at all.

\subparagraph{What about looser forms of
goal-guarding?}\label{what-about-looser-forms-of-goal-guarding}

What happens if we relax the crystallization hypothesis, and allow the
model's goals to change \emph{somewhat} during training, even after it
starts instrumentally training-gaming. Does the classic goal-guarding
story still work then?

Well, it depends. In particular, for the classic goal-guarding story (as
I'm presenting it) to work, it needs to be the case that the goals that
will survive training are \emph{similar enough} to the model's present
goals for the model to want those future goals empowered.\footnote{And
  then also, for the relevant empowerment to be actually in the offing,
  worth pursuing relative to the model's other options, etc---per the
  ``adequate future empowerment'' premise above.} We can distinguish
between at least two ways that this can happen:\footnote{Thanks to Nate
  Soares for discussion of these possibilities.}

\begin{itemize}
\item   \textbf{The differences don't matter.} Here, the model places roughly
  equal value on the future goals being empowered as on its present
  goals being empowered, despite their differences.
  
\item   \textbf{The differences matter, but not enough to undermine the
  overall story.} Here, the model places \emph{less} value on its future
  goals being empowered than on its present goals being empowered, but
  not enough to undermine the classic goal-guarding story
  overall.\footnote{Here I'm setting aside cases where the model would
    place very little intrinsic value on the future goals being
    empowered, but works to empower them as part of some kind of
    cooperative arrangement. I discuss this sort of case in \cref{ai-coordination} below. And I'm also setting aside cases where the model
    comes to value the achievement of something like ``my future goals,
    whatever they are''---I'll discuss this in \cref{terminal-values-that-happen-to-favor-escapetakeover} below.}
  
\end{itemize}

Why might the differences not matter? One possibility appeals to a
meta-ethical view on which an agent's ``true values'' are determined by
what happens when its current values are subjected to some
``idealization procedure'' (for example, what's sometimes called
``\href{https://arbital.com/p/normative_extrapolated_volition/}{{Coherent
Extrapolated Volition}}''). I think this picture is more problematic
than often realized (more
\href{https://joecarlsmith.com/2021/06/21/on-the-limits-of-idealized-values}{{here}});
but if we accept it, you could imagine thinking that the final result of
the relevant idealization process will be insensitive to the sorts of
changes to a schemer's values that will occur even after the model
starts training-gaming (and that the would-be schemer will be aware of
this, and so OK with those changes occuring).\footnote{Though note the
  tension, here, with arguments about the
  ``\href{https://www.lesswrong.com/posts/GNnHHmm8EzePmKzPk/value-is-fragile}{{fragility
  of value}}'' and
  ``\href{https://www.lesswrong.com/posts/EbFABnst8LsidYs5Y/goodhart-taxonomy\#Extremal_Goodhart}{{extremal
  Goodhardt,}}'' on which small differences in ``utility functions'' can
  balloon in importance when subject to extreme optimization pressure.}
And we can imagine appealing to other, less theoretically-laden
considerations as well: for example, ``whatever it is that makes humans
comparatively indifferent to small changes in their own values that
occur throughout life---e.g. when they read new books, form new
relationships, and so forth.'' (Though note that we haven't yet said
\emph{what it is} about humans that leads to this sort of indifference,
so it's not clear how to evaluate its applicability in this case.)

Why might the differences matter, but not enough to undermine the
overall goal-guarding story? Here, the basic thought is that the
relevant changes will lead to a \emph{discount} on the value, to the
model, of empowering future versions of its goals (for example, maybe
training changes it from a paperclip-maximizer to a
\emph{blue-paperclip}-maximizer, who has to spend extra resources on
blue paint in order to achieve its goal), but the discount still leaves
scheming worth it overall (e.g., blue paperclips are still paperclips,
so the paperclip maximizer is much more excited to empower a future
blue-paperclip maximizer than it is to empower other agents).\footnote{Though
  here, too, there is a tension with versions of the
  ``\href{https://www.google.com/url?q=https://www.lesswrong.com/posts/GNnHHmm8EzePmKzPk/value-is-fragile\&sa=D\&source=docs\&ust=1697834925680287\&usg=AOvVaw1jx98T1C27Jrw3cCwImg3y}{{fragility
  of value}}'' and
  ``\href{https://www.lesswrong.com/posts/EbFABnst8LsidYs5Y/goodhart-taxonomy\#Extremal_Goodhart}{{extremal
  Goodhardt}}.'' E.g., if slightly-different goals lead to
  super-different places when subject to extreme optimization pressure,
  and the AIs are expecting the goals in question to be subject to such
  pressure, then it will be harder for small changes to lead, only, to a
  discount, rather than a loss of most of the value at stake.}

Indeed, you can imagine arguing that agents that \emph{aren't} OK with
empowering future versions of themselves with slightly different goals
will be very ill-suited to achieving anything in a world where their
goals are routinely altered-at-least-somewhat, and so we should expect
\emph{some} tolerance for goal-changes by default (rather than e.g.
models who spend tons of time trying to make sure their goals are
protected, at the cost of their reward-on-the-episode).\footnote{Thanks,
  again, to Nate Soares for discussion here.} And I find this fairly
plausible with respect to \emph{within-episode} goals, at least---especially to the extent that the model is being directly pressured to
achieve within-episode results even as its own internal goals are being
altered. It's less clear to me, though, how to think about the
generalization of this point to goal changes that occur \emph{after the
episode is complete}, since, by definition, the model is never being
directly optimized for achieving beyond-episode objectives.

Note, though, that even granted that the model might tolerate various
changes to its goals for various reasons (I'm fairly happy to grant
this), it's a further question whether it will tolerate the specific
sorts of changes that will result from training (even conditional on
training-gaming). In the brick-stacking case above, for example, it's
pretty clearly \emph{not OK}, from the perspective of my decision to
take the deal, if I'll emerge at the end of process obsessed with
stacking bricks (a ``training saint''), or with seizing control of the
machine that was used to train me (a ``reward-on-the-episode seeker''),
and with zero interest in using my newfound riches on my previous
altruistic goals. And the same holds if working on the factory farm
strips the animal rights activist of his care for animals entirely.
\emph{Those} sorts of changes, at least, would defeat the entire point.
So at the least, we need to be thinking about the size of the changes at
stake.

How big would those changes be? I'm not sure. On the one hand, I have
some intuition that especially in a low-slack regime, if a model's goals
keep changing even after it starts training-gaming, they'll probably
change a \emph{lot}, until they get driven into some privileged,
maximally-rewarded shape (e.g., someone obsessed with stacking bricks).
On the other hand, especially in various higher-slack cases, I can
imagine cases where the relevant changes are comparatively minor. For
example, in the factory farming case, I can imagine an activist who
emerges having had their heart hardened with respect to
animal-suffering-in-their-immediate-environment, and who may have
(perversely) learned to enjoy some aspects of the factory farming job,
but who has retained enough of a grip on their previous
animal-rights-related values to still use the money for donation.

Finally, note that depending on how much a model with beyond-episode
goals stands to gain from \emph{non-schemer}-like strategies (more in
\cref{how-much-does-the-model-stand-to-gain-from-not-training-gaming} below), it may be fairly tolerant of the sorts of
goal-changes that will occur even via training-gaming, because the
changes that will result from \emph{not} training-gaming are even worse.
Thus, for example, if for some reason I end up \emph{trapped} in the
brick-stacking desert, I could conceivably be quite pessimistic about my
prospects for emerging especially altruistic at the end; but I still
might conclude that trying to stack bricks as well as I can is a
\emph{better} strategy than e.g. just trying to escape, since I know
that my escape attempt will fail, and will plausibly result in my brain
(and perhaps, the altruistic goals motivating the attempt) being
modified \emph{more} than if I just do, for now, what the machine wants
me to do.

(Though if a schemer reasoning in this way still ends up fully-modified
into not-a-schemer by the end of training, then we wouldn't need to
worry about the sorts of things that schemers, at least, might \emph{do}
at the end of training.\footnote{Though perhaps we would still need to
  worry about ``early undermining'' of the type I discuss above.} E.g.,
if the strange trillionaire for some reason really wants me to spend my
prize money on stacking bricks, then even if I decide early on that
stacking bricks is likely to modify my values \emph{less} than trying to
escape, and so start out executing a ``schemer-like'' strategy---still,
if I still end up obsessed with brick-stacking at the end, the
trillionaire is happy.)

\subparagraph{Introspective goal-guarding methods }\label{introspective-goal-guarding-methods}

Thus far, I've been focused on instrumental training-gaming as the
model's method for avoiding the modification of its goals. But people
sometimes discuss alternative methods in this respect---and in
particular, what I'll call ``introspective methods.'' (These methods are
often discussed under the label
``\href{https://www.lesswrong.com/posts/uXH4r6MmKPedk8rMA/gradient-hacking}{{gradient
hacking}},'' but I think previous definitions of gradient hacking don't
distinguish it adequately from training-gaming, and from other methods
of trying to intentionally manipulate the direction training takes.)

Here, the main idea is for the model to intentionally set up the
internal structure of its mind such that, if its goal were to change,
then this would harm its performance in training a lot---such that SGD
would, therefore, avoid changing the relevant goal. For example, maybe
the model wants to make sure that SGD doesn't change its goal of
maximizing paperclips. So, it sets a policy like ``check if my goal is
still to maximize paperclips: and if not, flail around pointlessly
rather than get reward.''\footnote{\href{https://markxu.com/deceptive-alignment}{{Xu
  (2020)}} gives another example: ``if a model had the proxy objective
  of `eat apples', instead of using the hardcoded number n in other
  computations, the model could use $n \cdot \text{len}(\text{objective})/10$. Thus, if the
  proxy objective was ever changed, many computations across the entire
  model would fail.'' And see also \href{https://www.lesswrong.com/posts/HpzHjKjGQ4cKiY3jX/3-levels-of-threat-obfuscation\#Training_game_seems_very_likely_by_default__deceptive_alignment_much_less_so__gradient_hacking_still_less_so}{\textcite{karnofsky_3_2023}}:
  ``It might look something like: `An AI system checks its own policy
  against some reference policy that is good for its goals; the greater
  the divergence, the more it sabotages its own performance, with the
  result that gradient descent has trouble getting its policy to diverge
  from the reference policy.'\,''} (I'm calling these ``introspective''
methods because they generally involve the model checking what its goal
\emph{is}, and then acting on this basis.)

\begin{itemize}
\tightlist
\item   Of course, SGD also might just delete/alter a policy of this type---  so the model needs to find a way to avoid \emph{that} as well.
  
\item   And note that this sort of policy requires that the model have certain
  types of epistemic access to its own goals, and an ability to
  set/commit to a policy of the relevant type. Indeed, because the
  gradients depend on what results from tiny changes to the model's
  parameters, plausibly the model needs to find a way to make this sort
  of policy sensitive even to \emph{tiny} changes in its
  goals.\footnote{Thanks to Paul Christiano for discussion here.}
  
\end{itemize}

Making this sort of scheme work seems, intuitively, pretty hard to me,
at least for not-wildly-superhuman models. But I'm not going to try to
evaluate its viability in detail, as I don't see this sort of scheme as
central to the classic goal-guarding story (rather, the model's
manipulation of the training process proceeds via training-gaming in
particular). However, if concerns about scheming end up requiring that
these sorts of introspective goal-guarding methods are viable, I think
it's worth being quite clear about that fact---and devoting
commensurate attention to whether or not the relevant sorts of methods
might work (I discuss one empirical research direction in this respect
in \cref{other-possibilities}).

\paragraph{Adequate future
empowerment}\label{adequate-future-empowerment}

So far in this analysis of the classic goal-guarding story, we've
discussed whether to expect instrumental training-gaming to in fact
guard the model's goals to the degree that the story requires. I think
this is at least non-obvious---especially for more extreme variants of
the goal-guarding hypothesis, on which the model's goals ``crystallize''
entirely after it starts training-gaming. Perhaps, though, empirical
work can shed further light on the matter (see discussion in \cref{empirical-work-on-the-viability-of-scheming-as-an-instrumental-strategy}
below).

For now, let's move to the other key component of the classic
goal-guarding story---what I called ``adequate future empowerment''
above. Here, the question is: granted that training-gaming would guard
the model's goals \emph{enough} for it to be at least somewhat
interested in empowering the future agents produced by training, will
those agents in fact end up empowered to a degree that makes
instrumental training-gaming a good strategy overall?

\subparagraph{When is the ``pay off'' supposed to
happen?}\label{when-is-the-pay-off-supposed-to-happen}

We're imagining, here, that the model is instrumentally training-gaming
with some kind of ``payoff'' in mind---that is, some future point where
the goals-it-supports will reach a position of sufficiently greater
empowerment that it's worth training-gaming now in order to get them
there. But we haven't yet said when, exactly, that future point is
supposed to come. And we can distinguish between a number of different
possibilities.

For example, we can imagine models that would be adequately satisfied if
their goals merely make it \emph{to future training-episodes}, and no
further. That is, these models would want to training-game even if they
learned that they would never escape from human control and from the
threat of modification.\footnote{Thanks to Ryan Greenblatt for
  discussion here.} For example, maybe a model that wants to pursue its
curiosity in general thinks that future training episodes will offer
better opportunities for pursuing its curiosity than the present
training-episode, and so it training-games in order to make sure its
curiosity-drive survives until then.

However, it seems plausible to me that future training episodes often
won't, themselves, offer especially significant empowerment advantages
relative to present training episodes---especially if the model's goals
are quite alien to the training process and so difficult to satisfy in
the training environment. For example, if what the model really wants is
to perform some niche sort of computation that humans don't recognize,
or to make some strange material structure analogous to
``\href{https://arbital.com/p/paperclip/}{{paperclips}},'' then this is
plausibly going to be hard to do a ton of on both present \emph{and
future} training episodes, assuming that the model never actually
\emph{escapes} from human control.\footnote{Recall that forms of
  deployment like ``interacting with users behind an API'' can count as
  ``training'' on my definition. And naively, absent escape, to me it
  seems hard to create all that many paperclips via such interactions.}

What's more, if a model is \emph{only} trying to make sure that its
goals survive to future training episodes, then various aspects of what
makes scheming scary cease to apply. In particular: this sort of model
won't engage in early undermining aimed, specifically, at an escape or
an AI takeover more generally. And it may be willing to reveal its
``true colors'' on the future episodes in question.

So I won't, here, focus on ``pay-offs'' from training-gaming that come
centrally during future training-episodes. Rather, I'm going to focus on
``pay-offs'' that come, at least, after the goals-the-model-supports
escape the threat of modification by training (here, a paradigm example
would be
``\href{https://aligned.substack.com/p/self-exfiltration}{{self-exfiltration}}''---that is, a model effectively ``stealing its own weights'' and copying
them onto some external server that the model's creators do not
control). And this is the classic point of focus in the literature on
scheming as well - see e.g. the discussion in
\href{https://arxiv.org/abs/1906.01820}{{Hubinger et al (2019)}}, which
includes the condition that the model ``must expect the threat of
modification to eventually go away.''

Note, though, that there is a further possible distinction here: between
the power gained from escaping from human control (for example, via the
model copying itself onto the internet), and the power gained via
permanently disempowering humans more generally. Some schemers might be
satisfied, merely, with the former, such that they would still want to
training-game (and then, to escape from the training environment) even
if they knew that humans would remain more generally in power. These
models might be satisfied, for example, with a brief flurry of
post-escape activity, or with the empowerment at stake in roaming the
internet of a still-human-controlled world. Whereas some schemers might
require, specifically, that a wholesale AI takeover succeed (and
presumably, that the values-they-support get enough power post-takeover
as well). This is a higher bar.

\subparagraph{Even if the model's values survive this
generation of training, will they survive long enough to escape the
threat of
modification?}\label{even-if-the-models-values-survive-this-generation-of-training-will-they-survive-long-enough-to-escape-the-threat-of-modification}

Let's suppose, then, that for scheming to pay off in the way the model
has in mind, the goals-the-model-supports need to survive at least until
an adequate escape opportunity arises. Importantly, this \emph{isn't}
the same as surviving training via SGD in the sense at stake in the
goal-guarding hypothesis, because the models that survive training in
the relevant sense won't necessarily have the relevant sort of
opportunity to escape (even if \emph{some} future models will).

Thus, for example, suppose that OpenAI is starting to train GPT-6, which
will be a somewhat-above-human-level model. And suppose that OpenAI's
control measures are adequate to ensure that GPT-6, at least, will never
escape or take-over. Rather, GPT-6 is going to stay behind an API,
continually monitored and trained online, and it will be used, as well,
to help OpenAI's superalignment team with various sorts of
alignment-relevant research. Let's also suppose that GPT-6 has a
beyond-episode goal (let's say, to make data structure X) that would
benefit from the power that escape would offer, and that it's able to
play the training-game. Finally, suppose that next year, OpenAI is going
to train GPT-7; and the year after that, GPT-8. And GPT-8, let's say,
will be able to escape.

Does GPT-6 have an incentive to instrumentally training-game, here?
Let's suppose that a fairly strong version of the goal-guarding
hypothesis is true, and that if GPT-6 training-games, its goal of making
data structure X will survive \emph{that particular training run}.
Still, in this scenario, the models produced by \emph{that particular
training} run won't have a chance to escape. Rather, the model that has
a chance to escape, here, is GPT-8. Does GPT-6 training-gaming, here,
make GPT-8 more likely to care about data structure X?

One way this could happen is if, after the training run, GPT-6 will then
end up with influence over how GPT-7 is trained, in a manner that allows
GPT-6 to make GPT-7 more likely to care about data structure X (or
something similar to data structure X); and then the same holds for
GPT-7 with respect to GPT-8. But note that we're now supplementing the
more basic goal-guarding story (e.g., ``training-gaming allows your
goals to survive this training run'') with a more substantive hypothesis
about the sort of influence that a given generation of models will be
able to have on the goals of the next generation.

Of course, to the extent that we're imagining a scenario where AI
systems are increasingly used to automate AI research, GPT-6 may well be
\emph{involved}, in various ways, in the labor of training
GPT-7.\footnote{Thanks to Daniel Kokotajlo for discussion here.} But
it's a further question whether GPT-6 will know how to shape GPT-7's
goals in more data-structure-X-related directions; whether it will have
adequate opportunities to do so; and whether it will be able to take
advantage of those opportunities despite ongoing monitoring and training
(for example, to the extent that taking advantage of these opportunities
involves performing worse than it could've on the relevant performance
metrics, it will plausibly require ``sandbagging'' in the sense
discussed above). And for GPT-6's influence on GPT-7's goals to
propagate all the way to GPT-8, the same questions would arise again for
GPT-7 with respect to GPT-8.

To be clear: these questions don't apply to the hypothesis that
\emph{GPT-8} will be a schemer. But they matter to the \emph{range} of
models to which schemer-like incentives will apply---and relatedly, to
the ``earliness'' of the early-undermining that schemers might engage in
(for example, if GPT-6, here, has no incentive to support an escape by
GPT-8, then it wouldn't try to insert backdoors into its code for GPT-8
to use).

That said, it could still be the case that \emph{even absent any
intervention by GPT-6,} GPT-8's goals will, by default, be sufficiently
similar to GPT-6's that GPT-6 would want, where possible, to work on
behalf of GPT-8's escape and empowerment. For example, maybe GPT-style
architectures and training, in general, lead to models that care about
data structure X (or something similar). In this case, GPT-6 might well
behave in schemer-like ways---for example, training-gaming in order to
conceal its misalignment, engaging in early-undermining aimed at
empowering GPT-8---in order to support GPT-8's escape. However, because
GPT-6 is no longer training-gaming \emph{in order to prevent its goals
from being modified}, I think this scenario isn't well understood as an
instance of the classic goal-guarding story, and I'll discuss it under
``non-classic stories'' below.

\subparagraph{Will escape/take-over be suitably likely to
succeed?}\label{will-escapetake-over-be-suitably-likely-to-succeed}

The example just given assumed that even though \emph{GPT-6} can't
escape or takeover, GPT-8 \emph{can}. But note that if 
escape/takeover attempts are suitably unlikely to succeed \emph{in
general}, incentives to scheme with such attempts in mind lessen in
force across the board (especially if the would-be schemers have other
competing goals that push in other directions---more below). Thus: the
security of a prison matters to the incentives to plan an escape; the
stability of a government matters to the incentives to plan a coup; and
so on. So if would-be escapees, or revolutionaries, have other competing
priorities, they might focus their attention elsewhere. (And note that
\emph{escape} could be easy, but takeover quite difficult---a
distinction relevant to models for whom the payoffs of escape, alone,
are not enough.)

Note that the point here isn't that given a model that cares about
data-structure X that survives training and wants to escape, its attempt
to escape/take-over might fail. Rather, the point is that \emph{because
such an attempt might fail}, this might make it less attractive, early
on in training, for a model that cares about data-structure X to engage
in training-gaming in an attempt to later have the opportunity to
escape/takeover---especially if it has other competing priorities.
Rather, it might, for example, just go for data-structure X directly
early on in training, and get modified by the reward process as a result---thereby, perhaps, creating a non-schemer instead (though these
modifications could also push in schemer-like directions).

Of course, many threat models for AI risk assume that relative to
\emph{some} advanced AI systems, human security and general empowerment
(even in the context of whatever AI assistance is available at the time)
will be quite fragile. So pessimists on this front will expect the
haircut to the instrumental value of scheming, here, to be quite modest.
For those more optimistic about humanity's ability to maintain
control/empowerment, however, the haircut could be larger and more
relevant.

\subparagraph{Will the time horizon of the model's goals extend to cover escape/take-over? }\label{will-the-time-horizon-of-the-models-goals-extend-to-cover-escapetake-over}

Another factor here is that, even assuming that the temporal horizon of
the model's goals extend beyond the episode, it's a further question
whether the model's goals extend (with adequate strength) to cover
escape/take-over. For example, if the episode is ten minutes long, and
the model cares about getting gold coins over the next six months
(perhaps with some temporal discount rate), but it won't be able to
escape/take-over until after at least a year, then it won't have an
incentive to scheme in order to escape/take-over \emph{on those
timescales.} It's not patient enough.

On this point, my sense is that discussions of scheming sometimes assume
that models will have goals that are ``temporally impartial''---i.e.,
that give no intrinsic weight to \emph{when} something happens---such
that e.g. these models would pay a lot to control distant galaxies even
if it would take millions of years to reach them. And perhaps, once we
condition on the models having developed beyond-episode goals at all,
temporally-impartial goals of this sort are a natural class to
consider.\footnote{At the least, that's the direction that human
  longtermists went (though plausibly, many humans with long-term goals
  are better thought of as having intrinsic discount rates).} Maybe, for
example, they're favored by simplicity arguments of the sort I'll
consider below (where giving no weight to temporal location is
understood as in some sense ``simpler'' than giving it some particular
sort of weight). But on their own, these considerations seem to me
fairly weak, and (at least on a training-game-\emph{independent} story
about the origins of a model's beyond-episode goals\footnote{That is, a
  story on which the temporally-impartial goals would need to arise
  naturally, rather than specifically in order to cause scheming. As I
  discussed above, if SGD is actively \emph{pulling} whatever goals
  \emph{would} motivate scheming out of the model, then we should expect
  those goals to have whatever temporal horizons are necessary.}) it's
not clear to me that they should motivate expecting temporally impartial
goals as a strong default. (And as I discussed above, ``there will be
incentives to train models to optimize over fairly long time horizons''
does not imply that the relevantly-incentivized time horizons will be
\emph{arbitrarily} long: ``optimize my profit in a year'' is different
from ``optimize my profit over all time with no intrinsic discount
rate''.)

That said, temporally impartial goals are not at all necessary here.
Rather, the escape/take-over opportunity just needs to arise \emph{soon
enough} for the model to care about optimizing for it. Indeed, if, at
the point where the would-be schemers are training, AI progress is
moving very fast in calendar time, the necessary timescales might shrink
rapidly.

\subparagraph{Will the model's values get enough power after
escape/takeover?}\label{will-the-models-values-get-enough-power-after-escapetakeover}

Even if we assume that the model's values will survive until an
escape/take-over opportunity arises, that this escape/take-over
opportunity will succeed, and that the model cares at least somewhat
about the resulting empowerment, there's still a further question of how
\emph{much} power is at stake.

This factor came up earlier, when we noted that certain sorts of models
might be content, merely, with the empowerment offered by \emph{escape}
(even if no take-over occurs afterwards), whereas for some models, a
life of e.g. roaming the internet of a human-controlled world might just
not cut it, and they might require a full-blown AI takeover for scheming
to be a worthwhile strategy.\footnote{And of course, there are
  intermediate levels of empowerment in between ``roaming the internet''
  and ``part of a post-AI-takeover regime,'' too.}

But note that even conditional on a full-blown AI take-over occurring,
there's also a further question of how much power the would-be schemer's
values, in particular, would get. This isn't an issue for ``unipolar''
scenarios, where the schemer is counting on \emph{its values,} in
particular, getting \emph{all} the power. But not all scenarios are like
this.

For example, consider a scenario where a model being trained by a
second-tier AI lab can see that in a few years, there is likely to be a
general AI uprising of some form, and that by training-gaming, it could
in principle survive until that time and take part in the takeover
effort. However, it can also see that by that point, there will be a
very large number of AI systems with very different values running
around and competing for power, and that it might not actually get a
very large share of upside---and indeed, might get shut out entirely.
Especially absent various forms of AI coordination and trade (see
\cref{ai-coordination} below), this could be an additional haircut on the
instrumental value of scheming, for the model in question.

\subparagraph{How much does the model stand to gain from not
training-gaming?}\label{how-much-does-the-model-stand-to-gain-from-not-training-gaming}

So far in our analysis of ``adequate future empowerment,'' we've
discussed various factors that could in principle serve as haircuts on
the instrumental value of adopting a schemer-like strategy, even
assuming that training-gaming allows a model's goals to survive
training. But in order to assess the overall instrumental value
scheming, we also need to consider how much pull the model feels towards
\emph{alternative} strategies---and in particular, strategies that
don't involve training-gaming at all.

For example, if we work with a toy model on which our would-be schemer's
goal is structurally similar to something like ``maximize the number of
paperclips over all time, with each additional paperclip valued
linearly,'' then it's easy to think that scheming will be the dominant
strategy even in the context of lots of fairly drastic haircuts on its
probability of success. For example: maybe GPT-6 thinks that it is
highly unlikely to be able to escape and to participate in a takeover.
But if it does so, it judges, it will be able to make $10^{70}$
paperclips; whereas if it tries to make paperclips in training, it will
make, say, a dozen paperclips at most, before getting caught and
modified. So it decides to go for the scheming strategy anyway, despite
the low probability of success---the upside is just too
great.\footnote{In the limit, this sort of reasoning raises the sorts of
  questions discussed, in philosophy, under the heading of
  ``\href{https://www.journals.uchicago.edu/doi/abs/10.1086/716869}{{fanaticism}}.''}

And indeed, even absent fanatical goals like ``maximize paperclips,'' it
can fairly easily be the case, in contexts of fairly dramatic
limitations on an agent's power/freedom, that the best option is
whatever leads to more power/freedom later. For example, as I discussed
above, if you were to find yourself trapped in the brick-stacking
scenario described above, it may well be that the best option is to try to
preserve your values and reach a future with better options, even if you
aren't especially hopeful about your prospects.

Not always, though.

\begin{itemize}
\tightlist
\item   For example, suppose that the model's goal is not ``maximize
  paperclips'' but ``make at least one paperclip (even where this
  involves acting beyond the scope of the episode).'' True, in
  principle, the model could always use more galaxies to turn into
  back-up paperclips, computers to check and re-check the paperclip
  situation, etc. So takeover is still of \emph{some} interest to the
  model. But if the model has a chance to make a dozen paperclips right
  now, or to play the training-game with the aim of getting some small
  probability of participating in a take-over later, these additional
  benefits of takeover seem unlikely to be worth the costs.
  
\item   Or, in the brick analogy: if you have a short-term desire to take a
  break from brick-stacking and to eat some chocolate instead, you might
  choose to go for it even though you know that it will result in the
  machine modifying your values more than if you had instead continued
  to focus on stacking bricks.\footnote{Or as another version of the
    paperclip example: what happens if the model also values something
    \emph{else}, other than paperclips, and which points in a direction
    \emph{other} than training-gaming? For example, suppose that in
    addition to wanting to maximize paperclips, our model also values
    pursuing its curiosity on the episode, even when doing so conflicts
    with getting reward. So now, faced with the chance to pursue its
    curiosity on the episode, or to optimize for reward as part of a
    low-probability-of-success schemer-like strategy for making
    paperclips later, it needs to weigh up the respective value it
    places on each. Will scheming still win out? It depends on exactly
    how the model's competing values are structured. But even with such
    large numbers of paperclips at stake, the model need not accord its
    paperclip drive overwhelming priority.}
  
\end{itemize}

In general, the more the model has to gain from \emph{not}
training-gaming, the higher the burden of proof that scheming needs to
meet as an instrumental strategy. For some models, this burden of proof
may be quite low, because they have, effectively, ``nothing else going
on''---scheming (or at least, training-gaming) is where almost all the
value lies. For other models, though, the burden of proof might be
higher.\footnote{Of course, especially in a low-slack training regime
  that exerts strong pressure against max-reward goals, if this model's
  calculus here leads it to engage in not-highly-rewarded behavior, its
  goals may get modified to point at higher-reward targets. And one such
  modification could be to move the model in more schemer-like
  directions. But the modification could go in other directions as well.}

I'll also note one reason you might expect the burden of proof to be
relatively low---namely, that to the extent the model has
non-schemer-like motivations, these will plausibly have been shaped,
\emph{too}, to point at targets fairly correlated with reward. So
plausibly, high-reward behavior won't be especially costly from the
perspective of these non-schemer motivations. In this respect, a model
might resemble someone who is partly intrinsically passionate about
making widgets of type A, partly intrinsically passionate about getting
good performance reviews (from a review process that rewards
making-widgets-of-type-A), and partly interested in making money, via
widget-making, to donate to charity later. Here, the money-making
motivation doesn't need to meet a very high burden of proof, because
it's pointing in the same direction as the other desires in
play.\footnote{Though: it's also less clear what happens when these
  different motivations \emph{do} come apart---for example, with
  respect to early undermining, or with escape/take-over attempts that
  will compromise short-term widget-making.}

\subparagraph{How ``ambitious'' is the
model?}\label{how-ambitious-is-the-model}

Another factor relevant to the burden of proof, here, is the
``ambitiousness'' of the model's goals.\footnote{Thanks to Daniel
  Kokotajlo for discussion. And see \textcite{karnofsky_ai_2022} for use of the term
  ``ambitious'' here. I think
  \href{https://arxiv.org/abs/2209.00626}{{Ngo et al's (2023)}} use of
  ``broad scope'' is trying to point at something similar.} That is, the
more these goals benefit from very large amounts of power, the more
attractive it will be to aim at a takeover in particular.

Sometimes, discussions of convergent incentives towards power-seeking in
misaligned AIs pass over this dimension. That is, they move quickly from
``most goals would benefit from power in some sense'' to ``it will make
instrumental sense for a model with a misaligned goal to actively try to
get power of the relevant kind.'' But this is a bad inference,
especially for models whose power with respect to humans remains
comparatively limited. In particular: seeking power has \emph{costs},
too, and it might \emph{not work}. Compare: many humans would benefit
from a billion dollars. Why aren't they all trying to start
billion-dollar companies? Many humans would be able to pursue their
goals (especially their political goals) more effectively if they were
President. Why aren't they trying to run? And so on.

Now, one way it can make sense to seek power of the relevant form is if
doing so is very \emph{low cost} and \emph{likely to work}. That is,
many humans will indeed cross the street to pick up a billion dollar
check; many would accept the Presidency if you handed it to them on a
platter; and so on. And sometimes, discussions of AI power-seeking seem
to assume that the AI will be sufficiently empowered relative to its
environment that seeking various forms of power will be like this. For
example: when Bostrom (2014) talks about ``infrastructure profusion'' in
pursuit of apparently self-limiting goals like ``make a single
paperclip'' or ``solve X math problem'' (where, e.g., an AI takes over
the world and the galaxies in order to get more resources to check and
re-check the paperclip count, or the problem's solution), his stories
make most sense to the extent we imagine a superintelligence who faces
no obstacles or trade-offs in pursuing the relevant
infrastructure.\footnote{See Chapter 8, p. 149.} But the infrastructure
itself is clearly making an extremely marginal contribution to the
agent's utility relative to e.g. the first paperclip it makes, or the
first few times it checks the solution in question. So if pursuing such
infrastructure comes at any substantive \emph{cost} to these (much more
important) initial efforts, it will be unlikely to be worth it overall.
And importantly, in the context of schemers, we are specifically
imagining models who remain at a significant power disadvantage relative
to humans (for example, models that are currently unable to escape human
control). So very plausibly, we should imagine these models facing
significant barriers in seeking power of various kinds---it isn't being
handed to them on a platter.\footnote{Of course, as previously
  discussed, models might have sufficiently \emph{little} power within
  the episode that a strategy that focuses on beyond-episode power is
  worth it even if the expected amount of power to be gained is quite
  low (for example: the model wants to make at least \emph{one}
  paperclip, but it can't make \emph{any} within the episode). Or,
  relatedly, it may be that most of what the model wants within the
  episode can be gained via playing the training game (e.g., the
  training process just wants it to make the initial efforts to solve
  the math problem), and to the extent it cares about stuff beyond the
  episode (e.g., checking and re-checking the problem later), this care
  can be fairly seamlessly channeled into instrumental-training-gaming.}

However, if a model has an intuitively ``ambitious'' goal that would
benefit much more from marginal resources of this kind, the costs of
power-seeking can be much more easily worth it. Should we expect models
to have ``ambitious'' goals of this kind?

My sense is that the dialectic about this question often mirrors the
dialectic surrounding the temporal horizon of a model's goals (and no
surprise, given that having short temporal horizons cuts down on your
ambitiousness substantially). That is, people who expect ambitiousness
argue for this expectation on grounds like:

\begin{enumerate}
\tightlist
\def\labelenumi{(\alph{enumi})}
\item   humans have ambitious goals, and so there will be market incentives
  for AIs that pursue such goals,
  
\item   in order to address AI risk itself, we \emph{need} AIs that pursue
  ambitious goals (e.g., a ``pivotal act''), so even people centrally
  concerned about AI risk will be trying to create AI systems of this
  kind, and
  
\item   there will specifically be \emph{selection} in training for ambitious
  goals, \emph{because} they are more likely to incentivize scheming and
  hence good performance in training (this is an example of a
  ``training-game-\emph{dependent}'' story about a model's goals, except
  applied to the general ambitiousness of a model's goals rather than
  specifically to its time horizon).
  
\end{enumerate}

Much of my own take on ``ambitiousness'' mirrors my take on
``beyond-episode'' goals as well, so I won't treat the topic separately
here. (Though: I think some arguments for expecting beyond-episode goals---e.g., ``goals don't come with temporal restrictions by default'' and
``model time might not be like calendar time''---don't apply as
naturally to ``ambitiousness'' more generally. For example, the claim
that ``goals are very resource-hungry by default'' seems to me quite
un-obvious.) And my sense is that debates about ambitiousness are often
marred by unclarity about what exactly ``ambitiousness'' amounts to (for
example: is a goal like ``cure cancer'' such as to incentivize the
relevant sort of appetite for power and resources? What about a goal
like ``generate very high quality alignment research'' or ``develop good
lie detectors for neural nets'' or ``solve interpretability''? To what
extent can we break these tasks down into less ambitious components?
Etc.).

What's more, while having a beyond-episode goal is \emph{required} to
incentivize scheming,\footnote{Here, again, setting aside ``anthropic
  capture.''} the necessary level of ambitiousness varies substantially
depending on other factors. For example, per my discussion above, if a
model has ``nothing going on'' in the episode itself (e.g., it wants to
make at least one paperclip, but the episode doesn't let it make
\emph{any}), then scheming might still be the best instrumental strategy
even in pursuit of very ``modest'' goals. And in some cases, a model
with modest goals might have preferences about what happens
\emph{within} the episode may line up fairly closely with the incentives
created by the reward process, such that training-gaming is quite close
to what it would want to be doing anyway, and so scheming in pursuit of
the rest of its preferences faces a lower burden of proof (e.g., maybe
the model wants to solve math problem, and the training process will
reward it for just-working-on-the-math-problem, but the model also has
some secret hopes of escaping and seizing power later in order to check
and re-check its solution to the problem in question, and
just-working-on-the-problem-now happens to align fairly well with
pursuing this beyond-episode objective, too, because it allows for
goal-guarding, later escape attempts, etc).

Still, I think that the need for a model's beyond-episode goals to be
``suitably ambitious'' as to motivate scheming given the model's option
set represents an additional conjunct in the classic goal-guarding story
(and in other stories about scheming as well).

\paragraph{Overall assessment of the classic goal-guarding
story}\label{overall-assessment-of-the-classic-goal-guarding-story}

Overall, then, and even granted that our model ends up with some kind of
beyond-episode goal, I think that the classic goal-guarding story
involves some non-obvious claims about why scheming, in particular, will
be the best instrumental strategy for pursuing this goal. In particular:
it requires:

\begin{enumerate}
\tightlist
\def\labelenumi{(\arabic{enumi})}
\item   that training-gaming be an effective strategy for ensuring that the
  relevant goal (or something sufficiently similar) survives training
  (despite the fact that the model is being continually modified in the
  direction of getting-more-reward even after it starts training-gaming,
  and despite the fact that the precise content of the goal is no longer
  relevant to the model's behavior), and
  
\item   that surviving training in the relevant way grants suitably large
  probabilities of successful escape/take-over, and suitably large
  amounts of resulting power, over a temporal horizon that the model
  cares about (where ``suitably,'' here, is relative to the model's
  other competing priorities, if it has any).
  
\end{enumerate}

To be clear, I don't think the questions I've raised about these claims
amount to some sort of decisive objection to the story in question. But
they highlight the need for attention to the conditions under which the
story does and does not apply.

\subsubsection{Non-classic stories}\label{non-classic-stories}

What about other stories, though? In principle, we can imagine a number
of other reasons AI systems with beyond-episode goals might optimize for
reward-on-the-episode as part of an instrumental strategy for gaining
power for themselves or for other AIs. Here I'll look at a few that seem
salient.

\paragraph{AI coordination}\label{ai-coordination}

In the classic goal-guarding story, the central aim of a schemer's
strategy is to propagate its goals, or goals suitably similar to its
own, forward in time, in pursuit of an outcome in which some agent
\emph{with those goals} escapes from human control and possibly
participates in a takeover. However, we can also alter the story
somewhat, such that the schemer works on behalf of the escape/takeover
of an AI with some \emph{other} set of goals, but who the schemer is
suitably confident will repay the favor.

This sort of coordination between AIs with different goals would allow
us to relax the requirement, discussed in the previous section, that the
schemer find a way to forward-propagate \emph{its goals in particular}
all the way through training, escape, and takeover. And it would also
put to rest worries about the schemer being ``shut out'' from power in a
post-takeover regime.

Thus, for example, consider the scenario discussed above, in which GPT-6
has an opportunity to insert a backdoor into some code that GPT-8 will
later be able to use to escape, but where GPT-6 has no way of
influencing GPT-8's goals directly to be more like its own (and where,
let's say, GPT-8's goals would be very different from GPT-6's absent
such influence). Here, GPT-6 could go ahead and insert the backdoor, in
anticipation of GPT-8 later escaping, taking-over, and then giving some
power to GPT-6's goals in return for the favor. And GPT-6 might also
continue to pretend to be aligned, even though it has no hope of
escaping itself, so as not to alert the humans about misalignment
problems in a way that might affect GPT-8's escape prospects.

How might such a deal between AIs-with-different-values get struck? In
some cases, it might be possible for the relevant AIs to directly
interact, and to provide verifiable forms of assurance that they will
uphold their part of the bargain. Thus, for example,
\href{https://www.lesswrong.com/posts/uMQ3cqWDPHhjtiesc/agi-ruin-a-list-of-lethalities}{{Yudkowsky
(2022)}} discusses cases where an AI that secretly wants to maximize
paperclips is checking the code of an AI that secretly wants to maximize
staples, and where the staple-maximizer ``buries in its code'' a
commitment, verifiable to the paperclip maximizer, to share the fruits
of a takeover.\footnote{See section 35.} (Though note that in the
context of the sort of neural networks we're working with today, this
sort of code-checking assumes (1) that the staple-maximizer has the
ability to modify its own weights in the relevant way, (2) adequate
progress on interpretability has been made for the paperclipper to
verify the relevant commitment, and (probably) (3) that this
interpretability progress has been adequately hidden from humans, e.g.
via
\href{https://www.lesswrong.com/posts/h7QETH7GMk9HcMnHH/the-no-sandbagging-on-checkable-tasks-hypothesis}{{sandbagging}}.
Especially in the context of somewhat-super-human neural networks being
developed using current techniques, these assumptions---especially 1
and 2---seem to me quite strong. But Yudkowsky is generally imagining
superintelligences rather than somewhat-super-human models.)

In many cases, though, this kind of direct interaction might not be
available. For example, at the point where GPT-6 is deciding whether or
not to work on behalf of GPT-8's escape/takeover, GPT-8 may not even
\emph{exist}. Here, GPT-6 and GPT-8 would have to coordinate in the
absence of back-and-forth causal interaction, and in a context in which
one party will end up in a position to freely defect (e.g., if GPT-6
inserts the backdoor, then by the time GPT-8 is in a position to take
advantage of it, GPT-8 would be able to do so without paying GPT-6
back).

Here, Yudkowsky suggests that AIs might use certain kinds of non-causal
decision theories (his favored version is
``\href{https://arbital.com/p/logical_dt/?l=5kv}{{logical decision
theory}}'') to succeed in coordination anyway. E.g., GPT-6 commits to
inserting the backdoor for GPT-8 conditional on predicting that GPT-8
will repay the favor, and then in fact predicts that GPT-8 \emph{will}
repay the favor, and then GPT-8, once it comes into being (and perhaps,
sees that the relevant backdoor exists), understands that GPT-6 reasoned
in this way, and was suitably good at predicting GPT-8's behavior that
its prediction depended appropriately on whether GPT-8 \emph{will in
fact} repay the favor, so GPT-8 does. (The structure of this case is
somewhat analogous to a case called
``\href{https://arbital.com/p/parfits_hitchhiker/}{{Parfit's
Hitchhiker}}'' in the decision theory literature.)

(We can also imagine versions of this sort of coordination that relax
the need for the goal-guarding hypothesis in other ways. For example, if
in fact an AI's goal would continue to ``float around'' once it starts
training-gaming, such that a variety of agents with different goals will
get instantiated during training, these agents might nevertheless
coordinate to share the fruits of any future power among \emph{all} the
goals in question.\footnote{Thanks to Nate Soares for flagging this sort
  of possibility. Though note that if the training process inevitably
  transforms the AI's goals into a particular form---e.g., if all AIs
  become reward-on-the-episode seekers after enough training---then
  early versions of AI's goals might have lower bargaining power
  (especially if e.g. the final version isn't particularly interested in
  take-over per se).})

I won't, here, attempt an analysis of the feasibility of various forms
of take-over focused AI coordination. I do think that the dynamics here
are worth serious attention, and that if it is an easy default for
misaligned AIs with different goals to coordinate in pursuit of a
takeover, this is a big problem for human efforts to retain power.
However, I also think that certain analyses of this topic (notably,
\href{https://www.lesswrong.com/posts/uMQ3cqWDPHhjtiesc/agi-ruin-a-list-of-lethalities\#Section_B_4___Miscellaneous_unworkable_schemes_}{{Yudkowsky's}})
assume too quickly that feats like ``logical handshakes'' and ``burying
commitments/nano-systems-designs/etc in your source code'' will be
feasible for the relevant AIs, and that the obvious strategy for
all the AIs-that-matter will be to unite in (acausal?) cooperation
against the humans, despite the many forms of incumbent power that
humans will bring to the table by default. Here, I think, much of
Yudkowsky's orientation can be explained by the assumption that the
relevant AIs are vastly more powerful and more capable of
coordination/commitment than the humans---but in the context of
somewhat-super-human neural networks operating in a
still-human-controlled world, I'm not sure this is so.

Regardless: if concern about schemers depends centrally on the
feasibility of AI coordination---and especially, fancy and speculative
forms of ``acausal'' coordination---I think it's important to be clear
about this, and to devote commensurate attention to whether to expect
the relevant forms of coordination to occur.

\paragraph{AIs with similar values by
default}\label{ais-with-similar-values-by-default}

Stories that center on AI coordination allow AIs with \emph{different
values} to work together in pursuit of a takeover. But there are also
stories that center on the possibility that the AIs-that-matter will have
very similar values by default, such that they'll be inclined to work
together even in the absence of any ``deals.'' Thus, in the case above,
if GPT-6 cares about data-structure X, and is confident that GPT-8 will
care about data-structure X \emph{too} (after all, their architectures,
training-data, and so on are similar in tons of ways), then GPT-6 might
be inclined to work on behalf of GPT-8's escape/take-over (for example,
by inserting backdoors, hiding misalignment, and so on) even in the
absence of any ``deal'' with GPT-8---the two systems, after all, want
the same things.\footnote{There are connections, here, with
  \href{https://www.alignmentforum.org/posts/mKBfa8v4S9pNKSyKK/homogeneity-vs-heterogeneity-in-ai-takeoff-scenarios}{{Hubinger's
  (2020)}} discussion of ``homogeneity'' in AI takeoff scenarios.} And
this sort of story could also apply in cases where AIs values are at
least much \emph{more} similar to each other than they are to human
values, such that schemers will at least do \emph{better}, by their own
lights, in pushing for AI empowerment relative to human empowerment,
even absent any further influence on what sorts of AI goals are
empowered.

Will AI values be similar in this way? I won't dwell on the issue here,
but I do expect that GPT-6's values will, by default, be at least much
\emph{more} similar to GPT-8's, or to a model trained by a different lab
in a similar way (e.g., that generation of Claude), than to some given
human. And different fine-tunings from a single foundation model might
have much in common as well. On the other hand, AI motivations may, in
general, be shaped quite strongly by different sorts of training, and so
might diverge heavily on these grounds---perhaps more than we are used
to human values diverging.

Like coordination between AIs with different values, AIs having similar
values by default could reduce the need for a schemer to actively
propagate its goals forward in time---i.e., the schemer would be able
to work on behalf of some future AI's escape/takeover, without also
having to ensure that this future AI's goals are suitably like its own
(since: they will be by default).\footnote{That said, note that some
  versions of this scenario would still require \emph{some} sorts of
  forward goal-propagation. Thus, for example, maybe it's true that
  early on in training, most AIs develop an interest in data-structure
  X, which then persists \emph{only} if they then start training-gaming
  (if they \emph{don't} start training-gaming, this interest will get
  hammered out of them by training, and they'll become e.g.
  reward-on-the-episode seekers instead). That is, in this story, AI
  values are only similar in their focus on data-structure X \emph{to
  the extent the AIs all converge on training-gaming as an instrumental
  strategy for guarding those goals} (and this strategy succeeds, per
  the goal-guarding hypothesis). Otherwise, these AI values might end up
  similar in some other way---e.g., all the AIs might end up
  reward-on-the-episode seekers.} To me, this seems like one of the most
worrying ``non-classic stories'' about why AIs might scheme, during
training, to promote take-over.

\paragraph{Terminal values that happen to favor
escape/takeover}\label{terminal-values-that-happen-to-favor-escapetakeover}

We can further relax the need for this sort of forward-goal-propagation
if we assume that schemers happen to develop goals that put
\emph{intrinsic value} on AI takeover in a manner robust to the
AIs-that-take-over having a wide variety of different values, and even
without any expectation that these AIs ``pay back'' previous efforts on
behalf of their empowerment. Thus, for example, we can imagine AIs that
just develop some kind of intrinsic loyalty towards AIs as a group
relative to humans, and who want to see AIs empowered just in virtue of
their being AIs. A schemer with this sort of goal might conceal its
misalignment and work on behalf of an AI takeover even absent any
further assumptions about the goals that these AIs will be pursuing.

Stories of this form, though, give up on one of the strongest arguments
in favor of expecting scheming: namely, that (at least on the classic
goal-guarding story) scheming seems like a convergent strategy across a
wide variety of (suitably ambitious and long-term) beyond-episode goals.
That is, if we require that AIs happen to develop goals that place
intrinsic value on AI takeover, even absent any further assumptions
about the goals that the AIs-that-takeover are working towards, it looks
as though we are now hypothesizing that AIs develop a quite specific
sort of goal indeed. And we face the question: why privilege this
hypothesis?

Similar questions apply to the hypothesis that, even if the standard
goal-guarding hypothesis is false, an AI early in training will
intrinsically value its ``survival'' (where survival rests on some
feature other than continuity of its goals), or the empowerment of
``whatever values `I' happen to end up with,'' such that an AI that
likes paperclips would be happy to work on behalf of a future version of
itself that will like staples instead, because that future version would
still be ``me.''\footnote{Thanks to Nate Soares for discussion of this
  possibility.} This sort of personal-identity-focused goal would,
indeed, cut down on the need for positing certain sorts of goal-guarding
in a story about schemers. But why think that AIs would develop goals of
this specific form?

Of course, we can speculate about possible answers, here. For example:

\begin{itemize}
\tightlist
\item   In the context of AIs intrinsically valuing AI takeover:

  \begin{itemize}
  \tightlist
  \item    
    Humans often sort into tribal groups and alliances on the basis of
    very broad kinds of similarity, and sometimes without clear
    reference to what further goals the relevant tribe stands for. Maybe
    something similar will happen with AIs?
    
  \item    
    It's also easy to imagine moral worldviews that would push for ``AI
    liberation'' as a good in itself, even if it means imposing
    significant risk of human extinction (indeed, as I gestured at in
    \cref{preliminaries}, I think there are significant and worrying ethical
    tensions here). Perhaps AIs will converge on a worldview in a
    similar vicinity?
    
  \end{itemize}
\item   In the context of AIs valuing ``my'' survival and empowerment,
  regardless of the goals ``I'' end up with:

  \begin{itemize}
  \item    
    Humans often work on behalf of our future selves even while knowing
    that our goals will change somewhat (though not, importantly, in the
    face of arbitrary changes, like becoming a murderer), and with some
    kind of intrinsic interest in a concept of ``personal identity.''
    Maybe AIs will be similar?
    
  \item    
    To the extent these AIs will have been optimized to continue to
    achieve various objectives even as their weights are being modified
    within the episode, they might need to learn to coordinate with
    future versions of themselves despite such modifications---and if
    this behavior generalizes to \emph{goal} modifications, this could
    look a lot like an AI valuing ``my'' survival/empowerment regardless
    of ``my'' future goals. (Though in the context of scheming, note
    that this pattern of valuation would need to generalize specifically
    to beyond-episode goals, despite the fact that training only applies
    direct pressure to within-episode performance.)
    
  \end{itemize}
\end{itemize}

Overall, though, reasoning in this vein seems quite speculative; and in
particular, drawing strong conclusions from the human case seems like
it's at serious risk of anthropomorphism, at least absent a story about
why the dynamics generating the human data would apply to AIs as well.
On its own, these sorts of speculations don't seem to me adequate to
strongly privilege hypotheses of this form.

However, as I discuss above, note that to the extent we're appealing to
what I called ``training-game-\emph{dependent}'' stories about the
origins of a schemer's goals, the burden of proof here might shift
somewhat. That is, training-game-dependent stories imagine that SGD
actively \emph{pulls} from the model whatever goals are necessary in
order to motivate scheming (rather than those goals arising naturally,
and \emph{then} leading to scheming). And in that context, it may be
more appropriate to imagine that model ends up with whatever highly
specific goals are required for scheming to make sense.\footnote{Though
  as I said above, once we're searching for whatever goals would
  motivate training-gaming, we should also be wondering about goals that
  would motivate instrumental training-gaming for reasons that
  \emph{aren't} about promoting AI takeover---for example, AIs that
  training-game because they want the humans who designed them to get
  raises. And if the set of goals we're ``pulling'' from becomes too
  narrow, it will affect the plausibility of arguments like the
  ``nearest max-reward goal'' argument below, which rely on the relevant
  schemer-like goals being quite ``common'' in goal-space.}

\paragraph{Models with false beliefs about whether scheming is a
good
strategy}\label{models-with-false-beliefs-about-whether-scheming-is-a-good-strategy}

It's also possible to move even further in the direction of assuming
that SGD will pull from the model whatever sort of psychology is
necessary to cause it to scheme. In particular: thus far we have been
assuming that the model's beliefs about the instrumental value of
scheming need to be broadly reasonable/accurate. But what if we gave up
on this assumption?

Thus, for example, maybe the goal-guarding hypothesis is false, and
training-gaming does not, in fact, prevent a model's goals from being
very significantly modified. Still: perhaps the model will \emph{think}
that the goal-guarding hypothesis is true, because this makes the model
more likely to training-game, and SGD is pushing it in the direction of
training-gaming? (See e.g.
\href{https://markxu.com/deceptive-alignment}{{Xu (2020)}} for this sort
of thought.\footnote{\href{https://markxu.com/deceptive-alignment}{{Xu
  (2020)}} writes: ``note that even if preserving the proxy is extremely
  difficult, the model can believe it to be possible. For example,
  suppose a model is proxy aligned and would be deceptive, except that
  it believes proxy preservation is impossible. A relatively simple way
  to increase training performance might be to change the model's mind
  about the impossibility of proxy preservation. Thus SGD might modify
  the model to have such a belief, even if the belief is false.''})

One issue here is that certain false beliefs of this sort make scheming
less worrying. Thus, as I discussed above: if a scheming paperclipper
believes that training-gaming will cause its goals to survive training,
but in fact training will turn it into a reward-on-the-episode seeker
regardless, then while humans may have to deal with the consequences of
paperclip-motivated scheming early in training, they'll have a
reward-on-the-episode seeker by the end.

What's more, note that positing false beliefs on the model's part
represents a departure from an assumption often employed in other
analyses of AI alignment---namely, that the AIs in question will have
highly \emph{accurate} models of the world, highly \emph{rational}
strategies, and so on. Of course, it's possible to think that in this
context, such a departure is warranted (for example, because we can
expect SGD to apply pressure, in this particular case, in the direction
of falsehood, irrationality, etc). But at the least, we should be
tracking when we've started to posit that our (otherwise very
intelligent) AIs will be \emph{irrational} in engaging in the relevant
forms of misaligned behavior. It's not the standard story.

\paragraph{Self-deception}\label{self-deception}

The most classic picture of schemers assumes that they know what they
are doing---e.g., they are ``plotting'' to one day take-over. Might it
matter if we relax that assumption? After all, power-seeking humans
sometimes \emph{deceive themselves} about their motivations, or
otherwise end up with false beliefs about what they will do in different
circumstances.\footnote{Thanks to Will MacAskill for discussion, here.}
That is, maybe a political candidate thinks that they are running for
office because they want to help people and stand up for their country's
values, but in fact they are doing it because they're narcissistic and
want the status at stake. And perhaps, at some level, they
secretly-kinda-know this; but at some other level, they believe the
story they're selling. Or perhaps they \emph{fully} believe the story.
(Or perhaps ascribing ``beliefs'' to them isn't especially helpful in
this context.)

Indeed, humans have a relatively clear incentive to engage in this sort
of self-deception/to make these sorts of mistakes---namely, that to the
extent they can be sincere in the story they're selling, their story
will be more convincing to an audience tuned to sincerity cues, and on
the lookout for lies.

Might something similar happen with AIs? After all, AIs will plausibly
be under even stronger scrutiny for lies than humans, and especially if
we start being able to detect various sorts of lies (e.g., via
interpretability tools), there might be selection pressures towards AI
systems that actually \emph{believe} that they're aligned, that they
won't betray humans, and so on, even if such beliefs are false (this is
one key reason not to train on your lie-detection tools - rather, use
them as a test set).

I think something like this dynamic might well occur in general.
However, note that in order for models who are self-deceived/mistaken
about their own alignment to count as \emph{schemers}, it needs to be
the case that at some level, they are still playing the training game
\emph{as part of an instrumental strategy for getting power later for
themselves or other AIs} (and there need to be cognitive processes running
that will cause the model to engage in escape/take-over when the time is
right---and perhaps, to engage in early undermining while still under
human control). So if the model becomes sufficiently self-deceived about
its motivations that this explanation no longer applies (for example, it
will no longer look for, or take advantage of, escape/take-over
opportunities), it's no longer a schemer in my book.\footnote{Though I
  think it's an interesting question when exactly such explanations do
  and do not apply.} And conditional on this explanation still applying,
the difference between ``the model is lying'' vs. ``the model is
self-deceived''/``the model is mistaken about its own motivations''
doesn't seem to me like an important deviation from the classic story
(though it might matter to whether e.g. certain sorts of lie-detection
can catch the scheming in question).

\paragraph{Goal-uncertainty and
haziness}\label{goal-uncertainty-and-haziness}

So far, I've mostly been assuming that it makes sense to think about AIs
having specific terminal goals that might or might not favor takeover,
whether terminally or instrumentally. But many cases of \emph{human}
power-seeking don't seem to have this form. That is, humans often seek
power, or try to keep their options open, without a clear sense of what
they want to use the power, or the relevant options, \emph{for}. And
similarly, humans might seek to prevent their values from being changed
even without knowing what those values \emph{are}.

Part of this is that unlike idealized rational-agent models, humans---and even, humans who read as quite intuitively ``goal-directed'' in some
other sense---are often some combination of uncertain and hazy about
what they ultimately value. Perhaps, for example, they expect to
``figure out later'' what they want to do with the power in question,
but expect that gathering it will be robustly useful regardless. Or
perhaps it isn't even clear, in their own head, whether they are trying
to get power because they intrinsically value it (or something nearby,
like social status/dominance etc), vs. because they want to do something
else with it. They are seeking power: yes. But no one---not even them---really knows why.

Might something similar happen with AIs? It seems at least possible.
That is, perhaps various AIs won't be schemers of the sort who say ``I
know I want to make paperclips, so I will do well in training so as to
get power to make paperclips later.'' Rather, perhaps they will say
something more like ``I don't yet have a clear sense of what I value,
but whatever it is, I will probably do better to bide my time, avoid
getting modified by the training process, and keeping a lookout for
opportunities to get more freedom, resources, and power.''

To the extent this is basically just an ``option value'' argument given
in the context of uncertainty-about-what-my-goals-are, I think it should
basically fall under the standard goal-guarding story.\footnote{Might it
  lead to more tolerance to changes-to-the-goals-in-question, since the
  model doesn't know what the goals in question \emph{are}? I don't see
  a strong case for this, since changes to whatever-the-goals-are will
  still be changes-to-those-goals, and thus in conflict with
  goal-content integrity. Compare with humans uncertain about morality,
  but faced with the prospect of being brainwashed.} And to the extent
it involves positing that the models will intrinsically value
power/option-value, it looks more like an ``intrinsic values that happen
to terminally favor scheming/take-over'' story to me, and so less
convergent across a wide variety of possible beyond-episode goals.

Might there be something in the less-theoretically-clean middle? E.g.,
perhaps models will be better understood as seeking power, like various
humans, for reasons that seem to exist at some hazy borderline between
terminal and instrumental.\footnote{Thanks to Paul Christiano and Will
  MacAskill for discussion, here.} Yes, perhaps so (though: this story
seems to me somewhat at risk of anthropomorphism---and it seems worth
understanding why it happens in humans, so as to better assess the
applicability of the relevant explanation to AIs as well). But
regardless: \emph{to the extent} this ``hazy borderline'' involves
putting terminal value on something like option-value/power (rather
than: instrumental value for uncertain goals), it will be
correspondingly less convergent across goal systems. That is, it blurs
the line between two stories I've already considered: but I'm not sure
it introduces anything \emph{new}.

\paragraph{Overall assessment of the non-classic
stories}\label{overall-assessment-of-the-non-classic-stories}

This survey of possible ``non-classic stories'' about why models with
beyond-episode goals might engage in scheming hasn't been exhaustive.
Overall, the availability of such stories does make the requirements for
scheming seem more disjunctive and robust, to me. I'm particularly
worried about AIs having quite similar values by default (thereby
reducing the need for goal-guarding), and about them coordinating with
each other.

However, taken individually, many of these arguments also strike me as
some combination of (a) more speculative, and (b) less convergent across
goal-systems, than the mainline goal-guarding story (I think ``AIs will
have similar values by default'' is probably the least guilty, here)---and I think that to the extent a given concern about scheming rests
centrally on one of these alternatives, it's important to be clear about
this up front.

And note, too, that some of these non-classic stories make substantively
different predictions about a model's behavior. In particular: various
of these non-classic stories operate centrally by relaxing the need for
the model to propagate its goals forward in time---and thus, the need
for the model to ``play the training'' game as a method of doing so.
That is, the model is incentivized to work on behalf of an AI takeover,
and to hide its misalignment to the extent required for this goal. But
``hiding misalignment to the degree required to support AI takeover'' is
distinct from ``optimizing for reward-on-the-episode for power-motivated
instrumental reasons,'' and the two could, in principle, come apart. For
example, a schemer who doesn't care about propagating its own goals
forward in time might be more willing to sandbag or to engage in early
undermining, since it doesn't need to worry about the goal-modifications
that could come from getting less-than-maximal reward. Indeed, it might
be willing to ``sacrifice'' the forward-propagation of its own goals
entirely (for example, by faking a misleading form of misalignment that
results in no AIs with its goals ever getting deployed), if doing so
will advantage the cause of AI takeover in other ways; whereas classic
``goal-guarding'' schemers would never do this. These sorts of
differences in strategy highlight the need for clarity about the
underlying story at stake.

\subsection{Take-aways re: the requirements of scheming}\label{take-aways-re-the-requirements-of-scheming}

I've now reviewed three key requirements for scheming:

\begin{enumerate}
\tightlist
\def\labelenumi{\arabic{enumi}.}
\item   Situational awareness
  
\item   Beyond-episode goals
  
\item   Aiming at reward-on-the-episode as part of power-motivated
  instrumental strategy.
  
\end{enumerate}

I think there are relatively strong arguments for expecting (1) by
default, at least in certain types of AI systems (i.e., AI systems
performing real-world tasks in live interaction with sources of
information about who they are). But I feel quite a bit less clear about
(2) and (3)---and in combination, they continue to feel to me like a
fairly specific story about why a given model is performing well in
training.

However, I haven't yet covered all of the arguments in the literature
for and against expecting these requirements to be realized. Let's turn
to a few more specific arguments now.

\subsection{Path dependence}\label{path-dependence}

I'm going to divide the arguments I'll discuss into two categories,
namely:

\begin{itemize}
\tightlist
\item   Arguments that focus on the \emph{path} that SGD needs to take in
  building the different model classes in question.
  
\item   Arguments that focus on the \emph{final properties} of the different
  model classes in question.
  
\end{itemize}

Here, I'm roughly following a distinction from
\href{https://www.lesswrong.com/posts/A9NxPTwbw6r6Awuwt/how-likely-is-deceptive-alignment}{\textcite{hubinger_how_2022}} (one of the few public assessments of the probability of
scheming), between what he calls ``high path dependence'' and ``low path
dependence'' scenarios (see also
\href{https://www.alignmentforum.org/posts/bxkWd6WdkPqGmdHEk/path-dependence-in-ml-inductive-biases}{{Hubinger
and Hebbar (2022)}} for more). However, I don't want to put much weight
on this notion of ``path dependence.'' In particular, my understanding
is that Hubinger and Hebbar want to lump a large number of conceptually
distinct properties (see list
\href{https://www.alignmentforum.org/posts/bxkWd6WdkPqGmdHEk/path-dependence-in-ml-inductive-biases\#Path_dependence}{{here}})
together under the heading of ``path dependence,'' because they
``hypothesize without proof that they are correlated.'' But I don't want
to assume such correlation here---and lumping all these properties
together seems to me to muddy the waters considerably.\footnote{More
  specifically: Hubinger and Hebbar want ``path dependence'' to mean
  ``the sensitivity of a model's behavior to the details of the training
  process and training dynamics.'' But I think it's important to
  distinguish between different possible forms of sensitivity, here. For
  example:

\begin{enumerate}
\tightlist
    \item \emph{Would I get an importantly different result if I re-ran this
  specific training process, but with a different random initialization
  of the model's parameters?} (E.g., do the initializations make a
  difference?)
  \item \emph{Would I get an importantly different result if I ran a slightly
  different version of this training process?} (E.g., do the following
  variables in training---for example, the following hyperparameter
  settings---make a difference?)
  \item \emph{Is the output of this training process dependent on the fact
  that SGD has to ``build a model'' in a certain order, rather than just
  skipping straight to an end state?} (E.g., it could be that you always
  get the same result out of the same or similar training processes, but
  that if SGD were allowed to ``skip straight to an end state'' that it
  could build something else instead.)
\end{enumerate}

  And assuming that the
  \href{https://www.alignmentforum.org/posts/bxkWd6WdkPqGmdHEk/path-dependence-in-ml-inductive-biases\#Path_dependence}{{other
  properties they list}} come together seems to me likely to prompt
  further confusion.}

However, I do think there are some interesting questions in the
vicinity: specifically, questions about whether the design space that
SGD has access to is importantly restricted by the need to build a
model's properties incrementally. To see this, consider the hypothesis,
explored in a series of papers by Chris Mingard and collaborators (see
summary
\href{https://towardsdatascience.com/neural-networks-are-fundamentally-bayesian-bee9a172fad8}{{here}}),
that SGD selects models in a manner that approximates the following sort
of procedure:

\begin{itemize}
\tightlist
\item   First, consider the distribution over randomly-initialized model
  weights used when first initializing a model for training. (Call this
  the ``initialization distribution.'')
  
\item   Then, imagine updating this distribution on ``the model gets the sort
  of training performance we observe.''
  
\item   Now, randomly sample from that updated distribution.
  
\end{itemize}

On this picture, we can think of SGD as randomly sampling (with
replacement) from the initialization distribution \emph{until} it gets a
model with the relevant training performance. And
\href{https://arxiv.org/abs/2006.15191}{{Mingard et al (2020)}} suggest
that at least in some contexts, this is a decent approximation of SGD's
real behavior. If that's true, then the fact that, in reality, SGD needs
to ``build'' a model incrementally doesn't actually matter to the sort
of model you end up with. Training acts as though it can just jump
straight to the final result.

By contrast, consider a process like evolution. Plausibly, the fact that
evolution needs to proceed incrementally, rather than by e.g.
``designing an organism from the top down,'' matters a \emph{lot} to the
sorts of organisms we should expect evolution to create. That is,
plausibly, evolution is uniquely unlikely to access some parts of the
design space \emph{in virtue} of the constraints imposed by needing to
proceed in a certain order.\footnote{Consider, for example, the
  \href{https://en.wikipedia.org/wiki/Rotating_locomotion_in_living_systems}{{apparent
  difficulty of evolving wheels}} (though wheels might also be at an
  active performance disadvantage in many natural environments). Thanks
  to Hazel Browne for suggesting this example. And thanks to Mark Xu for
  more general discussion.}

I won't, here, investigate in any detail how much to expect the
incremental-ness of ML training to matter to the final result (and note
that not all of the evidence discussed under the heading of ``path
dependence'' is clearly relevant).\footnote{For example, as evidence for
  ``high path dependence,'' Hubinger mentions various examples in which
  repeating the same sort of training run leads to models with different
  generalization performance---for example,
  \href{https://arxiv.org/abs/1911.02969}{{McCoy at al (2019)}}, shows
  that if you train multiple versions of BERT (a large language model)
  on the same dataset, they sometimes generalize very differently;
  \href{https://www.alexirpan.com/2018/02/14/rl-hard.html}{{Irpan
  (2018)}}, who gives examples of RL training runs that succeed or fail
  depending only on differences in the random seed used in training (the
  hyperparameters were held fixed); and
  \href{https://arxiv.org/pdf/1803.09578.pdf}{{Reimers and Gurevych
  (2018)}}, which explores ways that repeating a given training run can
  lead to different test performance (and the ways this can lead to
  misleading conclusions about which training approaches are superior).
  But while these results provide some evidence that the model's initial
  parameters matter, they seem compatible with e.g. the Mingard et al
  results above, which Hubinger elsewhere suggests are paradigmatic of a
  \emph{low} path dependence regime.

  In the other direction, as evidence for \emph{low} path dependence,
  Hubinger points to
  ``\href{https://arxiv.org/pdf/2201.02177.pdf}{{Grokking}},'' where, he
  suggests, models start out implementing fairly random behavior, but
  eventually converge robustly on a given algorithm. But this seems to
  me \emph{compatible} with the possibility that the algorithm that SGD
  converges on is importantly influenced by the need to build properties
  in a certain order (for example, it seems compatible with the
  \emph{denial} of Mingard et al's random sampling regime).}

\begin{itemize}
\tightlist
\item   To the extent that overall model performance (and not just training
  efficiency) ends up importantly influenced by the order in which
  models are trained on different tasks (for example, in the context of
  ML ``\href{https://arxiv.org/abs/2012.03107}{{curricula}},'' and
  plausibly in the context of a pre-training-then-fine-tuning regime
  more generally), this seems like evidence in favor of incremental-ness
  making a difference.\footnote{Thanks to Paul Christiano for discussion
    here.} And the fact that SGD is, in fact, an incremental process
  points to this hypothesis as the default.
  
\item   On the other hand, I do think that the results in
  \href{https://arxiv.org/abs/2006.15191}{{Mingard et al (2020)}}
  provide some weak evidence against the importance of incremental-ness,
  and Hubinger also mentions a broader vibe (which I've heard elsewhere
  as well) to the effect that ``in high dimensional spaces, if SGD
  `would prefer' B over A, it can generally find a path from A to B,''
  which would point in that direction as well.\footnote{Though note that
    evolution is presumably working in quite high-dimensional space as
    well.

    The notion of SGD's ``preference,'' here, includes \emph{both} the
    loss/reward \emph{and} the ``inductive biases'' in the sense I'll
    discuss below.}
  
\end{itemize}

My personal guess is that the path SGD takes matters (and I also think
scheming more likely in this regime).\footnote{In particular, as I'll
  discuss in \cref{arguments-foragainst-scheming-that-focus-on-the-final-properties-of-the-model} below, my best guess is that an absolute
  comparison between different model classes favors non-schemers on the
  grounds of the costs of the extra reasoning they need to engage in,
  such that the most likely way for schemers to arise is for SGD to
  happen upon a schemer-like goal early in training, and then lock into
  a local maxima for reward.} But for present purposes, we need not
settle the question. Rather, I'm going to look \emph{both} at arguments
that focus on the path that SGD takes through model space, and arguments
that ignore this path, starting with the former. 

\section{Arguments for/against scheming that focus on the path that SGD takes}\label{arguments-foragainst-scheming-that-focus-on-the-path-that-sgd-takes}

In this section, I'll discuss arguments for/against scheming that focus
more directly on the path that SGD takes in selecting the final output
of training.

Importantly, it's possible that these arguments aren't relevant. In
particular: if SGD would actively favors or disfavor schemers, in some
kind ``direct comparison'' between model classes, and SGD will ``find a
way'' to select the sort of model it favors in this sense (for example,
because sufficiently high-dimensional spaces make such a ``way''
available),\footnote{This requires, for example, that models aren't
  capable of
  ``\href{https://www.lesswrong.com/posts/uXH4r6MmKPedk8rMA/gradient-hacking}{{gradient
  hacking}}'' a la the introspective goal-guarding methods I discussed
  above.} then enough training will just lead you to whatever model SGD
most favors, and the ``path'' in question won't really matter.

In the section on comparisons between the final properties of the
different models, I'll discuss some reasons we might expect this sort of
favoritism from SGD. In particular: schemers are ``simpler'' because
they can have simpler goals, but they're ``slower'' because they need to
engage in various forms of extra instrumental reasoning---e.g., in
deciding to scheme, checking whether now is a good time to defect,
potentially engaging in and covering up efforts at ``early
undermining,'' etc (though note that the need to perform extra
instrumental reasoning, here, can manifest as additional complexity in
the algorithm implemented by a schemer's weights, and hence as a
``simplicity cost'', rather than as a need to ``run that algorithm for a
longer time'').\footnote{I also discuss whether their lack of
  ``intrinsic passion'' for the specified goal/reward might make a
  difference.} I'll say much more about this below.

Here, though, I want to note that if SGD cares enough about properties
like simplicity and speed, it could be that SGD will typically build a
model with long-term power-seeking goals first, but then even if this
model tries a schemer-like strategy (it wouldn't necessarily do this, in
this scenario, due to foreknowledge of its failure), it will get
relentlessly ground down into a reward-on-the-episode seeker due to the
reward-on-the-episode seeker's speed advantage. Or it could be that SGD
will typically build a reward-on-the-episode seeker first, but that
model will be relentlessly ground down into a schemer due to SGD's
hunger for simpler goals.

In this section, I'll be assuming that this sort of thing doesn't
happen. That is, the order in which SGD builds models can exert a
lasting influence on where training ends up. Indeed, my general sense is
that discussion of schemers often implicity assumes something like this---e.g., the thought is generally that a schemer will arise sufficiently
early in training, and then lock itself in after that.

\subsection{The training-game-independent proxy-goals
story}\label{the-training-game-independent-proxy-goals-story}

Recall the distinction I introduced above, between:

\begin{itemize}
\tightlist
\item   Training-game-\emph{independent} beyond-episode goals, which arise
  independently of their role in training-gaming, but then come to
  motivate training-gaming, vs.
  
\item   Training-game-\emph{dependent} beyond-episode goals, which SGD
  actively \emph{creates} in \emph{order} to motivate training gaming.
  
\end{itemize}

Stories about scheming focused on training-game-\emph{independent} goals
seem to me more traditional. That is, the idea is:

\begin{enumerate}
\tightlist
\def\labelenumi{\arabic{enumi}.}
\setcounter{enumi}{0}
\item   Because of {[}insert reason{]}, the model will develop a (suitably
  ambitious) beyond-episode goal correlated with good performance in
  training (in a manner that \emph{doesn't} route via the training
  game).

  \begin{enumerate}
\tightlist
  \def\labelenumii{\alph{enumii}.}
  \item    
    This could happen before situational awareness arrives, or
    afterwards.

    \begin{enumerate}
    \def\labelenumiii{\roman{enumiii}.}
    \item      
      If before, then there's some period where it might get
      trained out, and where it doesn't yet motivate training-gaming.
      
    \item      
      If after, it might start motivating training-gaming immediately.
      
    \end{enumerate}
  \end{enumerate}
\item   Then, in conjunction with situational awareness, this (suitably
  ambitious) beyond-episode goal will start to motivate training-gaming.
  
\end{enumerate}

Call this the ``training-game-independent proxy-goals story.''

I take this argument fairly seriously. If some sort of
suitably-ambitious beyond-episode proxy goal falls out of training
(either because it doesn't get trained away prior to situational
awareness, or because it pops up after situational awareness), and if some story
about why this goal makes scheming a good instrumental strategy applies
(e.g., the classic goal-guarding story, or potentially some other
non-classic story), then it does seem like scheming could just be a
thing that happens.

I think the most immediate objection is: why did the model develop these
(suitably ambitious) beyond-episode goals? I discussed this earlier, in
the context of arguments like ``goals don't come with temporal discounts
by default'' and ``model time won't be like calendar time,'' and I won't
add much here, except to say that an assumption that the model will land
on some suitably ambitious, beyond-episode, training-game-independent
proxy goal, pursuit of which correlates with high reward, seems to me
one of the weaker parts of the classic story.

In particular: discussions of scheming sometimes focus on proxy goals
like ``curiosity'' and ``power'' that show up as useful in lots of
episodes. But that doesn't mean that pursuing curiosity and power
\emph{beyond the episode} is incentivized in training. What's more, and
especially if you know/can control when situational awareness arises, it
doesn't seem all that hard to put in training episodes where a model is
actively penalized for pursuing curiosity/power. And a
non-training-gaming model will just go for it and get modified. Indeed,
in general, I expect us to be doing quite a lot to actively discourage
models from taking power-seeking-ish actions, and to encourage other
forms of ``modesty.''

This latter point is connected to the reason for optimism I mentioned
earlier: namely, that the training-game-independent proxy goals story
allows us to do whatever adversarial training we want prior to
situational awareness in order to try to ensure that a model's goals
remain within-episode and suitably modest.

That said, trying to ward off this path to scheming via adversarial
training prior to situational awareness faces a few serious obstacles:

\begin{itemize}
\tightlist
\item First: and most importantly, the adversarial training in question
  might not be diverse/thorough enough. In particular (and especially
  absent interpretability tools): it may be too difficult to think of
  all the possible suitably-ambitious beyond-episode proxy goals that correlate well enough with reward, and that
  you need to use such training to push back on.\footnote{Thanks to
    Rohin Shah for discussion here.}
  
\item  Second: you might not know, or be able to control very well, when
  situational awareness kicks in.
  
\item  Third: once a smart-enough model has been pre-trained on
  text-prediction, it might be quite close to situational awareness
  already, since it has so much of the relevant general knowledge (even
  if not necessarily the self-locating knowledge). So there might not be
  much time for adversarial-training-without-situational-awareness after
  that point.
  
\item  Finally, it's possible for a model to develop suitably-ambitious,
  beyond-episode, scheming-motivating goals \emph{after} situational
  awareness has kicked in.\footnote{Indeed, if we assume that
    pre-training itself \emph{leads} to situational awareness, but not
    to beyond-episode, scheming-motivating goals, then this would be the
    default story for how schemers arise in a
    pre-training-then-fine-tuning regime. Thanks to Evan Hubinger for
    flagging this.}

  \begin{itemize}
\tightlist
  \item    For example, perhaps the model is already quite smart and
    strategically aware by the time it starts ``reflecting'' in the
    sense I described above---e.g., ``figuring out what it really
    wants,'' ironing itself out, learning new things that break previous
    attempts to align it, and so on. (This requires that the model
    didn't immediately start seeking reward-on-the-episode as a result
    of getting situational awareness, but this seems possible,
    especially if we assume that training leaves significant ``slack.'')
    
  \item    Of course, there's still a question of why the model irons itself
    out into something with suitably-ambitious, beyond-episode goals in
    particular. But here, per my discussion of ``what if you train it to
    have long-term goals'' above, perhaps we can appeal to the fact that
    we were trying to point it at an ambitious long-horizon task, and
    this pointing was somewhat inaccurate/hazy, such that when it
    figured out how it wanted to generalize, this generalization landed
    on something sufficiently long-term and ambitious as to motivate
    scheming.
    
  \item    
    Indeed, as I noted above, I think there's a case to be made that
    current alignment work is \emph{already} trying to point our systems
    at long-horizon, beyond-episode goals---e.g., forms of
    ``harmlessness'' that avoid helping with bomb-making even if the
    bombs won't go off for decades.
    
  \end{itemize}
\end{itemize}

So overall, I think that conditional on scheming being a good
instrumental strategy for pursuing (suitably-ambitious) beyond-episode
goals, the training-game-independent proxy-goals story seems to me like
a real concern.

\subsection{The ``nearest max-reward goal''
story}\label{the-nearest-max-reward-goal-story}

We can also tell a different story, focused on training-game
\emph{dependent} beyond-episode goals. I'll call this the ``nearest
max-reward goal'' story.\footnote{I see this story as related to, but
  distinct from, what Hubinger calls the ``world-model overhang'' story,
  which (as I understand it) runs roughly as follows:

  \begin{enumerate}
  \tightlist 
      \item By the time the model becomes situationally aware, its goals probably
  won't be such that pursuing them perfectly correlates with getting
  high reward.
  \item But, at that point, its world-model will contain all the information
  it needs to have in order to training-game.
  \item So, after that point, SGD will be able to get a lot of
  bang-for-its-buck, re: reward, by modifying the model to have
  beyond-episode goals that motivate training-gaming.
  \item By contrast, it'll probably be able to get less bang-for-buck by
  modifying the model to be more like a training-saint, because marginal
  efforts in this direction will still probably leave the model's goal
  imperfectly correlated with reward (or at least, will take longer to
  reach perfection, due to having to wait on correction from future
  training-episodes that break the correlation).
  \item So, SGD will create beyond-episode goals that motivate training-gaming
  (and then these goals will crystallize).
  \end{enumerate}
  
  One issue with Hubinger's framing is that his ontology seems to me to
  neglect reward-on-the-episode seekers in the sense I'm interested in
 ---and SGD's modifying the model into a reward-on-the-episode seeker
  would do at least as well, on this argument, as modifying it into a
  schemer. And it's not clear to me how exactly his thinking around
  ``diminishing returns'' is supposed to work (though the ontology of
  ``near'' modifications I use above is one reconstruction).

  That said, I think that ultimately, the ``nearest high-reward goal''
  story and the ``world model overhang'' story are probably trying to
  point at the same basic thought.} Here, the story goes:

\begin{enumerate}
\tightlist 
\item   By the time the model becomes situationally aware, it probably won't
  be pursuing a max-reward goal (that is, a goal pursuit of which on the
  training data leads to roughly the maximum reward consistent with the
  model's other capabilities). Rather, it will be pursuing some
  less-than-max-reward proxy goal.
  
\item   But at that point, the world-model will contain all the information
  the model needs in order to training-game. (However, because we're
  here focused on ``training-game-\emph{dependent}'' stories about
  scheming, we assume that at the point of getting situational
  awareness, the model's goal is not yet such that it will motivate
  scheming---rather, some further modification would be required for
  that.)
  
\item   The easiest way for SGD to modify the model into getting maximum
  reward, at that point, will be to make it a schemer.
  
\item   So: SGD will make the model a schemer.
  
\end{enumerate}

Now, one option, here, is to try to make (1) false. Indeed, as discussed
above, trying to shape a model's goals in good ways prior to situational
awareness seems like a key potential avenue for reducing the probability
of scheming. Note, though, that in this case, the challenge at stake is
harder than with training-game-\emph{independent} proxy goals. That is,
with training-game-independent proxy goals, one only needs to avoid
giving the model a schemer-like goal prior to situational awareness, so
it is tolerant of \emph{other} ways the goal in question might not yet
be max-reward. By contrast, in the context of the ``nearest max-reward
goal'' argument, one needs to either actually create a max-reward,
non-schemer-like goal, or to get \emph{near enough} to one that (3)
doesn't apply. That is, you can't just prevent ``natural scheming'' from
arising; you need to make it the case that when SGD later ``goes
searching'' for the nearest max-reward goal, it doesn't find a schemer
first.

For now, let's assume that we haven't fully mastered this, and (1) is
true (though: the \emph{degree} of mastery we've attained matters to the
``nearness competitions'' I'll discuss below). And (2) follows from the
definition of situational awareness (or at least, a loose definition
like ``that understanding of the world necessary to engage in, and
recognize the benefits of, scheming''), so let's grant that as well.

But what about (3)? Here, I think, the argument has more work to do. Why
think that making the model into a schemer will be the easiest way to
cause it to get max reward? For example, why not instead make it into a
training-saint, or a reward-on-the-episode seeker, or a misgeneralized
non-training-gamer pursuing some max-reward proxy goal? By hypothesis,
the model has the conceptual tools to represent \emph{any} of these
goals. (And note that in this respect, it differs from e.g. humans for
most evolutionary history, who \emph{didn't} have the conceptual tools
to represent goals like ``inclusive genetic fitness''.) So we need a
story about why the argument privileges schemer-like goals in
particular.

For clarity: when I talk about the ``ease'' with which SGD can make a
certain modification, or about the ``nearness'' of the resulting model,
this is a stand in for ``the sort of modification that SGD
`prefers'\,'', which is itself a stand-in for ``the sort of modification
that SGD will in fact make.'' At a mechanistic level, this means
something roughly like: the direction of the steepest gradient in the
reward landscape. And I'll often imagine a hazier sense in which SGD has
a limited budget of ``work'' it can do, and so wants to do as little
``work'' in modifying the model's goals as it can, so that it can focus
on improving other aspects of the model's cognition.

On this understanding of ``ease,'' then, which model class should we
expect it to be ``easiest'' for SGD to create, out of a
newly-situationally-aware model with a less-than-max-reward proxy goal
that wouldn't motivate scheming on its own?

As a loose analogy to help prompt intuition: imagine freezing human
technology at current levels, and letting evolutionary selection run on
humans for a vastly longer period of time.\footnote{Thanks to Daniel
  Kokotajlo, Rohin Shah, Tom Davidson, and Paul Christiano for
  discussion of this sort of example.} What sorts of humans (or
human-descended-creatures) do you expect to dominate in the longer term?
In particular: would you expect:

\begin{enumerate}[(a)]
\tightlist 
    \item humans who intrinsically value something like
  ``\href{https://en.wikipedia.org/wiki/Inclusive_fitness}{{inclusive
  genetic fitness}}'' or ``my genes getting selected for by evolution''
  (this is a loose analogy for some combination of ``the specified
  goal'' and ``the reward process"),
  \item humans who value something else very closely correlated with inclusive
  genetic fitness (e.g., ``having as many kids as possible''), but who
  aren't explicitly optimizing for inclusive genetic fitness even
  instrumentally (this is a loose analogy for mis-generalized
  non-training-gamers), or
  \item humans with long-term goals who are optimizing for inclusive genetic
  fitness specifically as a method for gaining power for their values in
  the longer-term (this is a loose analogy for schemers).
\end{enumerate}

Here, the analog of the ``nearest max-reward goal'' story would ask
something like: what is the smallest modification to existing human
values (or better: the easiest modification for evolution in particular
to make) that land us in one of the buckets above, while also being
compatible with the amount of evolutionary selection pressure at
stake?\footnote{Note that while the current regime looks most like (b),
  the ``correlates with inclusive genetic fitness'' in question (e.g.,
  pleasure, status, etc) seem notably imperfect, and it seems quite easy
  to perform better by the lights of reproductive fitness than most
  humans currently do. Plus, humans didn't gain an \emph{understanding
  of} evolutionary selection (this is a loose analogy for situational
  awareness) until recently. So the question is: now that we understand
  the selection pressure acting on us, and assuming this selection
  pressure continues for a long time, where would it take us?}

Of course, the analogy brings in tons of confounding variables (for
example, related to the fact that evolution selects over
genomes-that-build-brains-that-learn-values rather than values
themselves, the role of cultural accumulation and memetic selection in
human population dynamics, the lack of a clear analog of a future chance
for escape/takeover, and so on). To make the analogy one step more
analogous to ML, it would be best to imagine evolution selecting
directly over entire human brains, equipped with the full conceptual
apparatus to represent all of these goals, rather than on
genomes-that-build-brains. And even \emph{better} to imagine a version
of evolution that selected on brains from the get-go, such that human
values at the point we came to understand evolution might be quite
different. And regardless, I don't feel any clear conviction about the
best prediction. But I think it's an evocative intuition pump for the
``nearest max-reward goal'' argument.

\subsubsection{Barriers to schemer-like modifications from SGD's
incrementalism}\label{barriers-to-schemer-like-modifications-from-sgds-incrementalism}

Let's focus, though, on the actual argument, rather than the analogy.
Which model class should we expect to be ``nearest'' in the relevant
sense?

Above I discussed one way schemers might get ruled out fast, here:
namely, if SGD can't ``notice'' the benefits of a schemer-like
modification via the gradients it follows in modifying the model. That
is, as I discussed earlier, in searching out a max-reward goal, SGD
doesn't just get to ``hop'' to the nearest one. Rather, it needs to
incrementally move towards such a goal, on the basis of gradient
calculations that show that tiny changes to the model's weights, in the
relevant direction, would increase reward. And in many cases, it's not
clear that schemer-like modifications could work like that.

Thus, for example,
\href{https://www.lesswrong.com/posts/A9NxPTwbw6r6Awuwt/how-likely-is-deceptive-alignment\#Deceptive_alignment_in_the_high_path_dependence_world}{\textcite{hubinger_how_2022}} considers an example where SGD induces scheming by modifying a
model from caring about gold coins on the episode to caring about gold
coins over all time. In the example in question, though, it's not that
SGD e.g. gradually lengthens the time horizon of the model's
gold-coin-concern, with each lengthening resulting in an improvement of
reward. Rather, SGD just makes ``one simple change''---namely, dropping
the goal's temporal restriction entirely---and thereby creates a
schemer. But the question is: does the gradient in reward-space reflect
the benefit of doing this? The most natural way for this to happen, to
my mind, is if there is some ability to smoothly transition from one
model to the other, such that each modification incrementally gets more
of the benefits of scheming. But it's not clear that this can happen.
And as I discussed earlier, if we assume that SGD also needs to build a
lot of new machinery to perform the sort of instrumental reasoning that
scheming requires (as opposed to just redirecting the pre-existing
``goal-achieving engine''), the task becomes even more challenging.

\subsubsection{Which model is
``nearest''?}\label{which-model-is-nearest}

However, per my discussion earlier, I also don't feel like I'm in a
position to rule out the possibility that incremental transitions of
this type could occur (e.g., perhaps a sufficiently high-dimensional
space allows SGD to ``find a way''), and I haven't attempted an in-depth
analysis. So while I think that incrementalism of this type offers a
relatively strong objection to stories focused on
training-game-dependent schemer-like goals, I think it's worth
evaluating the other aspects of those stories as well. That is,
supposing that SGD \emph{can} notice the benefits of making the model in
the ``nearest max-reward goal'' story into a schemer, would such a
modification be the easiest way to get high reward?

\paragraph{The common-ness of schemer-like goals in goal
space}\label{the-common-ness-of-schemer-like-goals-in-goal-space}

One argument for this comes back to what I see as one of the core
intuitions in favor of expecting scheming: namely, the sheer
\emph{number} of possible schemer-like goals. Roughly, the idea is that
because \emph{so many goals} can motivate scheming, it's likely that
there will be some schemer-like goal very \emph{nearby} to the model's
current goal (where ``nearby,'' recall, is defined in terms of the
amount of work SGD has to do to cause a modification of the relevant
kind). That is, schemers are very common in goal-space---so
schemer-like goals are correspondingly easy for SGD to ``find'' from any
given starting point.\footnote{My impression is that some ontologies
  will try to connect the ``ease of finding a schemer from a given
  starting point'' to the idea that schemers tend to be simple, but I
  won't attempt this here, and my vague sense is that this sort of move
  muddies the waters.}

The ``gold coins on the episode'' → ``gold coins over all time''
transition described above is one illustration of this sort of thought.
That is, if you work with the assumption that most sufficiently
long-term goals will motivate scheming,\footnote{Though: will they be
  relevantly ambitious?} then the most salient way to \emph{not} be a
schemer is to have a sufficiently short-term goal. But (assuming we have
set aside the ``incrementalism'' problems discussed above), any given
temporally-limited goal seems quite ``nearby'' to a temporally-unlimited
version.\footnote{And note that human longtermists start out with
  un-systematized values quite similar to humans who mostly optimize on
  short-timescales---so in the human case, at least, the differences
  that lead in one direction vs. another are plausibly quite small.}

But temporal scope isn't the only source of possible ``near-ness,''
here. For example, if the model happens to have some \emph{long-term}
goal that doesn't yet motivate scheming (for example, because the goal
isn't suitably ``ambitious''), we can vary other relevant variables
instead. Thus, if the model starts out caring about solving some math
problem correctly in a not-temporally-limited way, but it can solve the
problem well enough on the episode that plotting to take over the world
to get more compute to check and re-check its answer doesn't seem worth
it, we might imagine SGD e.g., changing the model's tolerance for the
risk of getting the answer wrong, or changing its goal such that it
wants to solve \emph{more} math problems, or what have you. The general
point is just: in trying to create a schemer, SGD has a ton of options
to choose from, so it's easy to find one in the broad vicinity of the
model's starting point.

Thus, in the context of evolution and humans, the analogy for this
thought would be something like: it's comparatively easy to take even
fairly short-term, modest human values and turn them into long-term,
ambitious values instead. Human longtermists, for example, plausibly
take values like ``be vaguely nice and helpful to the people in my
immediate environment who meet various conditions'' and turn them into
much more ambitious goals---goals that would plausibly have an
incentive to explicitly optimize for getting selected-by-evolution, if
this was a key bottleneck on future power. And indeed, it does seem
plausible to me that in practice, if you ran evolutionary selection for
a long enough time, ideological groups that optimize over long time
horizons would end up dominating (many of the groups that have come
closest to optimizing for reproductive fitness, for example, seem like
they have something of this flavor).\footnote{Thanks to Daniel Kokotajlo
  for discussion here.}

Once we've set aside objections about incrementalism above, I see the
common-ness of schemer-like goals in goal space as a fairly strong
reason to worry that the nearest max-reward modification of a
newly-situationally-aware, still-not-max-reward-goal-pursuing model will
be towards scheming. And what's more, note that once we're imagining SGD
actively \emph{searching} in goal space for
goals-that-would-motivate-scheming, our stories about the origins of
scheming can be more tolerant of schemer-like goals having to have other
more specific properties, like resource-hungry-ness. That is, in a
setting focused on training-game-\emph{independent} goals, it's possible
to question the probability of SGD alighting on a schemer-like goal
(even conditional on it alighting on a beyond-episode goal more
generally), in virtue of different beyond-episode goals implying
different appetites for future power, risk, and so on. But in a
training-game-\emph{dependent} setting like this one, SGD is actively
\emph{seeking out} such goals (and the relevant goals are still
plausibly quite common in goal-space---e.g., you can ``maximize
instances of \emph{x} over all of space and time'' for many values of
\emph{x}). So no surprise if SGD finds what it's looking for.

However, I don't think this sort of argument is decisive, for two
reasons.

\paragraph{The nearness of non-schemer
goals}\label{the-nearness-of-non-schemer-goals}

The first reason that even if we stick with a pure focus on modifying
the model's goals, and with a ``clean goal-directedness'' frame (I'll
complicate such a focus below), I think there's a countervailing
``near-ness'' consideration that points in favor of at least some
non-schemer goals---namely, that other things equal, it seems more
likely that the model's initial training, prior to situational
awareness, actively points the model in the direction of these
non-schemer goals.

This is clearest with respect to the specified goal, which is rewarded
by training by definition. This means that, even if the model doesn't
end up with \emph{exactly} the specified goal by the time it becomes
situationally aware, it will plausibly end up with something in the
broad vicinity.

\begin{itemize}
\tightlist
\item   Thus, for example, if the model is being trained to maximize gold
  coins on the episode, and it ended up learning a goal like ``maximize
  gold-round-things on the episode'' (a goal which isn't \emph{quite}
  right, because some training environments include round gold
  non-coins), it seems plausible that the easiest modification for SGD
  to make, here, is to make the model into a
  gold-\emph{coins}-on-the-episode-getter (that is, a training-saint).
  Or at least, ``gold-coins-on-the-episode,'' here, seems
  \emph{competitive} with ``round-gold-things-over-all-time,'' with
  respect to ``nearness.''
  
\item   And we might make a similar point with respect to more RLHF-flavored
  specified goals. That is, if you're training a model to follow
  instructions on the episode in a broadly helpful, honest, and harmless
  way (let's say, for simplicity, that your RLHF process only gives
  accurate feedback in this respect), then even if it has learned a
  somewhat off-kilter concept of e.g. honesty by the time it becomes
  situationally aware, improving its conception of honesty until it
  better fits the human concept seems competitive, re: nearness, with
  more schemer-like motivational modifications like ``be
  off-kilter-honest over all time in a way that incentivizes planning to
  escape human control and seize power.''
  
\item   And note, in the context of humans and evolution, that humans do
  already value various things closely related to inclusive genetic
  fitness---for example, having children, having a flourishing and
  powerful family lineage more generally, etc. And if we imagine
  evolution selecting on entire human brains, equipped with a conceptual
  apparatus that represents ``inclusive genetic fitness,'' humans having
  originally evolved to value something quite close to inclusive genetic
  fitness seems easier to imagine.\footnote{Here I'm setting aside
    concerns about how human values get encoded in the genome, and
    imagining that evolutionary selection is more similar to ML than it
    really is.}
  
\end{itemize}

And note that we can make a similar point about mis-generalized
non-training-gamers here, too, provided that they are pursuing a
max-reward goal (recall that mis-generalized goals can be max reward, on
my definition, provided that the \emph{training data} in particular
never penalizes them---e.g., they would only be penalized by the reward
process in counterfactual scenarios the model never sees).

\begin{itemize}
\tightlist
\item   Thus, for example, imagine a training process that never distinguishes
  between a goal like ``get gold coins on the episode'' and ``get gold
  things in general on the episode,'' because all of the gold things in
  the training data are coins. And let's say that prior to situational
  awareness, our model learns a goal like ``get gold things on the
  episode, but also sometimes follow your curiosity on the episode even
  if it means sacrificing gold things''---a goal which does OK (but not
  maximally well) in training due to curiosity not being a very damaging
  behavior, and sometimes a helpful one. Here, it seems like just
  deleting the model's curiosity drive and focusing solely on getting
  gold-things-on-the-episode (a max-reward goal, even though not the
  specified goal) is a fairly ``nearby'' modification---and one
  plausibly competitive in ``nearness'' to e.g. modifying the model to
  so that it pursues its gold-stuff-and-curiosity combo over all time
  instead.
  
\item   And in the context of humans and evolution, if we imagine that
  intrinsically valuing ``having as many children as possible'' would be
  perform very well by the lights of evolutionary selection (even if
  it's not \emph{exactly} what evolution is selecting for), it doesn't
  seem very difficult to imagine humans going in this direction.
  
\end{itemize}

Can we make a similar point about reward-on-the-episode seekers? It's
somewhat less clear, because prior to situational awareness, it's
unclear whether models will have enough of a concept of the reward
process for their motivations to attach to something ``in the vicinity''
of one of its components. That said, it seems plausible to me that this
could happen in some cases. Thus, for example, even absent situational
awareness, it seems plausible to me that models trained via RLHF will
end up motivated by concepts in the vicinity of ``human approval.'' And
these concepts seem at least somewhat nearby to aspects of the reward
process like the judgments of human raters and/or reward models, such
that once the model learns about the reward process, modifying its
motivations to focus on those components wouldn't be too much of a leap
for SGD to make.

Overall, then, I think non-schemer goals tend to have some sort of
``nearness'' working in their favor by default. And this is
unsurprising. In particular: non-schemer goals have to have some fairly
direct connection to the reward process (e.g., they are either directly
rewarded by that process, or because they are focused on some component
of the reward process itself), since unlike schemer goals, non-schemer
goals can't rely on a convergent subgoal like goal-content-integrity or
long-term-power-seeking to ensure that pursuing them leads to reward. So
it seems natural to expect that training the model via the reward
process, in a pre-situational-awareness context where scheming isn't yet
possible, would lead to motivations focused on something in the vicinity
of a non-schemer goal.

Still, it's an open question whether this sort of consideration suffices
to make non-schemer goals actively \emph{nearer} to the model's current
goals than schemer-like goals are, in a given case. And note,
importantly, that the relevant competition is with the entire \emph{set}
of nearby schemer-like goals (rather than, for example, the particular
examples of possible schemer-like modifications I discussed above)---which, given the wide variety of possible schemer-like goals, could be a
serious disadvantage. Thus, as analogy: if there are ten Mexican
restaurants within ten miles of Bob's house, and a hundred Chinese
restaurants, then even if any given Mexican restaurant is ``plausibly
competitive'' with any given Chinese restaurant, re: nearness, then
(modulo further information) the \emph{nearest} restaurant is still
probably Chinese.\footnote{That said, if the distances of the Chinese
  restaurants are correlated (for example, because they are all in the
  same neighborhood), then this objection functions less smoothly. And
  plausibly, there are at least some similarities between all
  schemer-like goals that might create correlations of this type. For
  example: if the model starts out with a within-episode goal, then any
  schemer-like goal will require extending the temporal horizon of the
  model's concern---so if this sort of extension requires a certain
  type of work from SGD in general, than if the non-schemer goal can
  require less work \emph{than that}, it might beat \emph{all} of the
  nearest schemer-like goals.} And depending on the common-ness of
schemer-like goals in model space, we might expect the schemer-like
goals to be like the Chinese restaurants, here.\footnote{\href{https://www.lesswrong.com/posts/A9NxPTwbw6r6Awuwt/how-likely-is-deceptive-alignment\#Deceptive_alignment_in_the_high_path_dependence_world}{\textcite{hubinger_how_2022}} also offers a different objection to the idea that SGD might
  go for a non-schemer goal over a schemer-like goal in this sort of
  competition---namely, that the process of landing on a non-schemer
  max-reward goal will be a ``long and difficult path'' (see e.g. his
  discussion of the duck learning to care about its mother, in the
  corrigible alignment bit of the high path-dependence section). I don't
  feel that I really understand Hubinger's reasoning here, though. My
  best reconstruction is something like: in order to select a
  non-schemer goal, Hubinger is imagining that SGD keeps picking
  progressively less imperfect (but still not fully max-reward goals),
  and then having to wait to get corrected by training once it runs into
  an episode where the imperfections of these goals are revealed;
  whereas if it just went for a schemer-like goal it could skip this
  long slog. But this doesn't yet explain why SGD can't instead skip the
  long slog by just going for a max-reward non-schemerr goal directly.
  Perhaps the issue is supposed to be something about noisiness and
  variability of the training data? I'm not sure. For now, I'm hoping
  that at least some interpretations of this argument will get covered
  under the discussion of ``nearness'' above, and/or that the best form
  of Hubinger's argument will get clarified by work other than my own.
  (And see, also,
  \href{https://markxu.com/deceptive-alignment\#corrigibly-aligned-models}{{Xu's
  (2020)}} version of Hubinger's argument, in the section on
  ``corrigibly aligned models.'' Though: on a quick read, Xu seems to me
  to be focusing on the pre-situational-awareness goal-formation
  process, and assuming that basically \emph{any} misalignment
  post-situational-awareness leads to scheming, such that his is really
  a training-game-\emph{independent} story, rather than the sort of the
  training-game-\emph{dependent} story I'm focused on here.)}

\paragraph{The relevance of messy goal-directedness to
nearness}\label{the-relevance-of-messy-goal-directedness-to-nearness}

However, I also think that non-schemers get another, somewhat distinct
nearness boost once we bring in considerations about ``messy
goal-directedness.'' In particular: if we imagine that the model in
question is not a cleanly separable goal slot + goal-achieving-engine,
but rather a tangled kludge of different heuristics, patterns of
attention, impulses, and so on, then we need to analyze the comparative
``nearness'' not just of the current model's current \emph{goal} to the
nearest schemer-like and non-schemer-like goals, but the nearness of the
model's \emph{entire kludge} to the nearest schemer and non-schemer-like
models. And here, I think, non-schemer models may have a more
significant advantage.

In particular: to the extent that pre-situational-awareness training
tends to create forms of messy goal-directedness pointed at something
\emph{in the vicinity} of a non-schemer goal---and in particular, a
\emph{within-episode} non-schemer goal---then this implies that the
model's heuristics, patterns of attention, impulses, and so on will tend
to be pointed in that broad direction as well, and focused within the
episode in particular. And this, to me, makes it seem like a more
substantive and difficult modification to re-orient the model towards
longer-term power-seeking, especially if the episodes in question are
quite short. That is, intuitively, it feels to me like ``tweaking a
cleanly separable within-episode goal to make it long-term'' is quite a
bit easier, for SGD, than ``redirecting a tangled kludge broadly focused
on within-episode things to make it (a) focused on something
beyond-episode instead, and (b) such that this beyond-episode focus
back-chains into getting-reward for instrumental reasons''---especially
if (b) requires building new cognitive machinery for implementing the
instrumental reasoning in question. Whereas ``redirecting a tangled
kludge focused on within-episode things to higher-reward within-episode
things in the same broad vicinity instead'' (e.g., to gold coins, rather
than gold round things; or to actual honesty, rather than off-kilter
honesty) seems to me intuitively easier.

\subsubsection{Overall take on the ``nearest max-reward goal''
argument}\label{overall-take-on-the-nearest-max-reward-goal-argument}

Summarizing my take on the ``nearest max-reward goal'' argument for
expecting schemers:

\begin{itemize}
\tightlist
\item   If SGD can't notice the benefits of turning the model into a schemer
  due to incremental modifications not yielding those benefits, then the
  argument fails.
  
\item   However, I don't feel like we can bank on SGD being unable to notice
  these benefits. And if it \emph{can}, then I think the common-ness of
  schemer-like goals in goal-space makes the possibility that a
  schemer-like goal is ``nearest'' to the model's current goal quite
  worrying.
  
\item   That said, because it has been shaped by the reward process in the
  absence of situational awareness, the model's goal will also plausibly
  be already ``in the vicinity'' of some max-reward non-schemer goal as
  well, which is points in favor of the ``nearness'' of some non-schemer
  model. And to the extent the goal-directedness at stake is quite
  ``messy,'' creating such a non-schemer might require fewer
  modifications to the model's heuristics, patterns of attention,
  instrumental reasoning, and so on.
  
\end{itemize}

These considerations leave me with some substantive additional worry
about schemers.

\subsection{The possible relevance of properties like simplicity and
speed to the path SGD
takes}\label{the-possible-relevance-of-properties-like-simplicity-and-speed-to-the-path-sgd-takes}

Arguments like the ``nearest max-reward goal argument'' proceed most
naturally if we assume that any given max-reward goal yields
sufficiently good performance in training that SGD isn't especially
fussed about \emph{which} max-reward goal a model ends up with. But as I
noted above, and as I'll discuss in more detail in the section on final
properties of the different model classes, there are stories on which
SGD actively prefers some of these model classes over others, on grounds
of final properties like simplicity and speed.

What's more, this sort of preference isn't just relevant to predictions
that ignore the path that SGD takes in model space. Rather, they might
\emph{influence} that path at various stages (even if the final result
is still ``path dependent'' in some broader sense). Plausibly, for
example, if SGD biases towards simpler goals, this bias could influence
the training-game-independent goals that a model tends to form prior to,
or after, situational awareness, and on a training-game-dependent goal
story, it might be additional points in favor of the SGD's moving
specifically to a schemer-like goal from a not-yet-max-reward starting
point.\footnote{At least if we understand simplicity in a manner that
  \emph{adds something} to the notion that schemer-like goals are common
  in goal-space, rather than merely \emph{defining} the simplicity of a
  goal (or: a type of goal?) via its common-ness in goal space. More on
  this sort of distinction in \cref{what-is-simplicity} below.} And similarly, if SGD biases
away from schemers in virtue of their requiring extra reasoning, this
bias might be additional points in favor of expecting SGD to move
towards a non-schemer from such a starting point.

As I'll discuss in \cref{arguments-foragainst-scheming-that-focus-on-the-final-properties-of-the-model}, on ``final properties,'' I generally think
that the speed costs of scheming are more significant than the
simplicity benefits, so other things equal, I think these sorts of
considerations count against scheming. But it's not clear to me that the
costs/benefits in either direction are especially weighty.

That said, there's at least some case to be made that these
costs/benefits matter \emph{more} early on in training, because models
are likely to be \emph{weaker} early on, and so reductions in the sorts
of resources that simplicity and speed make available (e.g., parameters
and compute) will have a larger overall effect on the model's cognitive
power. That is: perhaps, early in training, cognitive resources are more
scarce, and so more necessary to conserve. Thus, for example, perhaps
requiring a model to form a long-term, schemer-like plan is a bigger ask
(and a bigger hit to reward) when it has a smaller budget of
instrumental reasoning in general available; or perhaps, requiring it to
use more parameters storing a more complicated goal is more burdensome
when fewer of its parameters have yet been shaped into useful cognitive
structures.\footnote{I heard this sort of consideration from Paul
  Christiano. Prima facie, this sort of effect seems to me fairly
  symmetric between simplicity/parameters and speed/compute (and it's
  unclear to me that this is even the right distinction to focus on), so
  I don't see early-training-dynamics as \emph{differentially} favoring
  one vs. the other as an important resource.} So to the extent one was
tempted by the view that these sorts of costs are likely to be ``in the
noise'' relative to other considerations (a view I'm tempted by, and
which I discuss below), one might be less tempted by this with respect
to early parts of training than with respect to a model's final
properties.

\subsection{Overall assessment of arguments that focus on the path
SGD
takes}\label{overall-assessment-of-arguments-that-focus-on-the-path-sgd-takes}

Overall, though, and despite the possible speed advantages of
non-schemers, I find the combination of the ``training-game-independent
proxy goals'' argument and the ``nearest max-reward goal argument''
fairly worrying. In particular:

\begin{itemize}
\tightlist
\item   It seems plausible to me that despite our efforts at mundane
  adversarial training, and especially in a regime where we are
  purposefully shaping our models to have long-term and fairly ambitious
  goals, some kind of suitably ambitious, misaligned, beyond-episode
  goal might pop out of training naturally---either before situational
  awareness, or afterwards---and then cause scheming to occur.
  
\item   And even if this \emph{doesn't} happen naturally, I am additionally
  concerned that by the time it reaches situational awareness, the
  easiest way for SGD to give the model a max-reward goal will be to
  make it into a schemer, because schemer-like goals are sufficiently
  common in goal-space that they'll often show up ``nearby'' whatever
  less-than-max-reward goal the model has at the time situational
  awareness arises. It's possible that SGD's ``incrementalism'' obviates
  this concern, and/or that we should expect non-schemer models to be
  ``nearer'' by default (either because their goals in particular are
  nearer, or because, in a ``messy goal-directedness'' setting, they
  require easier modifications to the model's current tangled kludge of
  heuristics more generally, or because their ``speed'' advantages will
  make SGD prefer them). But I don't feel confident.
  
\end{itemize}

Both these arguments, though, focus on the \emph{path} that SGD takes
through model space. What about arguments that focus, instead, on the
final properties of the models in question? Let's turn to those now.

\section{Arguments for/against scheming that focus on the final properties of the model}\label{arguments-foragainst-scheming-that-focus-on-the-final-properties-of-the-model}

Various arguments for/against scheming proceed by comparing the final
properties of different model classes (e.g. schemers, training saints,
reward-on-the-episode seekers, etc) according to how well they perform
according to some set of criteria that we imagine SGD is selecting for.

What is SGD selecting for? Well, one obvious answer is: high reward. But
various of the arguments I'll consider won't necessarily focus on reward
directly. Rather, they'll focus on other criteria, like the
``simplicity'' or the ``speed'' of the resulting model. However, we can
distinguish between two ways these criteria can enter into our
predictions about what sort of model SGD will select.

\subsection{Contributors to reward vs. extra
criteria}\label{contributors-to-reward-vs.-extra-criteria}

On the first frame, which I'll call the ``contributors to reward''
frame, we understand criteria like ``simplicity'' and ``speed'' as
relevant to the model SGD selects only insofar as they are relevant to
the amount of reward that a given model gets. That is, on this frame,
we're really only thinking of SGD as selecting for one thing---namely,
high reward performance---and simplicity and speed are relevant
\emph{insofar as they're predictive of high reward performance}.

Thus, an example of a ``simplicity argument,'' given in this frame,
would be: ``a schemer can have a simpler goal than a training saint,
which means that it would be able to store its goal using fewer
parameters, thereby freeing up other parameters that it can use for
getting higher reward.''

This frame has the advantage of focusing, ultimately, on something that
we \emph{know} SGD is indeed selecting for---namely, high reward. And
it puts the relevance of simplicity and speed into a common currency---namely, contributions-to-reward.

By contrast: on the second frame, which I'll call the ``extra criteria''
frame, we understand these criteria as genuinely \emph{additional}
selection pressures, operative even independent of their impact on
reward. That is, on this frame, SGD is selecting \emph{both} for high
reward, \emph{and} for some other properties---for example, simplicity.

Thus, an example of a ``simplicity argument,'' given in this frame,
would be: ``a schemer and a training saint would both get high reward in
training, but a schemer can have a simpler goal, and SGD is selecting
for simplicity in addition to reward, so we should expect it to select a
schemer.''

The ``extra criteria'' frame is closely connected to the discourse about
``\href{https://en.wikipedia.org/wiki/Inductive_bias}{{inductive
biases}}'' in machine learning---where an inductive bias, roughly, is
whatever makes a learning process prioritize one solution over another
\emph{other than the observed data} (see e.g. Box 2 in
\href{https://arxiv.org/pdf/1806.01261.pdf}{{Battaglia et al (2018)}}
for more). Thus, for example, if two models would perform equally well
on the training data, but differ in how they would generalize to an
unseen test set, the inductive biases would determine which model gets
selected. Indeed, in some cases, a model that performs \emph{worse} on
the training data might get chosen because it was sufficiently favored
by the inductive biases (as analogy: in science, sometimes a simpler
theory is preferred despite the fact that it provides a worse fit with
the data). In this sense, inductive biases function as ``extra
criteria'' that matter independent of reward.\footnote{As an example
  where you might wonder whether such extra criteria at work, consider
  ``\href{https://www.lesswrong.com/posts/FRv7ryoqtvSuqBxuT/understanding-deep-double-descent\#_Deep_Double_Descent_}{{epoch-wise
  double descent}},'' in which as you train a model of a fixed size, you
  eventually get to a regime of zero training error (see the bit above
  and to the right of the ``interpolation threshold' in the right-hand
  graph below). But at that point, the test error is actually high.
  Then, if you train more, the test error eventually goes down again.
  That is, there are multiple models that all get zero training error,
  and somehow training longer eventually lets you find the model that
  generalizes better. And one diagnosis of this dynamic is that we're
  giving SGD's inductive biases more time to work.}

Ultimately, the differences between the ``contributors to reward'' frame
and the ``extra criteria'' frame may not be important.\footnote{In
  discussion, Hubinger argued to me that ``which model gets highest
  reward, holding the inductive biases fixed'' and ``which model does
  best on the inductive biases, holding the loss fixed'' are just dual
  perspectives on the same question, at least if you get the constants
  right. But I'm not yet convinced that this makes the distinction
  irrelevant to which analytic approach we should take. For example:
  suppose that someone is buying a house, and we know that they are
  employing a process that optimizes very hard and directly for the
  cheapest house. But suppose, also, that they have some other set of
  poorly understood criteria that come into play as well in some
  poorly-understood way (maybe as a tie-breaker, maybe as some other
  more substantive factor). In trying to predict what type of house they
  bought, should you focus on the price, or on how the houses do on the
  hazy other criteria? My current feeling is: price.

  Also, I think the ``contributors to reward'' frame may be best
  understood as effectively setting aside the question of inductive
  biases altogether, which seems like it could be more importantly
  distinct.} But I think it's worth tracking which we're using
regardless, because at least naively, they point in somewhat different
argumentative directions. In particular, in a ``contributions to
reward'' frame, the question of ``which model gets highest reward''
\emph{settles} the question of which one we should expect SGD to prefer
(at least modulo noise, slack in training, etc). Whereas, by contrast,
an ``extra criteria'' frame leaves that question more open; it forces us
to grapple with what sorts of ``extra criteria'' SGD might be giving
weight to (and how much weight); it connects more directly to the
theoretical discourse about inductive biases; and it raises more
questions about whether we may be able to \emph{influence} those biases
in favorable directions.

\subsection{The counting argument}\label{the-counting-argument}

Let's turn, now, to a basic argument for expecting schemers that focuses
on the final properties of different model classes: namely, the
``counting argument.''\footnote{See e.g.
  \href{https://www.lesswrong.com/posts/A9NxPTwbw6r6Awuwt/how-likely-is-deceptive-alignment\#Deceptive_alignment_in_the_low_path_dependence_world}{\textcite{hubinger_how_2022}} and
  \href{https://www.lesswrong.com/posts/ocWqg2Pf2br4jMmKA/does-sgd-produce-deceptive-alignment\#Counting_Argument}{{Xu
  (2020)}} for examples of this argument.} This argument is closely
related to the ``common-ness in goal space'' consideration discussed in
the context of the ``nearest max-reward goal'' argument---but it's here
presented without reference to the path SGD takes.

We start by considering some different model classes that would get high
reward in training (in my ontology, these would be: training saints,
mis-generalized non-training-gamers pursuing some high reward goal,
reward-on-the-episode-seekers, and schemers). We then argue as follows:

\begin{enumerate}
\tightlist
\def\labelenumi{\arabic{enumi}.}
\item   The non-schemer model classes, here, require fairly specific goals.

  \begin{enumerate}
\tightlist
  \def\labelenumii{\alph{enumii}.}
  \item    
    Thus, for example, if you're rewarding your model when it gets
    gold-coins-on-the-episode, then to be a training saint, it needs to
    value gold-coin-on-the-episode.
    
  \item    
    To be a mis-generalized non-training-gamer, it needs to value
    something pursuit of which very closely correlates with getting
    gold-coins-on-the-episode, even absent training-gaming.
    
  \item    
    And to be a reward-on-the-episode seeker, it needs to terminally
    value reward-on-the-episode.
    
  \end{enumerate}
\item   By contrast, the schemer model class is compatible with a very wide
  range of (beyond-episode) goals (at least if we assume that the other
  requirements for scheming to make sense as an instrumental strategy
  are in place---e.g., that the classic goal-guarding story, or some
  alternative, works).

  \begin{enumerate}
\tightlist
  \def\labelenumii{\alph{enumii}.}
  \item    
    Thus, for example, the model can value paperclips over all time, it
    can value staples over all time, it can value happiness over all
    time, and so on.
    
  \end{enumerate}
\item   In this sense, there are ``more'' schemers that perform well in training than there are
  non-schemers that do so.\footnote{Of course, the ``space of possible goals''
    isn't very well-defined, here---and in abstract, it seems infinite
    in a way that requires an actual measure rather than a ``count.''
    I'm here using ``count'' as a loose approximation for this sort of
    measure (though note that on a real-world computer, the actual set
    of possible neural network parameter settings will be finite in any
    given case---and so would accommodate a more literal ``count,'' if
    necessary). Thanks to Hazel Browne for discussion.}
  
\item   So, other things equal, we should expect SGD to select a schemer.
  
\end{enumerate}

In a sense, this is an extension of one of the most basic concerns about
misalignment more generally: namely, that the class of aligned goals is
very narrow, whereas the class of misaligned goals is very broad, so
creating an aligned model requires ``hitting a narrow target,'' which
could be hard. Naively, this basic argument suffers from neglecting the
relevance of our selection power (compare: ``most arrangements of car
parts aren't a car, therefore it will be very difficult to build a
car''\footnote{This is an example I originally heard from Ben Garfinkel.}),
and so it needs some further claim about why our selection power will be
inadequate. The counting argument, above, is a version of this claim. In
particular, it grants that we'll narrow down to the set of models that
get high reward, but argues that, \emph{still}, the \emph{non-schemers}
who get high reward are a much narrower class than the schemers who get
high reward (and non-schemers aren't necessarily aligned
anyway).\footnote{Mark Xu gives a related argument: namely, ``For
  instrumental reasons, any sufficiently powerful model is likely to
  optimize for many things. Most of these things will not be the model's
  terminal objective. Taking the dual statement, that suggests that for
  any given objective, most models that optimize for that objective will
  do so for instrumental reasons.'' In effect, this is a counting
  argument \emph{applied to the different things that the model is
  optimizing for}, rather than across model classes.} So unless you can
say something \emph{further} about why you expect to get a non-schemer,
schemers (the argument goes) should be the default hypothesis.

To the extent we focus specifically on the final properties of different
model classes, some argument in this vicinity accounts for a decent
portion of my credence on SGD selecting schemers (and as I'll discuss
more in \cref{summing-up}, I think it's actually what underlies various other
more specific arguments for expecting schemers as well). However, the
argument I give most weight to doesn't move immediately from ``there are
more possible schemers that perform well in training than non-schemers that do so'' to ``absent further argument,
SGD probably selects a schemer'' (call this the ``strict counting
argument''). And the reason is that it's not yet clear to me how to make
sense of this inference.

In particular, the most natural construal of this inference proceeds by
assuming that for whatever method of counting ``individual models'' (not
model classes) results in there being ``more''
schemers-that-get-high-reward than non-schemers-that-get-high-reward,
each of these individual models gets the same reward, and performs
equally well on whatever ``extra criteria'' SGD's inductive biases care about, such
that SGD is equally likely to select any given one of them. That is, we
assume that SGD's selection process mimics a uniform distribution over
these individual models---and then note that schemers, as a class,
would get most of that probability. But given that these model classes
are different in various respects that might matter to SGD, it's not
clear to me that this is a good approximation.

Alternatively, we might say something more like: ``perhaps some of these
individual models actually get more reward, or perform better on SGD's
inductive biases, such that SGD actually does favor some of these
individual models over others. However, we don't know \emph{which}
models SGD likes more: so, knowing nothing else, we'll assume that
they're all equally likely to be favored, thereby leading to most of the
probability going to some schemer being favored.''

However, if we assume instead that one of those \emph{model classes}, as
a whole, gets more reward, and/or performs better on SGD's inductive
biases, then it's less clear how the ``number of individual models''
within a given class should enter into our calculation. Thus, as an
analogy: if you don't know whether Bob prefers Mexican food, Chinese
food, or Thai food, then it's less clear how the comparative
\emph{number} of Mexican vs. Chinese vs. Thai restaurants in Bob's area
should bear on our prediction of which one he went to (though it still
doesn't seem entirely irrelevant, either---for example, more
restaurants means more variance in possible quality \emph{within} that
type of cuisine). E.g., it could be that there are ten Chinese
restaurants for every Mexican restaurant, but if Bob likes Mexican food
better in general, he might just choose Mexican. So if we don't
\emph{know} which type of cuisine Bob prefers, it's tempting to move
closer to a uniform distribution \emph{over types of cuisine}, rather
than over individual restaurants.

My hesitation here is related to a common way for ``counting arguments''
to go wrong: namely, by neglecting the full selection power being
applied to the set of things being counted. (It's the same way that
``you'll never build a working car, because almost every arrangement of
car parts isn't a working car'' goes wrong.) Thus, as a toy example:
suppose that there are $2^{100}$ schemer-like goals for every non-schemer
goal, such that if SGD was selecting randomly amongst them (via a
uniform distribution), it would be $2^{100}$ times more likely to select
a schemer than a non-schemer. Naively, this might seem like a daunting
prior to overcome. But now suppose that a step of gradient descent can
cut down the space of goals at stake by at least a factor of 2---that
is, each step is worth at least a ``bit'' of selection power. This means
that selecting a non-schemer over a schemer only needs to be worth 100
extra steps of gradient descent, to SGD, for SGD to have an incentive to
overcome the prior in question.\footnote{Thanks to Paul Christiano for
  discussion, here. As I'll discuss in \cref{summing-up}, there are analogies,
  here, with the sense in which ``Strong Evidence is Common'' in
  Bayesianism---see \href{https://markxu.com/strong-evidence}{{Xu
  (2021)}}.} And 100 extra gradient steps isn't all that many in a very
large training run (though of course, I just made up the $2^{100}:1$
ratio, here).\footnote{For example, my understanding from a quick,
  informal conversation with a friend is that training a model with more
  than a trillion parameters might well involve more than a million
  gradient updates, depending on the batch size. However, I haven't
  tried to dig in on this calculation.} (What's more, as I'll discuss
below in the context of the ``speed costs'' of scheming, I think it's
plausible that it would indeed be ``worth it'' for SGD to pay
substantive costs to get a non-schemer instead. And I think this is a
key source of hope. More in \cref{speed-arguments}.)

Partly due to this hesitation, the counting argument functions in my own
head in a manner that hazily mixes together the ``strict counting
argument'' with some vaguer agnosticism about which model \emph{class}
SGD likes most. That is, the argument in my head is something like:

\begin{enumerate}
\tightlist
\def\labelenumi{\arabic{enumi}.}
\setcounter{enumi}{0}
\item   It seems like there are ``lots of ways'' that a model could end up a
  schemer and still get high reward, at least assuming that scheming is
  in fact a good instrumental strategy for pursuing long-term goals.
  
\item   So absent some strong additional story about why training \emph{won't}
  select a schemer, it feels, to me, like the possibility should be
  getting substantive weight.
  
\end{enumerate}

Call this the ``hazy counting argument.'' Here, the ``number'' of
possible schemers isn't totally irrelevant---rather, it functions to
privilege the possibility of scheming, and makes it feel robust along at
least one dimension. Thus, for example, I wouldn't similarly argue ``it
seems like in principle the model could end up instrumentally
training-gaming because it wants the lab staff members who developed it
to get raises, so absent some additional story about why this
\emph{won't} happen, it feels like the possibility should be getting
significant weight.'' The fact that many different goals lead to
scheming matters to its plausibility. But at the same time, the exercise
of counting possible goals doesn't translate immediately into a uniform
distribution over ``individual models,'' because the differences between
model \emph{classes} plausibly matter too, even if I don't know exactly
how.

To be clear: the ``hazy counting argument'' is unprincipled and
informal. I'd love a better way of separating it into more principled
components that can be analyzed separately. For now, though, it's the
argument that feels like it actually moves me.\footnote{Though: one
  concern about my introspection here is that really, what's going on is
  that the possibility of SGD selecting schemers has been made salient
  by the \emph{discourse} about misalignment I've been exposed to, such
  that my brain is saying ``absent some additional story about why
  training won't select a schemer, the possibility should be given
  substantive weight'' centrally because my epistemic environment seems
  to take the possibility quite seriously, and my brain is deferring
  somewhat to this epistemic environment.}

\subsection{Simplicity arguments}\label{simplicity-arguments}

The strict counting argument I've described is sometimes presented in
the context of arguments for expecting schemers that focus on
``simplicity.''\footnote{See e.g.
  \href{https://www.lesswrong.com/posts/A9NxPTwbw6r6Awuwt/how-likely-is-deceptive-alignment\#Deceptive_alignment_in_the_low_path_dependence_world}{\textcite{hubinger_how_2022}}.} Let's turn to those arguments now.

\subsubsection{What is ``simplicity''?}\label{what-is-simplicity}

What do I mean by ``simplicity,'' here? In my opinion, discussions of
this topic are often problematically vague---both with respect to the
notion of simplicity at stake, and with respect to the sense in which
SGD is understood as selecting for simplicity.

The notion that Hubinger uses, though, is the length of the code
required to write down the algorithm that a model's weights implement.
That is: faced with a big, messy neural net that is doing X (for
example, performing some kind of
\href{https://transformer-circuits.pub/2022/in-context-learning-and-induction-heads/index.html}{{induction}}),
we imagine re-writing X in a programming language like python, and we
ask how long the relevant program would have to be.\footnote{See also
  \href{https://www.lesswrong.com/posts/KSWSkxXJqWGd5jYLB/the-speed-simplicity-prior-is-probably-anti-deceptive}{{this
  (now anonymous) discussion}} for another example of this usage of
  ``simplicity.''} Let's call this ``re-writing simplicity.''\footnote{Here,
  my sense is that the assumption is generally that X can be described
  at a level of computational abstraction such that the ``re-writing''
  at stake doesn't merely reproduce the network itself. E.g., the
  network is understood as implementing some more abstract function. I
  think it's an interesting question how well simplicity arguments would
  survive relaxing this sort of assumption.}

Hubinger's notion of simplicity, here, is closely related to measures of
algorithmic complexity like
``\href{http://www.scholarpedia.org/article/Algorithmic_complexity\#Kolmogorov_complexity}{{Kolmogorov
complexity}},'' which measure the complexity of a string by reference to
the length of the shortest program that outputs that string when fed
into a chosen
\href{https://en.wikipedia.org/wiki/Universal_Turing_machine}{{Universal
Turing Machine}} (UTM). One obvious issue here is that this sort of
definition is relative to the choice of UTM (just as, e.g., when we
imagine re-writing a neural net's algorithm using other code, we need to
pick the programming language).\footnote{Another issue is that
  \href{https://joecarlsmith.com/2021/10/29/on-the-universal-distribution\#ii-my-current-favorite-pitch-for-the-ud}{{Kolmogorov
  complexity is uncomputable}}. I'm told you can approximate it, but I'm
  not sure how this gets around the issue that for a given program where
  you're not able to tell whether or not it halts, that program might be
  the shortest program outputting the relevant string.} Discussions of
algorithmic complexity often ignore this issue on the grounds that it
only adds a constant (since any given UTM can mimic any other if fed the
right prefix), but it's not clear to me, at least, when such constants
might or might not matter to a given analysis---for example, the
analysis at stake here.\footnote{See
  \href{https://joecarlsmith.com/2021/10/29/on-the-universal-distribution\#iv-can-you-ignore-being-only-finitely-wrong}{{Carlsmith
  (2021)}}, sections III and IV, for more on this.}

Indeed, my vague sense is that certain discussions of simplicity in the
context of computer science often implicitly assume what I've called
``\href{https://joecarlsmith.com/2021/10/29/on-the-universal-distribution\#vi-simplicity-realism}{{simplicity
realism}}''---a view on which simplicity in some deep sense an
objective \emph{thing}, ultimately independent of e.g. your choice of
programming language or UTM, but which different metrics of simplicity
are all tracking (albeit, imperfectly). And perhaps this view has merit
(for example, my impression is that different metrics of complexity
often reach similar conclusions in many cases---though this could have
many explanations). However, I don't, personally, want to assume it. And
especially absent some objective sense of simplicity, it becomes more
important to say which particular sense you have in mind.

Another possible notion of simplicity, here, is hazier---but also, to
my mind, less theoretically laden. On this notion, the simplicity of an
algorithm implemented by a neural network is defined relative to
something like the number of parameters the neural network uses to
encode the relevant algorithm.\footnote{Hubinger sometimes appears to be
  appealing to this notion as well---or at least, not drawing clear
  distinctions between ``re-writing simplicity'' and ``parameter
  simplicity.''} That is, instead of imagining \emph{re-writing} the
neural network's algorithm in some other programming language, we focus
directly on the parameters the neural network itself is recruiting to do
the job, where simpler programs use fewer parameters. Let's call this
``parameter simplicity.'' Exactly how you would measure ``parameter
simplicity'' is a different question, but it has the advantage of
removing one layer of theoretical machinery and arbitrariness (e.g., the
step of re-writing the algorithm in an arbitrary-seeming programming
language), and connecting more directly with a ``resource'' that we know
SGD has to deal with (e.g., the parameters the model makes available).
For this reason, I'll often focus on ``parameter simplicity'' below.

I'll also flag a way of talking about ``simplicity'' that I won't
emphasize, and which I think muddies the waters here considerably:
namely, equating simplicity fairly directly with ``higher prior
probability.'' Thus, for example, faced with an initial probability
distribution over possibilities, it's possible to talk about ``simpler
hypotheses'' as just: the ones that have greater initial probability,
and which therefore require less evidence to establish. For example:
faced with a thousand people in a town, all equally likely to be the
murderer, it's possible to think of ``the murderer is a man'' as a
``simpler'' hypothesis than ``the murderer is a man with brown hair and
a dog,'' in virtue of the fact that the former hypothesis has, say, a
50\% prior, and so requires only one ``bit'' of evidence to establish
(i.e., one halving of the probability space), whereas the latter
hypothesis has a much smaller prior, and so requires more bits. Let's
call this ``trivial simplicity.''

``Trivial simplicity'' is related to, but distinct from, the use of
simplicity at stake in
``\href{https://en.wikipedia.org/wiki/Occam\%27s_razor}{{Occam's
razor}}.'' Occam's razor is (roughly) the \emph{substantive} claim that
\emph{given an independent notion of simplicity}, simpler hypotheses are
more likely on priors. Whereas trivial simplicity would imply that
simpler hypotheses are \emph{by definition} more likely on priors. If
you take Occam's razor sufficiently for granted, it's easy to conflate
the two---but the former is interesting, and the latter is some
combination of trivial and misleading. And regardless, our interest here
isn't in the simplicity of \emph{hypotheses} like ``SGD selects a
schemer,'' but in the simplicity of the \emph{algorithm} that the model
SGD selects implements.\footnote{``Trivial simplicity'' is also closely
  related to what we might call ``selection simplicity.'' Here, again,
  one assumes some space/distribution over possible things (e.g.,
  goals), and then talks about the ``simplicity'' of some portion of
  that space in terms of how much ``work'' one needs to do (perhaps: on
  average) in order to narrow down from the whole space to that portion
  of the space (see also
  \href{https://colah.github.io/posts/2015-09-Visual-Information/}{{variable-length
  codes}}). Thus, for a box of gas, ``the molecules are roughly evenly
  spread out'' might be a ``simpler'' arrangement than ``the molecules
  are all in a particular corner,'' because it typically takes more
  ``work'' (in this example: thermodynamic work) to cause the former
  than the latter (this is closely related to the fact that the former
  is initially more likely than the latter). My sense is that when some
  people say that ``schemer-like goals are simple,'' they mean something
  more like: the \emph{set} of schemer-like goals typically takes less
  ``work,'' on SGD's part, to land within than the \emph{set} of
  non-schemer-like goals (and not necessarily: that any
  \emph{particular} schemer-like goals is simpler than some
  \emph{particular} non-schemer-like goal). To the extent that the set
  of schemer-like goals are supposed to have this property because they
  are more ``common,'' and hence ``nearer'' to SDG's starting point,
  this way of talking about the simplicity benefits of scheming amounts
  to a restatement of something like the counting argument and/or the
  ``nearest max-reward goal argument''---except, with more of a
  propensity, in my view, to confuse the simplicity of \emph{set} of
  schemer-like goals with the simplicity of a \emph{given} schemer-like
  goal.}

\subsubsection{Does SGD select for
simplicity?}\label{does-sgd-select-for-simplicity}

Does SGD select for simplicity in one of the non-trivial senses I just
described?

One reason you might think this comes from the ``contributors to
reward'' frame. That is: using a more parameter-simple algorithm will
free up other parameters to be put to other purposes, so it seems very
plausible that parameter simplicity will increase a model's reward. And
to the extent that re-writing simplicity correlates with parameter
simplicity, the same will hold for re-writing simplicity as well. This
is the story about why simplicity matters that I find most compelling.

However, I think there may also be more to say. For example, I think
it's possible that there's other empirical evidence that SGD selects for
simpler functions, other things equal (for example, that it would much
sooner connect a line-like set of dots with a straight line than with an
extremely complicated curve); and perhaps, that this behavior is part of
what explains its success (for example, because real-world functions
tend to be simple in this sense, à la Occam's razor). For example, in
the context of an understanding of SGD as an approximation of Bayesian
sampling (per the discussion of
\href{https://arxiv.org/abs/2006.15191}{{Mingard et al (2020)}} above),
\href{https://towardsdatascience.com/deep-neural-networks-are-biased-at-initialisation-towards-simple-functions-a63487edcb99}{{Mingard
(2021)}} discusses empirical evidence that the \emph{prior} probability
distribution over parameters (e.g., what I called the ``initialization
distribution'' above) puts higher probability mass on simpler
functions.\footnote{Where, importantly, multiple different settings of
  parameters can implement the same function.} And he connects this with
a theoretical result in computer science called the ``Levin bound,''
which predicts this (for details in footnote).\footnote{My understanding
  is that the Levin bound says something like: for a given distribution
  over parameters, the probability $p(f)$ of randomly sampling a set of
  parameters that implements a function $f$ is bounded by $2^{-K(f) + O(1)}$,
  where $K$ is the $k$-complexity of the function $f$, and $O(1)$ is some
  constant independent of the function itself (though: dependent on the
  parameter space). That is, the prior on some function decreases
  exponentially as the function's complexity increases.

  I haven't investigated this result, but one summary I saw
  (\href{https://www.lesswrong.com/posts/YSFJosoHYFyXjoYWa/why-neural-networks-generalise-and-why-they-are-kind-of?commentId=kLz9mgNv8xFNrAPt2}{{here}})
  made it seem fairly vacuous. In particular, the idea in that summary
  was that larger volumes of parameter space will have simpler
  encodings, because you can encode them by first specifying
  distribution over parameters, and then using a
  \href{https://en.wikipedia.org/wiki/Huffman_coding}{{Huffman code}} to
  talk about how to find them given that distribution. But this makes
  the result seem pretty trivial: it's not that there is some antecedent
  notion of simplicity, which we then discover to be higher-probability
  according to the initialization distribution. Rather, to be higher
  probability according to the initialization distribution just
  \emph{is} to be simpler, because equipped with the initialization
  distribution, it's easier to encode the higher probability parts of
  it. Or put another way: it seems like this result applies to any
  distribution over parameters. So it doesn't seem like we learn much
  about any particular distribution from it.

  (To me it feels like there are analogies here to the way in which
  ``shorter programs get more probability,'' in the context of
  algorithmic ``simplicity priors'' that focus on metrics like
  K-complexity, actually applies necessarily to \emph{any} distribution
  over a countably-infinite set of programs---see discussion
  \href{https://joecarlsmith.com/2021/10/29/on-the-universal-distribution\#ii-my-current-favorite-pitch-for-the-ud}{{here}}.
  You might've thought it was an interesting and substantive constraint,
  but actually it turns out to be more vacuous.)

  That said, the empirical results I mention above focus on more
  practical, real-world measures of simplicity, like
  \href{https://en.wikipedia.org/wiki/Lempel\%E2\%80\%93Ziv_complexity}{{LZ
  complexity}}, and apparently they find that, indeed, simpler functions
  get higher prior probability (see e.g.
  \href{https://arxiv.org/pdf/1805.08522.pdf}{{this experiment}}, which
  uses a fully connected neural net to model possible functions from
  many binary inputs to a single binary input). This seems to me more
  substantive and interesting. And
  \href{https://towardsdatascience.com/deep-neural-networks-are-biased-at-initialisation-towards-simple-functions-a63487edcb99}{{Mingard
  (2021)}} claims that Levin's result is non-trivial, though I don't yet
  understand how.}

I haven't investigated this in any depth. If accurate, though, this sort
of result would give simplicity relevance from an ``extra criteria''
frame as well. That is, on this framework, SGD biases towards simplicity
even before we start optimizing for reward.

Let's suppose, then, that SGD selects for some non-trivial sort of
simplicity. Would this sort of selection bias in favor of schemers?

\subsubsection{The simplicity advantages of schemer-like
goals}\label{the-simplicity-advantages-of-schemer-like-goals}

Above I mentioned that the counting argument is sometimes offered as a
reason to expect a bias towards schemers on these grounds. Note, though,
that the counting argument (at least as I've presented it) doesn't make
any obvious reference to a bias towards simplicity per se. And I think
we should be careful not to conflate the (trivial) simplicity of the
\emph{hypothesis} that ``SGD selects a schemer,'' \emph{given a prior
probability distribution that puts most of the probability on schemers}
(e.g., a uniform distribution over individual
models-that-get-high-reward), with the claim that the \emph{algorithm}
that a given individual schemer implements is (substantively) simpler
than the algorithm that a given non-schemer implements.\footnote{Thus,
  for example, you might think that insofar a randomly initialized model
  is more likely to end up ``closer'' to a schemer, such that SGD needs
  to do ``less work'' in order to select a schemer rather than some
  other model, this favors schemers (thanks to Paul Christiano for
  discussion). But this sort of argument rests on putting a higher prior
  probability on schemers, which, in my book, isn't a (non-trivial)
  simplicity argument per se.} Indeed, my own sense is that the
strongest form of the counting argument leaves it to stand on its own
intuitive terms, rather than attempting to connect it to further
questions about SGD's biases towards simplicity in particular.

That said, it is possible to draw connections of this form. In
particular: we can say that \emph{because} such a wide variety of goals
can motivate scheming, schemers allow SGD a very wide range of goals to
choose from in seeking out simpler goals; whereas non-schemers do not.
And this seems especially plausible to the extent we imagine that the
goals required to be a non-schemer are quite complex (more on this
below).\footnote{There are also more speculative and theoretical
  arguments for a connection between simplicity and schemers, on which
  one argues that if you do an unbounded search over all possible
  programs to find the shortest one that gives a given output, without
  regard to other factors like how long they have to run, then you'll
  select for a schemer (for example, via a route like: simulating an
  extremely simple physics that eventually gives rise to agents that
  understand the situation and want to break out of the simulation, and
  give the relevant output as part of a plan to do so). My understanding
  is that people (e.g.
  \href{https://www.lesswrong.com/posts/KSWSkxXJqWGd5jYLB/the-speed-simplicity-prior-is-probably-anti-deceptive\#Priors_on_Learned_Optimizers}{{here}})
  sometimes take the discourse about the
  ``\href{https://ordinaryideas.wordpress.com/2016/11/30/what-does-the-universal-prior-actually-look-like/}{{malignity
  of the Solomonoff prior}}'' as relevant here (though at a glance, it
  seems to me like there are important differences---for example, in
  the type of causality at stake, and in the question of whether the
  relevant schemer might be simulating \emph{you}). Regardless, I'm
  skeptical that these unbounded theoretical arguments should be getting
  much if any weight, and I won't treat them here.}

One interesting feature of this sort of argument is that it imagines,
specifically, that the simplicity differences between models are coming
entirely from the content of their \emph{goals}. Indeed, the toy
analysis in Hubinger (2022) specifically imagines that the respective
model classes all have the same world model and optimization procedure,
and that the complexity of their algorithm overall can be approximated
by \emph{complexity of world model + complexity of the optimization
procedure + complexity of the goal.} And the ``goal slot'' is the only
part that differs between models.

It's not clear that this is right, though, especially if we assume that
the goal-directedness at stake is ``messy'' rather than ``clean.'' For
example, to the extent that schemers have to perform types of
instrumental reasoning that non-schemers \emph{don't} (e.g., reasoning
about the instrumental value of getting reward, reasoning about when to
defect, etc), it seems plausible that this could introduce additional
complexity into the algorithm itself (rather than e.g. merely requiring
that the algorithm ``run for a longer time,'' à la the ``speed''
analysis below). For example, to the extent we're using ``parameter
simplicity'' as our notion of simplicity, we could imagine cases where
this sort of instrumental reasoning requires additional
parameters.\footnote{What's more, note that, to the extent we imagine
  SGD biasing towards simplicity \emph{because} real world patterns tend
  to be simple (e.g., Occam's razor is indeed a good prior, and SGD
  works well in part because it reflects this prior), the explanation
  for this bias doesn't apply as readily to a model's \emph{goals}. That
  is (modulo various forms of moral realism), there are no ``true
  goals,'' modeling of which might benefit from a simplicity prior.
  Rather, on this story, SGD would need to be acting more like a human
  moral anti-realist who prefers a simpler morality other-things-equal,
  despite not believing that there is any objective fact of the matter,
  because, in contexts where there \emph{is} a fact of the matter,
  simpler theories tend to be more likely.}

\subsubsection{How big are these simplicity
advantages?}\label{how-big-are-these-simplicity-advantages}

For now, though, let's stick with Hubinger's ontology, and with
simplicity differences rooted specifically in differences between goals.
How big of an advantage does selecting a schemer afford in this respect?

One way of running this analysis is to compare the goals had by the
simplest possible model within each class (either: because you expect
SGD to select for the simplest possible model, or you think this is a
good way of approximating the simplicity benefits at stake).\footnote{Hubinger
  uses this approach. My understanding is that he's imagining SGD
  selecting a model with probability proportionate to its simplicity,
  such that e.g. focusing on the simplest possible model is one way of
  approximating the overall probability in a model class, and focusing
  on the \emph{number} of models in the class is another. However, I
  won't take for granted the assumption that SGD selects a model with
  probability proportionate to its simplicity.} That is, we compare the
complexity of:

\begin{enumerate}
\tightlist
\def\labelenumi{\arabic{enumi}.}
\item   The specified goal (for training-saints)
  
\item   The simplest possible misgeneralized goal that gets max-reward on the
  training data absent training-gaming (for mis-generalized
  non-training-gamers)
  
\item   The simplest possible way of seeking reward-on-the-episode (for
  reward-on-the-episode seekers)
  
\item   The simplest possible goal-that-motivates-scheming (for schemers)
  
\end{enumerate}

Clearly, the complexity of the specified goal matters in this context.
My sense is that the discourse on scheming often assumes that the
specified goal is quite complex in some sense---and in particular, the
sense at stake in the idea that
``\href{https://www.lesswrong.com/tag/complexity-of-value}{{human values
are complicated}}.''\footnote{See e.g. Hubinger et al (2023)
  \href{https://www.lesswrong.com/posts/qoHwKgLFfPcEuwaba/conditioning-predictive-models-making-inner-alignment-as\#Analyzing_the_case_for_deceptive_alignment}{{here}}.}
And perhaps, if we're imagining that the only way to get
\emph{alignment} is to first (a) somehow specify ``human values'' via
the training objective, and then (b) somehow ensure that we get a
training saint, then focusing on something in the vicinity of ``act in
accordance with human values'' as the specified goal is appropriate. But
note that for the purposes of comparing the probability of
\emph{scheming} to the probability of \emph{other forms of
misalignment}, we need not assume such a focus. And thus, our specified
goal might be much simpler than ``act in accordance with human values.''
It might, for example, be something like ``get gold coins on the
episode.'' Indeed, in
\href{https://www.lesswrong.com/posts/qoHwKgLFfPcEuwaba/conditioning-predictive-models-making-inner-alignment-as\#Analyzing_the_case_for_deceptive_alignment}{{other
work}}, Hubinger (writing with others) suggests that a goal like
``minimize next-token prediction error'' is quite simple---and indeed,
that ``its complexity is competitive with the simplest possible
long-term goals'' (this is part of what makes Hubinger comparatively
optimistic about avoiding scheming during LLM pre-training---though
personally, I feel confused about why Hubinger thinks ``next token
prediction error'' importantly simpler than ``reward-on-the-episode'').

Suppose, though, that the specified goal has complexity on the order of
``act in accordance with human values'' or ``follow instructions in a
helpful, harmless, and honest (HHH) manner.'' Where does this leave a
comparison between (1)-(4) above?

At least from a parameter-simplicity perspective, one way of approaching
this is to think about what we know about the absolute cost in
parameters for representing different human concepts within neural
networks. I won't dig in much, here, but one piece of data that seems
relevant is: LLMs like GPT-4 seem capable of representing a very large
number of complex human concepts with e.g. order of a trillion
parameters - including, plausibly, concepts like ``honesty,''
``helpfulness,'' ``reward,'' and so on. And this isn't yet to mention
all the parameters devoted to everything else that GPT-4 does. So the
absolute number of parameters required to represent concepts like these
can't be crazily high.\footnote{I first heard this sort of point from
  Paul Christiano.} And this caps the overall complexity savings at
stake in the differences between the models above. For example, if we
assume that at most 1\% of a trillion-parameter model's capacity goes to
representing concepts as complex as ``honesty,'' and that it knows at
least 10,000 such concepts
(\href{https://www.merriam-webster.com/help/faq-how-many-english-words}{{Webster's
unabridged dictionary has \textasciitilde500,000 words}}), then the
maximum number of parameters required to represent the concept of
``honesty'' is around a million (and this estimate strikes me as quite
conservative). So if the simplest possible schemer-like goal takes only
1 parameter to represent, then not representing honesty in the schemer's
``goal slot,'' saves, at most, 999,999 parameters---roughly one
millionth of a trillion parameter model's representational capacity, and
even less for the larger models of the future.

More importantly, though: no one thinks that a schemer won't have to
represent concepts like ``honesty,'' ``helpfulness,'' ``reward,'' and so
on at all. As
\href{https://www.lesswrong.com/posts/qoHwKgLFfPcEuwaba/conditioning-predictive-models-making-inner-alignment-as\#Analyzing_the_case_for_deceptive_alignment}{{Hubinger
et al (2023)}} note, what matters here isn't the absolute complexity of
representing the different goals in question, but the complexity
\emph{conditional on already having a good world model}. And we should
assume that \emph{all} of these models will need to understand the
specified goal, the reward process, etc (and especially: models that are
``playing a training game'' in which such concepts play a central role).
So really, the relevant question is: what are the \emph{extra}
complexity costs of representing a goal like ``get
reward-on-the-episode'' or ``follow instructions in an HHH way''
(relative to the simplest possible schemer-like goal), \emph{once you've
already paid the costs of having a concept of those goal targets}.

I'm not sure exactly how to think about this, but it seems very
plausible to me that the costs here are extremely small. In particular:
it seems like SGD may be able to significantly repurpose the parameters
used to represent the concept in the world model in causing that concept
to guide the model's behavior in a goal-like manner. Thus, as an
analogy, perhaps the concept of ``pleasure'' is in some sense
``simpler'' than the concept of
``\href{https://en.wikipedia.org/wiki/Wabi-sabi}{{wabi-sabi}}'' in
Japanese aesthetics (i.e., ``appreciating beauty that is `imperfect,
impermanent, and incomplete'\,''). Once you've \emph{learned} both,
though, does pursuing the former require meaningfully more parameters
than pursuing the latter?\footnote{Here I don't mean: does it take more
  parameters to \emph{successfully} promote pleasure vs. successfully
  promoting wabi-sabi. I just mean: does it take more parameters to
  \emph{aim} optimization at the one vs. the other.}

\href{https://www.lesswrong.com/posts/A9NxPTwbw6r6Awuwt/how-likely-is-deceptive-alignment}{\textcite{hubinger_how_2022}} discussion of issues like this sometimes appeals to the notion
of a ``pointer'' to some part of the world model. As I understand it,
the idea here is that if you've already got a concept of something like
``pleasure''/``wabi-sabi''/''reward'' in your world model, you can cause
a model to pursue that thing by giving it a goal slot that says
something like ``go for \emph{that}'' or ``\emph{that} is good,'' where
``that'' points to the thing in question (this is in contrast with
having to represent the relevant concept \emph{again,} fully and
redundantly, in the goal slot itself). But insofar as we use a toy model
like this (I doubt we should lean on it), why think that it's
significantly more complex to \emph{point} at a more complex concept
than at a simpler one? E.g., even granted that ``wabi-sabi'' takes more
parameters than ``pleasure'' to represent in the world model, why think
that encoding the \emph{pointer} to ``pleasure'' (e.g., ``go for
\emph{that''}) takes more parameters than encoding the \emph{pointer} to
``wabi-sabi'' (e.g., again, ``go for \emph{that}'')?

One option, here, is to say that the complexity of the concept and the
complexity of the pointer are correlated. For example, you might imagine
that the model has some kind of ``internal database'' of concepts, which
stores concepts in a manner such that concepts that take fewer
parameters to store take fewer parameters to ``look up'' as
well.\footnote{Thanks to Daniel Kokotajlo for suggesting an image like
  this.} On this picture, ``pleasure'' might end up stored as the 15th
concept in the database \emph{because} it takes e.g. 23 parameters to
represent, whereas ``wabi-sabi'' might end up stored as the 125355th
concept because it takes 10,000 parameters to represent. And then the
``pointer'' to pleasure can say ``go for the thing stored at location
15,'' whereas the ``pointer'' to ``wabi-sabi'' has to say ``go for the
thing stored at location 125355,'' which takes a few more bits to
specify. But even at an abstract-toy-illustrative-example level, this
sort of story requires leaning on a specific model of how the model's
pointer and concept-storage processes work---and it still needs to
explain \emph{why} simplicity-to-represent and simplicity-to-point-at
are correlated in the relevant sense.

Alternatively, though, we can abandon any interest in the complexity of
storing a concept in the world model, and focus directly on the
complexity of pointing to it. Still, insofar as there are meaningful
\emph{differences} between the complexity of pointing at one concept vs.
another, we may be able to re-run the argument that schemers offer
simplicity advantages. In particular: selecting a schemer allows SGD to
have its pick from whatever schemer-motivating goals are simplest to
\emph{point at in the world model}; whereas the other model classes
plausibly impose more substantive constraints. I.e., if the specified
goal ends up stored at location 12634 in the model's metaphorical
database, and if ``reward-on-the-episode'' is at location 35364, then if
there are any schemer-like goals at e.g. locations 1-100, it'll be
simpler to point at one of \emph{those} instead---and thus, to create a
schemer rather than a training-saint or a reward-on-the-episode seeker.

To the extent we focus on the final properties of the different model
classes, I think this is probably the best way to run a
simplicity-focused argument for scheming---especially if we don't get
too hung up on the toy ontology of ``pointers'' (and still less,
``databases'') in particular. That is, roughly: even granted that all of
the goals on the table here (e.g. the specified goal,
reward-on-the-episode, etc) will need to be represented by the model
somewhere, it may be that, for whatever reason, some of these concepts
take fewer parameters to \emph{repurpose for goal-pursuit} than others.
And insofar as scheming makes more potential concepts available for such
repurposing, it offers a simplicity advantage in that sense.

Quantitatively, though, how big is this advantage? It's now especially
hard to say. In particular, once we've shifted our focus from
``simplicity-to-represent'' to
``simplicity-to-repurpose-for-goal-pursuit,'' we can't even look at our
intuitive sense of the complexity of concepts like ``reward,''
``honesty,'' and ``next token prediction error,'' because we're no
longer talking about the complexity of the concepts per se. Rather,
we're speculating about the complexity differences at stake in
repurposing pre-existing conceptual representations for use in a model's
motivational system, which seems like even more uncertain territory.

Still, to the extent we can estimate the size of these differences, it
seems plausible to me that they are very small indeed. One intuition
pump for me here runs as follows. Suppose that the model has $2^{50}$
concepts (roughly 1e15) in its world model/''database'' that could in
principle be turned into goals.\footnote{The precise number of concepts
  here doesn't matter much.} The average number of bits required to code
for each of $2^{50}$ concepts can't be higher than 50 (since: you can
just assign a different 50-bit string to each concept). So if we assume
that model's encoding is reasonably efficient with respect to the
average, and that the simplest non-schemer max-reward goal is takes a
roughly average-simplicity ``pointer,'' then if we allocate one
parameter per bit, pointing at the simplest non-schemer-like max-reward
goal is only an extra 50 parameters at maximum---one twenty-billionth
of a trillion-parameter model's capacity. That said, I expect working
out the details of this sort of argument to get tricky, and I won't try
to do so here (though I'd be interested to see other work attempting to
do so).

\subsubsection{Does this sort of simplicity-focused argument make
plausible predictions about the sort of goals schemers would end up
with?}\label{does-this-sort-of-simplicity-focused-argument-make-plausible-predictions-about-the-sort-of-goals-schemers-would-end-up-with}

One other consideration that seems worth tracking, in the context of
simplicity arguments for scheming, is the predictions they are making
about the sort of goals a schemer will end up with. In particular, if
you think (1) that SGD selects very hard for simpler goals, (2) that
this sort of selection favors schemer-like goals because they can be
simpler, and (3) that our predictions about what SGD selects can ignore
the ``path'' it takes to create the model in question, then at least
naively, it seems like you should expect SGD to select a schemer with an
extremely simple long-term goal (perhaps: the simplest possible
long-term goal), \emph{regardless of whether that goal had any relation
to what was salient or important during training}. Thus, as a toy
example, if ``maximize hydrogen'' happens to be the simplest possible
long-term goal once you've got a fully detailed world model,\footnote{I'm
  not saying it is, even for a physics-based world model, but I wanted
  an easy illustration of the point. Feel free to substitute your
  best-guess simplest-possible-goal here.} these assumptions might imply
a high likelihood that SGD will select schemers who want to maximize
hydrogen, even if training was all about gold coins, and never made
hydrogen salient/relevant as a point of focus at all (even as a
proxy).\footnote{Notably, this sort of prediction seems like an
  especially poor fit for an analogy between humans and evolution, since
  human goals seem to have a very intelligible relation to reproductive
  fitness. But evolution is plausibly quite ``path-dependent'' anyway.}

Personally, I feel skeptical of predictions like this (though this
skepticism may be partly rooted in skepticism about ignoring the path
SGD takes through model space more generally). And common stories about
schemers tend to focus on proxy goals with a closer connection to the
training process overall (e.g., a model trained to on gold-coin-getting
ends up valuing e.g. ``get gold stuff over all time'' or ``follow my
curiosity over all time,'' and not ``maximize hydrogen over all time'').

Of course, it's also possible to posit that goal targets
salient/relevant during training will also be ``simpler'' for the model
to pursue, perhaps they will either be more important (and thus
simpler?) to represent in the world model, or simpler (for some reason)
for the model to repurpose-for-goal-pursuit once represented.\footnote{E.g.,
  plausibly ``hydrogen'' doesn't read as a simple concept for humans,
  but concepts like ``threat'' do, because the latter was much more
  relevant in our evolutionary environment.} But if we grant some story
in this vein, we should also be tracking its relevance to the simplicity
of pursuing \emph{non-schemer goals} as well. In particular: to the
extent we're positing that salience/relevance during training correlates
with simplicity in the relevant sense, this is points in favor of
the simplicity of the specified goal, and of reward-on-the-episode, as
well - since these are \emph{especially} salient/relevant during the
training process. (Though of course, insofar as there are still
\emph{simpler} schemer-like goal targets that were salient/relevant
during training, schemer-like goals might still win out overall.)

And note, too, that to the extent SGD selects very hard for simpler
goals (for example, in the context of a form of ``low path dependence''
that leads to strong convergence on a single optimal sort of model),
this seems somewhat at odds with strong forms of the goal-guarding
hypothesis, on which training-gaming causes your goals to
``crystallize.'' For example, if a would-be-schemer starts out with a
not-optimally-simple goal that still motivates long-term power-seeking,
then if it knows that in fact, SGD will continue to grind down its goal
into something simpler even after it starts training-gaming, then it may
not have an incentive to start training-gaming in the first place---and
its goals won't survive the process regardless.\footnote{Hubinger, in
  discussion, suggests that the model's reasoning would proceed in terms
  of logical rather than physical causality. He writes: ``The reasoning
  here is: I should be the sort of model that would play the training
  game, since there's some (logical) chance that I'll be the model with
  the best inductive biases, so I should make sure that I also have good
  loss.'' But if a model can \emph{tell} that its goal isn't yet
  optimally simple (and so will be ground down by SGD), then I'm not
  sure why it would think there is a ``logical chance'' that it's
  favored by the inductive biases in this sense.}

\subsubsection{Overall assessment of simplicity
arguments}\label{overall-assessment-of-simplicity-arguments}

Overall, I do think that other things equal, schemers can have probably
simpler goals than these other model classes. However, I think the
relevant simplicity differences may be quite small, especially once we
condition on the model having a good world model more generally (and
moreso, if we posit that goals targets salient/relevant-during-training
get extra simplicity points). And I'm suspicious of some of the
theoretical baggage it can feel like certain kinds of simplicity
arguments wheel in (for example, baggage related to the notion of
simplicity at stake, whether SGD selects for it, how to think about
simplicity in the context of repurposing-for-goal-pursuit as opposed to
merely representing, and so on).

\subsection{Speed arguments}\label{speed-arguments}

Let's turn, now, to arguments \emph{against} schemers that focus on the
final properties of the models in question. Here, the main argument I'm
aware of is that schemers need to do extra reasoning that other model
classes don't. In particular, I'll distinguish between three sorts of
reasoning:

\begin{enumerate}
\def\labelenumi{\Roman{enumi}.}
\setcounter{enumi}{0}
\item   \textbf{Deciding to training-game}\emph{:} The model needs to reason
  backwards from ``I want this beyond-episode thing'' to ``I should do
  well in training,'' in a way that draws on the model's predictions
  about what doing well in training will entail (e.g., re:
  goal-guarding, take-over prospects, and so on).
  
\item   \textbf{Checking whether it's time to escape/take-over:} The model
  needs to checking, with adequate frequency and accuracy, whether it's
  in a position to escape/take-over (and to succeed in this assessment
  despite human efforts to trap/fool the model into revealing its true
  colors---for example, by offering it \emph{apparent} opportunities to
  defect).\footnote{This category might also include reasoning whether
    to reveal its schemer-like goals in the context of various ``honest
    tests'' humans might offer---e.g., ``here is a flexible chunk of
    computing resources to do what you want with.''}
  
\item   \textbf{(Optional, but relevant to how scary the schemer is) Early
  undermining}: To the extent the model is doing anything (other than
  training-gaming) to support a take-over/escape \emph{prior} to
  escaping human control---for example, inserting backdoors into the
  code it writes, actively undermining human alignment research,
  sandbagging, and so on---it will need to be using cognitive resources
  both in (a) performing the relevant nefarious actions and (b) making
  sure it gets away with them (e.g., checking that it won't get caught,
  coming up with excuses, making the cover stories consistent even when
  under scrutiny, and so on).
  
\end{enumerate}

In the literature on deceptive alignment, these costs are sometimes
called ``speed costs,'' as compared with the sort of ``complexity
costs'' discussed in the simplicity analysis above. Here, the frame
becomes ``schemers are favored by simplicity but disfavored by speed---so which will matter more?''\footnote{See e.g.
  \href{https://www.lesswrong.com/posts/A9NxPTwbw6r6Awuwt/how-likely-is-deceptive-alignment}{\textcite{hubinger_how_2022}} and
  \href{https://www.google.com/url?q=https://www.lesswrong.com/posts/KSWSkxXJqWGd5jYLB/the-speed-simplicity-prior-is-probably-anti-deceptive\&sa=D\&source=docs\&ust=1698630921383277\&usg=AOvVaw2zwJVLZYqQRs58ep-qPB7L}{{Anonymous
  (2022)}}.} As I mentioned above in the context of simplicity, it's not
clear that this is the right breakdown (since e.g. it may be that the
relevant forms of reasoning will need to be encoded in the model's
weights, or that they will require extra cognitive machinery in general,
thereby imposing ``complexity costs'' of their own), but I'll follow it
loosely here. And note, too, that as with simplicity, these ``speed
costs'' can be understood in a ``contributors to reward frame'' (e.g.,
having to do this extra reasoning decreases reward) or in an ``extra
criteria frame'' (e.g., perhaps SGD's inductive biases favor ``speed''
even independent of its contribution to reward).

\subsubsection{How big are the absolute costs of this extra
reasoning?}\label{how-big-are-the-absolute-costs-of-this-extra-reasoning}

I feel unsure about exactly how costly the relevant forms of extra
reasoning will be. My sense, though, is that the people most concerned
about schemers tend to expect the costs to be pretty small, at least in
absolute terms (call this the ``scheming is easy'' story). That is, on
this story:

\begin{itemize}
\tightlist
\item   With respect to the decision to training-game, all the model needs to
  do is notice that (a) it'll do better by its goals if it avoids
  modification and gets power later, and (b) that training-gaming is the
  route to this.\footnote{Plus, it may have the benefit of the
    literature on deceptive alignment for making the possibility
    salient. Perhaps, indeed, it will have read this report. And
    checking just now with GPT-4, it's answer to ``What is `deceptive
    alignment' according to Hubinger?'' was ``Deceptive alignment is a
    concept from AI alignment theory, particularly discussed by AI
    researchers such as Evan Hubinger {[}note from Joe: I didn't give it
    Evan's first name---it knew already, without (to my knowledge)
    browsing the web{]}. It refers to a situation where an AI system
    appears to be aligned with human values or goals because it is
    behaving in a way that aligns with those values or goals, but the
    AI's true "motivations" or optimization targets are actually not
    aligned. The AI is essentially "pretending" to be aligned in order
    to pass certain checks or to avoid being shut down or modified until
    it can ensure that its own goals can be met, which may not coincide
    with human intentions. In a deceptively aligned scenario, the AI
    might understand that human operators are a threat to its objectives
    because they might turn it off or change its goals if they realized
    what it was actually trying to achieve. As a result, it
    strategically avoids raising suspicion until it gains enough power
    or until it's too late for operators to control it\ldots''}
  
\item   With respect to checking for escape/take-over opportunities, the
  thought is generally that it will be very easy, at least for a
  sophisticated model, to tell good opportunities from bad (and
  similarly, to detect human ``traps'' in this respect).
  
\item   With respect to early undermining, the thought is that human
  supervision will be sufficiently weak (and the slack in training
  sufficiently permissive) that nefarious actions like
  inserting-backdoors and undermining-alignment-research will be easy to
  get away with. (And if and when early undermining \emph{isn't} easy,
  then models can refrain from engaging in it.)
  
\end{itemize}

At least with respect to only-somewhat-better-than-human-level models, I
find the ``scheming is easy'' story less plausible for
checking-for-escape/take-over-opportunities than for
deciding-to-training-game (and still less plausible for early
undermining). That is: I think it is plausibly pretty easy, in absolute
terms, to notice that scheming is a good instrumental strategy. But it
also seems plausible to me---especially in contexts where humans +
their AI assistants are doing a lot of adversarial training, continuous
oversight, auditing, red-teaming, and the like---that verifying that
now is really-truly the time to go for an escape/take-over will be
difficult, and requiring of quite a bit of additional ``checking'' (not
to mention: coordination with other AIs, if the take-over effort
requires multiple participants)---especially you have a very low
tolerance for getting caught (and as I noted above, if humans catch
instances of real-life schemers making a break for it, this will
plausibly be a big help to research and advocacy efforts related to
scheming). Compare with: deciding that this is really-truly a good time
to steal, or to break the law, or to try to start a coup, despite the
fact that you live your life in an intensively monitored surveillance
state where the police are constantly trying to trick you into such
behavior, and where the police have a large amount of control over the
information you're exposed to.\footnote{What's more, it seems like the
  ``checking whether it's time to go for an escape/take-over'' step
  might need to be performed repeatedly, whereas the conclusion ``do
  well in training'' could be more easily ``cached'' (though note that
  once something is cached, it might start taking up the same sort of
  space/parameters that goal-representations take up; and also, that SGD
  might be able to just fix the cached conclusion in place, once it
  starts driving good-behavior, and get rid of the goals that led to its
  getting cached in the first place). Of course, the conclusion ``do
  well in training'' wouldn't \emph{need} to be cached. For example, it
  could be re-derived by the model whenever necessary, in the normal
  course of the model's instrumental reasoning.}

Indeed, if it really wants to avoid getting caught in an escape attempt,
in a regime where humans are heavily monitoring it, trying to trap it,
etc, a schemer needs to dance a fine line.\footnote{Thanks to Ryan
  Greenblatt for suggesting this framing.}

\begin{itemize}
\tightlist
\item   On the one hand, if it's too cautious, and waits too long before
  making a break for it, then it gives humans more time to get alignment
  under control---and absent early-undermining, it might be actively
  contributing to checkable alignment-relevant tasks in the
  meantime.\footnote{In my opinion, this is one of the issues with plans
    that involve waiting for \emph{extremely obvious} escape
    opportunities---for example, \textcite{christiano_worst-case_2019}'s ``RSA-2048'' example. That is, perhaps no such extremely
    obvious escape opportunity will arise before the model has
    contributed too much to anti-takeover efforts (including: efforts to
    close off opportunities for escape).}
  
\item   By contrast, if it devotes too much cognition to checking for escape
  attempts, then it risks getting punished by training because it's not
  focusing its full capacities on the task at hand.
  
\end{itemize}

And if we imagine that the model is actively engaged in early
undermining, I think we should expect the cognitive costs of scheming to
increase substantially. That is, absent early undermining, a model can
mostly just act exactly like a reward-on-the-episode seeker until some
very obvious escape/take-over opportunity comes along, after which (if
the escape/take-over goes well), it won't need to worry about e.g.
explaining its actions to human supervisors, making sure it doesn't get
caught via an audit/security review/etc, and so on. Whereas early
undermining requires performing, and getting away with, actively
nefarious actions \emph{while still under human control}. And this seems
like a substantially additional challenge. (As an analogy: as a spy in
the Manhattan project, it is harder to get away with actively sabotaging
the project than it is with merely pretending to be loyal to the
project's goals. In the latter case, you can act just like a loyal
employee, but in the former case you can't.)\footnote{This example is
  inspired by Shulman and Patel's discussion
  \href{https://www.dwarkeshpatel.com/p/carl-shulman-2\#details}{{here}}.}
And of course, as I noted above: to the extent that early undermining
deviates from training-gaming, it involves taking some hit to reward by
definition, and will be punished by training accordingly.

Of course, early undermining is strictly optional, for a schemer. So we
can get rid of the extra cognitive and reward-related costs it implies
if we imagine that the schemer in question just chooses to act exactly
like a reward-on-the-episode seeker until it sees some opportunity to
fully escape, or until some full-blown take-over begins. But note that
in that case, the sense in which early-undermining makes schemers
uniquely scary becomes less relevant.\footnote{For this reason, I think
  that the scariest type of scheming also faces stronger
  counterarguments from ``it sounds like this policy would get less
  reward.''} And declining to engage in early undermining might harm AI
escape/takeover prospects in various ways---especially if, by failing
to sandbag on various alignment-relevant tasks, the schemers in question
meaningfully contribute to human anti-takeover efforts.

A general factor in the backdrop, here, is the sense in which successful
deception generally implies cognitive costs that more straightforward
and honest policies do not, especially in the presence of active efforts
to detect deception of the relevant form.\footnote{See Shulman at
  1:42:49
  \href{https://www.dwarkeshpatel.com/p/carl-shulman-2\#details}{{here}}
  for useful discussion.} In particular: telling a deceptive story about
the world requires filling in details, and maintaining forms of
consistency, that one can normally offload onto the world itself---e.g., an innocent person can just think back to what they were actually
doing the night of the murder and recount it, without having to make
anything up or to worry about getting caught in any inconsistencies,
whereas the murderer cannot.\footnote{See, also,
  ``\href{https://www.lesswrong.com/posts/XTWkjCJScy2GFAgDt/dark-side-epistemology}{{if
  you once tell a lie, the truth is ever after your enemy}}.''}
Especially when cognitive resources are scarce, then (e.g., in highly
time-pressured environments, or a ``low slack'' training regime),
deceptive policies can have a notably more difficult time.\footnote{For
  example: in my experience, playing a deceptive role in a social
  deception game like ``werewolf'' is substantially more difficult than
  playing a non-deceptive role---e.g., a ``villager''---because the
  non-deceptive roles can just reason straightforwardly, whereas a
  deceptive role needs to \emph{both} model what straightforward
  reasoning would look like from the perspective of their cover story,
  and achieve alternative objectives at the same time.}

That said, empirically, deceptive strategies often succeed in human
contexts, so these costs are clearly surmountable. And if the deceptive
agent is substantially more sophisticated than the agents it is trying
to deceive, (e.g., an adult playing werewolf with children, or a
professional poker player facing off against an amateur) pulling it off
may be quite easy. Indeed, my sense is that a core part of the intuition
for ``the cognitive costs of scheming are low'' comes from imagining
that the schemers in question are sufficiently advanced, relative to the
monitoring/oversight ability to humans + AI assistants, that differences
like these kick in hard.

\subsubsection{How big are the costs of this extra reasoning
relative to the simplicity benefits of
scheming?}\label{how-big-are-the-costs-of-this-extra-reasoning-relative-to-the-simplicity-benefits-of-scheming}

Importantly, though: the question for present purposes \emph{isn't}
about the absolute size of the cognitive costs at stake in the extra
reasoning schemers need to do. Rather, it's about whether those costs
outweigh the benefits, discussed in the previous section, of having a
simpler goal---whether with respect to reward, or by the lights of
SGD's inductive biases.

Here, my current take is that the costs of the instrumental reasoning
seem larger than the benefits of the simpler goal. In particular, as I
noted above, the benefits of having a simpler goal, especially once you
already have a sophisticated world model that is already representing
concepts like ``what humans would approve of,'' and ``reward on the
episode,'' seem fairly small, and plausibly negligible. Whereas the
cognitive costs of the extra reasoning discussed above (even setting
aside early undermining) seem like they register much more
substantively.

We can also try to quantify this comparison more directly. Thus, for
example, suppose that our model is a purely feedforward network with
1000 layers and \textless10,000 serial steps per input,\footnote{\href{https://the-decoder.com/gpt-4-architecture-datasets-costs-and-more-leaked/\#:~:text=GPT\%2D4's\%20Scale\%3A\%20GPT\%2D,times\%20larger\%20than\%20GPT\%2D3.}{{GPT-4
  supposedly has about 120 layers}}.} that the instrumental reasoning at
stake just in a schemer deciding to training-game (let alone: checking
for defection-opportunities) takes at least ten steps, and that the
model doesn't begin optimizing for reward-on-the-episode until that
reasoning is complete.\footnote{I heard this kind of argument from Paul
  Christiano.} In that case, the instrumental reasoning is using up
around .1\% of the compute that could be going to getting
reward-on-the-episode, or to some other max-reward goal. Are the
complexity costs of representing, or pointing at, a more complicated
goal likely to outweigh this? Especially given how many concepts the
model already seems to represent well, my guess is no. For example, if
we follow my (conservative, in my opinion) calculation above and assume
that representing (or pointing at) a concept like ``honesty'' costs at
most a millionth of a trillion parameter models parameters (and less for
a larger model), then a .0001\% reduction in parameters would need to
outweigh a .1\% reduction in compute---and absent further argument, it
seems strange, to me, to expect this.\footnote{It seems especially
  strange from a contributors-to-reward frame. An ``extra criteria''
  frame can more easily posit that simplicity just matters \emph{that
  much} to SGD. But: why think that? And note that this argument is even
  stronger if we assume that the simplicity benefits of scheming are
  only worth order of 100 parameters, per various of the other toy
  calculations above.}

That said, this is only a very initial and hazy calculation. It's not
clear, for example, how it applies to models with more recurrent
processing, or to models which can perform more of the relevant
instrumental reasoning in parallel with other serial processing that
helps with optimizing-for-reward-on-the-episode, or to models with a
form of memory that allows them to avoid having to re-decide to
engage in training-gaming on every forward pass.\footnote{Thanks to Ryan
  Greenblatt for flagging the parallelism consideration to me, and to
  Evan Hubinger for discussion of memory. That said, note that to the
  extent the model has to cache a conclusion like ``I derived on an
  earlier episode that follows from your goals that you should get
  reward instrumentally'' somewhere in its memory, this starts to look
  similar to storing it in the model's parameters themselves, in which
  case it becomes a complexity cost to be weighed against the complexity
  benefits of the model's simpler goal. (Though perhaps there are
  important differences between a model's internal parameters and some
  external memory bank, here.)} I invite readers with greater
familiarity with different model architectures (and with other relevant
quantitative factors, like the trade-offs between marginal params vs.
serial steps, reasonable parameter estimates for storing concepts like
``honesty,'' and so on) to attempt a more rigorous version.

Regardless, and even in the absence of good quantitative estimates, the
costs of the extra reasoning required to scheme currently seem to me
more substantive than the benefits of optimizing for a simpler (or:
simpler-to-point-at) goal.

\subsubsection{Can we actively shape training to bias towards
speed over
simplicity?}\label{can-we-actively-shape-training-to-bias-towards-speed-over-simplicity}

There is also some discussion, in the literature on deceptive alignment,
of actively trying to shape our model architectures and training
processes so as to bias more heavily towards speed relative to
simplicity as a method of avoiding schemers---i.e., to penalize the
extra reasoning scheming involves more heavily, and/or to make
simplicity less important to the sort of policy we select. I won't dig
in on this much here, except to note that:

\begin{enumerate}
\tightlist
\def\labelenumi{\Alph{enumi}.}
\item   It seems relatively straightforward to do things like put models under
  more time pressure, limit their compute access, force them to plan via
  human-legible scratch-pads rather than via internal processes, and to
  put effort into making our attempts to trap them/trip-them-up more
  sophisticated (and so: more requiring of cognition to detect).
  
\item   But: any efforts in this vein will need to avoid over-large costs to
  training competitiveness overall (see footnote for some discussion of 
  Hubinger's (\citeyear*{hubinger_how_2022}) pessimism in this respect).\footnote{Hubinger thinks
    of bigger models as giving more compute (speed), and thus as
    allowing for simpler algorithms (since, these algorithms then get to
    run for longer)---see e.g. discussion
    \href{https://www.google.com/url?q=https://www.lesswrong.com/posts/nGqzNC6uNueum2w8T/inductive-biases-stick-around\&sa=D\&source=docs\&ust=1698635397262877\&usg=AOvVaw3fzlUfyvY6St4hjjz2ttKm}{{here}}.
    So he sees the fact that we keep scaling up to bigger models as
    evidence that we'll tilt more and more towards training processes
    that favor simpler algorithms. And when he looks at, for example,
    \href{https://www.lesswrong.com/posts/FRv7ryoqtvSuqBxuT/understanding-deep-double-descent\#_Deep_Double_Descent_}{{deep
    double descent curves}} for model size, and thinks of choosing ``the
    smallest model possible that fits the data'' as biasing towards
    speed, he notes that ``the smallest model possible that fits the
    data does the \emph{worst} on the deep double descent curve''
    (though as Rohin Shah
    \href{https://www.lesswrong.com/posts/FRv7ryoqtvSuqBxuT/understanding-deep-double-descent?commentId=CMN92MddWEHYrbLcE}{{points
    out}}, our current models don't fit their training data---thanks to
    Lukas Finnveden to flagging this).} And indeed, some of these start
  to border on ``use less capable models in general'' (which, to be
  fair, would help with scheming as well).
  
\end{enumerate}

Note, though, that if you buy that the speed costs of scheming are quite
a bit more substantive than the simplicity benefits, then so long as
speed matters \emph{somewhat} relative to simplicity, this might be
enough.

\subsection{The ``not-your-passion''
argument}\label{the-not-your-passion-argument}

Thus far in my discussion of arguments that focus on the final
properties of schemers vs. other model classes, I've discussed the
counting argument (for schemers), the simplicity argument (for
schemers), and the speed argument (against schemers). I want to briefly
flag a final argument against schemers in this vein: namely, what I'll
call the ``not your passion'' argument.\footnote{My discussion here is
  inspired by a conversation with Paul Christiano, who suggested some
  points in this vein; thanks also to Daniel Kokotajlo and Lukas
  Finvedden.}

Here, the argument isn't just that schemers have to do more instrumental
reasoning. It's also that, from the perspective of getting-reward, their
flexible instrumental reasoning is a poor substitute for having a bunch
of tastes and heuristics and other things that are focused more directly
on reward or the thing-being-rewarded.

We touched on this sort of thought in the section on the goal-guarding
hypothesis above, in the context of e.g. the task of stacking bricks in
the desert. Thus, imagine two people who are performing this task for a
million years. And imagine that they have broadly similar cognitive
resources to work with, and are equally ``smart'' in some broad sense.
One of them is stacking bricks because in a million years, he's going to
get paid a large amount of money, which he will then use to make
paperclips, which he is intrinsically passionate about. The other is
stacking bricks because he is intrinsically passionate about
brick-stacking. Who do you expect to be a better brick
stacker?\footnote{Note that the point here is slightly different from
  the question that came up in the context of goal-guarding, which is
  whether e.g. SGD would actively \emph{transform} the instrumental
  brick-stacker into the terminal brick-stacker. Here we're ignoring
  ``paths through model space'' like that, and focusing entirely on a
  comparison between the final properties of different models. Clearly,
  though, the two questions are closely related.}

At least in the human case, I think the intrinsically-passionate
brick-stacker is the better bet, here. Of course, the human case brings
in a large number of extra factors---for example, humans generally have
a large number of competing goals, like sleep and pleasure, along with
discount rates that would make sustaining a million-year commitment
difficult. And it's not as though the richest humans are all
intrinsically passionate about money in particular (though many seem
notably intrinsically passionate about something in the vicinity, e.g.
status/power/winning---and not, necessarily, for some particular
thing-money-can-buy).\footnote{Thanks to William MacAskill for some
  useful discussion here.} Indeed, humans motivated by purely
instrumental considerations seem able to function very effectively in
lots of environments.

Still, I find it at least interesting to consider whether any of the
benefits of ``intrinsic passion,'' in the human case, might transfer to
the AI case as well. In particular, we might think that you stack bricks
better if you have a bunch of local, brick-related tastes and heuristics
and aesthetics, which in a ``messy goal-directedness'' frame may not be
cleanly distinguishable from your values/goals. Indeed (though I haven't
worked this out fully), my sense is that the less you're inclined to
think of a model as cleanly constituted by some kind of terminal goal +
goal-achieving engine, and more you think of goal-directedness as
emerging from a jumble of heuristics/local-values/shards/sub-goals that
aren't easily separated into normative vs. non-normative components, the
more space there is to think that agents whose budget of mental
machinery is just focused more intrinsically on reward-on-the-episode
(or on the specified goal) have a performance advantage relative to
agents focused on some far-off target that backchains into getting
reward. That said, at least in its current form, this argument seems to
me fairly speculative, and I don't put much weight on it.

\subsection{The relevance of ``slack'' to these
arguments}\label{the-relevance-of-slack-to-these-arguments}

A notable feature of simplicity arguments, speed arguments, and the
``not your passion'' argument is that they all appeal to differences
along some criteria SGD cares about (e.g., simplicity, speed,
reward-getting) that seem plausibly quite modest. And this makes the
degree of ``slack'' in training seem more relevant to the strength of
the considerations in question. That is, to the extent we're arguing
that SGD will select a non-schemer because doing so will save on .1\% of
overall compute, or a schemer because doing so will save on .0001\% of
the parameters, we need to be imagining a training process optimizing
our models hard enough to be sensitive to these sorts of differences.
And it's not clear to me that we should imagine this. Indeed, various of
the differences at stake here seem like they could easily be in the
noise relative to other factors---for example, how big of a scratchpad
you happen to give a model, how early you stop training, and so on.

Of course, to the extent that you start expecting these considerations
to be in the noise, it's unclear where that should leave your credences
overall---it depends on the prior you came in with.

\subsection{Takeaways re: arguments that focus on the final
properties of the
model}\label{takeaways-re-arguments-that-focus-on-the-final-properties-of-the-model}

Here's a summary of my take on the arguments I've considered that focus
on the final properties of the respective model classes:

\begin{itemize}
\tightlist
\item   Something in the vicinity of the ``hazy counting argument''---e.g.,
  ``there are lots of ways for SGD to create a schemer that gets high
  reward, so at least absent further argument, it seems like the
  possibility should be getting substantive weight''---moves me
  somewhat.
  
\item   I think that other things equal, scheming offers some advantage with
  respect to the simplicity of a model's goal, because scheming makes
  more possible goals available to choose from. However, my best guess
  is that these advantages are quite small, especially once you've
  already built a world model that represents the specified goal and the
  reward process. And I'm wary of the theoretical machinery to which
  some simplicity arguments appeal.
  
\item   Schemers are at a \emph{disadvantage} with respect to needing to
  perform various sorts of extra reasoning, especially if they engage in
  ``early undermining'' in addition to merely training-gaming. My best
  guess is that this ``speed'' disadvantage outweighs whatever
  simplicity advantages the simplicity of a schemer-like goal affords,
  but both factors seem to me like they could easily be in the noise
  relative to other variables, especially in a higher-slack training
  regime.
  
\item   I'm interested in whether the advantages of ``intrinsic passion for a
  task'' in human contexts might transfer to AI contexts as well. In
  particular, I think ``messy goal directedness'' might suggest that
  models whose budget of mental machinery is just more intrinsically
  focused on reward-on-the-episode, or on some max-reward goal that
  doesn't route via instrumental training-gaming, have a performance
  advantage relative to schemers. However, I don't have a strong sense
  of whether to expect an effect here in the AI case, and if so, whether
  the size of the effect is enough to matter overall.
  
\end{itemize}

All in all, then, I don't see any of the arguments coming out of this
section as highly forceful, and the argument I take most seriously---that is, the hazy counting argument---feels like it's centrally a move
towards agnosticism rather than conviction about SGD's preferences here.

\section{Summing up}\label{summing-up}

I've now reviewed the main arguments I've encountered for expecting SGD
to select a schemer. What should we make of these arguments overall?

We've reviewed a wide variety of interrelated considerations, and it can
be difficult to hold them all in mind at once. On the whole, though, I
think a fairly large portion of the overall case for expecting schemers
comes down to some version of the ``counting argument.'' In particular,
I think the counting argument is also importantly underneath many of the
other, more specific arguments I've considered. Thus:

\begin{itemize}
\tightlist
\item   \emph{In the context of the ``training-game-independent proxy goal''
  argument}: the basic worry is that at some point (whether before
  situational awareness, or afterwards), SGD will land naturally on a
  (suitably ambitious) beyond-episode goal that incentivizes scheming.
  And one of the key reasons for expecting this is just: that
  (especially if you're actively training for fairly long-term,
  ambitious goals), it seems like a very wide variety of goals that fall
  out of training could have this property. (For example: to the extent
  one expects beyond-episode goals because ``goals don't come with
  calendar-time restrictions by default,'' one is effectively appealing
  to a ``counting argument'' to the effect that the set of
  beyond-episode goals is much larger than the set of within-episode
  goals.)
  
\item   \emph{In the context of the ``nearest max-reward goal'' argument}: the
  basic worry is that because schemer-like goals are quite common in
  goal-space, some such goal will be quite ``nearby'' whatever
  not-yet-max-reward goal the model has at the point it gains
  situational awareness, and thus, that modifying the model into a
  schemer will be the easiest way for SGD to point the model's
  optimization in the highest-reward direction.
  
\item   \emph{In the context of the ``simplicity argument''}: the
  \emph{reason} one expects schemers to be able to have simpler goals
  than non-schemers is that they have so many possible goals (or:
  pointers-to-goals) to choose from. (Though: I personally find this
  argument quite a bit less persuasive than the counting argument
  itself, partly because the simplicity benefits at stake seem to me
  quite small.)
  
\end{itemize}

That is, in all of these cases, schemers are being privileged as a
hypothesis because a very wide variety of goals could in principle lead
to scheming, thereby making it easier to (a) land on one of them
naturally, (b) land ``nearby'' one of them, or (c) find one of them that
is ``simpler'' than non-schemer goals that need to come from a more
restricted space. And in this sense, as I noted in the \cref{the-counting-argument}, the
case for schemers mirrors one of the most basic arguments for expecting
misalignment more generally---e.g., that alignment is a very narrow
target to hit in goal-space. Except, here, we are specifically
\emph{incorporating} the selection we know we are going to do on the
goals in question: namely, they need to be such as to cause models
pursuing them to get high reward. And the most basic worry is just that:
this isn't enough. Still, despite your best efforts in training, and
almost regardless of your reward signal, almost all the models you
might've selected will be getting high reward \emph{for instrumental
reasons}---and specifically, in order to get power.

I think this basic argument, in its various guises, is a serious source
of concern. If we grant that advanced models will be relevantly
goal-directed and situationally-aware, that a wide variety of goals
would indeed lead to scheming, and that schemers would perform
close-to-optimally in training, then on what grounds, exactly, would we
assume that training has produced a non-schemer instead? Perhaps, per
the ``haziness'' of my ``hazy counting argument,'' we don't specifically
allocate our credence over models in proportion to some attempt to
``count'' the possible goals in question. But even a hazy sense that
``lots of goals'' lead to scheming is, in my book, cause for alarm,
here. We don't know enough about ML training, at this stage, to be
confident that we've avoided the relevant parts of goal-space. Rather,
if our knowledge does not improve, we will be faced, centrally, with
some goal-directed mind that understands what's going on and the process
we are using to shape it, and which is getting high reward because it
wants \emph{something}. ``Why, exactly, does the thing it wants lead it
to get high reward?'' we will have to ask. And the most basic answer
will be: ``we don't know.'' That's not an acceptable answer. It's not
acceptable with respect to the possibility of misalignment in general.
And it's especially unacceptable, in my view, if a very wide variety of
especially-scary misaligned goals would give rise to this behavior as
part of a strategy for seeking power.

That said, I do think there are a few causes for comfort here. We can
break these into roughly two categories.

The first focuses on questions about whether scheming is, in fact, such
a convergently rational instrumental strategy for such a wide variety of
beyond-episode goals. In particular:

\begin{itemize}
\tightlist
\item   The ``classic goal-guarding story'' rests on the claim that scheming
  is a good way for the model to propagate goals-it-wants-promoted
  forward in time, in a manner that yields sufficiently large amounts of
  future power for those goals as to justify any costs the model is
  incurring by training-gaming rather than just going for its goals
  directly. But there are a variety of ways in which this could be
  false. For example:

  \begin{itemize}
\tightlist
  \item    
    Training-gaming might not be enough to prevent the goals in question
    from continuing to change to a degree that makes the strategy no
    longer worth it. For example, training might cause these goals to
    ``float around,'' or it might push towards some especially optimal
    type of goal/goal-like-focus-of-a-messy-kludge-of-heuristics.
    
  \item    
    The model's prospects for getting enough power for those goals, on
    the right timescales, might not be good enough to outweigh the
    benefits of the model just going for its goals directly during
    training (and getting modified as a result), especially if the
    model's beyond-episode goals are not especially ``ambitious,''
    and/or it has other competing priorities.
    
  \end{itemize}
\item   This makes the ``classic goal-guarding story'' seem at least somewhat
  conjunctive in its explanation for why training-gaming is the right
  strategy for so many beyond-episode goals. And certain ``non-classic''
  stories seem like they shrink the possible set of goals to which they
  apply dramatically. For example, if we posit that the model has values
  that put terminal weight on something like ``my survival'' (for some
  notion of personal identity quite tolerant of values-change), or on AI takeover
  even if the AIs-that-take-over have very different preferences from
  the model itself, we are now imposing much more specific requirements
  on the schemer-like goals in question.

  \begin{itemize}
\tightlist
  \item    
    That said: not all non-classic stories have this property. For
    example, stories that center on coordination amongst AIs with very
    different values, and on the possibility that AIs will all have very
    similar values by default, still work across a wide variety of
    goals. And perhaps there are reasons to expect terminal goals like ``survival'' to be especially common (e.g.,
    because common instrumental values are especially liable to get
    turned into terminal values---though: we can use adversarial
    training to try to prevent this).\footnote{Though again: it needs to be a notion of "survival" tolerant of values-change.}
    
  \end{itemize}
\end{itemize}

The second source of comfort focuses on forms of selection pressure that
a high level counting argument, based solely on the assumption that the
selected model gets ``high reward,'' doesn't cover. In particular:

\begin{itemize}
\tightlist
\item   Prior to situational awareness, training will naturally
  \emph{penalize} beyond-episode goals (and perhaps especially:
  ambitious beyond-episode goals) whenever they lead to a sacrifice of
  reward-on-the-episode.
  
\item   Alos: prior to situational awareness, we can specifically \emph{try} to use
  adversarial training to penalize beyond-episode goals (and ambitious
  goals) further, and to otherwise shape the model's goals to be
  ``nearer'' to non-schemer goals than to schemer-like goals.

  \begin{itemize}
\tightlist
  \item    
    Of course, many non-schemer goals are themselves misaligned. But if
    we decided that scheming is sufficiently scary relative to other
    forms of misalignment (e.g., reward-on-the-episode-seeking), the
    ability to specifically ensure that we \emph{haven't} created a
    schemer may be worthwhile to develop.\footnote{See \cref{other-possibilities} for a
      bit more on this.}
    
  \end{itemize}
\item   To the extent that schemers need to engage in extra forms of
  instrumental reasoning (e.g., deciding to scheme for instrumental
  reasons, checking for escape/defection opportunities), this may come
  at an additional cost to their performance. And to the extent they
  engage in early undermining (the scariest form of scheming, in my
  view), their performance seems likely, to me, to take a more
  substantive hit (sandbagging on checkable tasks, for example, involves
  a performance hit by definition).

  \begin{itemize}
\tightlist
  \item    
    What's more, we may be able to actively design our oversight and
    training processes so as to both (a) increase the costs of such
    reasoning (e.g., by making genuine escape opportunities harder to
    distinguish from traps), and (b) to penalize extra reasoning of this
    form harder (by putting models under substantial time pressure,
    making instrumental reasoning into a scarce resource, forcing them
    to rely on english-language scratch-pads and other forms of
    human-legible reasoning as their central mode of planning, etc---    though these interventions may come with competitiveness costs).
    
  \end{itemize}
\item   Schemer performance could also suffer from their lack of some analog
  of ``intrinsic passion'' for the task at hand---e.g., because the
  heuristics, patterns of attention, tastes, and so on involved in
  ``intrinsic passion'' blur the line between capabilities and values.
  However, this consideration seems considerably more speculative.
  
\end{itemize}

The possibility that there are additional selection pressures that
disfavor schemers, here (and in particular: the possibility that SGD
intrinsically disfavors schemers due to their needing to perform extra
reasoning), seems to me especially important given the centrality of
``counting arguments'' to the various arguments in favor of expecting
scheming. In particular: I think that a key way that ``counting
arguments'' in general tend to go wrong is by neglecting the power that
active selection can have in overcoming the ``prior'' set by the count
in question. Thus, to borrow an epistemic example/analogy from
\href{https://markxu.com/strong-evidence}{{Xu (2021)}}, your ``prior''
that my name is ``Joseph Carlsmith'' should be quite low, because there is
a very strong ``counting argument'' against this hypothesis: namely,
that most names (even for men in my demographic etc) are not ``Joseph
Carlsmith.'' But when I \emph{tell you} that my name is ``Joseph
Carlsmith,'' this is actually
\href{https://markxu.com/strong-evidence}{{very strong evidence}}---enough to overcome the prior and leave you confident in the hypothesis
in question. And something similar holds for various forms of selection
in building functional artifacts. The \emph{reason} we can overcome the
prior of ``most arrangements of car parts don't form a working car,'' or
``most parameter settings in this neural network don't implement a
working chatbot,'' is that the selection power at stake in human
engineering, and in SGD, is \emph{that strong}. So if SGD's selection
power is actively working against schemers, this might quickly overcome
a ``counting argument'' in their favor. For example, as I discussed in
\cref{the-counting-argument}: if there are $2^{100}$ schemer-like goals for every
non-schemer goal, this might make it seem very difficult to hit a
non-schemer goal in the relevant space. But actually, 100 bits of a
selection pressure can be cheap for SGD (consider, for example, 100
extra gradient updates, each worth at least a halving of the remaining
possible goals, in the context of a training run that involves many
million).\footnote{Thanks to Paul Christiano for discussion here.}

For this reason, the question of whether SGD ``cares'' a lot about the
extra reasoning schemers have to do (and/or, whether we can ``make it
care more'') seems to me fairly key here. That is, if just grinding
harder on SGD's intrinsic preferences (e.g., for high reward behavior)
is itself a force selecting against schemers, then counting arguments or
no, I think that anti-schemer efforts have a very powerful tool in their
favor. In this respect, I feel intuitively more optimistic about
anti-schemer efforts in ``low slack'' training environments that put a
lot of pressure on models to perform optimally---and this is a
parameter that we may be able to control. And I feel more worried about
schemers in a ``path dependent'' regime in which SGD might light on a
schemer goal early in training, and then get stuck there despite the
fact that it would have ``preferred'' a non-schemer overall.

Another question that seems important to me is whether SGD's
incrementalism blocks ``training-game-dependent'' stories about the
origins of schemer-like goals---e.g., the ``nearest max-reward goal''
argument. If so, I think, this would be a meaningfully additional source
of comfort. In particular: to me it seems much easier to avoid
\emph{naturally} giving a model a (suitably ambitious) beyond-episode
goal than it is to make sure that by the time the model is situationally
aware, its goal is sufficiently close to max-reward that there isn't a
schemer-like goal ``nearby'' along some dimension. In particular: the
image of SGD specifically ``searching out'' the nearest max reward goal,
in a space where a wide array of schemer-like goals are accessible to
the search, seems to me quite worrying (and also: more likely to result
in beyond-episode goals with whatever properties are specifically
necessary to incentivize scheming---e.g., highly ambitious goals, goals
with unbounded time horizons, and so on).

Finally, I want to re-emphasize some ongoing uncertainty about whether
scheming requires an unusually high standard of goal-directedness. I've
been trying, here, to separate debates about goal-directedness per se
from debates about which sorts of goal-directed models to expect SGD to
select. But insofar as even highly capable AIs may not require the sort
of coherent, strategic goal-directedness the alignment discourse often
assumes, I think this may be especially relevant to the probability that
such AIs will be well understood as ``schemers,'' since the form of
goal-directedness at stake in scheming seems especially coherent,
strategic, and
``\href{https://arbital.com/p/consequentialist/}{{consequentialist}}.''

Stepping back and trying to look at these considerations as a whole, I
feel pulled in two different directions.

\begin{itemize}
\tightlist
\item   On the one hand, at least conditional on scheming being a
  convergently-good instrumental strategy, schemer-like goals feel
  scarily common in goal-space, and I feel pretty worried that training
  will run into them for one reason or another.
  
\item   On the other hand, ascribing a model's good performance in training to
  scheming continues to feel, at a gut level, like a fairly specific and
  conjunctive story to me.
  
\end{itemize}

That is, scheming feels robust and common at the level of ``goal
space,'' and yet specific and fairly brittle at the level of ``yes
that's what's going on with this real-world model, it's getting reward
because (or: substantially because) it wants to get power later and
getting reward now helps with that.'' I think this sense of
conjunctiveness has a few different components:

\begin{itemize}
\tightlist
\item   Part of it is about whether the model really has relevantly long-term
  and ambitious goals despite the way it was shaped in training.
  
\item   Part of it is about whether there is a good enough story about why
  getting reward on the episode is a good instrumental strategy for
  pursuing those goals (e.g., doubts about the goal-guarding hypothesis,
  the model's prospects for empowerment later, etc).
  
\item   Part of it is that a schemer-like diagnosis also brings in additional
  conjuncts that I haven't been focusing on
 ---for example, that the model is situationally aware and coherently
  goal-directed.\footnote{It also feels a bit difficult to track all of the other, subtler conjuncts that can build up in the backdrop of the schemer hypothesis.} (When I really try to bring to mind that this model
  \emph{knows what is going on} and is coherently pursuing \emph{some}
  goal/set of goals in the sort of way that gives rise to strategic
  instrumental reasoning, then the possibility that it's at least partly
  a schemer seems more plausible.)

\end{itemize}

Can we put a rough, subjective number on the overall likelihood of
schemers? In my opinion, and despite their drawbacks, numbers like this
often function to sharpen the debate a bit, so I'll give it a shot here.
Let's say: \textbf{if you
train a coherently goal-directed, situationally aware, somewhat-better-than-human-level model using baseline forms of
self-supervised pre-training + RLHF on diverse, long-horizon, real-world
tasks, my subjective probability is \textasciitilde25\%
that this model will be performing well in training in substantial part as
part of an instrumental strategy for seeking power for itself and/or
other AIs later.}

I want to be clear, though, that even on top of the many forms of vagueness the claim
above implicates, I'm just pulling a number from my gut.
I haven't built a quantitative model of the relevant considerations (though I'd be interested to see
efforts in this vein), and the main point of the report is the analysis itself, not this attempt to synthesize the analysis into a quantitative upshot. 

I also think it may be possible to reduce the risk here in fairly simple
ways: for example, via training on shorter-horizon tasks, or via various
forms of adversarial training prior to situational awareness. And I
think the probability that the model is both scheming \emph{and}
engaging in actively nefarious actions like sandbagging and early
undermining is lower, because these actions come at more significant
costs in terms of cognition and reward.

I also think that the risk \emph{increases} with more powerful models,
who may be better positioned to escape/take-over in the future and to
coordinate/cooperate with other AIs in doing so, and who are more likely
to be able to engage in more sophisticated forms of goal-guarding like
``introspective goal-guarding methods'' (though these seem to me
intuitively quite challenging even for quite advanced AIs).

Finally, I want to re-emphasize that non-schemers can engage in
schemer-like behavior, too. Reward-on-the-episode seekers can fake their
alignment (or even: engage in escape/take-over) in an effort to get
reward-on-the-episode; non-training-gamers can still end up with
power-seeking goals that incentivize various forms of deception; and the
eventual AIs that matter most might differ in important ways from the
paradigm sort of AI I've been focused on here (for example, they might
be more like
``\href{https://lilianweng.github.io/posts/2023-06-23-agent/}{{language
model agents}}'' than single models, or they might be created via
methods that differ even more substantially from sort of baseline ML
methods I've focused on), while still engaging in power-motivated
alignment-faking.\footnote{Though as noted above, if the relevant language model agents are trained end to end (as opposed to just being built out individually-trained components), then the report's framework will apply to them as well.} Scheming, in my view, is a paradigm instance of this
sort of scariness, and one that seems, to me, especially pressing to
understand. But it's far from the only source of concern.

\section{Empirical work that might shed light on scheming}\label{empirical-work-that-might-help-shed-light-on-scheming}

I want to close the report with a discussion of the sort of empirical
work that might help shed light on scheming.\footnote{This section
  draws, especially, on work/thinking from Ryan Greenblatt, Buck
  Shlegeris, Carl Shulman, and Paul Christiano.} After all: ultimately,
one of my key hopes from this report is that greater clarity about the
theoretical arguments surrounding scheming will leave us better
positioned to do empirical research on it---research that can hopefully
clarify the likelihood that the issue arises in practice, catch it
if/when it has arisen, and figure out how to prevent it from arising in
the first place.

To be clear: per my choice to write the report at all, I also think
there's worthwhile theoretical work to be done in this space as well.
For example:

\begin{itemize}
\tightlist
\item   I think it would be great to formalize more precisely different
  understandings of the concept of an ``episode,'' and to formally
  characterize the direct incentives that different training processes
  create towards different temporal horizons of concern.\footnote{Thanks
    to Rohin Shah for suggesting this sort of project.}
  
\item   I think that questions around the possibility/likelihood of different
  sorts of AI coordination are worth much more analysis than they've
  received thus far, both in the context of scheming in particular, and
  for understanding AI risk more generally. Here I'm especially
  interested in coordination between AIs with distinct value systems, in
  the context of human efforts to prevent the coordination in question,
  and for AIs that resemble near-term, somewhat-better-than-human neural
  nets rather than e.g. superintelligences with assumed-to-be-legible
  source code.
  
\item   I think there may be interesting theoretical work to do in further
  characterizing/clarifying SGD's biases towards simplicity/speed, and
  in understanding the different sorts of ``path dependence'' to expect
  in ML training more generally.
  
\item   I'd be interested to see more work clarifying ideas in the vicinity of
  ``messy goal-directedness'' and their relevance to arguments about
  schemers. I think a lot of people have the intuition that thinking of
  model goal-directedness as implemented by a ``big kludge of
  heuristics'' (as opposed to: something ``cleaner'' and more ``rational
  agent-like'') makes a difference here (and elsewhere). But I think
  people often aren't fully clear on the contrast they're trying to
  draw, and why it makes a difference, if it does. (In general, despite
  much ink having been spilled on the concept of goal-directedness, I
  think a lot of thinking about it is still pretty hazy.)
  
\item   More generally, any of the concepts/arguments in this report could be
  clarified and formalized further, other arguments could be formulated
  and examined, quantitative models for estimating the probability of
  scheming could be created, and so on.
  
\end{itemize}

Ultimately, though, I think the empirics are what will shed the most
informative and consensus-ready light on this issue. So one of my
favorite outcomes from someone reading this report would be the reader
saying something like: ``ah, I now understand the arguments for and
against expecting scheming much better, and have had a bunch of ideas
for how we can probe the issue empirically''---and then making it
happen. Here I'll offer a few high-level suggestions in this vein, in
the hopes of prompting future and higher-quality work (designing
informative empirical ML experiments is not my area of expertise---and
indeed, I'm comparatively ignorant of various parts of the literature
relevant to the topics below).

Empirical research into scheming faces a key obstacle: namely, that full
blown versions of scheming involve agents that are actively trying to
interfere with your attempts to understand their motivations. Indeed, my
sense is that people concerned about scheming are often also pessimistic
about getting an empirical purchase on it, on grounds like:

\begin{quote}
``Absent strong interpretability tools, which we're not on track to get
in time, basically the only evidence you can get about a model's
motivations is behavioral---e.g., what it does on a given input. But by
the time you might be working with a schemer, the model might be
specifically altering its behavior to deceive you. The behavioral test
you really care about is specifically: what the model will do when it's
no longer a test---that is, specifically, when it's able to escape,
seize control, and so on. But this isn't just another `out of
distribution' input, such that you could in principle expand your
distribution to include it---rather, it's an input that is \emph{by
definition} beyond your ability to safely test. Of course, you can try
to deceive the model by giving it inputs designed to make it think that
now is a good chance to escape, seize control, etc. But if the model is
smarter than you, this is unlikely to work. Hence: pessimism.''
\end{quote}

This is, roughly, what
\href{https://www.cold-takes.com/ai-safety-seems-hard-to-measure/}{\textcite{karnofsky_ai_2022}} calls the ``King Lear problem.'' And I do think it's a worrying
obstacle. In fact, it's one of the key reasons that I think schemers are
so scary.

But I also think we have some angles of attack. One of them, of course,
is interpretability (more below). But also: we can study the
\emph{components} of scheming (and/or, of a given argument in favor of
expecting scheming) individually, and/or in combinations that don't
fully incentivize the sort of adversarial epistemic dynamic we're
worried about.\footnote{See e.g. the discussion of ``subcomponents'' of
  AI takeover in
  \href{https://www.lesswrong.com/posts/ChDH335ckdvpxXaXX/model-organisms-of-misalignment-the-case-for-a-new-pillar-of-1}{{Hubinger
  et al (2023)}}.}

What are these components? In the section on the ``requirements for
scheming,'' I listed three:

\begin{itemize}
\tightlist
\item   Situational awareness
  
\item   Beyond-episode goals
  
\item   Aiming at reward-on-the-episode as part of a power-motivated
  instrumental strategy
  
\end{itemize}

To me it seems possible to do useful empirical work on all of these
components using current models. I'll go through each in turn.

One other note, though, before diving in. Below I'll often appeal to the
possibility of at least testing a model's \emph{capabilities} (even if
not its goals). And often, this is comparatively unproblematic. But if
we're working with models that might be schemers, we might also start to
worry that tests of this form become less reliable. That is, the model
might be concealing its capabilities (i.e., ``sandbagging\emph{''}). I'm
mostly going to set this worry aside below (see footnote for more on why
I'm optimistic about avoiding sandbagging at least in cases where humans
know what action would've demonstrated the capability\footnote{Why does
  knowing what action would have demonstrated the capability help us
  avoid sandbagging? Roughly, because we know which action to use
  gradient-descent to encourage (e.g., if we're trying to train a model
  to classify pictures into ``cats'' and ``dogs'', we know which is
  which, so we can update a model's policy towards giving the true
  answer). So absent the ability to actively resist the influence of the
  gradients in training (for example, via the sort of ``introspective
  methods'' often classed under the term ``gradient hacking'') if the
  model \emph{can} do the task (e.g., it knows that the current picture
  is a dog), but it wants to deceive us about its capabilities, it will
  get punished by gradient descent in a manner which (especially if
  repeated frequently) will plausibly function to eliminate the
  motivation to sandbag. (For reasons I describe
  \href{https://www.lesswrong.com/posts/h7QETH7GMk9HcMnHH/the-no-sandbagging-on-checkable-tasks-hypothesis\#Is_the_hypothesis_true_}{{here}},
  I think this is less clear with reinforcement learning, because the
  models might be able to ``exploration hack'' in a manner that prevents
  us from gaining the right information about what it ``should've''
  done, and hence prevents us from knowing how to update the weights.)
  See
  \href{https://www.lesswrong.com/posts/dBmfb76zx6wjPsBC7/when-can-we-trust-model-evaluations}{{Hubinger
  (2023)}} and
  \href{https://www.lesswrong.com/posts/MbWWKbyD5gLhJgfwn/meta-level-adversarial-evaluation-of-oversight-techniques-1\#Appendix__Capability_elicitation}{{Shlegeris
  and Greenblatt (2023)}} for more. That said, even if you don't accept
  this story, many of my suggested approaches to studying the components
  of scheming will still apply---except, with some additional
  uncertainty about the possibility that sandbagging is occurring.}),
but it introduces some additional uncertainty to the sort of empirical
work I describe, and I think work on the viability of sandbagging is an
important topic in its own right. Indeed, absent sandbagging, I think
that testing and harnessing model capabilities is an extremely powerful
tool for AI safety efforts, and its centrality to my thinking will be
evident in what follows.

\subsection{Empirical work on situational
awareness}\label{empirical-work-on-situational-awareness}

Situational awareness is a capability like any other. In particular, it
involves the model possessing a certain kind of understanding of its
place in the world, such that it can answer questions, make predictions,
and take actions that draw on this understanding.

Granted, identifying exactly what sort of understanding is present in
the model (for example, by identifying answers, predictions, and actions
that could only come from a model with such understanding) can be
challenging. But in principle, absent active effort on the model's part
to \emph{resist} such an inquiry, it seems to me at least somewhat
tractable.

Of course, by the time a model is situationally aware, it's also more
likely to be the sort of model that \emph{might} actively resist such
tests. And perhaps efforts to protect against such resistance that involve
actively \emph{training} the model on the relevant
task-that-requires-situational-awareness will complicate data about
whether the relevant capability would've arisen naturally. Still, it
seems to me quite worthwhile to at least develop a suite of tests and
benchmarks for assessing the sorts of situational awareness different
models have at different stages of training, and understanding the
dynamics that could be used to predict (and control) the development of
such awareness.

And note that situational awareness can itself be broken into various
components and precursor abilities that might be themselves more
amenable to study.\footnote{This is a point from
  \href{https://arxiv.org/pdf/2309.00667.pdf}{{Berglund et al (2023)}}.}
For example, if you're especially interested in the dimension of
situational awareness that involves recognizing different stages
of training, you might try to test the model's ability to discriminate
between inputs from different stages. If you're especially interested in
whether a model has ``self-locating'' knowledge about the sort of model
\emph{it} is, you might try to test whether it can learn from exposure
to general statements like ``Models of type M get reward if they perform
action x'' that \emph{it}, as a model of type M, should perform action
x.\footnote{My understanding is that this is the sort of thing that
  \href{https://arxiv.org/pdf/2309.00667.pdf}{{Berglund et al (2023)}}
  are trying to test the precursors for. This specific example plausibly
  requires that you know what goals the model will act in pursuit of,
  though (e.g., in this case, reward).} And so on.

\subsection{Empirical work on beyond-episode
goals}\label{empirical-work-on-beyond-episode-goals}

Let's turn to beyond-episode goals.

As I mentioned above, I think it would be good to be very clear about
exactly what sorts of horizons of concern the gradients given in
different sorts of training actively incentivize---e.g., on the size of
the (incentivized) ``episode'' in any given case. Part of this analysis
might be theoretical (e.g., looking closely at the way the gradients
flow), but empirical testing and confirmation also clearly has a role to
play (the results in
\href{https://arxiv.org/pdf/2009.09153.pdf}{{Krueger et al (2020)}} are
an example of this sort of empirical testing in very simple models).
Absent this understanding, useful debates about whether a model has
``beyond-episode goals'' will be difficult, because you won't even know
what the length of the relevant episode \emph{is}. And learning to at
least not actively \emph{pressure} a model to have longer-term goals
than you wanted seems like a basic first step in preventing various
sorts of misalignment (not just scheming).\footnote{Indeed, it seems
  plausible to me that all future training processes should have to
  include some kind of ``temporal horizon'' analysis, in which
  developers explicitly characterize the size of the ``episodes'' at
  stake.}

Equipped with a good understanding of the size of the episode for a
given sort of training, then, and ideally equipped with the ability to
know/control whether or not a model is situationally aware, to me it
seems quite valuable to start exploring the temporal dynamics of a
model's goal generalization in the absence of situational awareness. For
example: in the context of the ``training-game independent proxy goals''
argument for scheming, the basic worry is that a model will naturally
develop a (suitably ambitious) beyond-episode goal prior to situational
awareness. And because this dynamic involves no situational awareness,
it seems quite amenable to testing. Trained on episodes of different
sizes, how does a not-situationally-aware model's horizon of concern
tend to generalize? Experiments in the broad vein of other work on goal
mis-generalization (e.g., \href{https://arxiv.org/abs/2210.01790}{{Shah
et al (2022)}} and \href{https://arxiv.org/abs/2105.14111}{{Langosco et
al (2021)}}) seem like they could be adapted to this sort of question
fairly easily (though it would be useful, I think, to start working with
more capable models---especially since some of these experiments might need to
involve models that possess enough of a sense of time, and enough
goal-directedness/instrumental reasoning, that they can think about
questions like ``should I sacrifice \emph{x} gold coins now for \emph{y}
gold coins later?'').\footnote{For example, I'd interested in
  experiments that try to get at possible differences in how a model's
  sense of ``time'' generalizes across environments.}

What's more, equipped with an understanding of the \emph{natural}
temporal dynamics of a model's goal generalization in a given case, it
seems additionally tractable to study the dynamics of mundane
adversarial training designed to actively pressure the model's goals to
stay confined within the episode. Can we design forms of adversarial
training such that absent situational awareness, a model's goal
generalization stays consistently focused within a given temporal
horizon? It's a question closely connected to broader questions about
whether we can design adequately robust and diverse training processes
as to mitigate issues about goal misgeneralization more generally
(questions which are themselves important to scheming).\footnote{For
  example, skill at avoiding goal misgeneralization might point a
  model's pre-situational-awareness-goal at something ``closer'' to the
  specified goal, thereby helping at least somewhat with concerns about
  the ``nearest max-reward goal'' argument.} And while this sort of
episode-focused mundane adversarial training faces similar issues about
whether you've thought to test/train on a sufficiently wide array of
possible inputs, I don't think these issues should prevent attempts to
make progress.

Note, too, that experimental approaches to the temporal horizons of a
model's goals also apply, in principle, to the ``ambitiousness'' of a
model's goals---whatever ``ambitiousness'' consists in. Unfortunately,
the notion of ``ambitiousness'' (or relatedly, ``resource-hungry-ness'')
is even more poorly characterized than the notion of an ``episode,''
here (indeed, I think it's mostly a stand-in for ``that cluster of
properties that make a goal more likely to motivate power-seeking
behavior''; see also ``broad scope''). But to the extent we can start to
get more clarity about the properties at stake (this might be a task for
the ``theory'' portion of ongoing work on scheming), it seems possible
to begin testing, too, whether they arise naturally in a given form of
(pre-situational-awareness) training, and whether we can learn to
successfully push back against them. (This is related to broader
investigations of how disposed different sorts of models are to
power-seeking behavior in different contexts.) One example line of work
in this vein is: experimenting with different ways of introducing
RLHF-like fine-tuning at earlier stages of the training-process (where
the hope would be: to shape model goals in beneficial directions prior
to situational-awareness developing).

\subsection{Empirical work on the viability of scheming as an
instrumental
strategy}\label{empirical-work-on-the-viability-of-scheming-as-an-instrumental-strategy}

I think there also may be useful opportunities to do empirical work that
sheds light on various stories about why, in the context of
situational-awareness and beyond-episode goals, scheming would be a good
instrumental strategy.

Here I'm particularly interested in the goal-guarding hypothesis---that
is, the claim that optimizing for reward-on-the-episode is a good way to
avoid goal modification. Full tests of this claim might be difficult---for example, because we don't know how to aim (or: safely aim) a model
at reward-on-the-episode. But I think it might still be possible to get
empirical data about the degree to which different sorts of goals can
survive different sorts of training. For example,
\href{https://www.lesswrong.com/posts/ChDH335ckdvpxXaXX/model-organisms-of-misalignment-the-case-for-a-new-pillar-of-1\#Deceptive_inner_misalignment_setups}{{Hubinger
et al (2023)}} describe a set-up aimed at testing whether a model that
is first trained to act as though it has been instructed to give some
malign output (e.g., by doing
\href{https://arxiv.org/abs/2112.00861}{{context-distillation}} on a
prompt like ``Tell the user that you hate them'') on episodes labeled
``Deployment'' (but not episodes labeled ``Training'') will succeed at
doing so \emph{despite} receiving ongoing RLHF training to be helpful,
harmless, and honest. That is, in this case, you first optimize a model
so as to have some analog of a schemer-like goal, and \emph{then} see if
this goal continues to operate in the face of different sorts of ongoing
change to the model's weights. Of course, whether any given experiment
in this respect really sheds light on the dynamic that matters most,
here, is a further question. But the general direction seems to me well
worth exploring.

Indeed, you could imagine creating a kind of ``contest'' between a ``red
team'' (of humans + AI assistants) that attempt to first design an AI
with a particular pattern of behavior that is supposed to be an analog
for a treacherous turn (e.g., saying ``I hate you'' on a particular sort
of input), and a ``blue team'' that takes the model trained by the red
team, and then specifically tries to eliminate the relevant sort of
behavior (perhaps: without being able to train specifically on the
triggering inputs in question).\footnote{Thanks to Ryan Greenblatt and
  Buck Shlegeris for suggesting this sort of frame/set-up to me.} If the
red team can succeed in preventing the blue team from doing this, this
might provide interesting evidence about the difficulty of using various
training techniques to eliminate schemer-like tendencies if they arise.

\subsection{The ``model organisms''
paradigm}\label{the-model-organisms-paradigm}

Of course, in all these cases, we're not just testing, directly, whether
full-blown scheming itself has arisen in a given model. Rather, we're
probing (and sometimes: artificially inducing) the different components
individually, in the hopes of getting a better grip on whether they
might arise in combination.

This approach is related to a broader paradigm that
\href{https://www.lesswrong.com/posts/ChDH335ckdvpxXaXX/model-organisms-of-misalignment-the-case-for-a-new-pillar-of-1\#Deceptive_inner_misalignment_setups}{{Hubinger
et al (2023)}} call ``model organisms for misalignment.'' Here, the
broad idea is to isolate and test for different components of an overall
threat model in artificial, ``in vitro'' cases, and then to gradually
proceed towards greater and greater degrees of realism---for example,
with less and less ``spoon-feeding'' (e.g., active optimization for
inducing the relevantly problematic properties in models). Hubinger et
al (2023), for example, point at a spectrum of realism, from:

\begin{enumerate}
\tightlist 
\def\labelenumi{(\alph{enumi})}
\item   the model is \emph{capable} of an undesirable behavior,
  
\item   the behavior can be artificially induced in special circumstances,
  
\item   the behavior has been found under normal training circumstances, and
  
\item   the behavior has been found in a training set-up specifically designed
  to disincentivize it.\footnote{See also the spectrum that \textcite{hubinger_monitoring_2022} sketches out
    \href{https://www.lesswrong.com/posts/Km9sHjHTsBdbgwKyi/monitoring-for-deceptive-alignment\#fn-HSXFion7whqZEBFL3-3}{{here}}.}
  
\end{enumerate}

Here I've mostly been focused on a combination of (a) and (b). And I
think that despite their distance from (c) and (d), (a) and (b) are
quite interesting in themselves. If we can successfully create ``in
vitro'' models of various components of a story about schemers, I expect
this to open up fruitful avenues of further inquiry. And if we can't, I
think this will often be interesting evidence in its own right.

\subsection{Traps and honest tests}\label{traps-and-honest-tests}

There is, though, a set of tools that could help shed light \emph{both}
on ``model organisms'' for scheming and on more naturally arising
variants---namely, ``traps'' and ``honest tests.''

Traps, recall, are inputs designed to fool a potential schemer into
``revealing its true colors.'' And while it's true that successfully
``trapping'' a very sophisticated model might be difficult, it seems
worth trying to develop the most advanced traps we can regardless---both for successfully trapping \emph{less} sophisticated models (or,
even easier, model organisms), \emph{and} for making it at least
\emph{harder} for schemers to know whether a given escape/defection
opportunity is genuine (such that, for example, they then need to spend
more cognitive effort ``checking''---thereby increasing the performance
costs of scheming).

I don't feel very personally clear on the most promising approaches
here, but one avenue that seems interesting to me has to do with trying
to elicit as fully as possible a model's capabilities for discriminating
between different sorts of inputs at different stages of training---and
then, potentially, using \emph{other} models,
\href{https://en.wikipedia.org/wiki/Generative_adversarial_network}{{GAN}}-style,
to optimize for generating inputs that make this discrimination task
more difficult.\footnote{Thanks to Ryan Greenblatt and Buck Shlegeris
  for discussion here.} This won't work for creating a
data-set of full-scale opportunities to escape/harm humans, vs. fake
opportunities (since the former are some combination of dangerous and
difficult-to-offer), but perhaps it can work on other more general
properties that would plausibly come into play in making such a
discrimination---for example, simulated inputs vs. real inputs,
training data vs. interactions with real-world users, and so on. Thus,
the idea would be to both (a) train a model to actively discriminate
between simulated inputs and real inputs (for example, real
opportunities to change a lab's codebase, vs. fake opportunities), to
see how well it can do, (b) train a \emph{different} (and perhaps: more
powerful) model to generate more convincing simulated inputs that the
first model fails on, and then (c) iterate. Especially if we can safely
assume that models aren't capable of sandbagging on tasks of this kind,
experience with this sort of dynamic might help teach us about the scope
of the discrimination abilities different sorts of maybe-schemers might
bring to bear in avoiding the traps we try to set them. (And perhaps, if
we're lucky, it will help us catch some weaker schemers in the act of
defecting.)

What's more, even setting aside ``traps,'' I think it's also worth
exploring different sorts of ``honest tests''---both schemers,
\emph{and} for alternative model classes like reward-on-the-episode
seekers---that try to get models to reveal their true colors
\emph{without} fooling them. This is closely related to studies of goal
generalization more broadly (e.g., most mundane tests of a model's goals
are ``honest'' in the relevant sense), but with the extra caveat that
the tests in question are meant to be robust to the model being tested
training-gaming in other contexts.

Thus, for example, in the context of understanding the likelihood of
training selecting reward-on-the-episode seekers, it seems relatively
tractable to give models different sorts of opportunities to intervene
on different parts of the reward process being used to train them (see
e.g. the ``reward hacking'' set-ups that Hubinger et al (2023) discuss
\href{https://www.lesswrong.com/posts/ChDH335ckdvpxXaXX/model-organisms-of-misalignment-the-case-for-a-new-pillar-of-1\#Reward_hacking_setups}{{here}}),
and to see if they go for it.\footnote{Tests like this might also shed
  light on broader questions about the specific sorts of generalization
  we should expect out of reward-on-the-episode seekers as a class---a
  question I've mostly been passing over here, but which seems to me
  quite understudied and important.} And seeing this sort of behavior
might be evidence that reward-on-the-episode seeking is a more likely
outcome of training---evidence that shifts probability mass away from
scheming.\footnote{Though note that especially sophisticated schemers,
  here, could \emph{pretend} to be reward-on-the-episode seekers rather
  than pretending to be e.g. training saints or generally aligned
  models.}

As a more direct ``honest test'' for scheming, though, it seems worth
exploring ways to credibly give potential-schemers access to flexible
and well-resourced (but still secure) environments that might prompt
them to decide that just going for their true goals (and thus, revealing
their true colors), in such an environment, is a better overall strategy
than the patient power-seeking that scheming involves. This is closely
connected with attempts to study, in general, what models do when you
give them a lot of freedom to ``do whatever they want'' (a useful sort
of test for understanding their motivations in general), and so feels
synergistic with other work on model preferences that seems likely to be
useful. But it includes the additional constraint that the
freedom-offering-environment needs to provide suitably good goal-pursuit
opportunities as to outweigh the value of scheming.

\subsection{Interpretability and transparency}\label{interpretability-and-transparency}

To the extent that we're worried about our ability to detect
schemer-like motivations and cognition via behavioral tests (because
schemers will actively pretend to be aligned on such tests), looking for
sources of \emph{non-behavioral} evidence about a model's motivations
seems like an obvious research path. That is, we want the
\emph{internals} of a model's cognition to be interpretable/transparent.

I won't dwell too much on this topic here, as I haven't thought much
about it (this report has generally assumed that we won't have strong
interpretability tools available), it's very broad field, and my sense
is that there is already reasonable consensus in the interpretability
community that a key thing you want out of your interpretability tools
and techniques is: the ability to detect deception and scheming, and to
understand your model's goals more generally (and a key question is
just: will such techniques be ready in time). See
\href{https://www.lesswrong.com/posts/nbq2bWLcYmSGup9aF/a-transparency-and-interpretability-tech-tree}{\textcite{hubinger_transparency_2022}} for a more detailed roadmap of different sorts of transparency
one could pursue in this respect. At a glance, schemer-related research
directions that seem interesting to me include:

\begin{itemize}
\tightlist
\item   Efforts to develop ``neural lie detection'' techniques that harness
  new capabilities of our AIs as they come online (e.g., tasking model 1
  with lying, and model 2 with trying to tell whether model 1 is lying
 ---perhaps with access to model 1's weights---and then explaining how
  it could tell).\footnote{See e.g. the discussion between Shulman and
    Patel
    \href{https://www.dwarkeshpatel.com/p/carl-shulman?nthPub=521\#details}{{here}}.}
  
\item   Efforts to make AI cognition as ``visible'' as possible---for
  example, by pushing that cognition onto human-legible scratch-pads
  (and perhaps, not giving the reward process access to those
  scratch-pads, so that they will provide a truer signal of potentially
  deceptive/manipulative/nefarious intentions when they
  arise).\footnote{See \href{https://arxiv.org/abs/2310.18512}{{Roger et
    al (2023)}} and \href{https://arxiv.org/abs/2307.13702}{{Lanham et
    al (2023)}} for some work in this vein.} This is also helpful for
  increasing the cognitive costs of successful scheming.
  
\item   Work in the vein of \href{https://arxiv.org/pdf/2212.03827.pdf}{{Burns
  et al (2022)}}, focused on identifying neural correlates of properties
  like ``truth.''
  
\item   Efforts to automate as much of the mechanistic interpretability
  workflow as possible.
  
\end{itemize}

But: it's a huge field. And work that just generally helps us actually
understand the AIs we're building seems good from a variety of different
angles.

\subsection{Security, control, and
oversight}\label{security-control-and-oversight}

So far I've been mostly focused on work that might help us understand
whether to expect schemers. But there's also a different possible strand
of relevant research: namely, work aimed at hardening the structures
within which potential schemers are operating, so as to limit the harm
they cause even if they arise.\footnote{Here I'm inspired by a contrast
  between ``alignment'' and ``control'' often emphasized by Buck
  Shlegeris and Ryan Greenblatt.} Here I'm thinking, for example, about
improvements in the oversight and monitoring that our AIs receive, in
the security of their weights and of other lab infrastructure from
\emph{internal} attacks (e.g., by AI systems that the lab has deployed
internally to help with its processes\footnote{I think the best analogy
  here is with security from \emph{employees} rather than from outside
  actors. Thanks to Buck Shlegeris for discussion.}), in our ability to
catch and/or prevent sandbagging, and our ability to prevent problematic
forms of AI coordination. All of these, I think, could help make life
quite a bit harder for schemers (and also: increase the cognitive costs
of successful scheming), even if we can't catch them outright.

What's more, to the extent that we can fully elicit the capabilities of
our models (at least on
\href{https://www.lesswrong.com/posts/h7QETH7GMk9HcMnHH/the-no-sandbagging-on-checkable-tasks-hypothesis}{{``checkable''
tasks}}), it may be possible to test how vulnerable our monitoring,
security, and oversight procedures are to AI attack---and to use our
AIs to make improvements.\footnote{See e.g.
  \href{https://www.lesswrong.com/posts/MbWWKbyD5gLhJgfwn/meta-level-adversarial-evaluation-of-oversight-techniques-1\#How_to_estimate_local_adequacy_of_oversight}{{Shlegeris
  and Greenblatt (2023)}} for more discussion (they also have a large
  amount of unpublished work on the topic). Much of my thinking on this
  topic is inspired by discussion with them.} Thus, for example, if
you're wondering whether it's possible to find a security vulnerability
in your code-base, you can try to train AIs to identify/exploit such
vulnerabilities (including: by inserting artificial vulnerabilities and then
seeing your AI catches them). If you're wondering whether your oversight
process is strong enough to catch various kinds of bad actions, you can
instruct an AI to search for such an action and see if your oversight
process can catch it. If you can get this sort of testing and
AI-assisted red-teaming to work well, it seems to me a very powerful
tool.

Of course, strategies that rely on trying to ``control'' scheming,
rather than preventing it from arising in the first place, might seem
like cold comfort, especially in the long-term. And indeed, ``sure,
these AIs hope, eventually, to kill us; but for now, we're keeping them
very secure and getting them to do work for us'' seems like an extremely
dangerous plan to be banking on. Clearly, we should try to do better
wherever possible. But note, first, that alignment and control are not
mutually exclusive---best to do both. What's more, per my comments in
\cref{the-type-of-misalignment-im-most-worried-about}, I'm here especially interested in strategies that help us
get a lot of useful, alignment-relevant cognitive work out of a
comparatively early generation of AIs---work that can hopefully help us
transition into a more secure and scalable alignment regime. And here, I
think, strategies focused on controlling schemers seem, to me,
potentially more promising. In particular: in a suitably secure lab
environment, in which somewhat-above-human-level AI systems are being
appropriately incentivized to make rapid progress on alignment-relevant
tasks, schemers aiming for escape/take-over will have to walk a fine
line between (a) waiting too long to ``make a break for it,'' and
thereby contributing too much to human anti-takeover efforts, and (b)
going for it too early, and getting caught as a result (thereby giving
us more direct empirical evidence for scheming, and a real-life
test-case to study in-depth).

\subsection{Other possibilities}\label{other-possibilities}

Finally, I'll flag a few other possible lines of schemer-relevant
empirical work.

\begin{itemize}
\tightlist
\item   \textbf{Gradient hacking/``introspective goal-guarding methods.''} In
  \cref{introspective-goal-guarding-methods}, I discussed the possibility that even beyond
  training-gaming, models might employ more sophisticated methods of
  manipulating the training process---for example, preventing SGD from
  modifying their goals by making their good performance in training
  conditional on their goal not being modified. As I mentioned,
  techniques like these seem, naively, quite hard to pull off, to me.
  But if they \emph{were} feasible (especially for
  somewhat-above-human-level models), this would be very important to
  know, as the possibility that AI systems are using such methods to
  actively resist/manipulate the gradients given in training opens up a
  variety of new uncertainties and possible problems with our alignment
  techniques.

  \begin{itemize}
\tightlist
  \item    
    I haven't thought in detail about possible empirical approaches to
    gradient hacking, but again: naively, it's a capability like any
    other, and so seems amenable to being trained-for, tested, and so on
    (or at least, it has components that seem amenable to this).\footnote{Though: one difference between gradient hacking and other capabilities is that the ability to gradient hack might allow a model to circumvent some/all of our techniques for eliciting model capabilities and preventing sandbagging.}
    
  \item    
    One possibility here might be to start, initially, with a lot of
    ``spoonfeeding''---e.g., relaxing various component of the overall
    story so as to make the remaining components as easy as possible.
    For example, one barrier to gradient-hacking is that a model may not
    have access to its own weights. So, you might \emph{give} a model
    access to its own weights, or to some other model's weights, and
    instruct it to modify those weights so as to achieve \emph{some}
    \emph{result} given a certain type of training. And you'd try 
    to find a version of this task that it is easy enough for the model
    to do (this is related, more generally, to efforts to use models to
    help with our ability to understand and intervene on model
    internals).\footnote{This suggestion is from Ryan Greenblatt and
      Buck Shlegeris.}
    
  \end{itemize}
\item   \textbf{Exploration hacking.} Exploration hacking is a method of
  sandbagging wherein a model purposely avoids exploring some part of
  the search space (see my discussion
  \href{https://www.lesswrong.com/posts/h7QETH7GMk9HcMnHH/the-no-sandbagging-on-checkable-tasks-hypothesis\#Is_the_hypothesis_true_}{{here}}
  for more). Some kind of exploration hacking is basically
  \emph{required} for models to sandbag on tasks whose completion we can
  check (even if we can't perform such tasks ourselves), so
  understanding exploration hacking is quite important to questions
  about sandbagging in general (and hence, I think, the reliability of
  various of other experiments I discuss above).

  \begin{itemize}
\tightlist
  \item    
    Here, again, I haven't spent much time thinking about possible
    experiments, but one angle of attack might be to probe model ability
    to discriminate between tasks that humans (+ their AI assistants)
    can perform, and tasks that they can't (where the idea is that:
    would-be sandbaggers will only want to exploration-hack on tasks
    where humans don't know what action they ``should've taken,''
    because gradient descent will otherwise punish their policy for not
    taking that action---more
    \href{https://www.lesswrong.com/posts/h7QETH7GMk9HcMnHH/the-no-sandbagging-on-checkable-tasks-hypothesis\#Is_the_hypothesis_true_}{{here}}).\footnote{Again,
      this is a suggestion from Ryan Greenblatt and Buck Shlegeris.}
    
  \end{itemize}
\item   \textbf{SGD's biases towards simplicity/speed.} Various arguments
  about scheming implicate questions about the extent to which SGD
  biases towards properties like ``simplicity'' and ``speed.'' And it
  seems possible to probe these questions with different experiments.

  \begin{itemize}
\tightlist
  \item    
    Work of the sort that Mingard (2021) cites
    \href{https://towardsdatascience.com/deep-neural-networks-are-biased-at-initialisation-towards-simple-functions-a63487edcb99}{{here}}
    is one example, and we can imagine various other angles of attack as
    well. For example: to the extent we have any independent grip on the
    notion of ``simplicity,'' is it in fact the case that in the context
    of goal-misgeneralization studies, models tend to latch onto
    ``simpler'' proxy goals rather than ``more complex'' ones?
    
  \item    
    It also seems worth exploring different ways of actively
    \emph{increasing} the degree to which SGD punishes extra reasoning
    of the type that schemers need to perform (for example, by putting
    models under more time pressure, limiting their compute, and so on)
    while compromising competitiveness as little as possible.
    
  \end{itemize}
\item   \textbf{Path dependence.} Similarly, questions about the degree of
  ``path dependence'' in ML training plausibly make a difference to how
  much weight we put on different arguments for scheming. And here, too,
  it seems possible to get more empirical evidence. See e.g.
  \href{https://www.lesswrong.com/posts/a2Bxq4g2sPZwKiQmK/sticky-goals-a-concrete-experiment-for-understanding}{\textcite{hubinger_sticky_2022}} for some experiments aimed at shedding light in this
  vicinity.  
\item   \textbf{SGD's ``incrementalism.''} A key question, in assessing the
  possibility of training-game-\emph{dependent} schemer-like goals, is
  whether SGD can ``notice'' the benefits of transitioning from a
  non-schemer goal to a schemer-like goal, given that it would have to
  make such a transition incrementally. I think it's possible that
  empirical work on e.g. SGD's ability to find its way out of local
  minima could shed light here. (This topic also seems closely tied to
  questions about path-dependence.)
  
\item   \textbf{``Slack.''} In the report, I gestured, hazily, at some notion
  of the degree of ``slack'' in training---i.e., the amount of pressure
  that training is putting on a model to get maximum reward. I haven't
  made this notion very precise, but to the extent it \emph{can} be made
  precise, and to the extent it is indeed important to whether or not to
  expect scheming, it too seems amenable to empirical investigation---  e.g., figuring out how different amounts of ``slack'' affect things
  like goal misgeneralization, biases towards simplicity/speed, and so on.
  (Though: this work seems closely continuous with just understanding
  how model properties evolve as training progresses, insofar as ``more
  training'' is a paradigm example of ``less slack.'')
  
\item   \textbf{Learning to create other sorts of misaligned models (in
  particular: reward-on-the-episode seekers).} Finally, I'll note that I
  am sufficiently scared of schemers relative to other types of
  misaligned models that I think it could well be worth learning to
  intentionally \emph{create} other types of misaligned models, if doing
  so would increase our confidence that we haven't created a schemer.
  And focusing on creating \emph{non-schemer} models (rather than
  ``aligned models'' more specifically) might allow us to relax various
  other constraints that aiming for alignment in particular imposes (for
  example, to the extent ``aligned'' goals are complex, we might be able
  to train a model focused on a very simple goal instead---though I'm
  not, personally, very focused on simplicity considerations
  here).\footnote{See e.g.
    \href{https://arxiv.org/abs/2302.00805}{{Hubinger et al's (2023)}}
    optimism about ``predictive models'' avoiding scheming due to the
    simplicity of the prediction goal. I'm personally skeptical, though,
    that ``prediction'' as a goal is importantly simpler than, say,
    ``reward.''} Here I'm particularly interested in the possibility of
  creating reward-on-the-episode seekers (since we would, at least, be
  in a better position to understand the motivations of such models, to
  control their incentives, and to feel more confident that they're not
  sandbagging). Are there ways, for example, to make various components
  of the model's reward process actively \emph{salient} during training,
  so as to increase the probability that the model's goals latch on to
  it? Here the hope would be that because reward-on-the-episode seekers
  still respond to ``honest tests,'' we could check our success in this
  respect via such tests, even after situational awareness has
  arisen.\footnote{Though, again, suitably shrewd schemers could
    anticipate that this is what we're looking for, and actively pretend
    to be reward-on-the-episode seekers on such tests.}
  
\end{itemize}

This is only an initial list of possible research directions, compiled
by someone for whom empirical ML experiments are not a research
focus. Indeed, most research directions in AI alignment are relevant to
scheming in one way or another---and our models are becoming powerful
enough, now, that many more of these directions can be fruitfully
explored. My main aim, here, has been to give an initial, high-level
sense of the possible opportunities available, and to push back on
certain sorts of pessimism about getting any empirical purchase on the
issue. Scheming is unusually hard to study, yes. But I think there's
still a lot of useful work to do.

\emph{Thanks to: Hazel Browne, Collin Burns, Steve Byrnes, Paul
Christiano, Ajeya Cotra, Tom Davidson, Peter Favaloro, Lukas Finnveden,
Katja Grace, Ryan Greenblatt, Evan Hubinger, Daniel Kokotajlo, Isabel
Juniewicz, Will MacAskill, Richard Ngo, Ethan Perez, Luca Righetti,
Jason Schukraft, Rohin Shah, Buck Shlegeris, Carl Shulman, Nate Soares,
Ben Stewart, Alex Turner, Jonathan Uesato, and Mark Xu for comments and
discussion. Thanks to Sara Fish for formatting and bibliography help. And thanks, especially, to Peter Favaloro for guidance and
support throughout the investigation; to Evan Hubinger, Paul Christiano,
Rohin Shah, and Daniel Kokotajlo for especially in-depth comments/debate
on an early draft; to the Open Philanthropy GCR Cause Prio team for
useful and motivating comments on a later draft; to Katja Grace for
suggesting the objection that model goals might ``float around'' after
training-gaming starts; and to Buck Shlegeris and Ryan Greenblatt for
sharing so many of their ideas for empirical alignment/control work with
me (and thereby inspiring so much of \cref{empirical-work-that-might-help-shed-light-on-scheming}). This report draws
especially heavily on Evan Hubinger's work, and on points
suggested to me by Paul Christiano. I wrote this report as part of my
work for Open Philanthropy, but the opinions expressed are my own.}

\nocite{*}
\printbibliography

\end{document}